\documentclass[
11pt, 
english, 
onehalfspacing, 
headsepline, 
consistentlayout, 
]{MastersDoctoralThesis} 

\usepackage[utf8]{inputenc} 
\usepackage[T1]{fontenc} 

\usepackage{mathpazo} 

\usepackage[autostyle=true]{csquotes} 

\usepackage{tikz}

\usepackage{multirow}

\usepackage{commath} 
\usepackage{accents} 
\newcommand*{\dt}[1]{{
	\accentset{\mbox{\large\bfseries .}}{#1}} }
\newcommand*{\ddt}[1]{%
	\accentset{\mbox{\large\bfseries .\hspace{-0.12ex}.}}{#1}} 

\usepackage{amsfonts}  

\usepackage{xcolor} 

\usepackage[toc,page]{appendix}

\newcommand{\overbar}[1]{\mkern 1.5mu\overline{\mkern-1.5mu#1\mkern-1.5mu}\mkern 1.5mu}

\usepackage{sidecap,amssymb} 

\usepackage[font={small},justification=justified]{caption}

\makeatletter
\renewcommand*\env@matrix[1][c]{\hskip -\arraycolsep
	\let\@ifnextchar\new@ifnextchar
	\array{*\c@MaxMatrixCols #1}}
\makeatother

\usepackage{changepage}

\thesistitle{} 
\supervisor{} 
\examiner{} 
\author{}

\keywords{}

\AtBeginDocument{
\hypersetup{pdftitle=\ttitle} 
\hypersetup{pdfauthor=\authorname} 
\hypersetup{pdfkeywords=\keywordnames} 
}

\begin{document}

\frontmatter 

\pagestyle{plain} 


\begin{titlepage}
\begin{center}
\vspace*{.06\textheight}

\rule{13.5cm}{.7pt}\\[0.7cm] 
{\huge \bfseries Analysis of some classical and quantum aspects of black hole physics \par}\vspace{0.7cm} 
\rule{13.5cm}{.7pt} \\[3cm] 

\Large\textsc{Luciano Gabbanelli}

\vfill

\vfill

\vspace{30pt}

\includegraphics[scale=3.2]{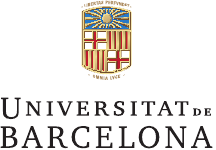} 

\vfill
\end{center}
\end{titlepage}


\begin{center}

\mbox{}
\thispagestyle{empty}

\vfill
©  2019\\[2ex]
All the content of this work is licensed under the Creative Commons

\end{center}

\cleardoublepage


\begin{adjustwidth*}{0.2cm}{0.2cm}

\begin{center}

\vspace*{.06\textheight}
\begin{center}
{\LARGE \bfseries Analysis of some classical and quantum aspects of black holes \ttitle\par}
\end{center}
\vspace{0.4cm} 

\vfill

by

\vfill

{\Large Luciano Gabbanelli\par}

\vspace{60pt}

\justifying

\noindent Thesis submitted in fulfillment of the requirements for the degree of {\it Philosophi\ae} {\it Doctor} in the Faculty of Physics of the University of Barcelona.

\vfill

\noindent PhD programme in Physics. Line of research: Particle Physics and Gravitation.

\vfill

\noindent This thesis was supervised and tutorized by 

\vfill

\centering
\vspace{13pt}

{\large Dom\`enec Espriu Climent}

\vspace{13pt}
\vfill

\justifying
\noindent Department of Quantum Physics and Astrophysics \& Institute of Cosmos Sciences of the University of Barcelona.

\vspace{40pt}
\begin{figure}[th]
	\centering
	\includegraphics[scale=2.8]{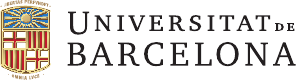}
	\decoRule
\end{figure}

\centering
\renewcommand{\today}{\ifcase \month \or January\or February\or March\or %
	April\or May \or June\or July\or August\or September\or October\or November\or %
	December\fi\ \number \year} 
{\large Barcelona, September 2019}\\[4cm] 

\end{center}

\end{adjustwidth*}

\cleardoublepage


\vspace*{50mm}

\begin{flushright}
	\textit{To my biological family, for being who they are.\\[3ex]
		To my chosen family, for being here.}
\end{flushright}

\thispagestyle{empty}





\begin{abstract}
\addchaptertocentry{\abstractname} 

For ninety years we have known that our universe is in expansion. Cosmological data favor an unknown form of intrinsic and fundamental uniform energy contributing approximately 68\% of the total energy budget in the current epoch. The simplest proposal in accordance with observations is the standard cosmological model consisting of a small but positive cosmological constant producing a gravitational repulsive effect driving the accelerated expansion. In standard cosmology general relativity is assumed as the theory for gravity, which in turn predicts that a sufficiently compact mass can deform spacetime and form a black hole. At a mathematical level, these objects are considered vacuum solutions described by very few parameters. For instance, a stationary black hole solution is completely described by its mass, angular momentum, and electric charge; and two black holes that share the same values for these parameters, are indistinguishable from one another.

On the basis of the usual metrics describing black holes, it is generally believed that all contained matter is localized in the center or, if rotating, on an infinitely thin ring. Recent approaches challenge this unintuitive assumption and consider matter just spread throughout the interior. Clearly, this begs for a quantum description in curved space. In past years, a novel approach established a new bridge between quantum information and the physics of black holes when an intriguing proposal was made: black holes could possibly be understood as Bose--Einstein condensates of soft interacting but densely packed {\it gravitons}. The aim of this thesis is to discuss how to construct a graviton condensate structure on top of a classical gravitational field describing black holes. A necessary parameter to be introduced for this analogy is a chemical potential which we discuss how to incorporate within general relativity. Next we search for solutions and, employing some very plausible assumptions, we find out that the condensate vanishes outside the horizon but is non-zero in its interior. These results can be extended easily to a Reissner--Nordstr\"om black hole. In fact, we find that the phenomenon seems to be rather generic and is associated with the presence of a horizon, acting as a confining potential. In order to see whether a Bose--Einstein condensate is preferred, we use the Brown--York quasilocal energy, finding that a condensate is energetically favourable in all cases in the classically forbidden region. The Brown--York quasilocal energy also allows us to derive a quasilocal potential, whose consequences allow us to suggest a possible mechanism to generate a graviton condensate in black holes. On the contrary, this is not the case for any kind of horizons; for instance, this mechanism appears not to be feasible in order to generate a quantum condensate behind the cosmological de Sitter horizon.

Furthermore, when a pair of black holes merge, an immense amount of energy should be given off as gravitational waves. Their wave forms have been recently confirmed to be perfectly described by general relativity. We discuss why for low frequency gravitational waves aimed to be detected by astrophysical PTA observations the fact that propagation should take place over an expanding (approximately globally de Sitter) spacetime should be taken into account. In this manner, harmonic waves produced in such mergers would become anharmonic when measured by cosmological observers. This effect is tiny but appears to be observable for gravitational waves to which PTA are sensitive. Therefore we have characterized modifications to the expected signal, and how it is related to the source and pulsar characteristics that are employed by the IPTA collaboration. If the cosmological constant were an intrinsic property, this experiment would be capable of confirming the relevance of $\Lambda$ at redshift $z<1$.

\end{abstract}



\hypersetup{allcolors=black}

\tableofcontents


\mainmatter 

\pagestyle{thesis} 

\chapter{Introduction}
\label{Introduction}




\section{From general relativity to black holes}

Even though gravity is ubiquitous, it is literally forty orders of magnitude weaker than all the other known forces. Gravity is the dominant force on the largest scales and the force we are most aware of in daily life and for longest; it literally grounds us. Macroscopically, matter is {\it neutral} with respect to the other three forces, whereas everything that has mass or energy experiences or generates gravity.

In 1915 Albert Einstein completed the general theory of relativity that changed the way of understanding the gravitational interaction \cite{EinsteinGR}. From this time on, spacetime would never be again a static place where {\it events} happen, but a dynamical structure in deep connection with those {\it events}. For instance, the local energy and momentum within the spacetime curve the very spacetime. 
Paraphrasing John Wheeler in \cite{MisnerThorneWheelerGravitation}:
"spacetime tells matter how to move; matter tells spacetime how to curve."
This interplay is seen at the level of the field equations where the local spacetime curvature, expressed by the Einstein tensor $G_{\mu\nu}$ on the left hand side, is altered because of the local energy and momentum within that spacetime given by the stress-energy tensor $T_{\mu\nu}$; this is
\footnote{The convention used for obtaining the Ricci tensor is by contracting the first and third indices of the Riemann tensor with the inverse metric as $R_{\mu\nu}=g^{\delta\lambda} R^{}_{\delta\mu\lambda\nu}$. The scalar curvature is the trace of the Ricci tensor $R=g^{\mu\nu}R_{\mu\nu}$ and $\kappa=8\pi G_N/c^4$ is the Einstein constant. $G_N$ is the Newtonian constant of gravitation.}
\begin{equation}\label{EEq}
	R_{\mu\nu}-\tfrac12\,R\,g_{\mu\nu} =\kappa\,T_{\mu\nu}\ .\end{equation}
A crucial guiding principle leading Einstein along the formulation of his theory was the equivalence between the gravitational and the inertial masses; to the point that he suggested that it should be ascended to the status of general principle. The equivalence principle states that "the movement of a body subjected to a gravitational field is physically indistinguishable to that of an accelerated body in flat spacetime". This means that locally, the spacetime is Minkowskian and the laws of physics exhibit local Lorentz invariance.
This universal characteristic is what urged Einstein to transfer the properties of the gravitational field to the structure of spacetime. The spacetime curvature and its fluctuations are a measure of the gravitational field.
As an immediate consequence, gravity could not longer be viewed as an instantaneous force. In the absence of external forces, any body immersed in a given spacetime must follow geodesics in accordance to the curvature of the dynamical spacetime it lives in (and it curves as well).

Since its formulation, general relativity has not only been exhaustively tested but also has predicted inconceivable phenomena. Having been born from a small contribution to the Mercury’s perihelion, an explanation of a small precession without the use of any arbitrary parameter, quickly changed the paradigm of how we understand the oldest known natural force.
Einstein jump to the fame soon after, during the total Sun eclipse of 1919, when Eddington confirmed the prediction for the deflection of starlight passing near the surface of the Sun. However it was not until 1959 that the first experiment (not observation) was performed to test this theory. Pound and Rebka measured the redshift of light when moving in the Earth's gravitational field. This research ushered in an era of precision tests of general relativity where more and more accurate tests began to be made. 
Some important examples are the Shapiro time delay of a signal when passing through a gravitational field (1966); or in a stronger-field regime the orbital decay due to gravitational wave emission by a Hulse--Taylor binary (1974), among many others. Nowadays, the first direct measurement of gravitational waves by the LIGO (2015), tests the theory in the very strong gravitational field limit, observing to date no deviations from the theory.
It is a common agreement in the relativistic community that we stand on the threshold of the gravitational-wave astronomy era.

Although Einstein's equations look elegantly simple, they are quite difficult to solve. As the curvature has an associated energy and the energy curves the spacetime, the equations have a strongly non-lineal character. In fact, Einstein itself used approximation methods in working out his initial predictions.
In was in 1916 when Schwarzschild found the first exact vacuum solution, other than the trivial flat solution \cite{Schwarzschild}.  It is a useful approximation describing slowly moving astronomical objects such as the Earth or the Sun. Known as the Schwarzschild metric\footnote{The $(-+++)$ signature has been used in chapters \ref{ChapterBHasBEC}, \ref{ChapterQLE} and \ref{ChapterBHasBECExtension}. We have changed to the $(+---)$ signature in chapters \ref{ChapterGW} and \ref{ChapterPTA}.},
\begin{equation}\label{MetricSchw}
	ds^2=g_{\mu\nu}\ dx^\mu dx^\nu=-\left(1-{\frac {r_S}{r}}\right)c^{2}\ dt^{2}+\left(1-{\frac {r_S}{r}}\right)^{-1}\ dr^{2}+r^{2}\ d\Omega^2\ ,\end{equation}
it is understood as the limit case of a static and spherically symmetric solution of Einstein's field equations. The variables range between $0\leq\theta\leq\pi$, $0\leq\phi<2\pi$ and the metric of the two-sphere is given by $d\Omega^2=d\theta^2+\sin^2\theta\ d\varphi^2$. The Schwarzschild radius depends only on the object mass, $r_S=2G_NM/c^2$.

The consequences of this solution were quite disconcerting and until today it is said that they hid in plain sight the breakdown of the classical description and the need of new physics to be understood.
In the following year, Hilbert presented an extensive study of the singularity structure and showed that the metric was non-regular both at $r=0$ and at $r=r_S$. For the former, there was general consensus about the {\it genuine} physical singularity behaviour of the curvature, which implies that some invariant (scalar) built upon the Riemann's tensor, such as curvature scalar or the Kretschmann invariant (the sum of Riemann's tensor components squared), diverges\footnote{In the original paper, the origin of coordinates has been placed over the Schwarzschild radius (where nowadays the event horizon of a black hole is situated).}. This means that the gravitational field becomes infinitely large at a gravitational singularity.
On the contrary, the nature for the latter remained unclear for many years.

It was Lema\^itre who showed in 1932 that this singularity was not physical (he used Lema\^itre coordinates and stated that it was an artifice introduced because of the coordinates that Schwarzschild had used); an illusion. Besides, seven years later, Robertson showed that a free falling observer {\it descending} in the Schwarzschild metric would cross the $r = r_S$ singularity in a finite amount of proper time $\Delta\tau$ even though this would take an infinite amount of time in terms of coordinate time $t$ employed in \eqref{MetricSchw}.
In any case, it was not until 1960 when new tools of differential geometry provided more precise definitions of what means for a Lorentzian manifold to be singular. This led to the definitive identification of the $r = r_S$ coordinate singularity as an event horizon: "a perfect unidirectional membrane: causal influences can cross it in only one direction" \cite{Finkelstein1958}.
Stated in other words, a null hypersurface from which points in the spacetime are causally disconnected from outside. When we choose a coordinate system where the metric does not diverge at the Schwarzschild radius, the following fact is observed \cite{TownsendBlackHoles}: in the inner region, $r<r_S$, everything, even photons directing radially outward, actually end up moving to decreasing values of $r$ and falling into the singularity. This indicates that even light remains trapped inside the surface given by $r=r_S$ and there is no longer a causal relationship with the exterior.
This is what allows us to interpret Schwarzschild's solution as a black hole.

The early 1970s brought a blossom in black hole physics when a close resemblance between the laws of black hole mechanics and the laws of thermodynamics were discovered. 
The second law of thermodynamics requires an entropy for black holes. In \cite{Entropy} Bekenstein conjectured that an entropy could be assigned to black holes and that it should be proportional to the area of its event horizon, $A_H=4\pi r_S{}^2$, divided by the Planck length, $\ell _P{}^2=\hbar G_N/c^3$; this is
\begin{equation}
	S=\frac{k_B\,A_H}{4\,\ell_P{}^2}\ .\end{equation}
$k_B$ is Boltzmann's constant. 
Bekenstein also noticed that there was a positive quantity, called the surface gravity $\kappa_H$, which should be an analogue of the temperature when comparing the laws of thermodynamics with that of black holes
\begin{equation}\label{TemperatureHawking}
	T=\frac{\hbar\,c\,\kappa_H}{2\,\pi\,k_B}\ .\end{equation}
In fact, it can be shown that having relations between the mass and the energy, the area and the entropy, the surface gravity and the temperature, 
black holes are consistent with the three laws of thermodynamics (see \cite{RossBHThermodynamics} for a nice review).

It is reasonable to think that the emergence of the singularity structure of black holes might be connected with the ignorance of the quantum mechanical effects.
It has been known for a long time that a quantum description of gravity is a problem and quantization of Einstein's equations gives rise to a non-renormalizable theory, uncontrollable at high energies.
It was Hawking who first attempted to put together quantum mechanics and black hole physics by semiclassical techniques without tackling the problem of curvature singularities. In certain regimes, quantum field theory (QFT) calculations can be performed in curved spacetime showing that black holes are indeed thermodynamic bodies with a temperature inversely proportional to its mass.
Both their temperature and the entropy depend only on asymptotic charges \cite{HawkingRadiation}.
The main prediction of his calculations points towards a slow evaporation of these objects due to the emission of thermal radiation with the same spectrum of a black body.
This led him to speculate that quantum information may be destroyed, and that it is not possible to trace back from the emitted Hawking quanta, the initial state that first formed a black hole \cite{HawkingBreakPredictability}.

This fundamental paradox is based on the fact that general relativity predicts the existence of surfaces at which time, as measured as a clock at infinity, stands still (event horizons). Gravity is well behaved at those surfaces (as we have already discussed free falling observers would pass through in a finite proper time), but this causes paradoxes in quantum mechanics. Other proposals apart from the Hawking breakdown of predictability are in favour of revising the laws of quantum mechanics \cite{tHooft1999}.  
This has turned into a fundamental problem in theoretical physics for more than 40 years. If the breakdown of general relativity is encoded in the singularity structure, the information paradox explicitly expresses an existing tension between general relativity and QFT. This evidences the need for a theory of quantum gravity.

\section{Self-completion via classicalization}

Alternative ideas have been proposed to provide a bottom-up picture reproducing the experimentally observed gravitational effects, but from a quantum field theoretic approach and without referring to general relativity at all.
The Newtonian potential models quite comprehensively most of the features of the solar system. Hence, given two masses, it is mandatory for the new theory to recover
\begin{equation}
	V_N(r)=-G\ \frac{m}{r}\end{equation}
in the non-relativistic limit. The similarities with the Coulomb potential are obvious, $V_C=Q/4\pi r$, and the difference as well, the gravitational potential is always attractive, whereas the photon also permits a repulsive behaviour.
This ensures that a spin 1 field would not be suitable for describing the gravitational interaction; nor would a spin 0. In a scalar field theory, the only allowed coupling between the hypothetical spin 0 gravitational mediator and the electromagnetic field is given by the Lagrangian density ${\cal L}_{int}\sim\psi T_\mu^\mu$. If we recall the electromagnetic stress-tensor \eqref{StressTensorElectro}, one immediately sees that $T_\mu^\mu=0$; and how it was stated by Eddington during the solar eclipse of 1919: "the light is drawn towards the sun".

Since fields higher than spin 2 are quite difficult to include consistently in a QFT, efforts were directed to the quantization of $h_{\mu\nu}$ emerging from the weak-field limit of the Einstein--Hilbert action 
\begin{equation}\label{ActionGR}
	S_{HE}=\frac{1}{2\kappa}\int d^4x\ \sqrt{-g}\ R\ .\end{equation}
From a {\it Wilsonian perspective}, low energy effective field theories (EFT) are perfectly well defined if we keep away from its cut-off scale.
Therefore, in their domain of applicability an EFT singles out the suitable (quantum) degrees of freedom that would serve as an adequate reference to describe a system at a given energy scale. The consideration of "suitable" refers to whether the system can be consistently described by means of a perturbative picture.
In the case of gravity, the cut-off scale is given by the Planck mass $M_P$, and in fact, it has been possible to compute explicit quantum corrections to classical spacetimes in the low-energy limit, $E\ll M_P$  \cite{DonoghueEFT}. 
The non-renormalizability of gravity does not represent a big deal for the theory {\it per se} at their applicability domain; however at energies beyond its domain the theory breaks down and gives rise to a non-renormalizable theory uncontrollable at high energies.
Let us now look at some orders of magnitude: neutron stars have masses lower than 2.16 M$_\odot$ inside a radius of the order of 10 km; hence an average density about the nuclear density. Some black holes have about the same mass, but they are about 10 times smaller and thus 1000 times more compact. Clearly, this begs for a quantum description in curved space.

Both approaches, a QFT description in terms of a massless spin 2 tensor field living in a Minkowskian space and Hawking's theory of quantizing fields on curved classical spacetimes, provides very interesting quantum mechanical effects. However, they also predict serious conceptual issues. As discussed above, the non-renormalizability and the information loss paradox point towards an unavoidable need of a more general theory. In what follows we will discuss an alternative perspective called {\it self-completion via classicalization}.
In order to motivate the discussion, let us briefly resume how the UV-completion of the Fermi's interaction is performed within the standard {\it Wilsonian approach} \cite{WilsonUVCompletion}. The UV-completion of the theory is achieved by integrating new degrees of freedom that allow the recovery of the weak coupling regime. This led to the replacement of the Fermi's contact interaction by a more complete theory, the electroweak theory.
On the contrary, {\it classicalization} proposed that gravity would require a {\it non-Wilsonian treatment}.

The Fermi theory was the precursor of the weak interaction. It consists of a Feynman diagram of four fermions directly interacting with one another at one vertex with the purpose of explaining the $\beta-$decay \cite{FermiBeta}. 
Roughly speaking, this theory is given by a Dirac Lagrangian together with a four-fermion interaction term
\begin{equation}
	{\cal L}_{int.} \sim G_F\ (\overline\Psi\Psi)^2\ ,\end{equation}
where $G_F$ is the Fermi coupling constant, which denotes the strength of the interaction. From dimensional analysis one obtains the following: the field's kinetic term implies $[\Psi]=$mass$^{1/2}\,\cdot\,$length$^{-1}$, then we have $[G_F]=$length$\,\cdot\,$mass$^{-1}$ for the coupling. In Planck units\footnote{Planck units are given by $\ c=G_N=k_B=\hbar=1$.}
the Fermi coupling constant dimensions are $[G_F]=$mass$^{-2}$.
It is common knowledge in QFT that interactions with coupling constants scaling as the inverse square of the mass are non-renormalizable and this is what happens in Fermi's four-fermion interaction.
While this theory describes the weak interaction remarkably well, this happens only in its {\it domain of applicability}. The cross-section grows as the square of the energy $\sim G_F{}^2\,E^2$, indeed once the center of mass energy of a certain process approaches the cut-off, all orders in the perturbative expansion become relevant, the cross-section grows without bound and unitary is violated. As a matter of fact, the theory is not valid at energies much higher than about $100$ GeV.
The fact that all terms become relevant is a consequence of the non-weak interaction between the chosen degrees of freedom; in fact they have rather entered in the strong coupling regime.
This sick behaviour of Fermi's theory claims the need of a UV--completion of the theory.

The electroweak theory represents the UV--completion of the Fermi weak interactions. The cure for the later is managed by introducing three mediators for the weak interaction; the vector bosons $W^\pm$ and $Z^0$. Consequently, the Fermi coupling constant \cite{Griffits} is usually expressed as
\begin{equation}
	G_F= \frac{\sqrt{2}}{8}\ (\hbar c)^{3} \left(\frac{g_W}{M_W\ c^2}\right)^2\ .\end{equation}
Notice that the coupling constant of the weak interaction $g_W$ and the mass of the $W$ boson which mediates the decay in question, $M_W$, do not appear separately; only their ratio occurs. At high energies it is better to describe the $\beta-$decay as a non-contact force field having a finite range, albeit very short. Now it is clearly seen that one can understand Fermi's theory as the low energy limit of the more general electroweak theory.

The findings in \cite{DvaliClassicalizationForEinsteinGravity, DvaliClassicalization} lead us to the conclusion that Einsteinan gravity is self-complete in a sense that is very different from the standard notion of the Wilsonian completeness.
The concepts of $\ell_P$ being a minimal length and the absence of trans-Planckian propagating quantum degrees of freedom \cite{LPMinimal}, as well as UV--IR connection through the black holes \cite{UVIRBH} have been around for some time. 
The authors argue that when the relevant degrees of freedom of our effective theory become strongly coupled, any attempt of probing the physics beyond, automatically bounces us back to the macroscopic distances due to black hole formation. 
The theory completes itself by producing some high-multiplicity states of the same particles that were already in the theory.
Each of the constituents would be in an extremely soft and weakly interacting regime with one another. The occupation number of the newly produced soft quanta is extremely high. The resulting state would approximately behave classically.
Hence, they postulate black holes as the real UV states of the theory.
As a natural consequence of this barrier, high energy scattering amplitudes are unitarized using black hole production.

\section{Black holes as self sustained quantum systems}

According to the usual metrics describing black holes, it is generally believed that all contained matter is localized in the center, or if rotating, on an infinitely thin ring.
Radically different approaches challenge this unintuitive assumption of localized matter and consider it spread throughout the interior \cite{BHMatterInterior}. The key point addressed is how matter can withstand the enormous pressures associated with such high densities. In \cite{Chapline} for the very first time it is assumed that matter near the horizon could be in a Bose--Einstein condensate phase. Soon after, the Bose--Einstein condensate (BEC) idea is extended to the whole interior and the "gravo-star" solution is discussed \cite{Mazur}.

The idea that many-particle states might represent the right path toward a quantum mechanical description of self-gravitating systems finds a natural explanation from the {\it classicalization} approach, encoded in the $2\rightarrow N$ scattering process \cite{Dvali2N}.
According to the Thorne's Hoop conjecture (see e.g. \cite{ThorneHoopClassic} for the classical formulation and \cite{ThorneHoopQuantum} for its quantum mechanical extension) black holes can form in a collision if the two colliding objects fall within their black disk. 
This acceptance is based on the fact that if a source with center of mass energy $\sqrt{s}$ is confined within its  gravitational (Schwarzschild) radius $R_g(s)=\sqrt{s}\,\ell_P{}^2$, must form a black hole.
Then it follows that when two high energy quanta are scattered with a center of mass energy $\sqrt{s}\gg M_P$ ,
the theory self-completes by producing {\it collective states} composed of a large number
\begin{equation}\label{GravitonNumber}
	N\simeq \frac{s}{M_P{}^2}\end{equation}
of soft virtual {\it gravitons}. Each of them is characterized by a typical quantum size
\begin{equation}\label{GravitonWavelength}
	\lambda\simeq \sqrt{N}\ \ell_P\end{equation}
and in the mean-field approximation ($N\gg1$), the semi-classical behavior of macroscopic black holes is recovered \footnote{Note that the weakly interacting {\it gravitons} are characterized by an energy
\begin{equation}\label{GravitonEnergy}
	\varepsilon=\frac{\hbar}{\lambda}=\frac{\hbar}{r_S}\ . \end{equation}
Energy conservation arguments implies that
\begin{equation}
	\sqrt{s}=N\ \varepsilon\ ,\end{equation}
from where the expression \eqref{GravitonNumber} it is derived. This clearly implies that
\begin{equation}
	N\gg1\ .\end{equation}}. 
Stated in other words, in the classicalization picture the role of the hard quanta is taken up by the collective effects of a self-bound state, composed of a large multiplicity of soft quanta produced in the process. 
The weakly-interacting constituents result in a quasi-classical final state that in a mean-field approximation ($N\gg1$) acquires some properties of classical black holes. This is the reason why it is said that such a theory self-completes in the UV by classicalization.

Starting from the previous discussion, it is possible to draw an interesting quantum picture for black holes without referring to any geometric arguments at all \cite{DG2,DvaliBHFormation}.
The microscopic corpuscular picture of black holes as $N-$graviton self-bound states at a {\it quantum critical point} \cite{DvaliQCP} enables us to translate the $N-$graviton production processes into the black hole formation, both in field and string theory scatterings.

Let us assume a $2\rightarrow N$ scattering process with a momentum exchange ${\bf p}$. Then the final state of $N$ gravitons each of them with a {\it dimensionless effective coupling}  $\alpha_g$.
According to the de Broglie relation $\lambda=\hbar/|{\bf p}|$, using \eqref{GravitonNumber} or \eqref{GravitonWavelength}, one can alternatively write
\begin{equation}\label{GravitonCoupling}
	\alpha_g=\frac{G_N\,|{\bf p}|^2}{\hbar}=\frac{\ell_P{}^2}{\lambda^2}\simeq\frac{M_P{}^2}{s}\simeq\frac{1}{N}\ .\end{equation}
It is explicitly seen that although each of the constituents interact very weakly among the other individual corpuscles ($\alpha_g\ll1$),
the system is globally in a strong coupling regime with a {\it collective dimensionless coupling} $g=\alpha_g\,N\sim1$. In a Bose--Einstein condensate language, non-perturbative effects due to the collective interaction among the $N$ gravitons resulting from the scattering process become extremely relevant and lead to the global (approximately) classical behaviour of the system.

Let us face this picture from another perspective and try to form a Bose--Einstein condensate of $N$ weakly interacting gravitons. 
Let us assume that each of them is characterized by a typical wavelength $\lambda \gg \ell_P$. It is obvious that each {\it graviton} would feel the collective binding potential by the other $N-1$ constituents.
At first order of approximation, it can be estimated to be $U\simeq-\hbar\,\alpha_g N/\lambda$.
Furthermore, the kinetic energy is $K\simeq|{\bf p}|\simeq\hbar/\lambda$. The resulting self-sustained state is expected to be in a marginally bound condition; then the energy balance yields $K+U\simeq0$. The direct implication is the quantum criticality condition
\begin{equation}
	\alpha_g\,N\simeq1\end{equation}
that coincides with the scaling found in the classicalization approach. 
This relation allows us to identify a black hole of mass $M$ with a graviton Bose--Einstein condensate on the verge of a quantum phase transition. 
The critical point separates two phases: for $\alpha N\ll1$, the system is in the phase in which collective {\it graviton}-{\it graviton} attraction is not enough to form a self-bound state and the graviton Bose-gas is essentially free. For $\alpha N=1$ the bound-state is formed, then corresponds to the point of black hole formation.
The typical wavelength of the {\it gravitons} constituting the condensate coincides with the Schwarzschild radius $\lambda\simeq r_S$.
Since \eqref{GravitonCoupling} holds, the {\it scaling laws of corpuscular black holes} are given by the equations \eqref{GravitonWavelength}, \eqref{GravitonCoupling} together with
\begin{equation}
	M=\sqrt{N}\ M_P\ .\end{equation}

\begin{figure}[!h]\begin{center}\vspace{10pt}
	\includegraphics[scale=0.3]{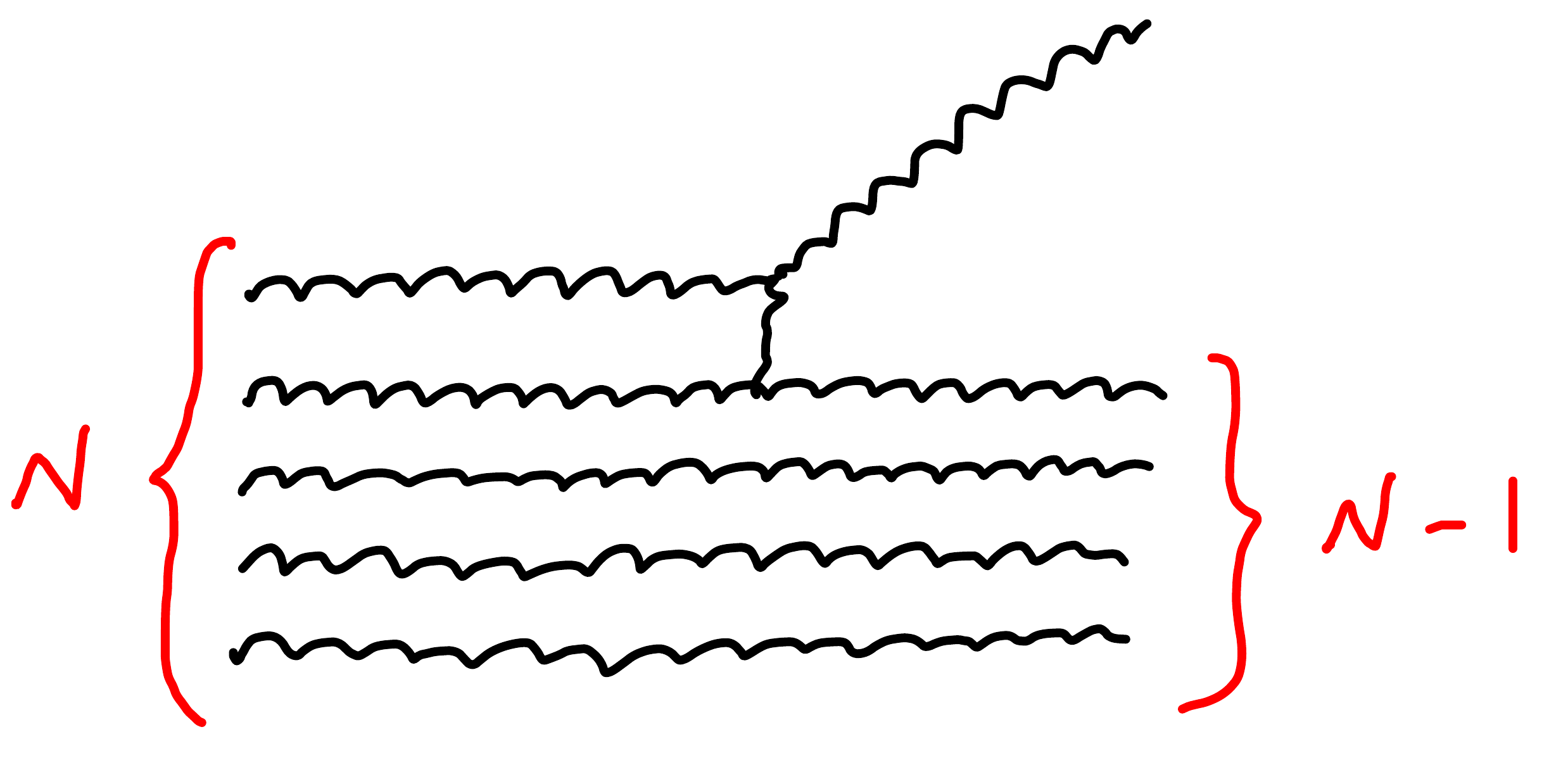}
	\captionsetup{width=0.85\textwidth}
	\caption{\small Leading order process responsible for quantum depletion of graviton condensate. Figure taken from \cite{DvaliBHFormation}.} \label{FigDepletion}\end{center}\end{figure}

Furthermore, at the critical point the system is characterized by collective Bogoliubov modes that become nearly gapless (up to $1/N$
resolution), and there is a maximal degeneracy of states. These degeneracy of states, at the same time represents the black hole entropy factor.
As one would expect for an interacting homogeneous Bose--Einstein condensate \cite{Bogoliubov}, the graviton wave-packet is "stuck" at the critical point. It cannot really hold on all its constituents, then it is slowly loosing {\it gravitons} through a {\it quantum depletion} process, reducing $N$, but maintaining the quantum criticality condition for each $N$. 
This leakage is a fully-quantum progenitor of what in a semi-classical treatment is called Hawking's radiation. Indeed, in \cite{DG2, DvaliBHFormation} the {\it depletion law}
\begin{equation}
	\frac{dN}{dt}\simeq-\frac{1}{\sqrt{N}\,\ell_P}+{\cal O}\left(\frac{1}{\ell_P\, N^{3/2}}\right)\end{equation}
is obtained. Figure \ref{FigDepletion} shows how depletion works, in the $2\rightarrow2$ scattering of two constituent {\it gravitons} one gains enough energy to leave the ground state and hence leaves the condensate. Considering the "scaling laws" previously discussed, the latter relation can be written in terms of the Arnowitt--Deser--Misner (ADM) mass $M$ as
\begin{equation}\label{DepletionMass}
	\frac{dM}{dt}\simeq- \frac{M_P{}^3}{\ell_P\,M^2}+{\cal O}\left( \frac{M_P{}^5}{\ell_P\,M^3}\right)\simeq-\frac{T^2}{\hbar}+{\cal O}\left(\frac{T^4}{\hbar\,M_P{}^2}\right)\ ,\end{equation}
where the temperature is given by $T =\hbar /G_N\,M = \hbar/\sqrt{N}\,\ell_P$. The result \eqref{DepletionMass} coincides with Hawking's arguments in the semiclassical limit given by $N\rightarrow\infty$, $\ell_P\rightarrow0$, but keeping $\sqrt{N}\,\ell_P$ and $\hbar$ fixed.

\section{Back to the geometry}

The corpuscular theory of gravity offers a way to unify most of the experimental observations in a single framework (from astrophysical to galactic and cosmological scales), solely based on gravity and matter.
This enables us to argue for the first time that the collective effects would be the correct explanation for the puzzling behavior of black holes; and by direct implication the apparent incompatibility of quantum mechanics and general relativity.
Unlike black holes, the new solution is thermodynamically stable and apparently bypasses the information loss paradox.
As an attempt to put together Dvali's proposal with general relativity solutions, the idea of black holes as graviton condensates is recovered \cite{AEG}.

The possibility of defining a quasilocal energy for different type of classical spacetimes presenting horizons is well established; therefore certain volume containing {\it gravitons}, each of them with an energy $\varepsilon$ given by \eqref{GravitonEnergy}, would have an associated energy density and a graviton number density. The depletion of the condensate must be associated to variations on both magnitudes in that volume. 
In this manner, it is thermodynamically consistent to associate a chemical potential $\mu$ to take into account the change of the particle number $N$ (according to Dvali's picture, in the phase transition). 

In Chapter \ref{ChapterBHasBEC} we present a simple way to introduce the chemical potential at the level of the Einstein's equations.
In this geometrical approach, this potential would regulate the components of a $2-$rank tensor $h_{\mu\nu}$, associated to perturbations of the causal structure of a black hole; for instance \eqref{MetricSchw}. 
The derived field equations resemble the non-linear Gross--Pitaevskii equation usually employed for modeling Bose--Einstein condensate. And the corresponding solution seems to correctly describe a self-sustained graviton condensate over a confining potential given by the metric itself. 
Besides, the condensate is intimately related to the classical field that sustains it and determines its characteristics. This suggests the possibility of interpreting the solution as a collective wave function in the framework of mean field approximation (usually a fairly good description for Bose--Einstein condensates). Remarkably enough, the characteristics of the resulting condensate are uniquely described by the Schwarzschild radius.

In \cite{AEGHorizons} analogous solutions for Reissner--Nordström and de Sitter metrics have been derived. In Chapter \ref{ChapterQLE} quasilocal-energy considerations, well established in black hole physics, have been extended to the obtained condensate-like solutions. In all cases the BEC structure is energetically favourable in the classically forbidden region. As mandatory, all solutions recover the classical characteristics in the visible side of the horizon, where the chemical potential vanishes.
Besides, the quasilocal energy can be used to define a quasilocal potential that may shed light upon the mechanism regulating the absorption and depletion of {\it gravitons}. The deep well structure around the horizon mediates the binding of matter falling into the black hole. Gravitational radiation produced by the infalling matter seems to feed the graviton condensate, then it is natural to interpret the chemical potential as the conjugate variable to the number of gravitons $N$. 

On the contrary, in Chapter \ref{ChapterBHasBECExtension} it is shown that it is not possible to define an analogous mechanism behind a cosmological de Sitter horizon. Then one is not able to generate a Newtonian potential as the result of a coherent state.

\section{The accelerating universe}\label{SecGWCC}

It is usual for imagination to go faster than technology and Einstein's equations \eqref{EEq} are not the exception. Only a year after Schwarzschild derived his first exact solution \eqref{MetricSchw}, theoretical consequences rapidly appear also in the large distances limit. Einstein himself applied his theory to the universe as a whole. 

At the time, the universe was thought to be static and the low relative velocities of the observed stars support this belief. The "greatest blunder" story is well know; Einstein knew that there was a missing piece for the universe's existence to take place. The dynamical nature of the field equations told him that the gravitational attraction would cause everything to collapse, whatever the initial equilibrium condition chosen. Hence, following the current observational presumptions, he added a universal constant $\lambda$ to counterbalance the effects of gravity and achieve the staticity; an attractive gravitational force being balanced by a repulsive cosmological constant force. 
Using the weak limit for the gravitational field and low velocities as a compass that guided him to the well establish Newtonian Poisson's equation, he proposed the following modification
\begin{equation}
	\nabla^2V(r)-\lambda\ V(r)=4\pi\ G_N\ \rho_m\ , \end{equation}
where the subscript $m$ stands always for matter, whatever its type; in this case the visible matter energy density. In such a manner, a constant gravitational potential $V\equiv-4\pi G_N \rho_m/\lambda$ would produce the desired static universe \cite{EinsteinCC}. From now on, we change to the terminology of $\Lambda$ for the cosmological constant.
In accordance to his proposal of modifying the Poisson’s equation, Einstein modified his field equations \eqref{EEq} introducing this new constant of nature\footnote{The sign of the cosmological term depends on the signature of the metric. Equation \eqref{EEqCC} corresponds to a $(+---)$ signature. On the contrary, for a signature $(-+++)$ the sign is a plus. See for instance equation \eqref{EEqCC-+++}.}
\begin{equation}\label{EEqCC}
	G_{\alpha\beta}-\Lambda\,g_{\alpha\beta} =\kappa\,T_{\alpha\beta}\ .\end{equation}
initiating the field of relativistic cosmology; another natural playground for Einstein's theory of gravity. He obtained the value $\Lambda=4\pi G_N\rho_m$ working in the nonrelativistic limit; and this would allow a static, homogeneous and spherically closed universe with radius of curvature given by $r_{curv.} \simeq \Lambda^{-1/2}$. 

This exact balance was completely {\it ad hoc}, and in fact
it was a thought that the cosmological constant spoiled the beauty and simplicity of general relativity. However, it was not until more than a decade after, in 1929, that Hubble's observations seemed to indicate that the universe was expanding \cite{Hubble}.
He confirmed experimentally the very famous Hubble law, the linear relationship between the distance of extragalactic sources and their recessional velocity $v=H_0\ D$. The Hubble constant is the Hubble's parameter at the time of observation 
\begin{equation}\label{HubbleConstantGeneric}
	H(T_0=0)=H_0\ .\end{equation} 
A significant tension between local and non-local measurements of the Hubble constant has been inferred from recent data. On the one side, the cosmic microwave background (CMB) measurements \cite{H0Plank} have reported a value $H_0 = (67.4\pm0.5)$ km s$^{-1}$ Mpc$^{-1}$ (Planck, 2018); contrary, direct measurements in the local universe using supernovae type Ia as standard candles \cite{Riess} have given $H_0 = (73.24\pm1.74)$ km s$^{-1}$ Mpc$^{-1}$ (SNe Ia, 2018b).
Latest measurements imply a discrepancy on the value for $H_0$ of more than $4\sigma$ and less than $6\sigma$ \cite{Verde}. This shows that the discrepancy does not seem to be dependent on the method, team, or source. Despite the controversy between late and early universe's techniques, the dynamical nature of our universe is commonly accepted. However, intriguing questions regarding the local behaviour of the Hubble constant triggered an intense debate that is still ongoing.

Coming back to the cosmological constant story \cite{Gamow}. 
Einstein failed to see that with or without cosmological constant, general relativity equations predict that the universe cannot be static\footnote{For a homogeneous and isotropic universe, Einstein's equations \eqref{EEqCC} take the structure written in \eqref{FriedmannEc}. Taking the time derivative of the fist equation and considering ordinary matter with an energy density obeying $\rho\sim a^{-3}$, one finds that for small perturbations  $a=a_0+\delta a$ the equilibrium is always unstable.}.
The person who told everybody to change the idea of space and time did not dare to predict what his theory was telling: the cosmic contraction or expansion with or without a cosmological constant, or even with the fine tuned value he had chosen despite of the instability problems he overlooked.
What's more, as pointed in \cite{RovelliCC}, it may not be even true that the introduction of $\Lambda$ was because of a matter of cosmology. Einstein knew about this term much earlier than his cosmological work. In \cite{Einstein1916} Einstein derived the gravitational field equations from a list of physical requirements. In a footnote he notice that their equations were not the most general possible ones; between other possible terms he mentioned the cosmological term. 
Hence $\Lambda$ was not an appendage added by Einstein to his theory to account for {\it observations}. Einstein seemed to know about $\Lambda$ being an integral and natural part of his theory. 
It was not a matter of "beauty" or "mystery", the cosmological constant seems to be one more constant in our fundamental theories.
It seems that the "blunder" was not a physically wrong prediction (the existence of a cosmological constant), but his silly prejudices that blinded him and made him loose the chance of doing an easy and spectacular prediction. Einstein missed an instability\footnote{Pointed out later on by Eddington \cite{Eddington}.} that could have led him to predict the cosmic expansion in 1917.

In the subsequent decades, much progress was made and cosmological data strongly spoke in favor of a positive cosmological constant at very large scales. The end of the 20$^{th}$ century brought the extraordinary discovery that our universe is expanding at accelerated rates at late eras. Two independent groups carried out accurate measurements of SNe Ia luminosity distances as a function of the redshift, measuring $\Lambda>0$ with a confidence level of more than $3\sigma$ \cite{AcceleratedExpansion}.
These results have been reinforced by different observations as CMB and the Baryon Acoustic Oscillations (BAO) among others. Today is widely accepted that the {\it easiest} explanation for this accelerated expansion is the dominance of $\Lambda$ over matter at late eras. Best fitting data says that over the 68\% of the energy density budget is in form of dark energy. These results are compatible with the existence of a cosmological constant in the current-epoch (see \cite{H0Plank,AbbotDES} for recent articles).
At the level of the equations this term counteracts the gravitational pull of matter with an {\it anti-gravity effect}, maintaining the general covariance of the theory; if small enough it passes the Solar System tests. However, the elusive nature of the cosmological constant still remains a mystery, and its {\it very small} value \cite{LambdaValue},
\begin{equation}\label{CC}
	\Lambda\sim 1.11 \times10^{-52}\ \text{m}^{-2}\ ,\end{equation}
does not make things easy. Until now the unobservable character remains at {\it not such large} distances. The cosmological constant problem is still an outstanding theoretical challenge: it is not so clear whether $\Lambda$ is an effective property valid only at {\it very large} scales or, on the contrary, a fundamental property of the spacetime. If the later is the case, it may be observable, as a matter of principle, at any scale. As can be seen, many questions concerning the ultimate nature of the cosmological constant are still unanswered.

Once $\Lambda$ is included in general relativity, its implications immediately surpass the cosmological arena. 
One of its specific characteristic is that it does not suffer any clustering effects (contrary for instance, to the scalar field dark energy).	
Interesting efforts turn to {\it local} observable effects, where {\it local} stands for sub-cosmological scales. For instance, whether it should affect the orbits of the Solar System and/or of double pulsars. The resulting effects seem to be too small for being detected \cite{LambdaLocalSolar}. At higher scales, effects are expected to be more relevant; corrections to the Virial relation are considered for extended galaxy clusters \cite{LambdaLocalGalaxy}. Other interesting possibility is assessing the presence of the cosmological constant for the case of light bending from distant objects\footnote{Even involving cosmological distances.}. Different studies derive different conclusions in which the order of magnitude of the effects of $\Lambda$ ranges from zero \cite{LambdaLensingZero} to appreciable values \cite{LambdaLensingBig}, moving through unobservably small ones \cite{LambdaLensingSmall}. All in all, these issues remain as an open debate.

\section{Gravitational waves propagating through an expanding universe}
It is well known that general relativity predicts the existence of gravitational waves. 
For describing the inspiraling, merger and final relaxation (ringdown) of black hole binaries one needs nothing other than vacuum spacetime dynamics.
The field equations simply become
\begin{equation}\label{RicciVacuum}
	R_{\alpha\beta}=0\end{equation}
and describe purely curved spacetime.
When a pair of black holes merge, an immense amount of energy is given off as ripples in the curvature of spacetime propagating at the speed of light $c$.

It is well established that the inclusion of a spread energy density immediately produces a non-trivial curvature of the spacetime. 
Therefore the logic would lead us to expect that the propagation of gravitational waves in our universe differs from that over a flat Minkowskian universe \eqref{RicciVacuum}. 
The usual treatment of gravitational waves is to perform small perturbations around a background, but in this case it should be a non flat background.
These issues and more have been reported in \cite{Bernabeu}, where linearized Einstein's equations in the presence of $\Lambda$ have been solved. For obtaining the solution, a coordinate choice is mandatory. 
The presence of a cosmological term adds new terms in the linearized equations. The relevance of these terms together with a careful discussion on the coordinate choice has been discussed in \cite{Espriu}. In fact, the easiest choice to solve the set of equations is the Lorenz gauge (or $\Lambda$ gauge) where the linearized equations are closely related to those of flat spacetime. These coordinates are nothing but different parametrizations of the Schwarzschild-de Sitter (SdS) spacetime.
However, the only coordinate system we can make sense of is one corresponding with the cosmological coordinates (FLRW, because the acronyms of Friedmann, Lema\^itre, Robertson and Walker); i.e. the one where the universe appears to be homogeneous and isotropic. For a de Sitter universe, the change of coordinates between both coordinate systems, from SdS to FLRW, is intricate but it can be worked out. Nonetheless, for the purpose of the topic under discussion we will see that it is enough only with its linearized version.

Once we know how to move between different coordinate systems, it is direct to transform the gravitational wave solution (easily found in SdS) to coordinates where we can make observable predictions. The conclusions of \cite{Espriu} are prominent: the transformed wave functions get modifications in their amplitude and in their phase of order ${\cal O}(\sqrt{\Lambda})$, because the propagation takes place over a dynamical background. We will see that the latter modifications would imply a possible way to measure {\it locally} the effect of the cosmological constant.
When waves travel they are redshifted but in a different way from the usual gravitational redshift usually used for electromagnetic radiation.
The dispersion relation is not linear anymore; it relates an effective frequency $w_{eff}$ with an effective wave number $k_{eff}$.
Of course the usual gravitational redshift for the frequency is recovered, but as it will be explained in Chapter \ref{ChapterGW}, there is much more to say about the consequences of $w_{eff}\neq k_{eff}$.
This chapter presents the results obtained in \cite{AEGgw}. Towards a realistic cosmology, we derived how to incorporate dark matter as a different background to the existent solution in de Sitter. This new type of matter has been modeled as a pressureless and non-interacting cold fluid.
Although mathematics get obscured because things get rapidly difficult when combining different types of matter, according to the previous discussion we can take advantage of the linearized arguments. Then a linearized version of the change of variables for these two components can be easily derived. One step further has been given in \cite{EGR}, where it is argued that this change of variables can be generalized to an arbitrary background, showing also the inclusion of radiation; however the validity of this goes only up to linear order. Certainly, the effect of the background over the gravitational wave propagation is ruled by the Hubble parameter today $(H_0)$, and it is not possible to disentangle the individual effect corresponding to each component; at least not at a linearized level.

\section{Gravitational wave detection}
Detection of gravitational waves is expected to be performed by pulsar timing array (PTA) observations sometime in the future. The experimental setup consists of an array of radiotelescopes for the accurate timing of a galaxy-sized net of rapidly-spinning neutron stars across the Milky Way.
These rapidly rotating and highly magnetized stars called pulsars emit beams of electromagnetic radiation. Among the thousands of existing pulsars, the millisecond pulsars (MSPs) are of particular interest given their very stable periods and the absence of glitches\footnote{Except for some very small glitches detected in some particular MSPs; see for instance \cite{Cognard}.}.

Three PTAs consortia carrying out such kind of observations have joined efforts in an international collaboration called the International Pulsar Timing Array (IPTA) \cite{Hobbs2010ManchesterIPTA2013}.
The aim of the collaboration is to observe variations in the time of arrival (ToA) of the electromagnetic signals coming from the pulsar array.  The differences between the predicted and measured ToAs are known as "timing residuals".
Pulsar-timing observations, unlike LIGO \cite{LIGO} or Virgo \cite{Virgo}, are sensitive to supermassive black hole (SMBH) binary mergers \cite{Shannon}. These binary systems composed by black holes with masses in the range $\sim(10^7 -10^{10}) M_\odot $, are amongst the primary candidates producing the kind of low-frequency gravitational waves (between the nanohertz to microhertz) aimed to be measured by IPTA. In \cite{Sopuerta} it is argued that the square kilometer array project (SKA) constitutes a significant improvement with respect to current PTAs projects. SKA will detect many more MSPs than what is currently possible and will allow to time them with very high precision, making them very sensitive to the small spacetime perturbations of gravitational waves.

Gravitational waves are emitted by very distant sources, then it is natural to expect a gravitational wave signature on the timing residuals modeled as a plain wave propagating in a approximately flat spacetime. 
Then the expansion of the universe is included as an {\it ad hoc} redshift in the frequency. Nonetheless, the propagation is over a nearly de Sitter universe, not a flat one, and we already know that there is more besides the known redshift.
In \cite{EGR} the propagation over non-flat spacetimes has been derived but from another point of view.
Chapter \ref{ChapterPTA} explicitly shows the wave equation in comoving coordinates. It is obtained from the linearized field equations for small perturbations around a FLRW expanding background.
As expected, the corresponding wave-like solution coincides with the solution obtained in Chapter \ref{ChapterGW} in the Lorenz gauge,  after performing the change of variables to cosmological coordinates. 
Analogously to the previous discussion, the aforementioned wave front suffers an amplitude and a phase modification. 
In Chapter \ref{ChapterPTA} we also show that the phase effect is reflected in an anomalous enhancement of the timing residual registered by pulsars, but only for the ones located at a particular angle subtended between the line observer-pulsar and the line observer-source. 

In summary, if waves are modeled bearing in mind that the propagation takes place in a dynamical universe, the local enhancement of the timing residual for certain pulsars could have an impact in PTA observations \cite{EspriuCC}.
The signal develops a peak-like pattern centered at a polar angle $\alpha$ with respect to the source direction as seen from Earth.
The location of the peak strongly depends on the distance to the gravitational wave source (and on the value of $\Lambda$, or correspondingly to a realistic cosmology, the value of $H_0$). According to the validity of the linearized analysis, the timing residual enhancement can be found at polar angles between $\alpha\sim10^\circ-\,40^\circ$ with respect to the source direction.
Even more, the theory states that the effect has a symmetry of revolution around the azimuthal angle $\beta$ perpendicular to the source direction.
This means that given a fixed source emitting gravitational waves, at a characteristic distance where pulsars of our galaxy are placed, one can find enhancement rings where the timing residual registers the discussed anomalous enhancement. As mentioned in \cite{E2}, where   the differential timing residual along the electromagnetic geodesic is examined, the peak corresponds to a {\it valley} where the phase is (nearly) stationary.

The height of the peak depends considerably on the location of the pulsar emitting the periodic electromagnetic beams; further away pulsars present significantly bigger signals which in some cases are even an order of magnitude bigger than if the computation is performed in a Minkowskian spacetime. 
We took a modest threshold as an example stating that we can discern pulsars with an enhancement magnitude of half the peak. Then any pulsar located inside the region bounded by the corresponding angles for this half peak  (i.e. a region $\alpha\pm\epsilon$) would experience an anomalous timing residual in their ToAs. This interval corresponds to the width of the previously defined enhancement ring, which may be up to tens of kpc.

Supporting the interesting detailed statistical analysis carried out in \cite{EspriuCC}, where a real array of pulsars taken from the ATNF catalog \cite{ATNF} has been used to compute the statistical significance of the intensified timing residual, the last chapter is also focused on the consequences of the characterization on the dimensions of peak-like part of the signal.
The linearized treatment has a range of validity. From our analysis it follows that the effect is presented for sources located between hundred of Mpc to a few Gpc away (which as mentioned before determines the $\alpha$ location). These distances might correspond to single sources SMBH merger observations for PTA and because of the strength of the derived effect, the increased value can maybe facilitate a first detection of gravitational waves by the IPTA collaboration.
If this observable effect is in fact present, we have an interesting method for choosing new pulsars apart from the nearly half a hundred already monitored by IPTA.
Also an hypothetical detection of gravitational waves taking into account these enhanced signals would imply experimental evidence in favour of {\it local} effects of the cosmological constant.


\chapter{Black holes as Bose--Einstein Condensates}
\label{ChapterBHasBEC}




\section{Summary of our current understanding}

In an interesting saga of papers, Dvali, G\'omez and coworkers \cite{DG2,DvaliBHFormation,DvaliQCP,DG1} have put forward an intriguing suggestion: black holes could perhaps be understood as Bose--Einstein condensates of gravitons. This proposal might be scientifically useful despite is based on a brutal approximation. Of course it might fail to capture aspects, but it might work to some extent, especially in the long wavelength regime and the implications can be conceptually staggering.

The condensate model of course cannot be a complete description of what happens in the black hole, but revealed a quite fruitful relation: it establishes a bridge between quantum information and black hole physics never discussed before. This would reveal new approaches on the deep quantum nature of such fascinating objects, and could lead to an alternative understanding of some of the possibly most striking features of black holes. For instance, Hawking radiation \cite{HawkingRadiation, Radiation} could be understood as being due to leakage from the condensate. Besides, this picture brings new ideas about the Bekenstein entropy \cite{Entropy}, the absence of hair \cite{NoHair}, as well as the quantum nature of information storage and the possible information loss in black holes \cite{InformationLoss}.

The main point of these works is that the physics of black holes can be understood in terms of a single number $N$, the number of (off-shell) {\it gravitons} contained in the Bose--Einstein condensate. These condensed {\it gravitons} have a wavelength $\lambda\sim \sqrt N\,\ell_P$, being $\ell_P$ the Planck length. They have a characteristic interaction strength $\alpha_g\sim 1/N$ and the leakage of the condensate leads to a Hawking temperature of order $T_H\sim1/(\sqrt{N}\,\ell_P)$, equal to the inverse of $\lambda$. The mass of the black hole is $M\sim\sqrt N\,M_P$ and its Schwarzschild radius therefore is given by $r_S\sim \sqrt N\,\ell_P$, thus agreeing with the Compton wavelength of the quantum {\it gravitons}, in accordance with the uncertainty principle that dictates $\lambda \simeq r_S$ in the ground state of the quantum system. Therefore, up to various factors of the Planck mass everything is governed by $N$, the number of intervening {\it gravitons}. Modulo some assumptions, all these results stem from the basic relation postulated in \cite{DG2} 
\begin{equation}
	r_g= \frac{N}{M}\end{equation}
that relates the number of gravitons $N$, the mass of the gravitating object $M$ and its gravitational radius $r_g$. For a Schwarzschild black hole $r_g=r_S$. 

We found these results intriguing and we tried to understand them in a different language, with the aim of being more quantitative. In \cite{AEG} the previous relations have been rederived from rather simple assumptions.

It is well known that a Bose–Einstein condensation of a spatially homogeneous gas with attractive interactions is precluded by a conventional phase transition into either a liquid or solid. Even when such a condensate can form, its occupation number is low. Repulsive forces act to stabilize the condensate against collapse, but gravitons do not have repulsive interactions, at least naively, and therefore one would conclude that a Bose--Einstein condensate is impossible to sustain, particularly as we expect $N$ to be very large. However, it is up to the equations themselves to establish whether such a condensate is possible or not. In addition, in theories of emergent gravity, the ultimate nature of gravitons may involve some kind of fermionic degrees of freedom (such as e.g. in the model suggested in \cite{AEP}). Then, repulsion is assured at some scale and fundamental collapse prevented.

The results in \cite{AEG} suggest that condensation is possible. However in our analysis we have separated slightly from original line of thought of Dvali {\it et al}. This approach is not geometric at all. There is no mention of horizon or metric. On the contrary, we adopt a more conservative approach; we assumed the preexistence of a classical gravitational field created by an unspecified source that generates the Schwarzschild metric. As it is known, nothing can classically escape from a black hole, so if we wish to interpret this in potential terms (which of course is not correct but serves us for the purpose of creating a picture of the phenomenon) it would correspond to a confining potential. On and above this classical potential one can envisage a number of quantum fields being trapped. For sure there is a gravity quantum field present; hence {\it gravitons} (longitudinal gravitons that is). It is expected that other quanta may get trapped by the black hole potential as well, along the following sections we will discuss such issue.

Continuing with our semiclassical analogy, these {\it gravitons}, and maybe other quanta, have had a long time to thermalize in a static (and "eternal") Schwarzschild black hole. It is therefore natural to expected that after cooling, a Bose--Einstein condensate of {\it gravitons} could eventually form. Of course these {\it gravitons} are in no way freely propagating transverse gravitons. They are necessary off-shell ($q^2\neq 0$) and thus have some sort of effective mass\footnote{It may help to get a picture of the phenomenon to think of them as quasiparticles.}.

In \cite{AEG} we give a set of equations that could describe a Bose--Einstein condensate constructed on top of a classical field created by a black hole. This set of equations would be identified as the counterpart of the Gross--Pitaevskii equation \cite{GPEquation}. 
Remarkably enough, the characteristics of the resulting condensate are uniquely described in terms of the Schwarzschild radius of the black hole and the value of a dimensionless parameter, interpreted as a chemical potential. 
A condensate appears not only to be possible but actually intimately related to the classical field that sustains it and determines its characteristics. It would therefore be tempting to go one step beyond and reverse the order of the logical implication; and eventually attempt to derive the classical field as a sort of mean field potential {\it\`a la Hartree--Fock}. However this will be left for future research.

\section{Building up a condensate over a Schwarzschild black hole background}\label{SectionBuildingCondensate}

Let us start with the Einstein field equations \eqref{EEq} describing the spacetime curvature. These equations are constructed with the metric tensor $g_{\alpha\beta}$ which in this context provides the geometric and causal structure of the built condensate and the spacetime where it is immersed.
We will denote by $\tilde g_{\alpha\beta}$ to the background metric that in this section will invariably be the Schwarzschild metric. Perturbations above this background metric will be denoted 
by $h_{\alpha\beta}$, so as 
\begin{equation}\label{MetricPerturbedGeneric}
	g_{\alpha\beta}= \tilde g_{\alpha\beta} + h_{\alpha\beta}\ .\end{equation}

To initiate our program we should, first of all, identify an equation (or set of equations) that could provide a suitable description of a Bose--Einstein condensate in the present context. In other words, we have to find the appropriate generalization of the Gross--Pitaevskii equation. The graviton condensate has necessarily to be described by a second-degree, symmetric tensor field $h_{\alpha\beta}$ that within our philosophy has to be connected necessarily with a perturbation of the classical metric. In order to keep things as simple as possible we will attempt to describe only condensates with quantum states having $l=0$. This is translated to spherically symmetric excitations of the metric of the form
\begin{equation}
\label{MetricSchwPert-h}
	ds^2=\Bigl[-\Bigl(1-\frac{r_S}{r}\Bigr)+h_{tt}\Bigr]\ dt^2+\Bigl[\Bigl( 1-\frac{r_S}{r}\Bigr)^{-1}+h_{rr}\Bigr]\ dr^2+r^2\ d\Omega^2\ .\end{equation}
The Einstein tensor derived from the previous metric will be expanded up to second order in $h_{\alpha\beta}$ to retain the leading non-linearities (self-interactions of the desired condensate). 

It is well known that Birkhoff’s theorem \cite{Birkhoff} guarantees the uniqueness of the solution of Einstein's equations in vacuum with the properties of having spherical symmetry and being static. As every dimensionful quantity can be expressed as a function of the Schwarzschild radius $r_S=2G_NM$, the difference between a given Schwarzschild metric and any perturbed solution built upon it must necessarily correspond to a change in the mass $M\rightarrow M+\delta M$. Even though this result is well known \cite{ReggeWheeler}, it is interesting to give a short review in a perturbative formalism. The relevant equations of the classical theory give the geometrical structure of the condensate, hence the corresponding part of the field equations. For the sake of simplicity these equations are included in Appendix \ref{AppxClassical}.

The fact that spherical symmetric perturbations are related to shifts in mass, indicates a correspondence between any perturbation $h_{\alpha\beta}$ (i.e. each {\it graviton}) and a certain amount of energy that is reflected in a change of the black hole mass.

\subsection{Einstein's equations as Gross--Pitaevskii equations}
The familiar Gross--Pitaevskii equation \cite{GPEquation} employed to describe Bose--Einstein condensates is a non-linear Schr\"odinger equation; i.e. an equation of motion that contains self-interactions (hence the non-linearity), a confining potential for the atoms or particles constituting the condensate, and a chemical potential, the conjugate thermodynamic variable of the number of particles or atoms contained in the condensate.

Among all these ingredients, perturbed Einstein's equations already contain most of them. They are already non-linear and while there is no an explicit confining potential (as befits a relativistic theory) they do confine particles, at least classically, because if the selected background corresponds to a Schwarzschild black hole, the strong gravitational field classically traps particles inside the horizon. However, there is one ingredient still missing, namely the equivalent of the chemical potential. Therefore we have to extend the formulation of perturbations around a classical black holes solution to the grand-canonical ensemble by adding to the appropriate action a chemical potential term.

As it is well known since the early days of quantum field theory \cite{PW} there are no conserved currents or continuity equations for fundamental fields that are chargeless (such as a real Klein--Gordon field). Therefore there is no way of defining a number operator for freely propagating gravitons or photons. 

However in the picture of \cite{DG2,DvaliBHFormation,DvaliQCP,DG1} the situation is different. If a Bose--Einstein condensate is present with {\it gravitons} acquiring all the same momenta $\sim 1/r_S$ and being weakly interacting, for macroscopic black holes (recall $\alpha_g\sim 1/N$) the total energy stored in the condensate should be $\sim N/r_S$ and this quantity would be a conserved one. Therefore, lest $r_S$ change, $N$ would be conserved too. The previous reasoning shows very clearly that the {\it gravitons} contemplated in the present scenario (of "maximum packing"), if realized,  have nothing to do with freely propagating gravitons as the number of massless on-shell gravitons is not a conserved quantity.

The energy contained in a given volume occupied by a non-interacting scalar field is 
\begin{equation}
	E= -\frac12 \int dV\ \phi\ \nabla^2\phi\,;\end{equation} 
by analogy, in the present case
\begin{equation}
	E=\frac12\int dV\ \varepsilon^2\ \hat h_{\alpha\beta}\,\hat h^{\alpha\beta}=\int dV\ \varepsilon \ \rho_{\hat h}\, ,\end{equation}
where we assume that the energy {\it per graviton} ($\,\varepsilon\,$) is constant and approximately given by $\varepsilon=1/\lambda$ with $\lambda=r_S$, and
\begin{equation}\label{Density}
	\rho_{\hat h}\equiv \frac12\ \hat h_{\alpha\beta}\ \frac{1}{\lambda}\ \hat h^{\alpha\beta}\ .\end{equation}
While there is no formally conserved current, the above quantity can be interpreted as a "graviton number density", which in principle is related to the perturbation tensor as $\hat h_{\alpha\beta}= M_P\, h_{\alpha\beta}$ to keep the correct dimensions. The Planck mass is used is favour of the universal gravitational constant, $M_P{}^2 =\hbar/G_N$.
The integral of the graviton density \eqref{Density} in the interior of the black hole has to be interpreted as the number of constituents of the condensate.

The above considerations can now be phrased in a Lagrangian language. The chemical potential term in the action should be related to the graviton density of the condensate inside a differential volume element $dV$. In order to respect the basic symmetry of the general theory of relativity, the simplest form of introducing such a term is by means of
\begin{equation}\label{ActionCPTerm}
	\Delta S_{chem. pot.}= -\frac12 \int d^4 x\ \sqrt{-\tilde g}\ \mu \ \hat h_{\alpha\beta}\,\hat h^{\alpha\beta}\ .\end{equation} 
$\tilde g$ stands for the determinant of the background metric. This term does of course resemble a mass term for the spin-2 excitation and indeed it is some sort of effective mass in practice as the {\it gravitons} in the condensate are quasi-non-interacting. However we will see that general relativity eventually requires for the quantity $\mu$ to transform as a scalar and to be position dependent, i.e. $\mu=\mu(r)$,  from requirements of self-consistency of the proposal.

The additional term it is invariant under the gauge group of diffeomorphism. Part of this statement is shown in \cite{Weinberg}: under an infinitesimal displacement in the coordinates of the form $\delta_{\scriptscriptstyle D} x^\alpha=\xi^\alpha$ the full metric changes as the lie derivative along the direction of the displacement, $\delta g_{\alpha\beta}={\cal L}_{\scriptscriptstyle D}g_{\alpha\beta}=\xi_{\alpha;\beta}+\xi_{\beta;\alpha}$. The same gauge transformation rules the background metric. As the perturbation is defined by $h_{\alpha\beta}= g_{\alpha\beta}-\tilde g_{\alpha\beta}$, under the same perturbation of the coordinate system, the same rule of covariant transformation is obtained. Together with the fact that the chemical potential $\mu$ behaves as a scalar under a general coordinate transformation, ensure automatically the general covariance of the theory. However, while the addition is diff invariant, it is {\em not} background independent. The separation $g_{\alpha\beta}=\tilde g_{\alpha\beta} +h_{\alpha\beta}$  leads to an action that depends on the choice of the background metric, which is the one of the chosen black hole. Likewise it is expected for $\mu$ to also depends on the background metric.

Therefore an appropriate action for the field $h_{\alpha\beta}$ is
\begin{equation}\label{ActionGRChemPot}
	S(h)={M_P}^2\int d^4x\ \sqrt{-g}\ R-\tfrac12\ {M_P}^2\int d^4x\ \sqrt{ -\tilde g}\ \mu\ h_{\alpha\beta}\,h^{\alpha\beta} \ .\end{equation}
Indices are raised and lowered using the full metric $g_{\alpha\beta}=\tilde g_{\alpha\beta}+ h_{\alpha\beta}$ in order to preserve diffeomorphism invariance, that is an exact invariance of the above action. A completely covariant expansion in powers of $h_{\alpha\beta}$ can be performed up to the desired order of accuracy.
By construction these equations will be non-linear, but we will keep only the leading non-linearities to maintain the formalism simple and analytically tractable. The left hand side is just the {\it Schr\"odinger part}, the additional piece is proportional to the chemical potential, both giving a Gross--Pitaevskii like equation(s).
It is worth noting again that the action \eqref{ActionGRChemPot} is as shown reparametrization invariant, but it is not background independent and it should not be. The fact that fluctuations take place above a black hole background (in whatever coordinates one chooses to describe it) does matter.

The action principle yields the two equations of motion for the perturbation field $h_{\alpha\beta}={\rm diag}(h_{tt}\,,\,h_{rr}\,,\,0\,,\,0)$; namely
\begin{equation}\label{EEqChemPot}
	G_{\alpha\beta}(\tilde g+h)= \mu\ \sqrt{\tfrac{\tilde g}{g}}\ \bigl(h_{\alpha\beta}-h_{\alpha\sigma}\,h^\sigma{}_\beta\bigr)\ .\end{equation}
that can be interpreted as the Gross--Pitaevskii equations describing the properties of a graviton condensate "sitting" in the black hole interior. The explicit derivation for these equations is included in Appendix \ref{AppxActionVariation}. 
It is important to keep in mind the following: we are working here in the grand canonical ensemble; this implies that the magnitude $\mu$ is an external field and does not vary in the action. In particular, for these equations of motion it is independent of $h_{\alpha\beta}$, so $\delta \mu/\delta h_{\alpha\beta}=0$. This should be the way of introducing the chemical potential. 
Otherwise, were $\mu$ not an external field, it would be necessary to take it into account when performing variations to derive the equations of motion. An equation of motion would be obtained for $\mu$ and this equation would nullify automatically the perturbation as well as the chemical potential itself, i.e. $h_{\alpha\beta}=0$ and $\mu=0$. Therefore, the external field $\mu$ is introduced in the theory as some kind of Lagrange multiplier. As Lagrange multipliers have constraint equations, the chemical potential of the theory may not be arbitrary at all. 
It must satisfy binding conditions with $h_{\alpha\beta}$. The main difference is that these constrains are implicit in a general theory of relativity. The restriction of $\mu$ is through the general covariance conditions for the action. In other words, the diffeomorphism covariance implies the covariant conservation of the Einstein tensor, and this in itself entails the same for the chemical potential term
\begin{equation}\label{DiffEqChemPot}
	\nabla^\beta G_{\alpha\beta}=0 \hspace{40pt}\Longrightarrow\hspace{40pt} \big[\ \mu\ \sqrt{\tfrac{\tilde g}{g}}\ \left(h_{\alpha\beta}-h_{\alpha\sigma}\, h^\sigma{}_\beta\right) \big]{}^{;\beta}=0\ .\end{equation}
The covariant derivative is defined using the full metric $g_{\alpha\beta}=\tilde g_{\alpha\beta} +h_{\alpha\beta}$. This differential equation of motion is valid up to every order in perturbation theory.

Subsequently, we will proceed to find acceptable solutions of these equations and interpret them. We will separate the problem into two regions: outside and inside of the black hole horizon in sections \ref{SectionOutsideHorizon} and \ref{SectionInsideHorizon} respectively. We certainly expect that the graviton condensate will disappear quickly in the outside region; imposing this as a boundary condition for $r\gg r_S$ we will see that in fact the condensate is identically zero on this side of the horizon. On the contrary, a unique non-trivial solution will be found in the interior of the black hole.

\section{Outside the horizon}\label{SectionOutsideHorizon}

In this section it will be shown that even after the inclusion of $\mu$ there is not other normalizable solution outside the black hole horizon than the trivial solution for the perturbation.

With the aim of motivating this solution it is practical to start with a perturbative analysis, retaining only the leading order in $h_{\alpha\beta}$ on the right hand side of \eqref{EEqChemPot} and neglecting possible terms of the form $\mu h^2$ on the grounds of being considered of $O(3)$. This treatment seems natural since the chemical potential (if present --despite of being null in this region--), far away from the condensate, should behave at most as a perturbative small parameter; and {\it small} changes in the chemical potential would be responsible of producing a {\it small} perturbation over the background metric. As a plus, the $O(2)$ system of equation takes a simple form, very similar to the familiar Gross--Pitaevskii equation, 
\begin{equation}\label{EEqChemPotO2}
	G^{\scriptscriptstyle(1)}_{\alpha\beta}\,(h)+ G^{\scriptscriptstyle(2)}_{\alpha\beta}\,(h^2)=\mu\ h_{\alpha\beta}\ .\end{equation}
The zero order corresponds to the vacuum solution and vanish. The left hand side has been already introduced in \eqref{EEqSchwPerturbed}.

\subsection{Asymptotic analysis}
If we place ourselves far away from the black hole, in the $r\to\infty$ limit, one is able to keep only dominant terms in the Einstein's equations. As we want a vanishing perturbation at infinity, the following ansatz is imposed in the faraway region:
\begin{equation}\left\{\hspace{15pt}\begin{split}
	h_{tt}&=A\ r^{\,-a}\\[2ex]h_{rr}&=B\ r^{\,-b}\\[2ex]\mu\ &=C\ r^{\,-c}\end{split}\right.\end{equation}
with $a,\,b,\,c >0$.

Before proceeding, an additional consideration is needed: since for us $h$ represents a localized Bose--Einstein condensate, in analogy with the wave function of a confined particle, the perturbation must be a square-integrable function. This is
\begin{equation}\label{PertSquareIntegrable}
	\int_{r_S}^{\infty}d^3x\ \sqrt{-\tilde g}\ {h}^2\equiv\int_{r_S}^{\infty}d^3x \sqrt{-\tilde g}\,\left(h_{tt}{}^2\,g^{tt\,2}+h_{rr}{}^2\,g^{rr\,2}\right) <\infty\,.\end{equation}
There are at least main two reasons that lead us to the previous requirement. Let us review first why one has to request square-integrability. Note that the solutions of \eqref{EEqChemPot} should not be understood as a quantum field, but rather as solutions of the Gross--Pitaevskii equation. Here we adopt Bogoliubov theory \cite{Bogoliubov} and its interpretation of the Gross--Pitaevskii wave function. This is the commonly accepted interpretation and essentially it boils down to the fact that in the large occupation limit ($N\to\infty$) the creation and annihilation operators of the ground state can be approximately treated as commuting c-numbers, and hence the many body quantum problem is described by a classical function called the "macroscopic wave function" or simply the "order parameter". Because the modulus square of this quantity is proportional to ${a}^\dagger a$ (we introduce the well known annihilation and creation operators $a$ and $a^\dagger$ to define the number operator), it is therefore proportional to $N$. Hence the order parameter itself is proportional to $\propto\sqrt{N/V}$. If the potential is not uniform, $N$ may of course depend on the coordinates. From this the need to require square-integrability follows. This interpretation is in the present case also supported by the dimensionality of $\rho_{\hat h}$ in \eqref{Density}.

Why then this precise condition? The leading role of the geometry has been to provide, in an indirect manner, a "confining potential" that traps {\it gravitons} inside the horizon. Therefore, from this point of view, when giving a physical interpretation to the chemical potential (see section \ref{SectionConnectionDG}) one implicitly assumes that the geometry of the spacetime is flat and that geometry acts via the external potential and the chemical one. In the present situation, it turns out that $\sqrt{-\tilde g} \propto r^2\sin\theta$ coincides with the one corresponding to a 3 dimensional spatial flat metric in spherical coordinates; therefore $d^3x\sqrt{-\tilde g}$ is reparametrization invariant in 3 dimensions. This interpretation is equivalent to defining a wave function normalized by the temporal component of the metric tensor, $\psi=\sqrt[4]{-\tilde g_{tt}}\ h$. 
Then, any magnitude computed by means of an integration over a 3-spatial volume $d^3V= d^3x\,\sqrt{\gamma}$ (where $\gamma$ corresponds to the determinant of the 3-spatial metric $\gamma_{ij}$) is effectively written in a fully covariant way as $d^3V\abs{\psi}^2=d^3x\sqrt{\gamma}\,\sqrt{-\tilde g_{tt}}\ h^2 =d^3x\sqrt{-\tilde g}\,h^2$, since the determinant of the 4-dimensional metric can be decomposed as $\tilde g=\tilde g_{tt}\gamma$, with $i,j=r,\theta,\phi$ if the metric is diagonal. Therefore, the volume element is coordinate dependent as well as the  square modulus of the macroscopic wave function $\psi$, but the combination of both is not; and we end up finally with the corresponding volume element to the previously discussed flat spacetime.

Whatever the case, in this perturbative limit the components of the Schwarzschild metric accomplish $\tilde g_{tt}{}^2,\ \tilde g_{rr}{}^2\longrightarrow 1$ in the limit of $r\rightarrow\infty$; then the integral in equation \eqref{PertSquareIntegrable} should fulfill
\begin{equation}
	4\,\pi\int_{r_S}^{\infty}dr\ r^2\ \left[{h_{tt}}^2+{h_{rr}}^2 +O(h^3)\right] <\int_{r_S}^{\infty} \frac{dr}{r} =\infty \hspace{20pt} \Bigl(\ \begin{gathered}{\small \text{logarithmic}} \\[-0.75ex]{\small\text{divergence}}\end{gathered}\ \Bigr)\ .\end{equation}
For this to happen, if a power law ansatz at infinity is imposed, the exponents must obey $a,\,b>3/2$.

After the inclusion of the chemical potential term, two of the there linearly independent equations coming from \eqref{EEqChemPotO2} becomes nonzero, the temporal and radial due to the new term, while the angular remains unchanged,
\begin{eqnarray}\label{EEqSchwInfinity}\begin{split}
	\left(1-b\right)\,\frac{B}{r^{b+2}}-\left(1-2\,b\right)\,\frac{B^2}{r^{2b+2}}&=\frac{A\,C}{r^{a+c}}&\\[4ex]
	\left(\frac{r_S}{r}+a\right)\frac{A}{r^{a+2}}-\frac{B}{r^{b+2}}+\left(\frac{r_S}{r}+a\right)\frac{A^2}{r^{2a+2}}-\left(\frac{r_S}{r}+a\right)\frac{A\,B}{r^{a+b+2}}+{\frac{B^2}{r^{2b+2}}}&=\frac{B\,C}{r^{b+c}}& \\[4ex]
	\left(\frac{r_S}{r}+a^2\right)\frac{2\,A}{r^{a+2}}+\left(\frac{r_S}{r}-b\right)\frac{2\,B}{r^{b+2}}+\left(\frac{2r_S}{r}+3\,a^2\right)\frac{A^2}{r^{2a+2}}\hspace{63.5pt}& &\\-\left({\frac{r_S}{r}}-b\right)\frac{4\,B^2}{r^{2b+2}}-\left[\left(2+b\right)\frac{r_S}{r}+2\,a^2+a\,b\right]\frac{A\,B}{r^{a+b+2}} &=\ \ 0\ .& \end{split}\end{eqnarray}
The requirement of the solution being square-integrable exclude that $a$ and $b$ could be zero. This implies automatically the neglection of the terms proportional to the Schwarzschild radius. 
Even more, with the previous considerations we can also neglect all terms coming from second order in perturbations, as they vanish faster than the linear ones. It is natural for the Einstein tensor to become Minkowskian after these requirements. After all, \eqref{EEqSchwInfinity} reduce to the following asymptotic identities
\begin{eqnarray}\label{EEqSchwInfinityO1}\begin{split}
	\left(1-b\right)\frac{B}{r^{b+2}}\quad&=&\frac{A\,C}{r^{a+\rho}}& \\[3ex] \frac{A\,a}{r^{a+2}} -\frac{B}{r^{b+2}}\quad&=&\frac{B\,C}{r^{b+c}}&\\[3ex]
	\frac{2\,A\,a^2}{r^{a+2}} -\frac{2\,B\,b}{r^{b+2}}\quad&=& 0\hspace{5pt}&\ \ .\end{split}\end{eqnarray}
The third equation, the angular component $G_\theta{}^\theta$, make explicit how the components of the perturbation behaves between each other;
it is mandatory for both terms to contribute at infinity. If this is not the case, this would nullify $A$ (or $B$), and then, via the two fist equations in \eqref{EEqSchwInfinityO1}, fix $B=0$ (or $A=0$) and $C=0$; i.e. $h_{\alpha\beta}=\mu=0$. 
The competition of both terms is possible if and only if $a=b$. This condition modifies the angular equation in \eqref{EEqSchwInfinityO1} as
\begin{equation}
	\left(A\,a-B\right)\frac{2\,a}{r^{a+2}}=0 \hspace{40pt}\Longrightarrow\hspace{40pt} A\,a=B\ .\end{equation}
Then, from the second equation in \eqref{EEqSchwInfinityO1} the following relationship can be read
\begin{equation}
	\left(A\,a-B\right)\frac{1}{r^{a+2}}\quad=\quad0\quad=\quad\frac{B\,C}{r^{a+c}}\ .\end{equation}
This automatically leads to a null chemical potential as the only possible solution for this region with the proposed ansatz.

\subsubsection*{Faster decays}
We have also explored the possibility of exponentially vanishing chemical potential when $r\to\infty$ with analogous conclusions. 
If we change the ansatz and impose an exponential decreasing solution for the perturbations at infinity,
\begin{eqnarray}\left\{\hspace{15pt}\begin{split}
	&h_{tt}&\hspace{-5pt}=\ A\ e^{a\,r}\\[2ex]&h_{rr}&\hspace{-5pt}=\ B\ e^{b\,r}\\[2ex]&\mu &\hspace{-5pt}=\ C\ e^{c\,r}\end{split}\right.\end{eqnarray}
with $a,\,b,\,c<0$, the linear order dominates in front of the higher ones. As decreasing solutions are expected, the parameters $a$, $b$ and $c$ must be not null. If one use this ansatz in the Einstein's equations partially evaluated at infinity, one finds that in the angular equation there is no way for both leading terms to compete between each other
\begin{equation}
	G_\theta{}^\theta\ :\hspace{15pt} -2\,A\,a^2\,e^{a\,r}-2\,B\,b\,\frac{e^{b\,r}}{r}=0\ .\end{equation}
In conclusion, depending on the fact whether if $a$ is bigger or not than $\beta$, $A=0$ or $B=0$. For the first case, if $A=0$, the leading terms of the temporal equation
\begin{equation}
	G_t{}^t\ :\hspace{15pt} B\,b\,\frac{e^{b\,r}}{r}=A\,C\,e^{\left(a+c\right)\,r}\end{equation}
fixes $B=0$. Being both, $A$ and $B$ null, the chemical potential disappears from the theory in this region. 
For the second case, if $B=0$, the same equation $G_t{}^t$ nullifies $A$ (as no null solutions for $C$ are expected). 
No perturbation makes the chemical potential senseless. To sum up, this ansatz implies as well that the only solution 
at infinity is a null perturbation, $h_{\alpha\beta}=0$, and the disappearance of the chemical potential from the theory, 
$\mu=0$. 

Likewise it can be seen that a much faster Gaussian-like decay
\begin{eqnarray}\left\{\hspace{15pt}\begin{split}
	&h_{tt}&\hspace{-5pt}=\ A\ e^{a\,r^2}\\[2ex]&h_{rr}&\hspace{-5pt}=\ B\ e^{b\,r^2}\\[2ex]&\mu &\hspace{-5pt}=\ C\ e^{c\,r^2}\end{split}\right.\end{eqnarray}
is also excluded with analogous calculations. Null perturbation and null chemical potential is the only solution in this zone.

\subsection{Numerical analysis}

Before concluding this section we will give additionally a numerical argument that confirms the vanishing of the condensate and the chemical potential in the outer region when the perturbations are null at infinity.

The following change of variables allows rewriting the relevant components of the Einstein tensor \eqref{EEqSchwPerturbed} in terms of dimensionless quantities
\begin{eqnarray}\label{ChangeVariablesXz}\left\{\hspace{15pt}\begin{split}
	&z& \hspace{-5pt}=&\quad\frac{r_S}{r}\\[2ex] 
	&X(z)&\hspace{-5pt}=&\quad \mu(r)\, r^2	\ .\end{split}\right.\end{eqnarray}
This quantities are universal as $r_S$ and $\mu$ are the only physical parameters. This redefinition maps the exterior part into a compact interval: $r=\infty$ is transformed into $z=0$ and $r=r_S$ into $z=1$. It is well known that Schwarzschild spacetime is asymptotically flat; therefore, at infinity ($z=0$) it is expected for the perturbation to vanish ($h_{\alpha\beta}=0$). Nevertheless equations are difficult to deal with because they are singular at $z=0$, i.e. it is not possible to isolate the higher order derivatives.

Let us see how we can proceed. As mentioned before, we have to settle for a point  $z\sim0$ (but $z\neq 0$) to set up boundary conditions  for the numerical integration procedure to start. Likewise setting $h_{rr}=0$ at any arbitrary value of $z$ outside the event horizon, immediately triggers divergences at the first step of the routine. Therefore we take a "small" initial value for the perturbation in a point near infinity ($z\sim0$), and then decrease this initial value $\epsilon$; this is
\begin{equation}
	h_{\alpha\beta}\ {\small(r\rightarrow \infty)}\quad\longrightarrow\quad \epsilon\ .\end{equation}
for a decreasing $\epsilon\ll1$.

\begin{figure}[!h]\begin{center}
	\includegraphics[scale=0.4]{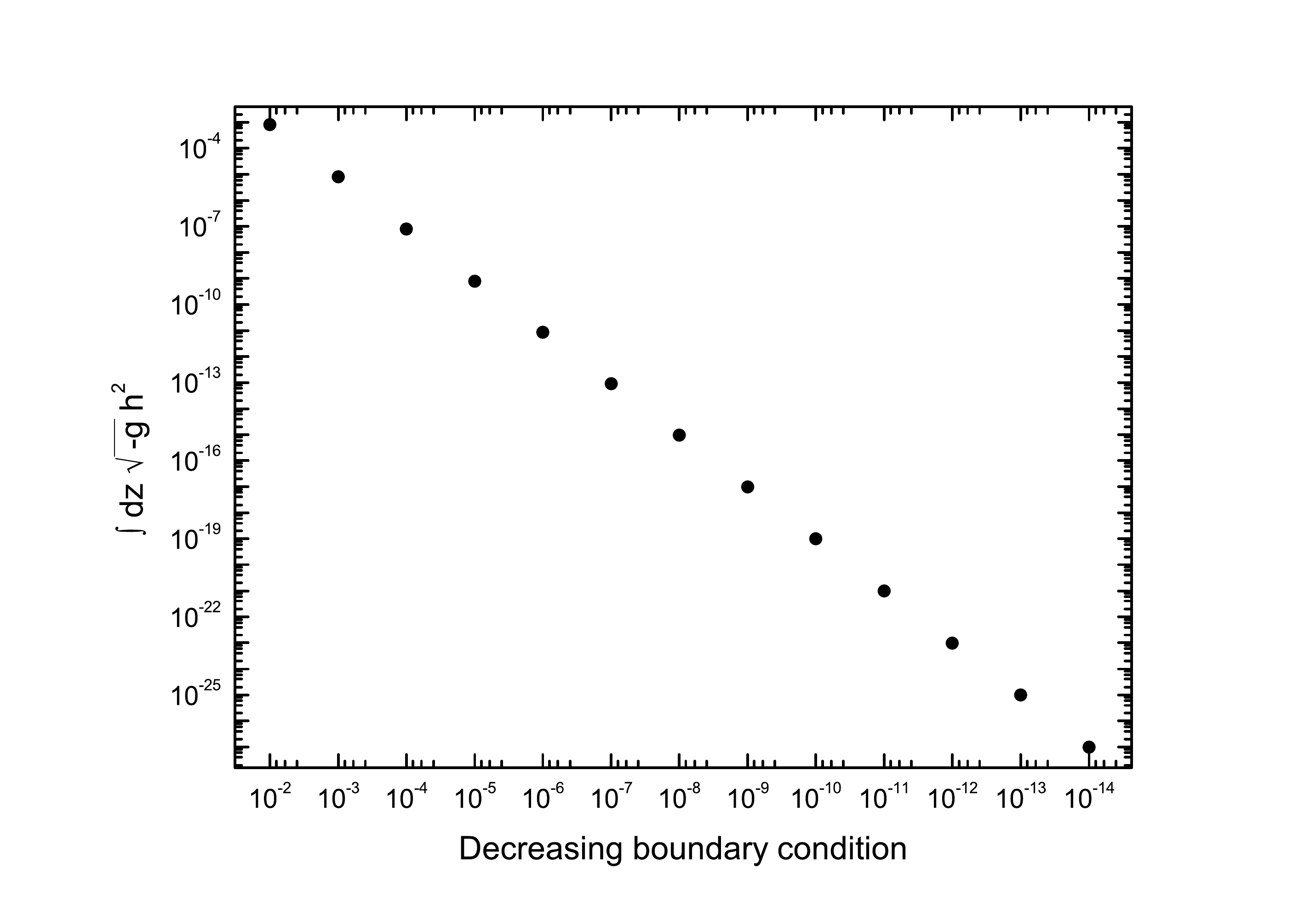}
	\captionsetup{width=0.85\textwidth}
	\caption{\small Each point represents a magnitude proportional to the integral of $h^2$ outside the event horizon, computed for a particular initial condition for the perturbation placed near infinity, $h_{\alpha\beta}(\infty)$. The graph is presented in a log--log scale. A linear fit gives a slope that asymptotically approach $2$ as the boundary condition is nullified; this implies that the integral vanishes quadratically. The closer the limit to an asymptotically flat spacetime is (i.e. decreasing the initial condition near infinity), the smaller this integral becomes.} \label{NOut}\end{center}\end{figure}

With the set of solutions for the components of $h_{\alpha\beta}$ obtained for each of the initial condition imposed, it is possible to compute the quantity $\int dV \, h^2$ defined in equation \eqref{PertSquareIntegrable}. The integral extends over the exterior of the black hole; this is $z\subset(0\,;\,1)$.
In Figure \ref{NOut} the behaviour of the integral of $h^2$ outside the black hole is presented when the initial condition for the integration approaches zero at a fixed point far away from the condensate (near infinity or $z\sim0$). It is observed that this integral is invariably small. The numerical analysis results very stable and the initial condition can be reduced as many orders of magnitude as desired. The graph is presented in a log--log scale, hence the linear behaviour. The points are fitted by a linear function with a slope approaching $2$ as more points with lower initial condition are added. This implies that the integral converges quadratically to zero as the initial condition for $h_{\alpha\beta}$ nullifies. This seems to confirm ---also numerically--- that there is no condensate, i.e. $h_{\alpha\beta}=0$, and no dimensionless chemical potential, $X=0$, in the outer region of the black hole.

\section{Inside the horizon}\label{SectionInsideHorizon}

In order to study the behaviour of our equations when the horizon is crossed, it is convenient to keep working in the $z=r_S/r$ and $X=\mu\,r^2$ variables introduced in \eqref{ChangeVariablesXz} from the previous section and redefine the perturbations as
\begin{eqnarray}\label{PertRedefinition}\left\{\hspace{15pt}\begin{split}
	&h_{tt}(r)&=&\quad\left(1-z\right)\gamma_{tt}(z)\\[2ex]&h_{rr}(r)&=&\quad\left(1-z\right)\gamma_{rr}(z)\ .\end{split}\right.\end{eqnarray}
With the new perturbations the Einstein's equations \eqref{EEqSchwPerturbed} become more compact and can be dimensionalized. The three relevant components of the nondimensional tensor $ r^2G_{\alpha}{}^\beta$ read as
\begin{equation}\label{EEqGO2}\begin{split}
	-(2z+1)(z-1)^2\gamma_{rr} -z(z-1)^3{\gamma_{rr}}' +(4z+1)(z-1)^4{\gamma_{rr}}^2&\\ +2z(z-1)^5\gamma_{rr} \,{\gamma_{rr}}' &\\[4ex] 
	-(z-1)^2\gamma_{rr} -z(z-1){\gamma_{tt}}' -z(z-1)\gamma_{tt}\,{\gamma_{tt}}' +z(z-1)^3\gamma_{rr}\,{\gamma_{tt}}'&\\ +(z-1)^4{\gamma_{rr}}^2& \\[4ex]
	-\big[2z(z-2)(z-1)\gamma_{rr}+z(z-2)(z-1)^2{\gamma_{rr}}'-z(5z-2){\gamma_{tt}}'&\\-2z^2(z-1){\gamma_{tt}}''-4z(z-2)(z-1)^3\,\gamma_{rr}{}^2-z^2(z-1){{\gamma_{tt}}'}^2&\\+z(7z-2)(z-1)^2\gamma_{rr}{\gamma_{tt}}'+z^2(z-1)^3{\gamma_{rr}}'\,{\gamma_{tt}}'&
	\\-2z(z-2)(z-1)^4\gamma_{rr}{\gamma_{rr}}'-z(5z-2)\gamma_{tt}\,{\gamma_{tt}}'&\\-2z^2(z-1)\gamma_{tt}\,{\gamma_{tt}}''+2z^2(z-1)^3\gamma_{rr}\,{\gamma_{tt}}''\big]/4&\ .\end{split}\end{equation}
Of course with the change of variables, now the prime stand for a derivative with respect to the dimensionless inverse radial coordinate $z$.

\subsection{Linearization near the event horizon}
In order to get a feeling for possible solutions to these equations we consider their linearized approximation. Only terms linear in $\gamma_{tt}$, $\gamma_{rr}$ and their derivatives are kept in the region $z\to 1$. Beforehand no restriction for the dimensionless chemical potential $X$ is imposed, therefore $X\gamma_{\alpha\beta}$ is a priori considered as a linear contribution in the perturbation. That is
\begin{equation}\begin{split}
	3(z-1)^2\gamma_{rr}+(z-1)^3{\gamma_{rr}}'=X\ \gamma_{tt}& \\[3ex] (z-1)^2\gamma_{rr}+(z-1){\gamma_{tt}}'= -X\ (z-1)^2\gamma_{rr} &\\[3ex] 2(z-1)\gamma_{rr}+(z-1)^2{\gamma_{rr}}'+3{\gamma_{tt}}'+2(z-1){\gamma_{tt}}''=0&\ .\end{split}\end{equation}
Let us now make the following ansatz for the new perturbations
\begin{eqnarray}\left\{\hspace{15pt}\begin{split}
	&\gamma_{tt}&\sim&\quad A (z-1)^a \\[2ex] &\gamma_{rr}&\sim&\quad B (z-1)^b\end{split}\right.\end{eqnarray}
while we leave the chemical potential $X=X(z)$ as a free function. The three equations take respectively the following form
\begin{equation}\begin{split}
	B(3+b)(z-1)^{b+2}=X\ A(z-1)^a& \\[3ex] B(z-1)^{b+2}+Aa(z-1)^{a}=-X\ B(z-1)^{b+2} &\\[3ex] B(2+b)(z-1)^{b+1}+Aa(1+2a)(z-1)^{a-1}=0&\ .\end{split}\end{equation}
It is worth noting that if $a=b+2$, all terms contribute as we get closer to the event horizon and $X$ behaves as a constant. In this situation, we obtain three equations for the coefficients 
\begin{equation}\begin{split}
	B(1+a)=X\ A&\\[3ex] B+a A= - X\ B& \\[3ex] a[B+A(1+2a)]=0&\ .\end{split}\end{equation}
The system of equations is algebraic for the variable $X$ (that as we have seen should behave as a constant as $z\to 1$); therefore, it is possible to eliminate $X$ by combining the temporal and radial equation. In the present ansatz this leads to
\begin{equation}
	B^2(1+a)+AB+A^2a=0\ .\end{equation}
Together with the angular equation, this determines the solutions up to a single constant. There are two possible solutions for this system
\begin{itemize}
\item[(i)]	\hfill \makebox[0pt][r]{%
	\begin{minipage}[b]{\textwidth}\begin{eqnarray}\label{PertO2}\quad\begin{split}
	&a\ =&\hspace{-7pt}0&\\[2ex]&b\ =-&\hspace{-7pt}2&\end{split}
	\hspace{40pt}A\ =-B\hspace{30pt}\Longrightarrow \hspace{30pt}\left\{\quad\begin{split}
	&\gamma_{tt}&=&\quad A \\[1ex] &\gamma_{rr}&=&-\frac{A}{(z-1)^2}\end{split}\right.\end{eqnarray}\end{minipage}}

\item[(ii)]	\hfill \makebox[0pt][r]{%
	\begin{minipage}[b]{\textwidth}\begin{eqnarray}
	\qquad\begin{split}
	&a\ =-&\hspace{-7pt}1&  \\[2ex] &b\ =-&\hspace{-7pt}3& \end{split} \hspace{40pt} A\ =\quad B  \hspace{30pt} \Longrightarrow\hspace{30pt} \left\{\quad\begin{split}
	&\gamma_{tt}&=&\quad \frac{A}{(z-1)} \\[1ex] &\gamma_{rr}&=&\quad \frac{A}{(z-1)^3}\ \ .\end{split}\right.\end{eqnarray} \end{minipage}}\end{itemize}
In any case, at least one of the perturbations is divergent over the event horizon. Nonetheless, the first solution 
appears to be integrable, while this is not the case for the second one. In the case of (i), we get $X=-1$; while if (ii) is taken as solution, $X=0$ (i.e. no chemical potential at all)\footnote{Solution ii) represents however a volume-preserving fluctuation at the linear order.}.

However, because these solutions do not vanish when $z\to 1$, justified doubts can be casted on the relevance of the 
linearized equations. Let us examine this point taking into consideration only the solution (i), which is integrable according to the considerations of the previous section.

In order to see if this solution is modified when non-linearities are switched on, we substitute it back in the $O(2)$ equations system where self-interactions matters, and see if the solution survives or how would it get modified. For a compete $O(2)$ picture, we need to expand the right hand side of the field equations \eqref{EEqChemPot}. Therefore, the Einstein's equations corresponding to the non trivial components of the chemical potential tensor
\begin{equation}\label{EEqChemPotO2Full}\begin{split}
	r^2G_t{}^t(\tilde g+h)&= X\ \tilde g^{tt}\sqrt{-\tilde g^{tt}\, \tilde g^{rr}}\left[1-\tfrac52\ \tilde g^{tt}\ h_{tt}-\tfrac12\ \tilde g^{rr}\ h_{rr}\right]h_{tt}\\[4ex]
	r^2G_r{}^r(\tilde g+h)&= X\ \tilde g^{rr}\sqrt{-\tilde g^{rr}\tilde g^{tt}}\left[1- \tfrac{1}{2}\ \tilde g^{tt}\ h_{tt}-\tfrac{5}{2}\ \tilde g^{rr}\ h_{rr}\right] h_{rr}\ .\end{split}\end{equation}
In view of the foregoing, near $z\sim 1$, the left hand side, given by \eqref{EEqGO2}, together with the right hand side, changed to the $\gamma_{\alpha\beta}$ perturbation variable defined on \eqref{PertRedefinition}, give the temporal, radial and angular equations
\begin{equation}\begin{split}
	-3(z-1)^2\gamma_{rr} -(z-1)^3{\gamma_{rr}}' +5(z-1)^4{\gamma_{rr}}^2 +2(z-1)^5\gamma_{rr}\,{\gamma_{rr}}' \hspace{61pt}&\\ =-X\gamma_{tt}\left[1+\tfrac{5}{2}\ \gamma_{tt} -\tfrac{1}{2}\left(z-1\right)^2\gamma_{rr}\right]&\\[4ex] 
	-(z-1)^2\,\gamma_{rr}-(z-1){\gamma_{tt}}'-(z-1)\gamma_{tt}\,{\gamma_{tt}}'+(z-1)^3\gamma_{rr}\,{\gamma_{tt}}'\hspace{74pt}&\\+(z-1)^4{\gamma_{rr}}^2=X\left(z-1\right)^2\gamma_{rr}\left[1+\tfrac{1}{2}\gamma_{tt}-\tfrac{5}{2}(z-1)^2\gamma_{rr}\right]\\[4ex]
	\big[2(z-1)\gamma_{rr} +(z-1)^2{\gamma_{rr}}' +3{\gamma_{tt}}' +2(z-1){\gamma_{tt}}'' -4(z-1)^3{\gamma_{rr}}^2\hspace{51pt}&\\ -2\,(z-1)^4\gamma_{rr}\,{\gamma_{rr}}'+3\gamma_{tt}\,{\gamma_{tt}}'-5(z-1)^2\gamma_{rr}\,{\gamma_{tt}}'-(z-1)^3{\gamma_{rr}}'\,{\gamma_{tt}}'\hspace{41pt}&\\+(z-1){{\gamma_{tt}}'}^2+2(z-1)\gamma_{tt}\,{\gamma_{tt}}''-2(z-1)^3\gamma_{rr}\,{\gamma_{tt}}''\big]/4=0&\ ,\end{split}\end{equation}
respectively. Replacing the possible solution, $\gamma_{tt}=A$ --this eliminates any derivative of $\gamma_{tt}$--
and  $\gamma_{rr}=-A/(z-1)^2$ into these equations, we obtain the following relations
\begin{equation}\begin{split}
	3\,A-2\,A+5\,A^2-4\,A^2=(1+A)A=-X\ A\left(1+3\ A\right)&\\[3ex]
	A+A^2=(1+A)A=-X\ A\left(1+3A\right) &\\[3ex]
	-\frac{2\,A}{(z-1)} +\frac{2\,A}{(z-1)} -\frac{4\,A^2}{(z-1)} +\frac{4\,A^2}{(z-1)} =0&\ .\end{split}\end{equation}
Therefore, quite surprisingly, the linear solution is still an exact solution of the non-linear quadratic differential equations. Thus, we conclude that \eqref{PertO2} is solution of the second order system of equations. 

In addition, this exercise gives an interesting result:
\begin{equation}\label{ChemPotO2}
	X\simeq -\frac{1+A}{1+3A}\sim-1+2A+\dots \ ,\end{equation}
where $A$ is so far arbitrary, also in sign. Note that at linear level we got $X=-1$ and the fact that the quadratic equation gives an $O(A)$ correction to this result is consistent as we are implicitly assuming that $|h_{\alpha\beta}|\ll |\tilde g_{\alpha\beta}|$, i.e. $|A|\ll 1$. The constant $A$ itself is arbitrary and is not determined by the structure of the equations. Changes in $A$ appear as an overall factor in the solution.

\subsection{Numerical analysis}
So far we have seen that the solution i) given in (\ref{PertO2}) satisfies not only the linearized approximation but also the full quadratic equations. Could it happen that the previous analysis misses some solution? To try to answer this question, we have performed a numerical integration of the basic equations retaining the leading order on the right hand side of the equations, i.e. the set of equations \eqref{EEqChemPotO2} which are much more easier to approach and would give a hint of the collective modes of the trapped graviton condensate. Although the structure of these field equations resembles the Gross--Pitaevskii equation, where the chemical potential term enters together with a linear factor in the perturbation, we are really interested in solving the equations given by \eqref{EEqChemPot}; as they can be derived from an action principle.

Of course no matter at which order the truncation is taken, the dimensionless character of the main equations remains; it is useful to introduce a nondimensional radial variable $r/r_S$. The integration is performed in a compact spacetime domain $[\epsilon\,;\,1-\epsilon]\subset(0\,;\,1)$.
\footnote{It is impossible to start integrating from either 0 (the black hole center) or 1 (its event horizon) exactly because as can be seen from \eqref{EEqSchwPerturbed} the system of equations is singular at $r=0$ and $r=r_S$; i.e. one cannot determine in either point the value of the higher order derivatives because precisely at the origin and at the horizon they are multiplied by factors that vanish or diverge.}
It is consistent to use boundary conditions according to the asymptotic solution derived in (\ref{PertO2}). Parametrized as 
\begin{eqnarray}\begin{split}
	&\gamma_{tt}(1-\epsilon)\ =\hspace{10pt}A\\[2ex]&\gamma_{rr}(1-\epsilon)\ =-A\ \frac{(1-\epsilon)^2}{\epsilon^2}\\[2ex]&\gamma_{tt}{}'(1-\epsilon) \ =\hspace{10pt} 0\end{split}\end{eqnarray}
it can be seen that the constant $A$ itself is arbitrary and is not determined by the equations. Besides, changes in $A$ appear as an overall factor in the solution. 

\begin{figure}[!b]\begin{center}
	\includegraphics[scale=0.85]{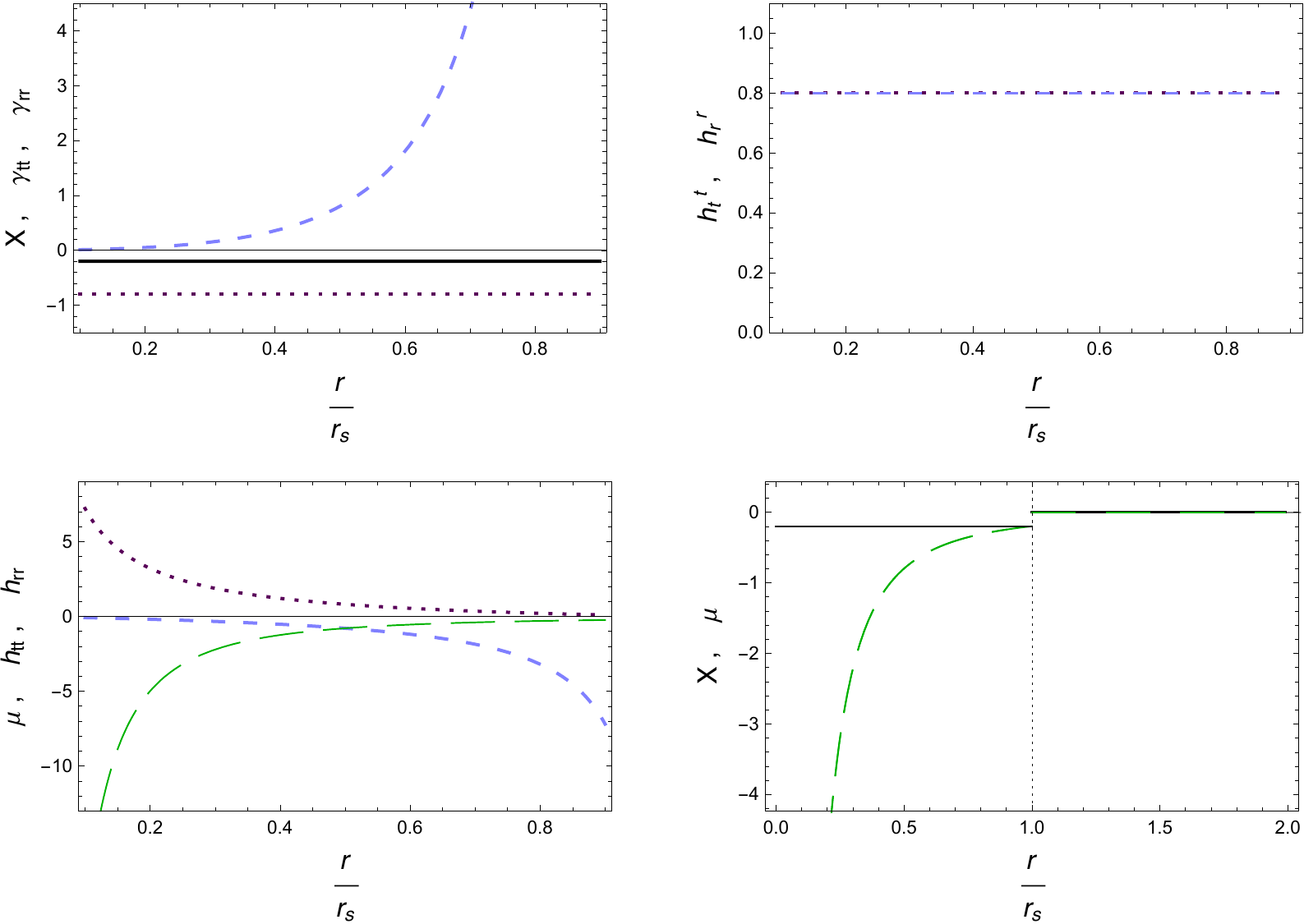}
	\captionsetup{width=0.85\textwidth}
	\caption{\small In this plot $A=-0.8$ has been taken; this implies $\gamma_{tt}=-0.8$ and $X=-0.2$ inside the black hole. On the first plot, $\gamma_{tt}$ (dotted) and $\gamma_{rr}$ (dashed) are represented together with the dimensionless chemical potential $X$ (solid). On the second plot, the numerical integration shows that the solution of the $O(2)$ equations has constant and equal components $h_t{}^t$ and $h_r{}^r$ (same markers are used for each component). On the third graphic, the components of the perturbation $h_{tt}$ (dotted), $h_{rr}$ (dashed) and the dimensionful chemical potential $\mu$ (long dashed) are plotted. Finally, the fourth graphic gives the behaviour of the chemical potential, both $\mu$ and $X$. Across the black hole horizon, this variable experience a jump that in this case is $\Delta X=\Delta\mu\simeq0.2$. As already emphasized, a whole family of solutions are obtained, parametrized by a real arbitrary constant $A$. One can simply trade this constant for the value of $\varphi$.}
	\label{FigSolutionO2}\end{center}\end{figure}

For certain fixed values of $A$, varying the {\it small} value of $\epsilon$, numerical integrations give curves with common structures as the one presented in Figure \ref{FigSolutionO2}. The conclusion from the analytical study is reinforced after performing this numerical analysis of the basic $O(2)$ equations. We found only one solution with the same general features; in the first plot of Figure \ref{FigSolutionO2} it is seen that $\gamma_{tt}$ as well as the function related to the chemical potential $X$, turns out to be constant in the interior of the black hole. No other solutions are found.

Along this section, we have seen how the equations simplifies in such a way that they look simpler if expressed in terms of the functions $\gamma_{\alpha\beta}$. However this is not the case for the solution; to understand what the solution \eqref{PertO2} means is better to undo the redefinition of the components of the wave function (i.e. $\gamma_{\alpha\beta}\rightarrow h_{\alpha\beta}$) by means of \eqref{PertRedefinition}. It is of interest to raise one index of the components of the perturbation with the inverse full metric. In the second plot of Figure \ref{FigSolutionO2} the results of the numerical integration for $h_t{}^t$ and $h_r{}^r$ are shown; as can be seen, the obtained functions are revealed to be constant and equal throughout the interior of the black hole 
\begin{equation}\label{PertCoContraConstant}
	h_t{}^t = h_r{}^r =\ {\rm constant}\ .\end{equation}

The solution for the initially defined variables, the perturbation $h_{\alpha\beta}$, and the chemical potential $\mu$ are presented in the third plot. The components of the perturbation turns out to be proportional to the corresponding metric element where each of the two belongs; i.e. $h_{tt}=h_t{}^t g_{tt}$ and $h_{rr}=h_r{}^r  g_{rr}$ with constant factors \eqref{PertCoContraConstant}. The angular degrees of freedom remain unchanged, $h_\theta{}^\theta=h_\phi{}^\phi= 0$. The chemical potential is negative and presents a jump at the horizon; this is shown in the fourth plot of Figure \ref{FigSolutionO2}. Before drawing conclusions from the solution let us move to the complete equations; we will see that it is possible to derive an analytic solution from the system of equations that coincides with our previous asymptotic and numerical perturbative analysis.

\subsection{Exact solution}

In turn to study the full equation of motion we are interested in, equation \eqref{EEqChemPot}, let us recover the exact chemical potential term for the right hand side of Einstein's equations
\begin{equation}\label{ChemPotTerm}
	X_{\alpha\beta}=\mu\ \sqrt{\tfrac{\tilde g}{g}}\ \bigl(h_{\alpha\beta}-h_{\alpha\sigma}h^\sigma{}_\beta\bigr)\ .\end{equation}
We will abandon the perturbative analysis and show an analytic solution for the field equations.
Inspired by the previous analysis, both perturbative and numerical of the main equations derived at $O(2)$, we reformulate our picture starting from a constant metric fluctuation with mixed indices (one covariant, one contravariant).
Whereas the numerical integration have proved to be extremely precise in giving the relation \eqref{PertCoContraConstant} between components, the solution presents instabilities at some {\it very close} scale. Consequently let us write first
\begin{equation}
	h_t{}^t=\varphi_t\ ; \hspace{40pt} h_r{}^r=\varphi_r\ ; \hspace{40pt} h_\theta{}^\theta=h_\phi{}^\phi=0 \end{equation}
where $\varphi_t$ and $\varphi_r$ are $r-$ independent. With such condition, the perturbation takes the following form
\begin{equation}\label{PertCoContra}
	h_{\alpha\beta}=\text{diag}\left(\frac{\varphi_t}{1-\varphi_t}\ \tilde g_{tt}\,;\,\frac{\varphi_r}{1-\varphi_r}\ \tilde g_{rr}\,;\,0\,;\,0\right)\ .\end{equation}
This implies for the total metric to become
\begin{equation}\label{MetricCondensate}
	g_{\mu\nu}={\rm diag}\Bigl(\,\frac{1}{1-\varphi_t}\ \tilde g_{tt}\,,\,\frac{1}{1-\varphi_r}\ \tilde g_{rr}\,,\, \tilde g_{\theta\theta}\,,\,\tilde g_{\phi\phi}\,\Bigr)\end{equation}
and seems to impose an upper limit for the constant values of $\varphi_t$ and $\varphi_r$  by the requirement of a regular metric along the whole interior\footnote{The upper bound for the wave function will become clear when computing the number $N$ of constituents of the condensate; such value for the wave function corresponds to the limit $N\rightarrow\infty$.}.  

The exact Einstein's equations of motion for the theory simplifies enormously with respect to the perturbative cases, and reduces to the following two
\begin{equation}\begin{aligned}\label{EEqhCte2}
	&-\frac{\varphi_r}{r^2} =\mu\ \varphi_t \left(1-\varphi_t\right)^{3/2}(1-\varphi_r)^{1/2}\\[3ex]
	&-\frac{\varphi_r}{r^2} =\mu\ \varphi_r\left(1-\varphi_t\right)^{1/2}(1-\varphi_r)^{3/2}\ .\end{aligned}\end{equation}
Of course if we expand these two equations up to $O(2)$ in the components of the perturbations we recover the equations \eqref{EEqChemPotO2Full} when changing to the picture with mixed indices, i.e. for $h_t{}^t=\varphi_t$ and $h_r{}^r=\varphi_r$.
These equations are valid also for the external solution; however, in the case of the external region, we know that Minkowskian metric must be recovered far away from the sources. This condition implies that $\varphi_t=\varphi_r=0$ and the same for the chemical potential $\mu=0$, as explained in section \ref{SectionOutsideHorizon}. 

Among the two possible solutions of the latter algebraical system, only one is compatible with an acceptable limit for small perturbations, namely $\varphi_t=\varphi_r=\varphi$ (we expect $0<\varphi<1$). All things considered, the two equations in \eqref{EEqhCte2} are equal, therefore the resulting equation of motion for a constant $\varphi$ becomes
\begin{equation}\begin{aligned}\label{EEqhCte}
	&-\frac{\varphi}{r^2} =\mu\ \left(1-\varphi\right)^2\varphi\ .\end{aligned}\end{equation}
Within our philosophy (\ref{EEqChemPot}) is understood as a Gross--Pitaevskii equation for a condensate wave function $\varphi$. Because the solution is constant within the chosen formulation, the {\it kinetic} term drops and one is left with a purely algebraic, mean-field-like equation\footnote{Gross--Pitaevskii equations containing no derivative terms are widely known. For instance when describing a uniform gas of interacting atoms, the corresponding equation is $$g\ \varphi^2=\mu\ ,$$ where $g$ is the (repulsive) coupling constant, $\varphi$ the order parameter for the Bose--Einstein condensate, and $\mu$, needless to say, is the chemical potential.}. Added to this, it can be derived from an action principle.
This unique non-linear equation produces a unique branch of solutions that relate the value of the condensate with the chemical potential directly from \eqref{EEqhCte}, and is described by 
\begin{equation}\label{ChemPotSolution}
	\mu= \frac{\mu_0}{r^2}\ ,\hspace{30pt}\text{with} \hspace{ 30pt} \mu_0= -\frac1{(1-\varphi)^2}\ .\end{equation}
For completeness, let us mention that the picture of a Gross--Pitaevskii like equation \eqref{EEqhCte} providing a dimensionless constant $\mu_0$ given by \eqref{ChemPotSolution} should be consistent with the analytical and perturbative analysis performed to the $O(2)$ system. And of course it is so; we are retrieving the $O(2)$ solution previously found, with $\varphi=-A(1+A)$ playing the role of the free constant discussed along this section. This relation between $\varphi$ and $A$ allows to write $\mu_0\sim-1+2A+\dots$ which coincides with \eqref{ChemPotO2}.

Before moving on, it is mandatory to make a comment on the covariant conservation of our equations. At the end of section \ref{SectionBuildingCondensate} we have pointed out that the diffeomorphism invariance entails a differential equation for the chemical potential, namely (\ref{DiffEqChemPot}). Bianchi identities entails the covariant conservation of the Einstein tensor, and this implies automatically the same for the left hand side of our equations of motion
\begin{equation}
	\Bigl(\ \mu\,\tfrac{\sqrt{-\tilde g}}{\sqrt{-g}}\ \Bigr){}^{,\,\nu}\ \bigl(h_{\mu\nu}-h_{\mu\sigma}h^\sigma{}_\nu\bigr)+\mu\ \tfrac{\sqrt{-\tilde g}}{\sqrt{-g}}\ \bigl(h_{\mu\nu}- h_{\mu\sigma}h^\sigma{}_\nu\bigr){}^{;\,\nu}=0\ .\end{equation}
The latter equation is a set of four equations for $\alpha=t,\, r,\, \theta,\, \phi$, but only one is non trivial; this equation is the radial one, $\alpha=r$. When the components of the perturbation are equal and constant, i.e. $h_t{}^t=h_r{}^r=\varphi$, the general covariance condition yields the following differential equation for the chemical potential
\begin{equation}
	\Bigl(\varphi-\frac{\varphi^2}{2}\Bigr)\ \Bigl(\partial_r\mu+\frac{2\mu}{r}\Bigr)=0\ .\end{equation}
The integration is direct, $\mu=\mu_0\,r^{-2}$, and the only degree of freedom is a boundary condition when integrating this differential equation governing the chemical potential. This value should coincide with the value of our constant dimensionless chemical potential in \eqref{ChemPotSolution}; this is $\mu_0=-1/(1-\varphi)^2$.

The qualitative result is the existence of a normalizable solution that can be interpreted as the collective wave function of a graviton condensate. A unique relation is obtained between this (constant) wave function and a (also constant) dimensionless chemical potential.
It is to be expected that other quanta may form condensates too, being described by Gross--Pitaevskii like equations similar in spirit to the one considered for {\it gravitons}. This would lead to the introduction of chemical potentials conjugate to their respective number of particles. However it is not clear {\it a priori} which equations they obey; in particular which potential energy one should use in these cases. We will advocate here to use a potential derived from the quasilocal energy and investigate some possible consequences of this in next chapter. However let us first discuss in which manner our research connects with other proposals on corpuscular black holes.

\section{Consequences of the results and connection with previous proposals.}\label{SectionConnectionDG}

In order to present a physical interpretation for the obtained results, let us compute the number of particles (bosons) present in the condensate. It is immediate to see that the solution found for the wave function is of finite norm. Taking into account that the perturbation $h_{\alpha\beta}$ is null from the event's horizon onwards (we have seen that the condensate is confined behind the event horizon), the endpoint on the integration limit can be fixed at the Schwarzschild radius. This way, the number of particles is related to the following integral
\begin{equation}\label{Integralh2}
	\int_0^{\infty} d^3x \sqrt{-\tilde g}\ h_\alpha{}^\beta{h_\beta}^\alpha =4\pi\int_0^{r_S} dr\ r^2\ \bigl(h_t{}^{t\,2} + h_r{}^{r\,2}\bigr)= \frac{8\pi}{3}\ {r_S}^3\ \varphi^2\end{equation}
which states that the integral of the square modulus of the wave function has a constant value. Because of the definition of the chemical potential term, the volume element for the background metric $d^3x\sqrt{-\tilde g}$ has been used. 

The fact that this magnitude is constant automatically ensures a constant behaviour for the probability density of the wave function defined in \eqref{Density}, as ${h_\alpha}^\beta{h_\beta}^\alpha =h_{\alpha\beta}h^{\alpha\beta}$. Then, we are able to compute the total number of gravitons of the condensate as
\begin{equation}\label{NumberOfParticles}
	N=\int d^3x\ \sqrt{\tilde g} \ \rho_{\hat h} \end{equation}
where the density has been introduced in equation \eqref{Density}. Assuming the "maximum packaging" condition $\lambda=r_S$ explained in \cite{DG2} and retrieving the missing constants, we can relate the quantity in (\ref{Integralh2}) to the integral of the density $\rho_{\hat h}$. Then, with the arguments of section \ref{SectionBuildingCondensate}, the number of particles \eqref{NumberOfParticles} is expressed as
\begin{equation}\label{mistery}
	N=\frac{4\pi}{3}\ {M_P}^2\ \varphi^2\ {r_S}^2 \hspace{40pt}\Longrightarrow\hspace{40pt} r_S= \sqrt{\frac{3}{4\pi\varphi^2}}\  \sqrt{N}\ \ell_P \ .\end{equation}
These relations agrees nicely with the proposal of Dvali and G\'omez. The rest of relations of their work can basically be derived from this.

Possibly our more striking results are that the dimensionless chemical potential $\mu_0= \mu(r) r^2$ stays constant and non-zero throughout the interior of the black hole, and that so does the quantity $h_\alpha{}^\alpha= \varphi$ previously defined and entirely determined by the value of $\mu_0$, and viceversa. Therefore, it is totally natural to interpret $\mu_0$ as the variable conjugate to $N$, the number of gravitons.

Of course the solution $\varphi=0$ is also possible. If $\mu_0=0$ the only solution is $\varphi=0$, $N=0$ reproducing the familiar Schwarzschild metric in the black hole interior. However, the interesting point is that solutions with $\varphi\neq 0$ exist and the value of $\mu_0$ is uniquely given by \eqref{ChemPotSolution}. Assuming continuity of the solution the limit $\varphi\to 0$ would correspond to $\mu_0=-1$. This dimensionless number would then be a unique property of a Schwarzschild black hole.

As seen above the dimensionless chemical potential has a rather peculiar behaviour. As $\mu_0=\mu\,r^2$ is a constant function, then $\mu\propto 1/r^2$ it is not null inside and over the event horizon. Outside it appears to be exactly zero. Let us now for a moment forget about the geometrical interpretation of black hole physics and let us treat the problem as a collective many body phenomenon. It is clear why {\it gravitons} are trapped behind the horizon: the jump of the chemical potential at $r=r_S$ would prevent the {\it particles} inside from reaching infinity. From this point of view it is quite natural to have a lower chemical potential inside the horizon than outside (where is obviously zero) as otherwise the configuration would be thermodynamically unstable. In the present solution particles ({\it gravitons} in our case) cannot escape.

However this is not completely true as the picture itself suggests that one of the modes can escape at a time without paying any energy penalty if the maximum packaging condition is verified. Let us do a semiclassical calculation inspired by this picture; using $M\sim M_P\sqrt N$:
\begin{equation}
	\frac{dM}{dt}\simeq \frac{M_P}{2\sqrt N}\ \frac{dN}{dt}= \frac{1}{2r_S}\frac{dN}{dt}\,.\end{equation}
To estimate $dN/dt$ (which is negative) we can use geometrical arguments to determine the flux. If we assume that for a given value of $r_S$ only one mode can get out (as hypothesized above) and that propagation takes place at the speed of light, elementary considerations\footnote{To determine the rate of variation of $N$ we have to multiply the flux times the surface $4\pi r_S^2$,
$$\frac{dN}{dt}=\Phi_{\scriptscriptstyle G}\ A= -\frac3{r_S}\ .$$
The flux is negative and can be computed as the density of the mode times the velocity, where we have assumed that $c=1$ in our units. Since the density of the mode is constant in the interior, it is just $3/4\pi r_S^3$.} lead to
\begin{equation}
	\frac{dM}{dt}\simeq -\frac{3}{2}\frac{1}{r_S{}^2}\,.\end{equation}
This agrees with the results of \cite{DG2} --for instance Eq. (35)-- and yields $T\simeq 1/r_S$. Within this picture, several questions concerning the long-standing issue of loss of information may arise as the outcoming state looks thermal \cite{HawkingBreakPredictability} but apparently is not; or at least not totally so. However we shall refrain of dwelling on this any further at this point.

The main result from the previous rather detailed analysis is that the black hole is able to sustain a graviton Bose--Einstein condensate (and surely similar condensate's made of other quanta). But is that condensate really present? Our results do not answer that question, but if we reflect on the case where the limit $N\to 0$ is taken, without disturbing the black hole geometry (i.e. keeping $r_S$ constant) this requires taking $\varphi\to 0$ in a way that the ratio $\sqrt{N}/\varphi$ is fixed. Then one gets a dimensionless $\mu_0=X=-1$. This value appears to be universal and independent of any hypothesis. The metric is 100\% Schwarzschild everywhere. We conclude that a black hole produces necessarily a (trapping) non-zero chemical potential when the physical system is expressed in terms of the grand-canonical ensemble. Outside the black hole, $\mu=0$ ($X=0$).

Another way of reaching this conclusion is by taking a closer look to our exact equations in the previous section. If one sets $\mu=0$ then necessarily $\varphi=0$ and one gets the classical Schwarzschild solution everywhere, but the reverse is not true. One can have $\varphi=0$ but this does not imply $\mu=0$. Let us emphasize that these results go beyond the second order perturbative expansion used in parts of this article.

As {\it gravitons} cross the horizon and are trapped by the black hole classical gravitational field they eventually thermalize and  form a BEC. The eventual energy surplus generated in this process is used to increase the mass and therefore the Schwarzschild radius. $\varphi$ is now non-zero; it is directly proportional to $\sqrt{N}$, the number of gravitons, and the dimensionless chemical potential departs from the value $\mu_0=X=-1$, presumably increasing it in modulo. As soon as $\varphi\neq 0$ the metric inside the black hole is not anymore Schwarzschild (but continues to be Schwarzschild outside). Note that the metric is destabilized and becomes singular for $\varphi=1$, so surely this is an upper limit where $N\rightarrow\infty$.  

Yet another possible interpretation, that we disfavour, could be the following. Each black hole has associated a given constant value of $\mu_0=X$, hence of $\varphi$. Then equations \eqref{mistery} would imply that after the emission of each {\it graviton}, the value of $r_S$ is readjusted. The problem with this interpretation is that it would require a new dimensionless magnitude ($\mu_0$ or $X$) to characterize a Schwarzschild black hole; something that most black holes practitioners would probably find hard to accept. In either case the metric in the black hole interior is different from the Schwarzschild one.

\chapter{Quasilocal energy for condensate structures}
\label{ChapterQLE}



\section{Quasilocal energy definitions in general relativity}

It is a fundamental tenet of general relativity that there is no such concept as local energy of the gravitational field. 
This can be understood as a consequence of the equivalence principle, implying that there is no proper definition for the energy density in gravitating systems.
One of the greatest achievements has certainly been the possibility of defining a global gravitational energy and momentum, both at spatial and null infinity (Arnowitt--Deser--Misner, in short ADM, or Bondi masses \cite{ADMBondi} between other formulations).

More ambitious efforts to associate an energy to extended, but finite, spacetime domains have resulted in quasilocal definitions. 
Initial attempts in this direction involved pseudotensor methods, where part of the "gravitational energy" stands as a source term in the form of a stress energy momentum. However this separation is not true for all observers, and there is no general definition for obtaining it. Einstein was the first one to derive a stress tensor from a variational principle. From a perspective of decades of research in this field, his definition, while not being diffeomorphism invariant, was quite successful. In the 60s, M\o{}ller discovered a bulk action, instead of being quadratic on the Christoffel symbols, $\Gamma_{\mu\nu}^\sigma$, as it was in the $\Gamma\Gamma-$action proposed by Einstein \cite{Einstein1916}, it was quadratic on the tetrad connection coefficients, $\omega_\mu^{\alpha\beta}$; the $\omega\omega-$action \cite{Moller}. This proposal gained the property of being diffeomorphism invariant; however it was not fully invariant under "internal" transformations of the tetrad and it was not constructed purely with the metric as Einstein's formulation. 
A major step forward was made when it was possible to generalize the quasilocal treatment into the language of two-spinors (see for instance the works of Szabados \cite{SzabadosP}). From this powerful formalism numerous interesting results have been obtained.
Although several approaches have led to some interesting results, no definitive consensus for the expression of the quasilocal energy seems to have emerged (the most important proposals are reviewed in \cite{Szabados} and references therein).

In the present study we will make use of the canonical quasilocal formalism based in the so-called Trace $K-$action \cite{YorkKAction}.
Quasilocal properties can be derived by means of a field–theoretic generalization of the Hamilton--Jacobi theory, without inquiring into the detailed structure of the quasilocal (Hamiltonian) phase space. 
In fact, it provides a formal basis for deriving from the action a stress energy momentum tensor with a purely metric formalism (as the $\Gamma\Gamma-$action), differing from the standard Hilbert action by metric dependent boundary terms. It is worth noting that (contrary to the $\Gamma\Gamma-$action) the Trace$K-$action is manifestly invariant under diffeomorphism and independent on any background structures\footnote{Any action admits subtraction terms, function of the fixed boundary data. In some cases, as the one we will deal with, it is convenient to fix this ambiguity by introducing a background spacetime; however, this is not a required feature of the canonical quasilocal formalism.}. 
Besides, although it is not clear how to apply the Noether theorem as this action is not constructed from a bulk Lagrangian, it is possible to avoid the use of this theorem by means of the Hamilton--Jacobi theory.

In what follows we will give the basic ingredients of this formalism. Then, in section \ref{SectionQLEBY}, we will review how to define a consistent quasilocal energy for spacetime exhibiting horizons. In the remaining sections, following \cite{AEGHorizons}, first we will analyze the consequences of applying this formalism to the zone beyond the horizon. Finally, it will be shown that with the quasilocal energy, one  can associate also a quasilocal binding-like potential. In this manner, a consistent picture can be obtained to constrain the behaviour of a graviton condensate structure gravitationally.

\section{Hamilton–Jacobi Formalism}\label{SectionQLEHJ}
There is no specific construction for the Hamilton–Jacobi equations. The starting point is to define a Hamiltonian making use of the action, which is well known in gravity. The mathematical formulation of this idea is to take a compact region of the spacetime $M$. 
Time is globally defined inside this region and provides a smooth foliation into a family of spacelike slices $\Sigma$. This foliation admits for the metric to introduce the usual ADM decomposition \cite{ADM},
\begin{equation}\label{MetricADMDecomposition}
	g_{\mu\nu}\ dx^\mu dx^\nu=-N^2dt^2+\gamma_{ij}(dx^i+V^idt) (dx^j+V^jdt)\ .\end{equation}
$N$ is the lapse function and $V^i$ the shift vector. In the static and spherically symmetric cases under consideration in section \ref{ChapterBHasBEC}, this metric gets the simpler structure; for convenience it will be written as
\begin{equation}\label{MetricSphSym}
	ds^2=-\epsilon N^2\ dt^2+\epsilon f^{-2}\ dr^2+r^2\ d\Omega^2\ ;\end{equation}
with such definition $N(r)$ and $f(r)$ are always positive functions of the coordinate $r$ and $\epsilon$ is either $+1$ or $-1$. If we are placed outside the horizon, $t$ is a timelike coordinate, then the temporal metric potential $g_{tt}$ is negative and in this region $\epsilon=+1$. At the horizon, $g_{tt}$ and $g_{rr}$ exchange sign, hence in the interior $\epsilon=-1$. Even more, when a condensate is present the inverse relation between $g_{tt}$ and $g_{rr}$ is broken.

In this manner the decomposition $M=[t';t'']\times\Sigma$ can be thought of as if it were a family of hypersurfaces defined by a timelike and future pointing unit normal $u_\mu=-\epsilon N\, \delta_\mu^t$. 
The  metric tensor induced on each leaf $\Sigma_t$ is
\begin{equation}
	\gamma_{\mu\nu}=g_{\mu\nu}+u_\mu u_\nu\ .\end{equation}
Hence $\gamma_\mu^\nu$ works as a projector tensor onto $\Sigma$.
For space coordinates $x^i$ adapted to the foliation $\Sigma:\ i=(r,\theta,\phi)$, we can define a spacial metric $\gamma_{ij}$, such that contracted with its inverse gives the Kronecker delta $\gamma_i^j=\delta_i^j$.

Also curvature tensors for the intrinsic curvature $R_{ijkl}$ and the extrinsic curvature $\Theta_{ij}$ can be defined. It is interesting for us the latter one, which is defined as usual
\begin{equation}\label{ExtrinsicCurvatureSigma}
	\Theta_{ij}=-\frac12\,\mathcal{L}_u\gamma_{ij}\ .\end{equation}

\begin{figure}[!b]\begin{center}
	\includegraphics[scale=0.2]{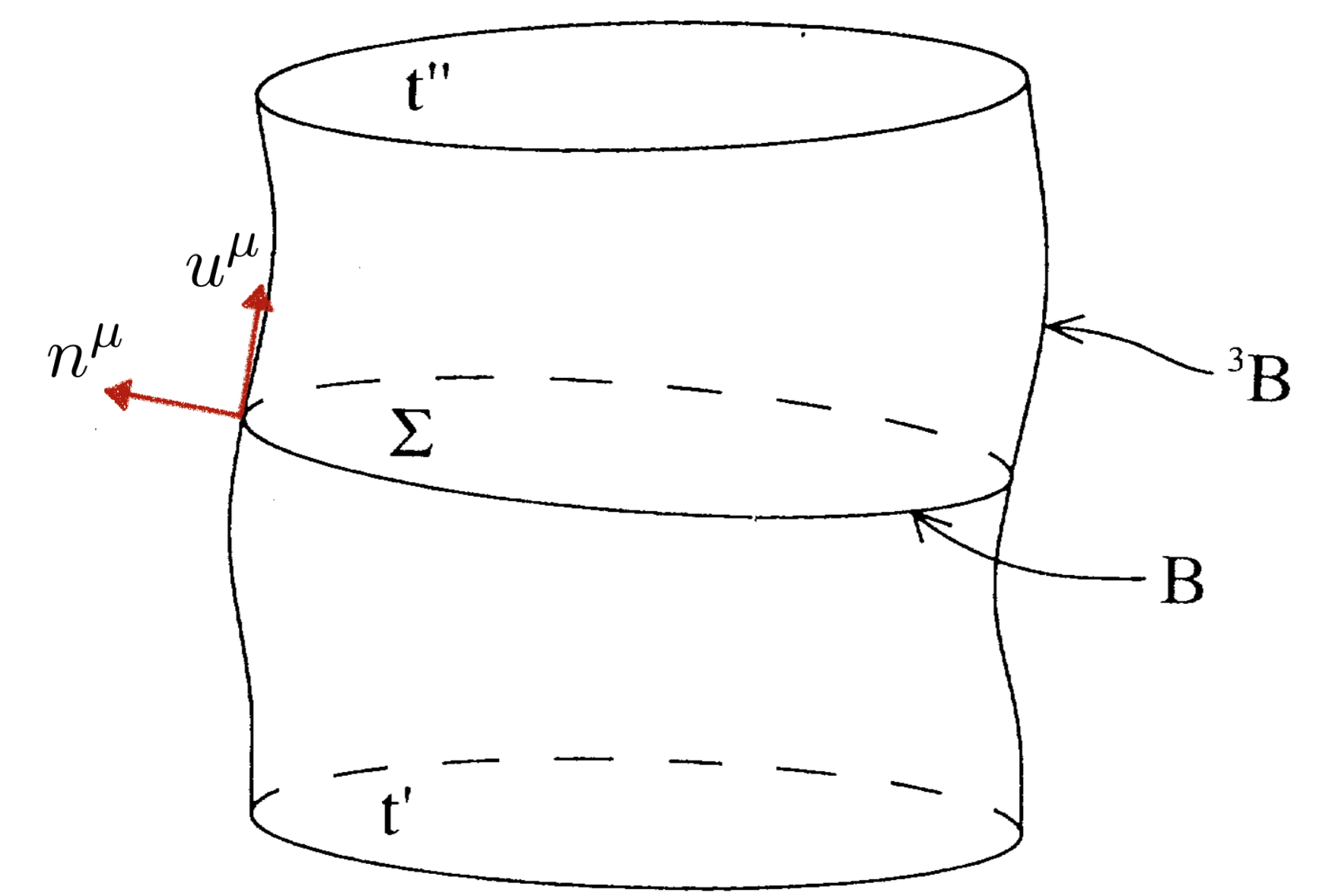}\captionsetup{width=0.85\textwidth}\caption{\small A generic spacelike slice $\Sigma$ has two-boundary $B$. The spacetime $M$ has as boundaries the time history of $B$, a timelike three-surface $^3B$, and the initial and final spacelike hypersurfaces defined at $t'$ and at $t''$.} \label{FigureManifold}\end{center}\end{figure}

Each generic leaf $\Sigma$ has a spatially closed two-boundary $\partial\Sigma=B$. The time history of these two-surface boundaries $B$ is the timelike three-surface boundary $^3B=[t';t'']\times B$, characterized by a spacelike normal $n^\mu$ pointing outwards as it is depicted in Figure \ref{FigureManifold}. An overlined notation will be used for tensors living in this submanifold, let us note that the boundary $B$ and its history $^3B$ need not be physical barriers. The induced metric is
\begin{equation}\label{Metric3B}
	\overline\gamma_{\mu\nu}=g_{\mu\nu}-n_\mu n_\nu\ ,\end{equation} 
$\overline\gamma_\mu^\nu$ projects onto $^3B$.
Analogously, intrinsic coordinates $i=(t,\theta,\phi)$ defines tensors on $^3B$: an intrinsic metric $\overline\gamma_{ij}$, the projector becomes the delta $\overline\gamma_i^j=\delta_i^j$, an extrinsic curvature
\begin{equation}\label{ExtrinsicCurvature3B}
	\overline\Theta_{ij}=-\frac12\,\mathcal{L}_n\overline\gamma_{ij}\ ,\end{equation}
and so on and so forth.

As we will see, when $\Sigma$ is taken to intersect $^3B$ orthogonally, the Brown--York quasilocal energy arises from the boundary term in the action. This orthogonality implies for the three-boundary normal $n^\mu$
to satisfy $(u\cdot n)\lvert_{^3B}=0$.
Therefore, the unit normal $n^\mu$ in the spacetime region to the three-boundary $^3B$ 'coincides' with the outward unit normal in $\Sigma$ to the two-boundary $B$. The normal vector to constant $r$ surfaces is
\begin{equation}\label{NormalB}
	n^i=\epsilon f\,\delta_r^i \ ;\end{equation}
with unit length in the hypersurface three-metric, i.e. $n^j\gamma_{ij}n^j\vert_{^3B}=1$.
This way of defining objects, factorizing the signs via $\epsilon$, makes the extension of the definition of the quasilocal energy to the interior zone, beyond the event horizon, straightforward. Details can be found in \cite{Lundgren}, where it is argued on physical grounds why normal vectors change sign at $r=r_S$. Basically they must keep pointing outwards and to the future.

Let now $N$ and $N^a$ be the lapse and the shift, this orthogonal restriction allows the metric on $^3B$ to be decomposed as
\begin{equation}\label{Metric3BGeneric}
	\overline\gamma_{ij}\ dx^idx^j= -N^2dt^2+\sigma_{ab}(dx^a+V^adt)(dx^b+V^bdt)\ .\end{equation}
Now $a,b=\theta,\phi$ are the coordinates on $B$ and $\sigma_{ab}$ is the two-metric. These two-boundaries $B$ can be thought as the intersection of the family of leaves $\Sigma$ with the three-boundary $^3B$ as shown in Figure \ref{FigureManifold}. The metric can alternatively be written as
\begin{equation}\label{MetricB}
	\sigma_{\mu\nu}=\gamma_{\mu\nu}-n_\mu n_\nu=\overline\gamma_{\mu\nu}+u_\mu u_\nu=g_{\mu\nu}-n_\mu n_\nu+u_\mu u_\nu\ .\end{equation}
In the case of spherically symmetric and static spacetimes from the metric decomposition \eqref{Metric3BGeneric} it is seen that the induced two-metric of $B$ has clearly the structure for a sphere; i.e. $ds^2=r^2\ d\Omega^2$.
If the connection is metric compatible as in the cases of interest, the usual definition for the extrinsic curvature of being proportional to the Lie derivative $\mathcal{L}$ along the corresponding unit normal \eqref{NormalB}, an analogous definitions as the ones written in \eqref{ExtrinsicCurvatureSigma} or \eqref{ExtrinsicCurvature3B}, in this case for the two-boundary $B$ as embedded in $\Sigma$, can be equivalently written as the covariant derivative in the spacelike slice of the normalized tangent vector to this hypersurface as defined on $\Sigma$
\begin{equation}\label{ExtrinsicCurvatureB}
	K_{ij}=-\sigma_i^k\,\nabla_k\,n_j \end{equation}
along a direction of the projector operator $\sigma_i^j$ constructed with \eqref{MetricB}.

The purpose of this decomposition is to establish an analogy with this method deduced in the context of nonrelativistic mechanics. To make the generalization more tractable, let us summarized the notation in Table \ref{TableDecomosition}. Also some technical expressions will be included in the Appendix \ref{AppxExtrinsicCurvature}.

\vspace{13pt}
\begin{table}[!h]\begin{center}\begin{tabular}{lcccc}\hline
			&Metric&\begin{tabular}{c}Unit\\Normal\\\end{tabular}&\begin{tabular}{c}Extrinsic\\curvature\end{tabular}&\begin{tabular}{c}Conjugate\\momentum\end{tabular} \\\hline\hline Spacetime $M$&$g_{\mu\nu}$&&&\\ Hypersurfaces $\Sigma$&$\gamma_{ij}$&$u^\mu$&$\Theta_{ij}$& $P^{ij}$\\Three-boundary $^3B$&$\overline\gamma_{ij}$&$n^\mu$& $\overline\Theta_{ij}$ &$\pi^{ij}$\\Two-boundary $B$ &$\sigma_{ab}$ &$n^i$&$K_{ab}$\\\hline\end{tabular}\end{center}\vspace{-15pt}\captionsetup{width=0.85\textwidth}\caption{\small Summary of notation for the spacetime decomposition.} \label{TableDecomosition}\end{table}

 \subsection{Nonrelativistic mechanics review}

Given a classical mechanical system of $n$ degrees of freedom with configuration manifold $Q$ and a Lagrangian ${\cal L}:TQ\times\mathbb{R}\rightarrow\mathbb{R}$ that is assumed to be first order and may depend on time explicitly.
In this mathematical formalism, given two configurations $(q'_a,t')$ and  $(q''_a,t'')$, the corresponding action functional is
\begin{equation}
	S^{(nr)}[q(t)]=\int_{t'}^{t''}{\cal L}\,[q_a(t);\dt{q}_a(t);t]\ dt=\int_{t'}^{t''}\bigl(\,\sum_{a}\dt q_a\,p_a-H\,[q_a,p_a,t]\,\bigr)\ dt\ ;\end{equation}
where $q_a(t)$ are the generalized coordinates of the state space (phase space $+$ time)\footnote{The pairs $(q_a,t)$ may be called a history or world line in the ‘spacetime’ $Q\times\mathbb{R}$ for proceeding to a relativistic formulation.}, from $q'_a(t')$ to $q''_a(t'')$, with tangent $\dt{q}_a(t)$ at the parametrization parameter $t$.
A Legendre transformation of the Lagrangian provides the action in canonical form and gives the usual definition for the Hamiltonian of the system in terms of the generalized coordinates $q_a$ and the generalized canonical momenta $p_a$.
It is expected that a generalization of this theory provides a formal basis for identifying from the action a stress-energy-momentum tensor for general relativity. 

Varying this action with the two fixed points, one gets
\begin{equation}\label{ActionVariationNR}\begin{split}
	\delta S^{(nr)}=&\ \Big(\begin{gathered}\text{ equations of motion }\\[-1ex] \text{terms}\end{gathered}\Big)+\sum_{a}p_a\,\delta q_a\  \big\vert_{t'}^{t''}-H\,\delta t\ \big\vert_{t'}^{t''}\ . \end{split}\end{equation}
Variations among classical solutions extremize $S^{(nr)}$; the terms giving the equations of motion vanish. The requirement that the action has null variation leads to the Hamilton--Jacobi equations
\begin{equation}\label{HJEqNR}
	p_a=\frac{\partial S}{\partial q_a}\ \left(=\frac{\partial{\cal L}}{\partial\dt{q}_a}\right)\ ; \hspace{50pt} H=-\frac{\partial S}{\partial t}\ .\end{equation}
Clearly, the action is not unique. Subtraction terms $S^{(nr)}[q(t)]-S^0[q(t)]$ of the form $S^0[q(t)]=\int_{t'}^{t''}(dh/dt)\,dt$, with any arbitrary smooth function $h = h(q_a(t),t)$ is an equally good action for obtaining the same dynamics. 
Obviously, the subtraction term $S^0[q(t)]$ modifies both magnitudes, the canonical momenta and the energy according to $p_a \rightarrow p_a-(\partial h/\partial q_a)$ and $E\rightarrow E+(\partial h/\partial t)$, respectively.

There is therefore an intimate connection between the classical Hamiltonian (the energy), the change in time of the classical action and the fact that choice of the action is not unique.

\subsection{Covariant Hamiltonian formalism}

The guiding principle for obtaining the quasilocal energy is the analogy between the action variation in nonrelativistic mechanics and variation in general relativity.
Now that we have defined the tensors needed at the beginning of section \ref{SectionQLEHJ} and in the Appendix \ref{AppxExtrinsicCurvature}, let us move to generalize the Hamilton--Jacobi method to general relativity.

In order for the action to provide a well-defined variational principle for Einstein theory, one fixes the intrinsic three-metrics over the boundaries (rather than all the $g_{\mu\nu}$) \cite{YorkKAction}. Then we have a bulk Hilbert action over $M$, with boundary three-metric $\overline{\gamma}_{ij}$ fixed on $^3B$ and the hypersurface metric $\gamma_{ij}$ fixed on the end points of the world-lines $t'$ and $t''$, on where to apply variational principles
\begin{equation}\begin{split}
	S=&\frac1{2\kappa}\int_Md^4x\sqrt{-g}\ R+S_m\\
	&+\frac1{\kappa}\int_{\Sigma}d^3x\sqrt{\gamma}\ \Theta\,\big\vert_{t'}^{t''} -\frac1{\kappa}\int_{^3B}d^3x\sqrt{-\overline\gamma}\ \overline\Theta+\frac1\kappa\int_{^3B}d^3x\sqrt{-\overline\gamma}\ \overline{\overline\Theta}\ .\end{split}\end{equation}
$S_m$ may include the matter action as well as the chemical potential or the cosmological constant terms when corresponds. The last term is the Gibbons--Hawking normalizing factor \cite{Gibbons}; $\overline{\overline\Theta}$ corresponds to the extrinsic curvature of the three-boundary but as embedded in a flat background four-geometry (the spacetime in which we want to obtain zero quasilocal quantities). This term is usually interpreted as our freedom to shift the zero point of the energy.

Varying the action with respect to arbitrary variations in the metric and matter fields, we get
\begin{equation}\label{ActionVariationGR}\begin{split}
	\delta S=&\ \Big(\begin{gathered}\text{ equations of motion }\\[-1ex] \text{terms}\end{gathered}\Big)+\int_\Sigma d^3x\ P^{ij} \delta \gamma_{ij}\Bigr\vert_{t_i}^{t_f}+ \int_{^3B} d^3x\ \left(\pi-\pi_0\right)^{ij}\delta\overline\gamma_{ij} \ . \end{split}\end{equation}
The Hilbert action minimally coupled to matter is stationary ($\delta S=0$) if Einstein's equations are satisfied. When restricted to classical solutions, the equations of motion vanish and we are left with the terms given by the two momenta: when matter fields are minimally coupled to gravity, the gravitational momentum canonically conjugate to $\gamma_{ij}$ for the hypersurfaces $\Sigma$ is defined as
\begin{equation}\label{MomentumTensorsP}
	P^{ij}=\frac1{2\kappa}\ \sqrt{\gamma} \left(\Theta\,\gamma^{ij}-\Theta^{ij}\right)\ ;\end{equation}
and the boundary momentum $\pi^{ij}$ conjugate to $\overline\gamma_{ij}$, as
\begin{equation}\label{MomentumTensorsPi}
	\pi^{ij}=-\frac1{2\kappa}\ \sqrt{-\overline\gamma}\left(\overline\Theta \,\overline\gamma^{\,ij}-\overline\Theta^{\,ij}\right)\ .\end{equation}
The $\pi_0^{ij}$ term is conjugate to the flat metric and is simply a shift due to the Gibbons--Hawking normalizing factor; fixing this term means to choose a ‘reference configuration’. The analogues of the Hamilton--Jacobi equations \eqref{HJEqNR} are the following two relations (the equations of the nonrelativistic case are recalled on the right for making the correspondence straightforward)
\begin{equation}\label{HJEqGR}
	\left\{\begin{split}&P^{ij}=\frac{\delta S}{\delta\gamma_{ij}}\\[2ex]& (\pi-\pi_0)^{ij}=\frac{\delta S}{\delta\overline\gamma_{ij}}\end{split}\right. \qquad\iff\qquad\left\{\begin{split}&p_a=\frac{\partial S}{\partial q_a} \\[2ex] &H=-\frac{\partial S}{\partial t}\ .  \end{split}\right.\end{equation}

The generalization is quite direct. The three-metric $\overline\gamma_{ij}$ provides the metrical distance between all spacetime intervals in the boundary
manifold $^3B$ (including time between spacelike surfaces); therefore the notion of energy (the one equation for the Hamiltonian) in nonrelativistic
mechanics is generalized to a stress energy momentum defined on $^3B$ that characterizes the entire system (gravitational field, chemical potential fields and any matter fields and/or cosmological constant). In accordance with the standard definition for the matter stress tensor 
\begin{equation}
	T^{\mu\nu}=\frac{2}{\sqrt{-g}}\ \frac{\delta S_m}{\delta g_{\mu\nu}} \ ,\end{equation}
the following surface stress tensor is conveniently defined
\begin{equation}\label{StressTensorSurface3B}
	\tau^{ij}\equiv\frac2{\sqrt{-\overline\gamma\,}}\ \frac{\delta S}{\delta\overline\gamma_{ij}}=\frac2{\sqrt{-\overline{\gamma}\,}}\left(\pi^{ij}-\pi^{ij}_0\right)\ .\end{equation}

An important feature of this stress tensor is apparent when considering two concentric spherical surfaces $B_1$ and $B_2$. In the limit where $B_1$ and $B_2$ approach each other, the total surface stress energy momentum result in
\begin{equation}
	\tau^{ij}=\frac{2}{\sqrt{-\overline\gamma}}\left(\pi_2^{ij}-\pi_1^{ij}\right)\end{equation}
since $\tau_1{}^{ij}$ is computed using the outward normal to $B_2$ and reference terms $\pi_0{}^{ij}$ cancel between each other. This tensor embodies the well-known result of Lanczos and Israel
in general relativity \cite{Lanczos-Israel}, which relates the jump in the momentum $\pi^{ij}$ to the matter stress tensor of the surface layer.
In the infinitesimally thin layer limit, the three-geometries of each `side' of the layer coincide and there is no gravitational contribution to $\tau^{ij}$.
The direct physical implication of this result is the widely known absence of a local gravitational energy momentum \cite{MisnerThorneWheelerLibro}.

\section{Brown--York quasilocal energy}\label{SectionQLEBY}

Let us give one step further and rederive the quasilocal energy elaborated by Brown and York \cite{BrownYorkMath} in order to analyze its behaviour for several spacetime configurations where horizons are present, with or without the possible graviton condensates just discussed in Chapter \ref{ChapterBHasBEC}. This magnitude is one of the promising candidates for obtaining the quasilocal energy in general relativity. According to this proposal, a suitable fixing of the metric over the boundaries \cite{YorkKAction} provides a well-defined action principle for gravity and matter, and dictates a natural choice for the definition of the quasilocal energy contained in the region of interest without the need for any other geometric structures. 
Besides, its properties can be obtained through a Hamilton--Jacobi type of analysis involving the canonical action as it has been recently describes in the previous section. Among these properties we could mention that the proposal of \cite{BrownYorkMath} agrees with the Newtonian limit for spherical stars, it is applicable to thermodynamic problems \cite{Braden,York} and the asymptotic limit at infinity is the Arnowitt--Deser--Misner expression \cite{ADM} for the energy in asymptotically flat spacetimes \cite{Regge}. 

Physically, the Brown and York proposal \cite{BrownYork} is meant to represent the total energy enclosed by a space-like surface. This information is encoded in the surface tensor $\tau^{ij}$ defined on \eqref{StressTensorSurface3B}.
As it has been stated in \eqref{MetricADMDecomposition},
the 3+1 decomposition of the spacetime, together with the orthogonal condition for the normal vectors $u^\mu n_\mu=0$,
yields the 2+1 decomposition \eqref{Metric3BGeneric} of the metric $\overline\gamma_{ij}$ on $^3B$ as well. Hence the normal and tangential projections of $\tau^{ij}$ on the two-surface $B$ defines the proper energy, momentum, and spacial-stress surface densities, given by
\begin{equation}\label{TensorSurface}\begin{split}
	&\varepsilon=u_iu_j\tau^{ij}\\[2ex]
	&j_a=-\sigma_{ai}u_j\tau^{ij}\\[2ex]
	&s^{ab}=\sigma_i^a\sigma_j^b\tau^{ij}\ \end{split}\end{equation}
respectively.

We are interested in the first magnitude; $\tau^{ij}$ is proportional to the boundary momentum \eqref{MomentumTensorsPi}. Written as in \eqref{MomentumBoundaryDecomposition}, it is immediate to see that
\begin{equation}
	\frac{2}{\sqrt{-\overline\gamma}}\ u_iu_j \pi^{ij}=\frac{2}{N\sqrt{\sigma}}\ \frac{N\sqrt{\sigma}}{2\kappa}\ K=\frac K\kappa\ .\end{equation}
At this stage, as the action is not unique and arbitrary functions of the fixed boundary data can be included without modifying the equations of motion, the reference tensor $\pi_0^{ij}$ contributes with a trace curvature term $K_0$ as embedded in Minkowski. 
Integrating the energy density along a slice defining a two-boundary $B$, leads us directly to the definition of the quasilocal energy
\begin{equation}\label{QLEnergy}
	E=\int_Bd^2x\ \sqrt{\sigma}\ \varepsilon=\frac1\kappa \int_Bd^2x\ \sqrt{\sigma}\,\left(K-K_0\right)\ . \end{equation}
The energy result in the subtraction of the total mean curvature of $B$ es embedded in $\Sigma$ with the total mean curvature of $B$ es embedded in a Minkowskian reference frame, times the inverse of the Einstein's gravitational constant.

Notice that the separation of the four dimensional space in a product $M=\Sigma\times t$ implies that the quasilocal energy is invariant under general coordinate transformations on $\Sigma$ but not under all four dimensional general coordinate transformations, which also involve time. Therefore quasilocal energy will be dependent on the proper time of the observer \cite{WuChenLiu}.

Let us now focus this definition over arbitrary static and spherically symmetric spacetimes generically written as in \eqref{MetricSphSym}.
For the calculation it is only needed the following connection coefficients
\begin{equation}
	\Gamma_{rr}^r=-\frac{f'}{f} \hspace{80pt} \Gamma_{r\theta}^\theta=\Gamma_{r\phi}^\phi=\frac{1}{r}\ ; \end{equation}
the prime denotes $r$ derivatives. The trace of the extrinsic curvature needed for computing the quasilocal energy is
\begin{equation}
	K=-2\,\epsilon\ \frac{f(r)}{r}\end{equation}
and the subtraction term is obtain embedding the sphere $B$ in a Minkowskian spacetime by setting $f(r)=\epsilon=1$, this is
\begin{equation}
	K_0=-\frac{2}{r}\ .\end{equation}
The next step is straightforward, putting everything together in \eqref{QLEnergy} with the determinant of the induced two-metric given by $\sqrt{\sigma}=r^2\sin^2\theta$, the quasilocal energy is 
\begin{equation}\label{QLEnergyIntegrated}
	E=\frac{r}{G_N}\left[1-\epsilon\,f(r)\right]\ . \end{equation}

Apart from the intrinsic interest of defining an energy, using a least-energy principle could allow us to determine whether a graviton condensate is in some sense preferred to the standard no-condensate situation. One might after all expect that matter falling into a black hole will eventually thermalize and the formation of a BEC may not seem such an exotic possibility then, but in the case of gravitons the formation of the condensate is a lot less intuitive. Minimization of the quasilocal energy may be a convenient tool to answer whether a BEC for gravitons exists or not.

\section{Quasilocal energy for a graviton condensate black hole}\label{SectionQLESchw}

Let us now consider the change in the quasilocal energy, as defined above, when a condensate with spherical symmetry, such as the one previously described, exists in a static and spherically symmetric black hole. The significant metric elements in \eqref{MetricSphSym}, both within and beyond the event horizon, are obtained particularizing \eqref{MetricCondensate} to the Schwarzschild solution
\begin{equation}\label{MetricSchwBECFactors}
	N(r)=\sqrt{\frac{\epsilon}{1-\varphi}\left(1-{\frac{r_S}{r}}\right)}\ ; \hspace{30pt}f(r)=\sqrt{\epsilon\left(1-\varphi\right)\left(1-{\frac{r_S}{r}}\right)}\ .\end{equation}
In the outer region we get the known solution by setting $\epsilon=1$ and $\varphi=0$. In the interior $r$ is a timelike coordinate, $\epsilon=-1$ and $\varphi$ may be different from zero. The following expression describes the quasilocal energy \eqref{QLEnergyIntegrated} in both cases
\begin{equation}\label{QLEnergySchw}
	E=\frac{r}{G_N}\,\left[1-\epsilon\sqrt{\epsilon\left(1-\varphi\right)\left(1-{\frac{r_S}{r}}\right)}\right]\ .\end{equation}

We have written the expressions above in such a way that taking $\varphi=0$ and $\epsilon=1$ for $r>r_S$ we get the quasilocal energy of the black hole as a whole as seen from outside\footnote{That is from the reference system of an observer at rest at $r=\infty$.}. The
quasilocal energy inside an $r=$ constant surface placed outside the event horizon is given by
\begin{equation}\label{QLESchwOut}
	E= \frac{r}{G_N}\,\left(1-\sqrt{1-\frac{r_S}{r}}\right)\ .\end{equation}
Assuming the existence of a BEC with $\varphi\neq0$ and that inside the horizon the $r$ coordinate becomes timelike with $\epsilon=-1$
\begin{equation}\label{QLESchwIn}
	E= \frac{r}{G_N}\,\left( 1+\sqrt{1-\varphi}\sqrt{\frac{r_S}{r}-1}\right)\  .\end{equation}
These expressions were derived in \cite{BrownYork} in the case $\varphi=0$.

\begin{figure}[!t]\begin{center}
	\includegraphics[scale=1.3]{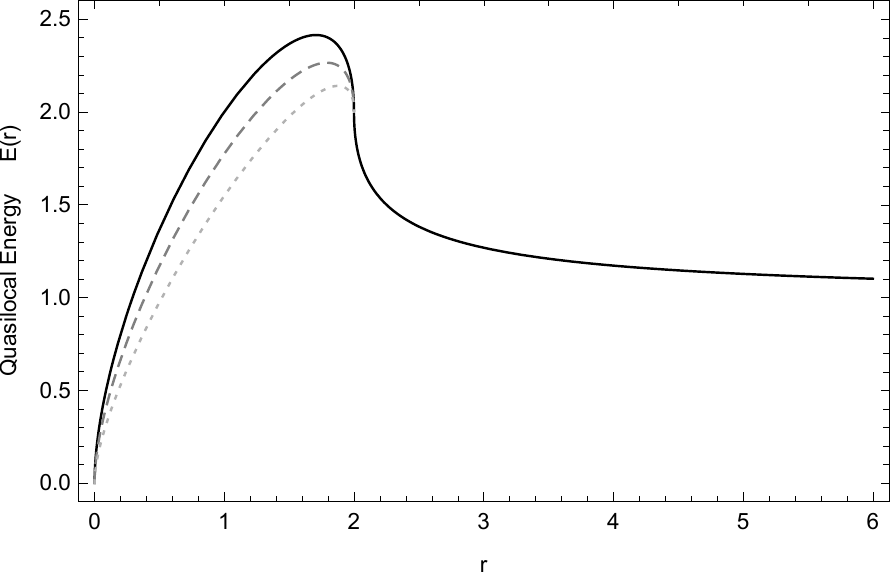}\captionsetup{width=0.85\textwidth}\caption{\small Quasilocal energy for a Schwarzschild black hole computed inside as well as outside the event horizon. Both axes are in units of the mass $M,\,G_N=1$. The event horizon is located at $2M$. In the inner region we compare a black hole in classical theory of general relativity where $\varphi=0$ (solid line), and various values for the condensate parameter where we understand that a BEC structure is preferred minimizing the resulting energy of the object. We use as a comparison two values for the condensate: $\varphi=0.4$ dashed line and $\varphi=0.7$ dotted line.} \label{QLEPlotSchw}\end{center}\end{figure}

The quasilocal energy of the entire Schwarzschild metric is plotted in Figure \ref{QLEPlotSchw}.
This shows that a condensate structure lowers the energy with respect the familiar Schwarzschild metric. The higher the value for $\varphi < 1 $ chosen, the lower the quasilocal energy will be. Figure \ref{QLEPlotSchw} tells us that the total energy inside the event horizon (including gravitational energy as well as the energy of the $r = 0$ singularity) is twice the black hole mass
\begin{equation}
	E(r=r_S)=2M \end{equation}
and twice the energy of the total spacetime as can be seen from the asymptotic behaviour for the quasilocal energy at infinity.
An expansion of \eqref{QLESchwOut} around infinity yields	
\begin{equation}
	\lim_{r\rightarrow\infty} E(r)\simeq \frac{r}{G_N}\,\left\{1-\left[1-\frac{r_S}{2\,r}+{\cal O}\left(\frac{1}{r^2}\right)\right]\right\}=\frac{r_S}{2G_N}\equiv M\ ,\end{equation}
which is certainly an expected result.

The reinterpretation of a black hole as describing a graviton BEC does not modify three remarkable features of the quasilocal energy presented in \cite{Lundgren} that are perhaps not widely acknowledged. The first one is that the quasilocal energy is zero at the singularity. The infinity of the energy of the gravitational field at $r=0$ predicted by Newtonian gravity is removed. The second striking fact comes from the energy distribution inside the black hole. Inside the horizon, the quasilocal energy has its maximum value at
\begin{equation}
	r_{max}=G_N\,M\left(1+ \frac{1}{\sqrt{2-\varphi}}\right)\ .\end{equation}
In the limit of the classical theory, this maximum value is located at a radius equals to $(1+1/\sqrt{2})G_N\,M$.
As the condensate parameter $\varphi$ increases, the maximum moves towards the event horizon and for the limiting value $\varphi=1$, it lies {\em exactly} at the horizon.
All this means that the black hole looks like an extended object where most of the energy seems to be 'stored' not far from the horizon.
The third feature  is that the derivative of the quasilocal energy matches across the horizon, but is infinite there regardless of the value of $\varphi$.

Note that expanding the quasilocal energy and subtracting the black hole mass M one gets for $r_S\ll r$
\begin{equation}
	E(r)-M = \frac{G_N\,M^2}{2r}+\dots\end{equation}
The leading term is the classical gravitational self-energy of a shell of matter with total mass $M$ and radius $r$.

\section{Gravitational binding}\label{GravitationalBinding}

For $r\gg r_S$, if we denote by $E(r)\vert_M $ the quasilocal energy of a black hole of mass $M$, 
\begin{equation}\begin{split}
	E(r)\vert_{M+m}-E(r)\vert_{M}&\ \simeq\ m+G_N\,\frac{M\,m}{r}+\frac32\, G_N{}^2\,\frac{M^2\,m}{r^2} +\dots\\[2ex]
	&\ =\ m\,\frac{\partial}{\partial_M} E(r) + {\cal O}(m^2)\ .\end{split}\end{equation}
The first term corresponds to the gravitational potential energy of masses $m$ and $M$ separated by a distance $r$. It is also the energy (for large values of $r$) that it takes to separate the small mass $m$ from $M$ and take it to infinity. This leads us to interpret the quantity
\begin{equation}\label{potentialquasilocal}
	V(r)= 1-\frac{\partial}{\partial M}\left(\frac{r}{G_N}\sqrt{1-\frac{2G_NM}{r}}\right)=1-\frac{1}{\sqrt{1-\frac{2G_NM}{r}}}\end{equation}
as the binding potential energy associated to the quasilocal energy outside the horizon. Now, if we take one step forward and extend these ideas to the black hole interior, we will get 
\begin{equation}
	V(r)= 1-\frac{\partial}{\partial M}\left(\frac{r}{G_N}\sqrt{1-\varphi} \sqrt{\frac{2G_NM}{r}-1}\right)=1-\frac{\sqrt{1-\varphi}}{\sqrt{\frac{2G_NM}{r}-1}}\end{equation}
as the analogous potential energy inside the horizon. As occurs with the quasilocal energy, there are different ways to derive a quasilocal potential as well (both inside and outside black holes). For instance, in \cite{Blau} the authors define a quasilocal potential coming from the Brown--York quasilocal energy that does not coincide with our definition. They compare their definition with the Misner--Sharp potential, derived from its respective quasilocal energy too \cite{MisnerSharp}.

\begin{figure}[t]\begin{center}
	\includegraphics[scale=1.3]{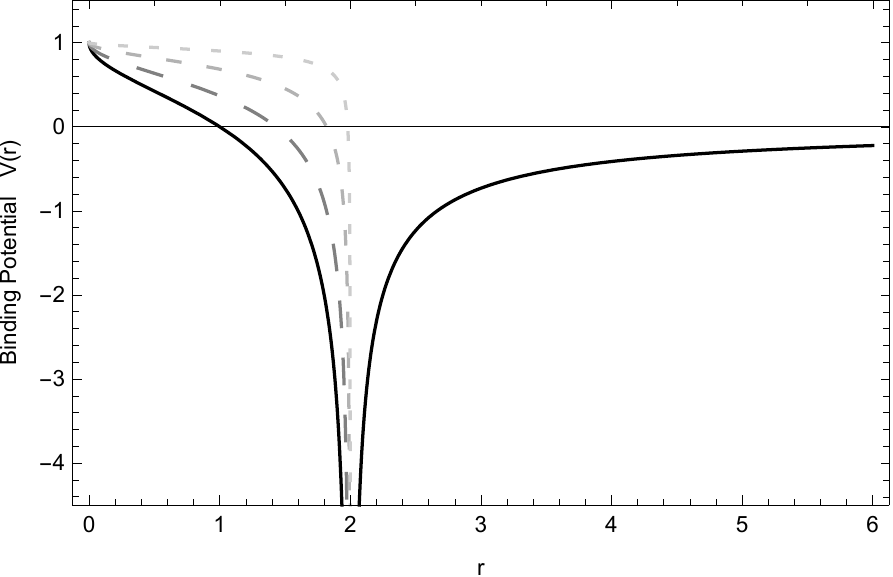}\captionsetup{width=0.85\textwidth}\caption{\small Potential profile derived from Brown--York energy for Schwarzschild spacetime, with $M=1$ and various values for the condensate parameter $\varphi$. The solid line corresponds to $\varphi=0$. As $\varphi$ approaches the limiting value $\varphi=1$ the inner part of the potential tends to a square potential.}\label{Potential}\end{center}\end{figure}

We note that $V(r)$  is singular at $r=r_S$, but we also note that this singularity is integrable. The following limits are readily found:
\begin{equation}
	\lim_{r\to 0}V(r)=1 \ ,\hspace{80pt}\lim_{r\to \infty}V(r)= 0\ .\end{equation}
A plot of the profile is presented in Figure \ref{Potential}. There we can see the potential that seems to emerge from the Brown--York prescription for the quasilocal energy in the case where the background metric is 
Schwarzschild.
As we observe, the binding takes place in a region that is relatively close to the horizon, which is not at all unexpected after having seen the shape of the black hole quasilocal energy profile.

There are several considerations that may follow from this potential profile if we accept the previous interpretation. Notice that a massive particle of mass $m$  initially at rest at infinity follows the trajectory described by (see e.g. \cite{rindler})
\begin{equation}
	\dt r\,^2= \left(1-\frac{r_S}{r}\right)^2\frac{r_S}{r}\ ,\end{equation}
where the dot means derivative with respect the coordinate time $t$ (i.e. not proper time). The radial momentum will be  $p_r= m\gamma \dt r$. As the particle approaches $r=r_S$ from the outside, $p_r$ tends to zero as it is well known. If we take Figure \ref{Potential} at face value the particle falls in the potential valley while emitting gravitational radiation. However, quantum mechanics should stabilize it at some point. Let us use a simple argument to make some energy considerations (a similar argument works fine in the case of the hydrogen atom).

Indeed, assuming that $\Delta p_r \sim p_r$, according to the uncertainty principle $\Delta r > 1/p_r$. This leads to the following order of magnitude estimate:
\begin{equation}
	\Delta r \simeq \sqrt{\frac{r_S}{m}}\ .\end{equation}
This gives us the characteristic width of the potential and the approximate energy level. Introducing the relevant units, for a black hole of about 30 solar masses and considering the fall of an electron, this gives $\Delta r \simeq 1 $mm. This result is surprising as it would imply that falling matter accumulates in a very thin shell on both sides of the horizon. The associated potential level would approximately be $\varepsilon/m \simeq 1-\sqrt{r_S\,m}$.

\begin{figure}[b!]\begin{center}
	\includegraphics[scale=1.3]{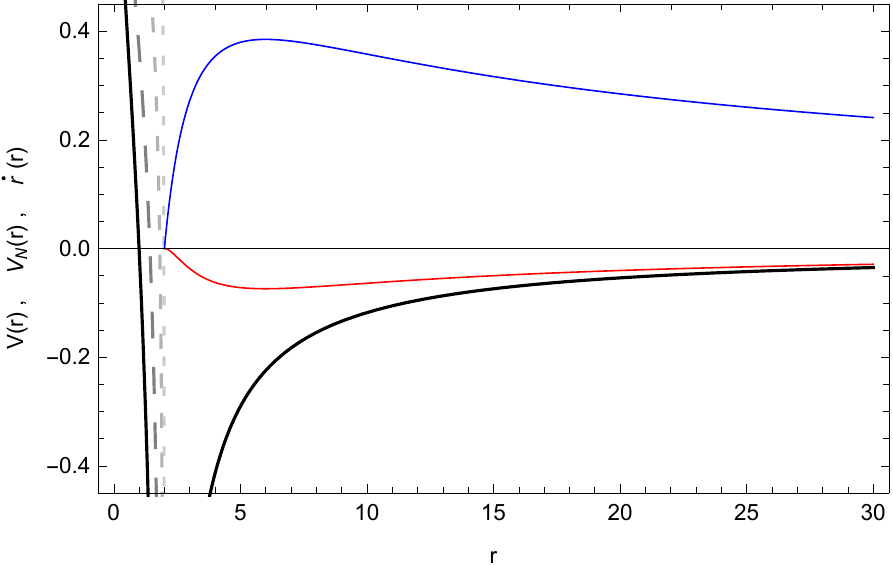}\captionsetup{width=0.85\textwidth}\vspace{5pt} \caption{\small Evolution of the velocity of a falling object as seen from a distant observer (blue curve). The maximum is $2c/(3\sqrt{3})$ and is attained at $r=3\,r_S$. We also plot the (fictitious) Newtonian potential that such an observer would `measure' from the fall of the object (red curve). In the same plot we show a zoom of the quasilocal potential profile derived from Brown--York energy for Schwarzschild spacetime, with $M=1$ presented in Figure \ref{Potential}. In the interior we plot the evolution of this magnitude as $\varphi$ gets close to its limiting value $\varphi=1$ and the potential approaches a step function.} \label{ParticleFalling}\end{center}\end{figure}

Figure \ref{ParticleFalling} may help us to understand the situation. In this figure we plot the velocity of a particle falling in a black hole of mass $M=1$ following a radial geodesic, as seen from an observer at infinity. We also plot the Newtonian potential that such an observer would deduce from measuring the motion of the falling particle along its geodesic motion. This (fictitious) potential agrees for the most part of the trajectory with the one derived from the quasilocal energy, which however is much deeper close to $r=r_S$. The difference between the two potentials should be observed by the distant reference point as gravitational wave emission. The total energy emitted should be approximately $m \sqrt{r_S\,m} \gg m$.
Obviously, only part of this energy is emitted outside the black hole and eventually reaches a remote observer. It is to be expected that a good part of it remains inside the black hole horizon. This is a natural explanation for the graviton condensate, present when $\varphi\neq 0$, whose properties have been discussed in the previous sections.

Assuming for the sake of the discussion that (a) {\it gravitons} in the condensate are in a state close to the maximum packing condition, i.e. have an energy $\sim 1/r_S$ and (b) that all the gravitational energy emitted by the falling particle is eventually stored in the black hole interior, the capture by the black hole of a particle of mass $m$ would create $n= (m\,r_S)^{\frac32}$ gravitons. 
This is a fairly large number so it is not unexpected that the continuous capture of matter by a black hole brings the value of the graviton condensate close to 1, its limiting value,  $N\sim (M/M_P)^2$. At that point, most of the mass of the black hole is actually due to the graviton condensate: gravitational (quasi)local energy has turned into black hole mass. This simple argument gives strong plausibility to the thesis sustained by Dvali and G\'omez.

Certainly, along the process just described, $M$ and consequently the shape of the potential does not stay constant, so a detailed dynamical analysis would be required to understand this process more accurately.  Of course all the previous considerations hinge on the validity of the hypothesis behind the quasilocal energy principle.

It is actually interesting to see what happens when the value of the graviton BEC approaches the limiting value $\varphi=1$ beyond which the metric becomes intrinsically singular. The quasilocal potential becomes a constant (the quasilocal energy is a linearly rising function of $r$ in the black hole interior in this limit). The result would apparently be a barrier right at the horizon location for falling particles. In this limit it would seem that gravitationally trapped objects would adhere to the black hole external surface.

\chapter{Condensates beyond horizons}
\label{ChapterBHasBECExtension}

\section{Bose--Einstein condensate extensions}

In Chapter \ref{ChapterBHasBEC}, we have derived a collective solution representing a bosonic condensate constructed with {\it gravitons} on top of a confining classical Schwarzschild potential.
The ground state of the system is dictated by a gravitational Gross--Pitaevskii-like equation and is strongly background dependent.
The characteristics of the solution present some interesting similarities with the usually employed approximated schemes such as Hartree--Fock. 
A main characteristic is that the total wave-function of the system of $N$ bosons is taken as a product of single-particle functions in the same quantum state.
When this is so, and the single-particle wave function satisfies the Gross--Pitaevskii equation, the total wave-function minimizes the expectation value of the Hamiltonian under a normalization condition giving the total number of constituents $N$. In the gravitational counterpart this aspect is translated into the maximum packaging condition for the {\it gravitons}. As shown in Chapter \ref{ChapterQLE} the energy is minimized for the condensate structure and is preferred over vacuum solutions.

In the present chapter we will give the first two simplest generalizations to other self-bounded classical field configurations: Reissner--Nordstr\"om and de Sitter background metrics.
We have found that condensate-like solutions may be present in all these cases too, and its structure is strongly dependent on the trapping geometry. Here again, its constituents are regulated by some sort of chemical potential. 
We then proceed to study the energy associated to these solutions and we found that they are in fact energetically favourable.
In the nonrelativistic case the mean-field theories provide a correct qualitative picture of the phase transition. Analogous implications are expected here and it is plausible to interpret the different static gravitational potential with our effective quantum description.

\section{Bose--Einstein condensate in Reissner--Nordström}\label{RRNN}\label{SectionRRNN}

In the present section, we shall extend the Bose--Einstein condensate formalism to a Reissner--Nordstr\"om geometry \cite{ReissnerNordstrom}.
When a black hole is coupled both to the gravitational and to the electromagnetic fields, the geometry is given by
\begin{equation}\label{MetricRN}
	\tilde{g}_{\alpha\beta}\,dx^\alpha dx^\beta=-\left(1-{\frac{r_S}{r}}+ \frac{r_Q{}^2}{r^2}\right)dt^2+\left(1-{\frac{r_S}{r}}+\frac{r_Q{}^2}{r^2}\right)^{-1}dr^2+r^2\,d\Omega^2\ .\end{equation}
This is a static solution to the Einstein–Maxwell field equations that describes an spherically symmetric electrically charged body of mass $M$ and charge $Q$ \footnote{In the weak gravitational field limit we have $-g_{tt} \simeq 1 + 2\Phi_N$, with $\Phi_N$ the Newtonian potential. Then it is see that the electromagnetic field produces a gravitational repulsion by means of the term $+r_Q{}^2/r^2$. This is not an effect of the electromagnetic interaction, since it forms part of the metric, and hence, also acts on neutral particles.}. 
In order to introduce the electromagnetic interaction into the condensate formalism at the level of the action, \eqref{ActionGRChemPot} should be modified by adding the following Lagrangian density
\begin{equation}
	\Delta S_{em}=\int d^4x\sqrt{-g}\,{\cal L}_{em}\ ;\end{equation}
where ${\cal L}_{em}=-F_{\alpha\beta}F^{\alpha\beta}/4$. The corresponding electromagnetic stress–energy tensor is expressed in terms of the field strength $F_{\alpha\beta}=\partial_\alpha A_\beta-\partial_\beta A_\alpha$ as usual
\begin{equation}\label{StressTensorElectro}
	T_{\alpha\beta}=F_{\alpha\sigma}F_\beta{}^\sigma-\frac{1}{4}F_{\rho\sigma}F^{\rho\sigma}\,g_{\alpha\beta} \ .\end{equation}
In the above metric $r_Q{}^2=G_N\,Q^2$ is a characteristic length. The tensor \eqref{StressTensorElectro} describes the electromagnetic field in the outer region and is constructed by means of the electric potential $A_\alpha=\left(Q/r,0,0,0\right)$.

Equivalent arguments as the one discussed in \eqref{SectionOutsideHorizon} leads us to the classical theory in this zone. No perturbations nor chemical potential are present in our visible region; therefore, the Einstein--Maxwell equations are the usual for describing the electromagnetic field in curved spacetime all the way from $r=\infty$ up to the (outer) event horizon located at
\footnote{As long as $2r_Q \leq r_S$; otherwise there can be no physical event horizon. Objects with a charge greater than their mass can exist in nature, but they cannot collapse down to a black hole unless they display a naked singularity.} 
\begin{equation}\label{EventHorizonRN}
	r_+=\frac12\left(r_S+\sqrt{r_S{}^2-4\,r_Q{}^2}\right),\end{equation}
where the metric becomes singular in these coordinates. There is another point where the metric presents a singularity; this is
\begin{equation}\label{EventHorizonRNCauchy}
	r_-=\frac12\left(r_S-\sqrt{r_S{}^2-4\,r_Q{}^2}\right).\end{equation}
This is called the inner horizon. In the region between horizons $r_-<r<r_+$ the metric elements $\tilde g_{tt}$ and $\tilde g_{rr}$ exchange signs.

Let us entertain the possible existence of a condensate beyond the outer event horizon. In this region the chemical potential term should be considered, hence the set of equations to solve are
\begin{equation}\label{EEqRN}
	G_{\mu\nu}(\tilde g+h)=\kappa T_{\mu\nu}+X_{\mu\nu}\ ,\end{equation}
where $X_{\mu\nu}$ is the chemical potential term defined in \eqref{ChemPotTerm}. We shall explore the existence of solutions of the type \eqref{MetricCondensate}. For doing so, we redefine the electromagnetic tensor of the charged black hole in \eqref{StressTensorElectro} with a new unknown factor
\begin{equation}
	F_{\mu\nu}\quad\rightarrow\quad C\,F_{\mu\nu}\ \end{equation}
(this can be understood as a renormalization of the charge $Q$).

It can be seen that a family of solutions describing a condensate exists. Indeed from the angular equations of motion we get
\begin{equation}\label{EinEqRNAngular}
	(1-\varphi)\,\frac{Q^2}{r^4}=-\, C^2\,g^{tt}g^{rr} \,\frac{Q^2}{r^4}\ ,\end{equation}
where the effective metric ($g=\tilde g +h$) is used to raise and lower indices. Because $\tilde{g}^{\,tt}\,\tilde{g}^{\,rr}=-1$
this equation defines the relation of the new  constant $C$  to the value of the condensate
\begin{equation}\label{C}
	C^2=\frac1{1-\varphi}\ .\end{equation}
On the other hand, we have the sector where the chemical potential acts, given by the $(t,r)-\,$sector. Both equations entail the same information and the resulting equation is
\begin{equation}\label{RNEinEqTR}
	-\left(1-\varphi\right)\frac{Q^2}{r^4}-\frac{\varphi}{r^2}=\mu\,\left(1-\varphi\right)^2\varphi+C^2\,g^{tt}g^{rr}\,\frac{Q^2}{r^4}\ .\end{equation}
The value for $C$ in \eqref{C} makes the terms with $Q$ to cancel and this equation exactly reproduces the one
obtained in the Schwarzschild case in equation \eqref{EEqhCte}; hence the chemical potential yields
\begin{equation}
	\mu r^2=-\frac{1}{(1-\varphi)^2}\ .\end{equation}
As we have seen in previous cases, the value of the BEC is in principle arbitrary $0<\varphi<1$ and 
its relation to the chemical potential 
remains unchanged with respect to the Schwarzschild case.

All obvious limiting cases behave as expected. If we turn off the charge, this is $Q=0$, the field equations and their corresponding solutions reduces to the uncharged Schwarzschild case. Also, if we make the chemical potential to vanish, the condensate disappear and we are left with the classical Reissner–Nordström vacuum solution.  
Maxwell equations in a curved vacuum are still satisfied in this solution for any arbitrary $Q$
\begin{equation}\begin{split}
	&{\cal D}^\nu F_{\mu\nu}=0\ ,\\[2ex]&F_{\nu\rho,\mu}+F_{\rho\mu,\nu}+ F_{\mu\nu,\rho}=0\ .\end{split}\end{equation}
The electromagnetic energy $(E^2+B^2)/2$ is scaled by the factor $C^2= 1/(1-\varphi)$ if $\varphi\neq 0$.
This is indeed equivalent to scaling the charge. Notice that the redefinition of the charge does not change the background metric, nor, consequently, the location of the two horizons.

Reissner--Nordstr\"om is not a particularly physically relevant solution as the existence of a macroscopic electrically charged black hole is very unlikely, so its interest is mostly theoretical as it would imply an accumulation of charged matter. Taking this into account one can possibly understand the modification of the charge implied by the redefinition $Q\to CQ$ as the consequence of the fact that the condensate density increases too. This redefinition makes $Q$ to diverge when $\varphi \to 1$.

\subsection{Quasilocal energy of a Reissner--Nordström black hole}\label{SectionQLEScw}

We shall use again the results of section \ref{ChapterQLE}. In this case, the radial metric element is 
\begin{equation}\label{fRN}
	f(r)=\sqrt{1-\varphi}\ \sqrt{\epsilon\left(1-\frac{r_S}{r}+\frac{r_Q{}^2}{r^2}\right)}\ .\end{equation}
In the outer horizon the same change of sign of $\epsilon$ as in the Schwarzschild case takes place. As we move inwards and cross the inner horizon the signature of the $t$ and $r$ coordinates are exchanged again, and $t$ recovers its timelike character.

Including the electromagnetic field into the theory will not change the definition of the quasilocal energy. The quasilocal energy is meant to measure the gravitational energy associated to a specific geometry, and so only the gravitational action is important. Of course, the addition of a new field changes the metric and this is the way that adding the electromagnetism influences the quasilocal energy.

The quasilocal energy in this case is given by 
\begin{equation}
	E(r)=\frac{r}{G_N}\left[1-\epsilon\sqrt{1-\varphi}\ \sqrt{\epsilon\left(1-\frac{2m}{r}+\frac{Q^2}{r^2}\right)}\right]\ ,\end{equation}
with $\epsilon=1$ when $t$ is timelike and  $\epsilon=-1$ when the temporal coordinate is spacelike. A striking feature is that the energy becomes negative for $r< Q^2/2M$. This radius is always inside the inner horizon. The value of the quasilocal energy at the singularity is $E(r=0)=-\abs{Q}$. The singularity at $r=0$ has the electric field of a point charge, and so, using just classical electromagnetism, its self energy should diverge.

\begin{figure}[h]\begin{center}
	\includegraphics[scale=1.1]{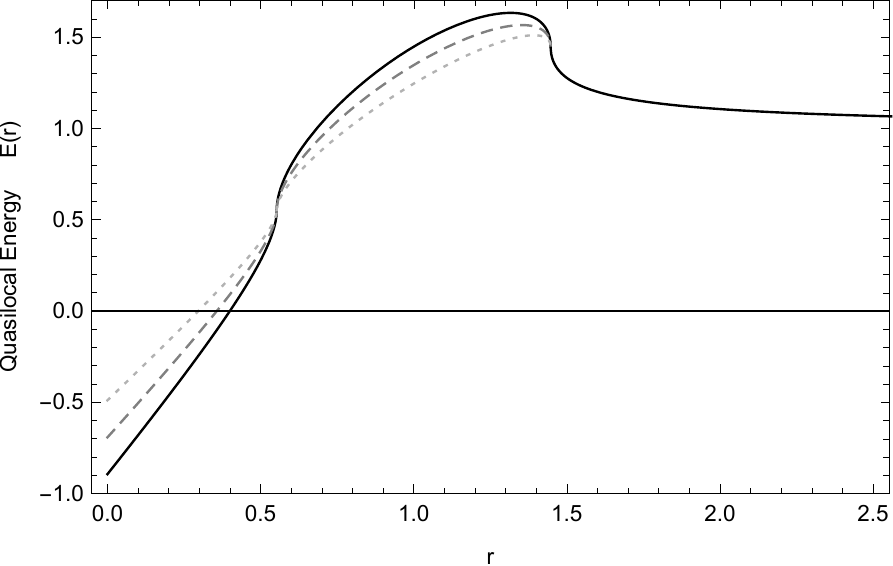}
	\captionsetup{width=0.85\textwidth}
	\caption{\small Quasilocal energy of a Reissner–Nordström black hole, with a charge given by $Q^2=0.8M^2$. Both horizons are clearly identified by the divergence of the derivative of the energy. Inside the external event horizon we compare the classical theory where the wave function is null, $\varphi=0$, with a condensate structure with the same values for $\varphi$ used in the Schwarzschild case. As we can see from the plot, inside the Cauchy horizon (where the $t$ and $r$ coordinates exchange signs again), the condensate structure is not favoured any more; hence the theory dictates a ring condensate between both horizons. Both axes are in units of the mass $M$ of the black hole.}\label{QLE.RN}\end{center}\end{figure}

In Figure \ref{QLE.RN}  the quasilocal energy for these type of black holes is presented. As we see, in the regions $r>r_+$ and $r<r_-$ the usual (no condensate) solution is energetically preferred, while in the region between horizons the opposite behaviour is taken place and a condensate is viable. In particular, inside the inner horizon there should not be a condensate. We conclude that for this case the condensate has the structure of a spherical shell. According to the proponents of the quasilocal energy, this energy includes the contribution of all fields.

The conclusions that can be drawn are in line with those discussed in the previous section for a Schwarzschild black hole.

\vspace{13pt}

\section{Bose--Einstein condensate in de Sitter}\label{dS}

It is quite peculiar for reasons that we will discuss below that also in this case the presence of an event horizon seems associated to a possible Bose--Einstein condensate. Let us repeat the previous calculations. In a de Sitter universe a cosmological term must be included to the Hilbert--Einstein action \eqref{ActionGR}
\begin{equation}
	S_{dS}=M_P{}^2 \int d^4x\sqrt{-g}\,(R-2\Lambda)\ . \end{equation}
Including also the chemical potential action \eqref{ActionCPTerm}, variational principles lead us to the Einstein field equations with a cosmological constant and a chemical potential term which are 
\begin{equation}\label{EEqCC-+++}
	G_{\mu\nu}=-\Lambda g_{\mu\nu} + X_{\mu\nu}\ .\end{equation}
The following maximally symmetric solution
\begin{equation}\label{MetricdS}
	ds^2=-\left(1-\frac{\Lambda}{3}\,r^2\right)\ dt^2+ \left(1-\frac{\Lambda}{3}\,r^2\right)^{-1}\ dr^2+r^2\ d\Omega^2\end{equation}
is obtained in our part of the universe. The reasoning goes as in section \ref{SectionOutsideHorizon}, but using $r=0$ as the Minkowskian location where it is mandatory for the condensate to vanish. As it vanishes at $r=0$ it remains null up to the cosmological horizon, which is located at 
\begin{equation}
	r_\Lambda=\sqrt{\frac3\Lambda}\ .\end{equation}

However, when we go thorough this horizon, it is expected a condensate to be present. Searching for solutions with a nonzero chemical potential, we redefine the cosmological constant term with the help of a new constant $C$: $\Lambda \rightarrow C\,\Lambda$, while the metric is modified to
\begin{equation}\label{MetricFullGenericDS}
	g_{\mu\nu}=\text{diag}\left(\frac{1}{1-\varphi}\,\tilde g_{tt}\,;\,\frac{1}{1-\varphi}\,\tilde g_{rr}\,;\,\tilde g_{\theta\theta}\,;\,\tilde g_{\phi\phi}\right), \end{equation}
as in the Schwarzschild and Reissner--Nordstr\"om cases.

Any of the two angular components of the Einstein's equations fixes this constant with respect to the condensate as
\begin{equation}
	C=1-\varphi \ .\end{equation}
This election for $C$ makes the remaining field equations to take the same form as in \eqref{EEqhCte}; the cosmological contribution inside the Einstein tensor cancel out with the modified cosmological term. As the equation is the same, the structure for the chemical potential remain unchanged with respect to the Schwarzschild case \eqref{ChemPotSolution}. That is
\begin{equation}
	\mu(r) = \frac{\mu_0}{r^2}\ .\end{equation}
In the visible part of our universe there is no condensate. This case is fairly obvious as at $r=0$ the metric is Minkowskian. At $r=r_\Lambda$ there is a discontinuity in the chemical potential. There is a peculiar situation here: contrary to the Reissner--Nordstr\"om case, where the electromagnetic field is enhanced by the condensate structure, the cosmological constant term is screened by the condensate structure. The presence of the graviton condensate decreases the value of the cosmological constant which becomes zero for the limiting value $\varphi=1$.

\subsection{Quasilocal energy analysis over a de Sitter horizon}\label{SectionQLEdS}

\begin{figure}[t!]\begin{center}
	\includegraphics[scale=1.1]{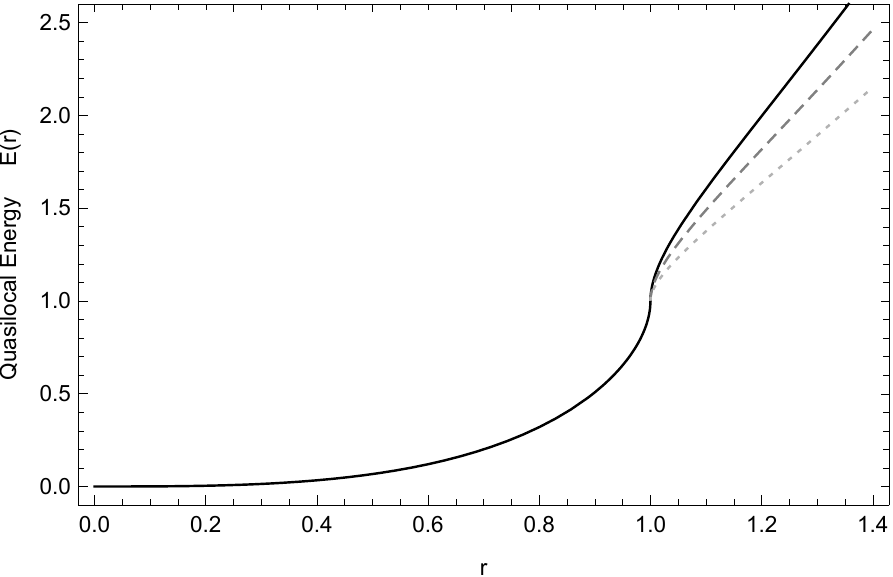}\captionsetup{width=0.85\textwidth}\caption{\small Quasilocal energy for de Sitter spacetime. The axes are in units of the cosmological horizon, i.e. $r_\Lambda=1$. Beyond the horizon we present the curves for the classical solution in solid line and two different condensate values, $\varphi=0.2$ and $\varphi=0.4$, with dashed line and dotted respectively.} \label{QLEdeSitter}\end{center}\end{figure}

The fundamentals of the previous derivations of the quasilocal energy do not change if a cosmological constant is considered. Now $f(r)$ in \eqref{MetricSphSym} is given by
\begin{equation}
	f(r)=\sqrt{1-\varphi}\ \sqrt{\epsilon\left(1-\frac{\Lambda}{3}\,r^2\right)}\ ,\end{equation}
where $\varphi$ is zero in our visible universe. When $\epsilon=1$, $g_{tt}$ is negative and $t$ is a timelike coordinate. At a horizon, $g_{tt}$ and $g_{rr}$ exchange signs and so $\epsilon=-1$. Then
\begin{equation}\label{QLEdS}
	E(r)=\frac{r}{G_N}\left[1-\epsilon\ \sqrt{1-\varphi}\ \sqrt{\epsilon\left(1-\frac{\Lambda}{3}\,r^2\right)}\right]\ .\end{equation}
The quasilocal energy for a de Sitter metric is shown in Figure \ref{QLEdeSitter}. 

As expected, the energy continually grows with increasing $r$. The horizon forms when the energy at the surface is larger than $E(r_c)=r_\Lambda/G_N$. As in the Schwarzschild case, the formation of a condensate diminishes the energy with respect to the vacuum case, so it seems that the condensate structure is favorable against the vacuum itself.

\begin{figure}[!t]\begin{center}
	\includegraphics[scale=1.1]{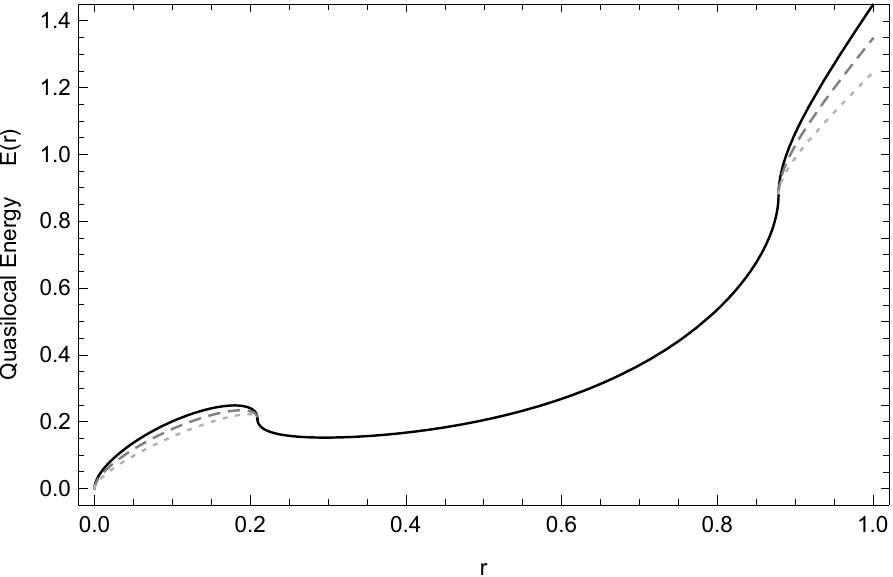}\captionsetup{width=0.85\textwidth}\caption{\small Quasilocal energy for a Schwarzschild black hole embedded in a de Sitter space. The units are such that $r_\Lambda=1$. The black hole mass is unrealistically large, $r_S=0.1\,r_c$ to make details notorious. Beyond both horizons we plot a condensate structure for comparison, with $\varphi=0.2$ and $\varphi=0.4$ in dashed and dotted lines respectively.} \label{QLESchwdeSitter}\end{center}\end{figure}

Just for completeness the solution combining the Schwarzschild and de Sitter spacetimes can be easily computed too. The quasilocal energy is plotted in Figure \ref{QLESchwdeSitter}.
Both horizons, the black hole horizon and the cosmological horizon, can be directly seen in the curve. In both points even though the derivative of the energy matches across the horizon, it is divergent. Also in both points, the horizon forms when the quasilocal energy crossed the value $r/G_N$.

The above picture provides a criterion to discern whether a condensate can be formed or not. Any time the radial coordinate transform its character from spacelike to timelike, when crossing a horizon, a condensate structure is energetically favored with respect to the vacuum solution.

\subsection{The case of the de Sitter horizon}

Before entering in a more detailed discussion about possible physical interpretations in the case
of the cosmological horizon in de Sitter, let us 
analyze this spacetime but using an inverse radial coordinate
\begin{equation}\left\{\begin{split}
	&\quad r=\frac{1}{z} \\[2ex]&\quad dr=-\frac{1}{z^2}\ dz\ .\end{split}\right.\end{equation}
With this change, the metric reads as
\begin{equation}\label{MetricdSz}
	ds^2=-\left(1-\frac{\Lambda}{3}\,\frac{1}{z^2}\right)\ dt^2 +\left(1-\frac{\Lambda}{3}\,\frac{1}{z^2}\right)^{-1}\,\frac{1}{z^4}\ dz^2 +\frac{1}{z^2}\ d\Omega^2\ .\end{equation}
This change of variables maps the cosmological horizon to a concentric sphere centered at $z=0$.
The Einstein tensor is constant in $r$ coordinates, then it is not changed by this change of variables 
and the equations of motion remains the same as well.
Let us now consider the geodesics in this spacetime for straight trajectories characterized by constant angles
$\theta$ and $\phi$. Some Christoffel symbols used to compute geodesics are
\begin{equation}\begin{split}
	&\Gamma_{tt}^z=\frac{3\,\Lambda\,z^2-\Lambda^2}{9\,z}\\[2ex] &\Gamma_{tz}^t=\frac{\Lambda}{3\,z^3-\Lambda\,z}\\[2ex] &\Gamma_{zz}^z=-\frac{6\,z^2-\Lambda}{3\,z^3-\Lambda\,z}\ .\ \end{split}\end{equation}
The geodesic equation is 
\begin{equation}
	\frac{d^2z}{dt^2}=-\frac{3z^2-\Lambda}{9z}\ \Lambda +\frac{ 6z^2+\Lambda}{3z^3-\Lambda z}\left(\frac{dz}{dt}\right)^2\ .\end{equation}
This equation is virtually identical in structure to the one corresponding to a particle moving towards the horizon in a Schwarzschild metric with vanishing angular momentum and, exactly like in that case, the observer sees the particle to slow down as it approaches the horizon that ---at least classically--- never crosses (according to the observer at infinity in the case of Schwarzschild, at $r=0$  in the case of de Sitter, i.e. at $z=\infty$).

Continuing with this analogy we derive the quasilocal energy using the $z$ coordinate where the space beyond the de Sitter horizon is mapped onto a sphere of radius $z_\Lambda= 1/r_\Lambda$. It is given by
\begin{equation}\label{quasilocaldesitterz}
	E(z)=\frac{1}{z}\,\left(1-\sqrt{1-\frac{\Lambda}{3}\frac{1}{z^2}}\right)\end{equation}
for $z> z_\Lambda$ and
\begin{equation}
	E(z)=\frac{1}{z}\,\left(1+\sqrt{1-\varphi}\,\sqrt{\frac{\Lambda}{3}\frac{1}{z^2}-1}\right)\ \end{equation}
for $z< z_\Lambda$.
The derivation for this expression was made directly using the inverse $z$ coordinate. If we go back to the radial coordinate  $z\rightarrow r$ we obtain the expression \eqref{QLEdS} but with a difference of a global minus sign.

Radial geodesics (in the usual coordinate $r$) can be obtained easily from the expression
\begin{equation}
	\dot r^2= \left(1-\frac{\Lambda r^2}{3}\right)^2\frac{\Lambda r^2}{3}\ ,\end{equation}
where $\Lambda r^2/6$ would be the analogous of the Newtonian potential. Note that the usual interpretation that one element of vacuum repeals each other, making the expansion of the universe obvious, is very intuitive in FLRW coordinates, but is not so obvious in the static coordinates used in this section.

In spite of the formal similarity between the expressions for the quasilocal energy \eqref{quasilocaldesitterz} and \eqref{QLESchwOut}, the situation is different because it does not seem possible to derive a potential such as the one in \eqref{potentialquasilocal}. The reason is that \eqref{quasilocaldesitterz} is mass-independent. 

As a consequence, the argument that was presented in the case of a Schwarzschild black hole, where it was argued that the presence of the potential well implied a substantial amount of gravitational radiation that could be stored in the inner part of the black hole in form of a graviton condensate does not hold. In the present case the potential well appears to be absent and consequently there is no substantial energy loss in form of gravitational radiation.

Intuitively, the de Sitter horizon appears to be somewhat different from the one associated to a Schwarzschild black hole (or to a Reissner--Nordstr\"om for that matter). To the external observer at infinity, the horizon at $r=r_S$ has a very clear meaning and objective existence. On the contrary, in the de Sitter spacetime the horizon is dependent on the observer situation (i.e. where the $r=0$ point is assumed to be). However as discussed in some detail in the seminal paper of Gibbons and Hawking \cite{GibbonsHawking} this might not pose any conceptual problem in principle. It is only  measurable effects by an observer what matters. 

As it is well known, there is a temperature associated to the de Sitter horizon, given by \cite{GibbonsHawking}
\begin{equation}
	T=\frac{1}{2\pi}\sqrt{\frac{\Lambda}{3}}\ .\end{equation}
This raises a puzzling possibility: matter falling into the de Sitter horizon in the inverse coordinate $z$; actually accumulating close to the horizon in the static coordinates $(r,t)$. In the Schwarzschild case we concluded that this would most likely imply a growth in the value of the graviton condensate $\varphi$. If this were the case, we immediately notice that because of the change in the cosmological constant beyond the horizon $\Lambda \to C\Lambda$ with $C= 1-\varphi$ the vacuum energy would decrease with time. However, this does not affect neither the position of the horizon, as the change in the metric is merely a multiplicative factor in the time and radial components, nor the value of the cosmological constant in the visible part of the universe.

However, the fact that there is no gravitational binding potential of the form (\ref{potentialquasilocal}), obtained in the Schwarzschild case, would imply that there is no way to create of a gravitational BEC and therefore in spite of being energetically favourable, the value of the condensate should be zero and consequently there would be no change in the cosmological constant neither inside nor outside the de Sitter horizon.

\chapter{Gravitational waves propagating over nonempty backgrounds}
\label{ChapterGW}



\section{The standard cosmological Model}

The $\Lambda$CDM (Lambda cold dark matter) is the simplest parametrization of the Big Bang cosmological model. The three main contributions are the cosmological constant, which is associated with dark energy, the cold dark matter (CDM) and ordinary matter (SM). This model is {\it "almost universally accepted by cosmologists as the best description of the present data"} \cite{Lahav} consistent with the history of our universe, since a very small fraction of a second after the Big Bang until the undergoing accelerated phase, providing a reasonably good description of the various processes occurred along its 13.8 billion years of existence. 
This concordance model describes the cosmic dynamics using the theory of general relativity with its field equations, together with very different branches of physics, from astronomy and thermodynamics to nuclear and particle physics. The success of the model lies in the capability of properly fitting a large variety of observational data.

One of the fundamental pillars upon which the concordance model is built upon is the cosmological principle, which postulates that our location in the universe is not special; the large scale isotropy (it looks the same in all directions) and homogeneity (from every location) are the two key features \cite{Book}. These two constraints can of course only be checked in the region where we have observational access and only in our world line. However no world line is  special: neither ours nor any other for any galaxy. Therefore, isotropy about all galaxies implies homogeneity, and the cosmological principle can be deduced from the isotropic hypothesis, which is exceedingly sustained by radiation backgrounds\footnote{For instance, CMB observations shows very faint anisotropies of the order of $\Delta T /T \sim10^{-4} -10^{-5}$.}. Theoretical predictions of the $\Lambda CDM$ model state that the matter distribution in our universe becomes homogeneous at scales larger than ${\cal R}_H=62.0\,h^{-1}$ Mpc ($\sim100$ Mpc), in close agreement with data. For instance, in \cite{Ntelis} a transition value to cosmic homogeneity of ${\cal R}_H= (63.3\pm0.7)\, h^{-1}$ Mpc is inferred from large scale structures. The values are given in terms of the reduced Planck constant $h = H_0/[100$ km s$^{-1}$ Mpc$^{-1}]$; recall the different values for the Hubble constant discuss below equation \eqref{HubbleConstantGeneric}.

Different collaborations studying various cosmological phenomena \cite{LambdaValue}, from temperature fluctuations measurements to high-redshift supernovae observations, CMB and BAO measurements, weak lensing techniques (such as cosmic shear and redshift space distortion), etc. agrees on a structure for the metric tensor accomplishing the cosmological principle. The model uses the FLRW metric\footnote{At this point we change to the $(+---)$ signature to follow the usual standards in cosmology.}, which is written as
\begin{equation}\label{MetricFLRWGeneric}
	ds^2=dT^2-a^2(T)\left(\ \frac{dR^2}{1-k\ \tfrac{R^2}{R_0{}^2}}+R^2\ d\Omega^2\,\right)\ .\end{equation}
The coordinate $T$ refers to the cosmic time and with the radial coordinate ranging from $0\leq R<R_0$, we get the well known comoving coordinates; we will refer to them with capital letters: $(T,R,\theta,\phi)$. $a(T)$ is the scale factor, a dimensionless quantity which gives a mapping between the theoretical predictions and the observations; i.e. between the (constant) comoving distance and the (physical) proper distance at a given cosmic time. The usual election of $a(T_0)=a_0=1$ at the present age of the universe has been taken as the reference time. The curvature parameter $k$ can take three different values, $+1$, $-1$ or $0$, depending on the geometry of the spacetime, whether it is closed, open or flat respectively.

Suppose that the history of the universe is reasonably well described by the Einstein's equations with a cosmological constant \eqref{EEqCC}. At a background level we can treat the components of the universe as {\it some kind of fluids} uniformly distributed along the whole space. In thermodynamic equilibrium, the energy momentum tensor for each component can be expressed as
\begin{equation}\label{StressTensorPF}
	T^{\alpha\beta}=\left(\rho+P\right)U^\alpha U^\beta-P\,g^{\alpha\beta}\ ,\end{equation}
in the time-positive metric signature tensor notation. $U^\alpha$ is the 4-velocity of the fluid with respect to the observer; the energy density $\rho$ and the isotropic pressure $P$ are measured in the rest frame of the fluid. The recently introduced FLRW comoving coordinates $(T,R,\theta,\phi)$, belong to the so-called rest frame and implies $U^{T}=U_{T}=1$, whereas $U_{R}=U_\theta=U_\phi=0$. The normalization condition $g_{\alpha\beta}U^\alpha U^\beta =1 $ is fulfilled. In order to determine the time evolution of the scale factor, Einstein field equations \eqref{EEqCC} are required together with some equations of state that relate the pressure and the density for every single component present; these are
\begin{equation}\label{EoS}
	P_i= \omega_i\ \rho_i\ .\end{equation}

Using the recently introduced geometric properties of homogeneity and isotropy with the metric \eqref{MetricFLRWGeneric} and  the aforementioned energy-momentum tensors \eqref{StressTensorPF}, the field equations are
\begin{equation}\label{FriedmannEc}\begin{split}
	\left(\ \frac{\dt a}{a}\ \right)^2+\frac{k}{a^2}-\frac\Lambda3=\frac\kappa3\ \rho_m&\\[3ex]\frac{\ddt a}a-\frac\Lambda3 =-\frac\kappa6\ ( \rho_m+3\,P_m&)\ .\end{split}\end{equation}
They are the so-called Friedmann equations. The dot stands for derivatives with respect to the cosmic time $T$. It is usually useful to work with the Hubble function defined as $H=\dt a/a$. At this point when we talk about matter, we are not distinguishing between nonrelativistic matter (baryonic or dark matter) and relativistic matter (radiation). Further on we will study each component separately.

\begin{figure}[!t]
	\begin{minipage}[c]{0.45\textwidth}
	\includegraphics[scale=0.72]{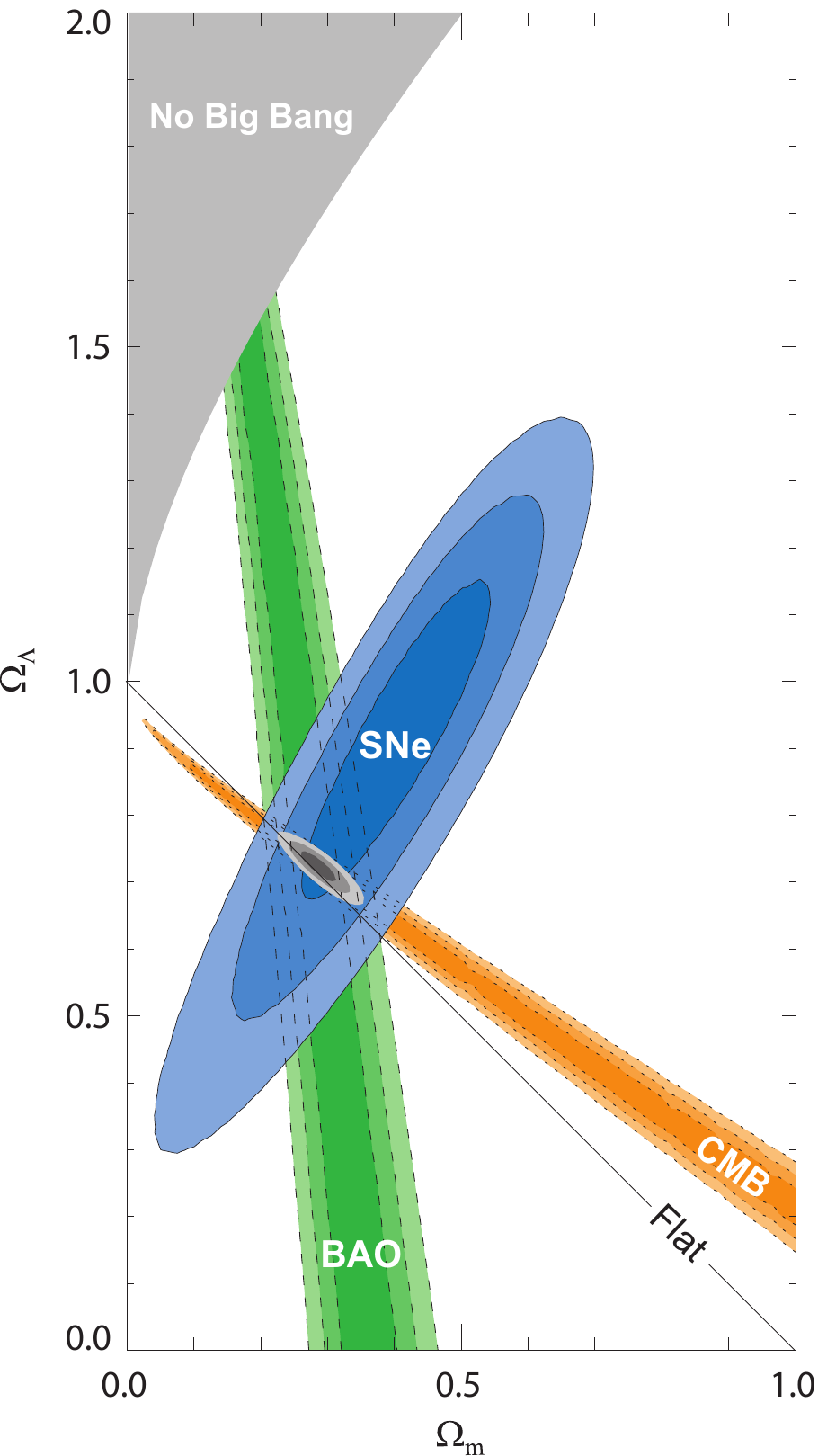}\end{minipage}\hfill
	\begin{minipage}[c]{0.64\textwidth}\vspace{-10pt}
	\caption{\small $\Omega_\Lambda-\Omega_{m}$ plane from the CMB, BAO and the Union SNe Ia set, together with their combination \cite{Kowalski}. The confidence level contours of 68.3\%, 95.4\% and 99.7\% are shown in color gradient. The values above the "Flat" line indicate a closed universe. These observations appear to point towards $\Omega_\Lambda+\Omega_m\sim1$ within a small percentage, meaning that the universe seems to be approximately flat on the scale of our horizon. More precisely, the {\it spatial curvature density} today \eqref{DensityCurvature} indicates that the radius of the spacial universe is at least about one order of magnitude larger than the radius of the visible universe. 
	Clearly, we are surely influenced by our experience in Earth, which is flat within our visible horizon. The impossibility of conceiving a flat infinite Earth is surely a well-founded prejudice on our unnatural perspective for a perfectly flat universe. } \label{FigFlat}\end{minipage}\end{figure}

It is customary to divide the first equation in \eqref{FriedmannEc} by the square of the Hubble parameter today, $H(a_0)=H_0$, and define the following density parameters
\begin{align}\label{DensityParamMatter}
	&\Omega_m=\frac{\kappa\,\rho_m}{3\,H_0{}^2}\\[2ex]\label{DensityParamCC}
	&\Omega_{\scriptscriptstyle\Lambda}=\frac\Lambda{3\,H_0{}^2}\\[2ex]\label{DensityCurvature} &\Omega_k=-\frac{k}{a^2\,H_0{}^2}\ . \end{align}
The sum of the three equals unity. Each density parameter corresponds to the ratio between the actual (or observed) energy density $\rho_i$ and the critical energy density $\rho_c=3H_0{}^2/\kappa$; this is $\Omega_{i}=\rho_i/\rho_c$. From Figure \ref{FigFlat} it is seen that observations favour a flat universe within a small amount $\Omega_k\sim0$. These analysis assumed $\omega_{\scriptscriptstyle\Lambda}=1$ for the equation of state parameter for the cosmological constant \cite{Peebles}, along this lines the same assumption has been made. While current observations are consistent with $k=0$, they are very far from selecting out only this value.

For the sake of simplicity and in accordance with future approximations, we assume $k=0$ and consider the remaining two parameters \eqref{DensityParamMatter} and \eqref{DensityParamCC}. 
At this point, we can distinguish different types of matter characterized by different cosmological parameters $w_i$ in \eqref{EoS}. Each contribution has a weight in the effective matter density parameter
\begin{equation}\label{DensityParamNR}
	\Omega_m\ =\ \Omega_{nonrelativistic\ matter}\ +\ \Omega_{relativistic\ matter}\ .\end{equation}
Several independent techniques like cluster mass-to-light ratios \cite{Carlberg-97}, baryon densities in clusters \cite{BaryonDensity}, weak lensing of clusters \cite{WeakLensing}, the existence of massive clusters at high redshift \cite{ClusterHR} constrain the model such that baryons account for the $\sim 5\%$ energy content of the universe, whereas dark matter for the $\sim 25\%$.
Relativistic matter encompasses the measure of the present mass in the relativistic $3\ K$ thermal cosmic microwave background radiation that almost homogeneously fills the space, together with the accompanying low mass neutrinos, accounting for an almost negligible contribution of
\begin{equation}\label{DensityParamR}
	\Omega_{rel}\ \sim\ 10^{-4}\,-\,10^{-5}\ .\end{equation}
The remaining component corresponds to dark energy, which is roughly 30 times bigger than the observed matter energy \cite{Hinshaw}.
We live in the precision Cosmology era and the $\Lambda$CDM model is still resisting the new generation of galaxy surveys and CMB probes.

The equation of state \eqref{EoS} for the main components of the universe provides enormous information. The accepted values for the state parameters of the three main contributions of the model are
\begin{eqnarray}\label{StateParameters}\begin{split}
	&\omega_{\scriptscriptstyle\Lambda}&=&&-1&\\[2ex] &\omega_{nr}&=&&0&\qquad(\text{presureless})\\[2ex] &\omega_{rel}&=&&\frac13&\ .\end{split}\end{eqnarray}
The possibility of relating the pressure and the density separately for each components states that the second equation in \eqref{FriedmannEc} can be rewritten as
\begin{equation}\label{FriedmannEc2}\begin{split}
	\frac{\ddt a}a=-\frac\kappa6\ \sum_{i=1}\ (\rho_i+3\ P_i)=\frac\kappa6\ (2\rho_{\scriptscriptstyle\Lambda}- 2\rho_r-\rho_{nr}&)\ .\end{split}\end{equation}
We favour the use of the density of the cosmological constant, rather than the cosmological constant itself. We also distinguish the relativistic from the nonrelativistic matter (either baryonic or non-baryonic). Interestingly, this equation that describes the acceleration of the universe clearly brings that $\rho_{\scriptscriptstyle\Lambda}$ speeds up the expansion, whilst the matter content slows it down. Generally, any background component satisfying $\rho_i+3\ P_i<0$ contributes positively to the acceleration.

With the purpose of discussing a novel effect coming from the fact that gravitational waves propagate over an expanding background. Next section reviews previous results about this effect obtained in a theoretical de Sitter universe. This line of research argues that this effect seems to have a window to be measured in Pulsar Timing Array (PTA) observations; in fact a detailed characterization of the expected signal in these observations is presented in Chapter \ref{ChapterPTA}. In section \ref{SectionLinGrav} we employ some basic ingredients of the $\Lambda$CDM cosmology recently reviewed to study the linearization of Einstein's equations and argue why this is the proper conceptual framework. Next, in section \ref{CosCte}, we give the main issues related to the choice of the coordinate system for the de Sitter case. This discussion actually contains the essential ingredient of the present study. 
Later we consider the case where there is no cosmological constant but only non-relativistic matter; and finally in section \ref{SectionGWLCDM}, we extend our considerations to the combined  $\Lambda$CDM case of two main components (with known analytic solution): dark matter and dark energy. We have derived the coordinate transformation between spherically symmetric coordinates centered at a coalescence of two black holes where gravitational waves are produced, and FLRW expanding coordinates, where gravitational waves propagate and are measured. One step further is given in section \ref{SectionRelMatt} where the same procedure is performed for relativistic matter. This allows one to infer a linearized picture in terms of the Hubble parameter $H_0$ which would take into account all constituents of the universe as a background entity in the last section.

In this manner, we have found a direct link to see whether this effect has a realistic correlation with current measurements and describe our universe in a realistic way. In this chapter we focus on the more fundamental aspects concerning the definition of and relation between the different coordinate systems, while we leave the discussion of their impact in PTA data for next chapter.

\section{The importance on the choice of coordinate systems} \label{SectionChoiceCtes}

In \cite{EspriuCC,E2} the influence of a non-vanishing cosmological constant on the detection of gravitational waves in PTA was discussed\footnote{That is, beyond the frequency redshift due to the expansion of the universe.}. It was found that the value of the timing residuals observed in PTA for a given range of angles subtended by the source and the pulsars were dependent on the value of $\Lambda$. Given that the gravitational waves potentially observed in PTA would correspond to the final phase of the fusion of two very massive black holes lurking at the center of two colliding galaxies at relatively low redshift, this result opens the possibility of `local' measurements of the cosmological constant; namely at subcosmological distances.

To simplify things, imagine a large mass $M$ and a much smaller mass $m$ orbiting around it.  For the time being let us assume that there is no cosmological constant and no matter density. In the situation just described, the geometry is very approximately described by the well known Schwarzschild metric corresponding to a mass $M$ \cite{Schwarzschild}. This metric is expressed in terms of the radial coordinate $r$ and a time coordinate $t$. This time coordinate $t$ is just the time ticked by a clock located at $r\to \infty$, i.e. an observer at rest at infinity.

Next let us assume that a cosmological constant $\Lambda \neq 0$ is present. In this case the relevant metric is also well known \cite{Kottler}; the Schwarzschild--de Sitter (SdS) metric:
\begin{equation}\label{MetricSdS}
	ds^2=\left(1-\frac{2G_NM}{r}-\frac{\Lambda}{3}\,r^2\right)\ dt^2-\left(1-\frac{2G_NM}{r}-\frac{\Lambda}{3}\,r^2\right)^{-1}\ dr^2-r^2\ d\Omega^2\ . \end{equation}
This metric is unique once the requirements of spherical symmetry and time independence of the metric elements are imposed \cite{Birkhoff}. It exhibits two horizons at
\begin{equation}
	r_S=2G_NM\hspace{40pt} \text{and} \hspace{40pt}r_\Lambda=\sqrt{\tfrac{3}{\Lambda}}\ .\end{equation}
Here $r$ has the same physical realization as in the case of the Schwarzschild metric mentioned above, but the metric 
is not anymore Minkowskian when $r \to \infty$ so one 
must understand carefully what the meaning of the time coordinate $t$ is. In this work we will consistently denote with lower case letters, i.e. $(t,r,\theta,\phi)$, the coordinates relevant in the location where
gravitational waves are emitted.

In the situation just described, if gravitational waves are produced it is obvious that they will be periodic in the time coordinate $t$ (see also the discussion in section \ref{SectionLinGrav}), and at a distance sufficiently large from the source but very small compared to cosmological distances they will be described by (a superposition of) approximately harmonic functions of the form
\begin{equation}\label{gw1}
	h_{\mu\nu}(w;t,r)= \frac{e_{\mu\nu}}{r}\,\cos\left[w\,(t-r)\right]\ ,\end{equation}
with $e_{\mu\nu} $ being the polarization tensor. This is so because the metric is dominated by the strong gravitational field created by the mass $M$, and in its vicinity the contribution of the $\Lambda r^2$ term (that produces very small corrections in the orbit parameters) is totally negligible. On the other hand, determining the precise form of the front wave (namely the actual superposition of harmonic functions) is a non-trivial task that in general requires a detailed analysis. Here we assume that the gravitational wave emission is strongly dominated by a single harmonic with a frequency that is governed by the period of rotation. There would be no problem to include in our considerations a dependence of the amplitude on $t$, particularly relevant in the last stages of the spiraling down process, but we do not consider it here for simplicity.

If at long distances from the source the universe is approximately Minkowskian, as it would be the case if the metric is globally Schwarzschild, this wave front would be seen by a remote observer with exactly the same functional form. This is not the case when a cosmological constant or dust are present. The universe cannot be considered to be globally Minkowskian although at very short distances this is a good approximation, except in the case of strong gravitational sources being present. In the vicinity of the very massive object of mass $M$, Einstein's equations together with symmetry considerations dictate the form of the solution.

However, as discussed in section \ref{SecGWCC} our spacetime is not Minkowskian far away from the gravitational wave source source and we know that the distribution of matter and energy governs the way clocks tick. Exactly as clocks in the vicinity of a strong gravitational field created by a very massive object signal a time $t$ to a remote observer, at cosmological scales the universe being homogeneous and isotropic dictates at what rate comoving clocks, far from strong gravitational sources,  tick. We know that the latter physical situation is described by a FLRW metric described by coordinates $(T,R,\theta,\phi)$
\begin{equation}\label{MetricFLRWGenericK0}
	ds^2=dT^2-a^2(T)\left(\ dR^2+R^2\ d\Omega^2\ \right)\ .\end{equation}
Then a cosmological observer located very far away from the black hole merger will not see the same functional dependence in the wave front simply because $T\neq t$ and $R\neq r$. 

It is particularly interesting to consider the case where only a cosmological constant is present because then the coordinate transformation between the two coordinate systems is precisely known when one is placed far away from the source of the gravitational field \cite{Espriu}, as is obviously the case if $r \gg 2MG$. The universe is globally de Sitter; SdS and FLRW coordinates correspond to slicing the same spacetime in two different ways. Both are solutions very far from the source. Clocks tick with time $t$ near the large mass $M$ and with time $T$ at the cosmological distances where we observe the phenomenon. In both cases the distribution of matter and energy dictates what time is. 

The cosmological time and comoving space coordinates will in fact be non-trivial functions of the SdS coordinates i.e. $T(t,r)$, $R(t,r)$. These transformations are well known in the case where the universe is globally de Sitter, and will be reviewed in the next sections. Adding matter further modify these transformations.  As a consequence of the different coordinate systems, some anharmonicities appear when (\ref{gw1}) is written in terms of the observer coordinates $T$, $R$. Given that the cosmological constant is small, these anharmonicities are numerically small too, but the results of \cite{EspriuCC} suggest that nevertheless they could be measurable in a realistic PTA observation.

Conventionally, the expansion of the universe is taken into account by replacing $w$ by its redshifted counterpart in \eqref{gw1}. If $z$ is small
\begin{equation}\label{Frequency}
	w^\prime \simeq w\,(1-z)\ ,\end{equation} 
with $z$ being the redshift factor. We will see that this effect on the frequency is indeed reproduced, but there is more than this. 

Let us justify why in PTA observations the coordinates that an astronomer uses to study the phenomenon are (up to some modifications described in Chapter \ref{ChapterPTA}) $T$ and $R$ to a high degree of accuracy. Galaxies are following the cosmic expansion with respect the source of the gravitational waves and therefore the relevant change of coordinates for an observer sitting at, say, the center of mass of a galaxy would be the one relating $(t,r)$ with $(T,R)$. Needless to say, gravitational fields inside the galaxy locally modify this relation. These modifications are truly negligible, except for those that will be considered in the next chapter. Therefore in this work we will consistently denote the coordinates used  by an Earth-bound observer by $(T,R)$. Naturally, the Earth and pulsars are gravitationally bound to the Galaxy and the distance of a given pulsar to an observer on the Earth is a constant distance $L$ in coordinate $R$. On the other hand, the gravitational wave source is not gravitationally bound, so it feels the expansion of the universe. In short, the effect that will be discussed and analyzed in this work has nothing to do with an hypothetical expansion of the distance between the pulsar and the observer (which is not present) but rather with the non-trivial transformation relating coordinates $(t,r)$ and $(T,R)$.       

The  results obtained in \cite{EspriuCC,E2} were not yet directly applicable to a realistic measurements of PTA timing residuals because at the very least non-relativistic dust needs to be included to get a realistic description of the cosmology. This is in fact the main purpose of the following part of this thesis.

\section{Linearized gravity framework}\label{SectionLinGrav}

The linearized theory of gravity is nothing else than perturbation theory around Minkowski spacetime. This is, starting with a small deviation  $|h_{\mu\nu}|\ll1$ from a flat metric tensor,
\begin{equation}\label{MetricetricPertMink}
	g_{\mu\nu}=\eta_{\mu\nu}+h_{\mu\nu}\,.\end{equation}
The usual procedure for obtaining the linearized Einstein tensor can be found in any textbook on general relativity (see e.g. \cite{Cheng}). However, a gauge choice is mandatory to obtain a solution to these equations in order to avoid redundancy under coordinate transformations $x^\mu\rightarrow x^\mu+\xi^{\mu}(x)$. 
Although the following discussion is valid in any gauge, it is simplest in the familiar Lorenz gauge
\begin{equation}\label{LorenzGaugeCond1}
	\partial_\mu h^\mu{}_\nu - \frac{1}{2}\partial_\nu h=0 \,,\end{equation}
or defining the trace-reversed perturbation variable
\begin{equation}\label{LorenzGaugeCond2}
	\tilde h_{\mu\nu}=  h_{\mu\nu} - \frac{1}{2}\,h\,\eta_{\mu\nu} \quad \Rightarrow \quad \partial_\mu \tilde h^\mu{}_\nu =0\,.\end{equation}
In this gauge the linearized Einstein's equations, including the cosmological constant and the energy momentum tensor read
\begin{equation}\label{FullLin}
	\Box \tilde h_{\mu\nu}=-2\,\Lambda\eta_{\mu\nu} -2\,\kappa\,T_{\mu\nu}\ .\end{equation}
As discussed in \cite{Espriu}, up to first order in $\Lambda$ the perturbation $h_{\mu\nu}$ in equation. \eqref{MetricetricPertMink} can be decomposed into a gravitational wave perturbation $h_{\mu\nu}^{\scriptscriptstyle (\mathrm{GW})}$ and a background modification due to the cosmological constant $h_{\mu\nu}^{\scriptscriptstyle (\Lambda)}$, being both {\it small} in magnitude. It is straightforward to propose a new contribution due to dust $h_{\mu\nu}^{\scriptscriptstyle (\mathrm{dust})}$. 
This component would contribute as an independent new background that will not interact with the other one, the one 
corresponding to $\Lambda$, and that will be also {\it small} enough so that the linearized approximation to be meaningful. 
Therefore, the total metric would be written as
\begin{equation}
	g_{\mu\nu}=\eta_{\mu\nu}+h_{\mu\nu}^{\scriptscriptstyle (\mathrm{GW})}+h_{\mu\nu}^{\scriptscriptstyle (\Lambda)}+h_{\mu\nu}^{\scriptscriptstyle (\mathrm{dust})}\,.\end{equation}
The first two contributions have been already discussed in \cite{Bernabeu}. The two independent background would satisfy each a simpler equation of motion
\begin{equation}\label{LinLam}
	\Box \tilde h_{\mu\nu}^{\scriptscriptstyle (\Lambda)}=-2\,\Lambda\eta_{\mu\nu}\end{equation}
for the cosmological constant and 
\begin{equation}
	\Box \tilde h_{\mu\nu}^{\scriptscriptstyle (\mathrm{dust})}=-2\,\kappa T_{\mu\nu}\end{equation}
for a dark matter background uniformly distributed along the spacetime. Finally the perturbation due to
the gravitational wave itself satisfies the usual homogeneous wave equation
\begin{equation}
	\Box \tilde h_{\mu\nu}^{\scriptscriptstyle (GW)}=0\,.\end{equation}
In the previous discussion we have not included the modification due to the gravitational field created by the very
massive object of mass $M$ as this is negligible at large distances, but the perturbation is also additive as is easily
seen by expanding Schwarzschild in powers of $G_N$.
Thus the discussion is easiest in linearized gravity. The contribution of the different terms simply adds to construct the total metric. And because the deviations with respect to Minkowski spacetime are small away from the gravitational wave source, linearization is a priori well justified. 

If we stress the importance of the linearization process it is because it is not always possible  to accomplish in all coordinate systems. For instance, it is impossible in cosmological coordinates. This will be discussed in more detail in section \ref{CosCte}. As a consequence {\em there are no harmonic gravitational waves waves} in FLRW coordinates\footnote{This statement remains true even after taking into account the $R-$dependent redshift.}. In SdS coordinates, however, linearization is possible \cite{Espriu,EspriuCC}, and linearization remains valid after including dust, which is one of the main results of this chapter.

The solution of equation \eqref{LinLam} as explained in \cite{Espriu} should correspond to the linearization (i.e. expanding in $\Lambda$ at first order) of equation \eqref{MetricdS}; i.e. the metric \eqref{MetricSdSCCAprox}. In fact this statement is not totally correct because in \eqref{FullLin} the Lorenz gauge condition is assumed, while the SdS metric does not fulfill this gauge condition, as it can be easily checked.
In fact, as explained in \cite{Espriu} a trivial coordinate transformation turns the SdS metric into one that complies with the Lorenz gauge. This coordinate transformation is (a) time-independent and (b) is of order $\Lambda$. The discussion on gravitational waves and the separation in background and waves is however simplest in the Lorenz gauge and it is the one just presented. In fact, the relevant effects to be discussed later are all of order $\sqrt{\Lambda}$ and order $\Lambda$ corrections are safe to be neglected.

\section{Gravitational wave propagation in de Sitter}\label{CosCte}

De Sitter geometry models a spatially flat universe and neglects matter, so the dynamic of the universe is dominated by the cosmological constant $\Lambda$. Obviously there exist a number of useful coordinate choices for this spacetime. These consist in picking a convenient time choice and thus defining a family of spacelike surfaces. Some slices make explicit a cosmologically expanding space with flat curved spatial part, whereas some other do not.

There is as well a unique choice of coordinates in which the metric does not depend on time at all; as is described in equation \eqref{MetricSdS} if the black hole contribution were omitted
\begin{equation}\label{MetricdS+---}
	ds^2=\left(1-\frac{\Lambda}{3}\,r^2\right)\ dt^2 -\left(1-\frac{ \Lambda}{3}\,r^2\right)^{-1}\ dr^2-r^2\ d\Omega^2\ .\end{equation}
The mere existence of such a choice is enough to tell us that there is no fundamental sense in which this is an expanding cosmological spacetime. The notion of expansion is a concept that is linked to a coordinate choice. Yet, the choice of coordinates is not totally arbitrary. Observers' clocks tick at a given rate, and they can measure the local space geometry at a fixed time. 

Cosmological observers detect that the universe as we see it is spatially flat and that gravitationally unbound objects separate from each other at a rate that is governed by the cosmological scale factor. $a(T)$ defines the metric \eqref{MetricFLRWGenericK0} in cosmological coordinates and the first --flat, i.e. with $k=0$ in \eqref{DensityCurvature}-- equation of motion from \eqref{FriedmannEc} with $\rho_m=0$, is a sufficient condition to get
\begin{equation}\label{ScaleFactorDE}
	\frac{\dt a(T)}{a(T)}\equiv H=\sqrt{\frac{\Lambda}{3}} \hspace{40pt}\Longrightarrow\hspace{40pt} a(T)=a_0\ e^{\sqrt{\frac{ \Lambda}{3}}\ \Delta T}\ .\end{equation}
The cosmological constant sets the expansion rate, given by the Hubble parameter $H$. $\Delta T=T-T_0$ is the time interval and $T_0$ is some initial time which is arbitrary and associated with the initial choice of the scale factor such that $a(T_0)=a_0$.

Then the physical coordinate system for discussing cosmology is given by the FLRW metric with coordinates $(T,R,\theta,\phi)$, as this coordinate system encodes the observed homogeneity and isotropy of the matter sources summarized in the cosmological principle \cite{Book}. Of course, space redefinitions of the metric are innocuous. It is of no consequence to choose to describe the world around us using Cartesian or polar coordinates. As long as coordinate transformations do not involve time in any essential way, any coordinate choice will be equally good to describe a universe conforming to the cosmological principle.

Time and space will however look very different to an observer in the vicinity of a black hole or, for that matter, in the vicinity of any strong gravitational source that is spherically symmetric, centered at one point that we conventionally denote by $r=0$. The spacetime there is isotropic, but not homogeneous. If the black hole is static, we know that the solution is unique and it is given by the Schwarzschild metric. If in addition, there is a cosmological constant, the corresponding metric will be the one given in equation \eqref{MetricSdS}.

The SdS metric approximates a Schwarzschild space for {\it small} $r$; and for {\it large} $r$ the space approximates a de Sitter space. We are focusing right now on effects on gravitational wave propagation, hence we are in a {\it large} $r$ regime. Far away from a SMBH with a typical Schwarzschild radius of $\sim 10^{11}$ m, the term $r_S/r$ can be safely neglected from our considerations and only the $\Lambda$ dependence needs to be taken into account; i.e. under this regime the metric is given by \eqref{MetricdS+---}. In this case the transformation relating the SdS coordinates to the FLRW ones are well known (see e.g. \cite{Espriu})
\begin{equation}\label{ChVarCCExact} \left\{\begin{split} \hspace{7pt}
	t(T,R)&=T-\sqrt{\frac{3}{\Lambda}}\,\log\sqrt{1-\frac{\Lambda}{3}\,a^2(T)\,R^2}\\[2ex]\hspace{7pt}r(T,R)&=a(T)\,R=a_{0}\,e^{\sqrt{\frac{\Lambda}{3}}\,\Delta T}\,R \ ;\end{split}\right.\end{equation}
the $\theta$ and $\phi$ coordinates do not transform. Note that even though for $r\gg r_S$ the influence of the black hole is negligible, its presence sets up a global coordinate system, and $r$ and $t$ have a well defined physical meaning.

The SdS metric in the {\it far away from the source} limit can be linearized  expanding \eqref{MetricdS+---} in powers of $\Lambda$. For $r\to \infty$, but $\Lambda r^2 \ll 1$ (i.e. well before the cosmological horizon)
\begin{equation}\label{MetricSdSCCAprox}
	ds^2=\left(1-\frac{\Lambda}{3}\,r^2\right)\ dt^2-\left(1+\frac{ \Lambda}{3}\,r^2\right)\ dr^2-r^2\ d\Omega^2\ .\end{equation}
This metric satisfies a linearized version of Einstein field equations \eqref{EEqCC}. On the contrary, the FLRW metric does not fulfill any linearized version of Einstein's equations\footnote{This is easy to understand by realizing that in the FLRW metric $\Lambda$ appears in a non-analytic form.}; in fact, no metric that depends only on time can be solution of the linearized Einstein's equations with a cosmological constant, whatever the gauge choice selected.

If linearization were possible in FLRW coordinates, gravitational waves could be harmonic in those coordinates; i.e. they could obey a (flat) wave equation. But this as we have seen is impossible. As a consequence, gravitational waves are {\em necessarily} anharmonic in FLRW coordinates. Therefore the proper procedure is to find a coordinate system where linearization is possible and the wave perturbation and the background can be treated additively. Luckily this is accomplished in SdS coordinates because they are analytic in $\Lambda$ and have a smooth Minkowski limit. Linearization and, consequently, harmonic waves remain valid when dust is included. The change from SdS to FLRW is necessarily not a "soft" one --it must have a singularity somewhere, and it does. 

If one attempts to solve directly the wave equation with a non-zero cosmological constant in FLRW using perturbation theory in $\Lambda$, one generates secular terms that grow with $(T,R)$ and invalidate perturbation theory.    

The coordinate transformations relating SdS and FLRW can be expanded for small values of $\Lambda T^2$  or $\Lambda R^2$ (this will always be the relevant situation in the ensuing discussions)
\begin{equation}\label{ChVarCCLinear} \left\{\begin{split} \hspace{7pt}
	t(T,R)&\approx T+a_{0}^2\left(\frac{R^2}{2}\sqrt{\frac{\Lambda}{3}} +\Delta T\,R^2 \,\frac{\Lambda}{3}\right)+\dots\\[2ex]
	r(T,R)&\approx a_{0}\left[1+\Delta T\,\sqrt{\frac{\Lambda}{3}}+\Delta T^2\,\frac{\Lambda}{6}\right]R+\dots\end{split}\right.\end{equation}
The change is of course still non--analytical in $\Lambda$. Applying this approximate transformation to the SdS metric, one ends up with an approximate version of the FLRW metric
\begin{equation}
	ds^2\approx dT^2-a_{0}^2\left(1+2\,\sqrt{\frac{\Lambda}{3}}\,\Delta T+2\,\frac{\Lambda}{3}\,\Delta T^2\right) \left(dR^2+R^2\ d\Omega^2\right)\end{equation}
which is the expansion of (\ref{MetricFLRWGeneric}) for small values of the cosmological constant. In practice, the leading $\sqrt\Lambda$ term will be the relevant one for our purposes. The extension of these considerations to the case where a cosmological constant and non-relativistic matter coexist comes next.

\section{Non-relativistic matter}\label{SectionNoRelMatt}

First we would like to repeat the discussion previously done in section \ref{CosCte} in the case where there is no cosmological constant, just dust. Namely we ask ourselves what would be the equivalent of equation \eqref{ChVarCCExact} in such a case.

According to the discussion around \eqref{DensityParamNR}, we will model the non-relativistic matter background as a pressureless {\it dust-like} perfect fluid, with a state parameter given in \eqref{StateParameters}. The gravitational field is produced entirely by matter characterized by a positive mass density, given by the scalar function $\rho_{\scriptscriptstyle dust}$, but with the absence of pressure. The stress-energy tensor is 
\begin{equation}\label{EMTensorDust}
	T^{\mu\nu}=\rho_{\scriptscriptstyle dust}\,U^\mu U^\nu\,.\end{equation}

As said above, in FLRW comoving coordinates the four--velocity is given by $U^\mu=(1,0,0,0)$; so the stress-energy tensor reduces to $T^{\mu\nu}=diag(\rho_{\scriptscriptstyle dust},0,0,0)$. The Einstein's equations derived for this metric structure together with the equation of state corresponding to $\omega_{\scriptscriptstyle dust}=0$, give the solution \cite{Weinberg} 
\begin{eqnarray}\label{ScaleFactorDM}\begin{split}
	&a(T)&=&\quad a_{0}\left(\frac{T}{T_{0}} \right)^{2/3}\ , \\[2ex] &\rho_{\scriptscriptstyle dust}(T)&=&\quad \frac{4}{3\,\kappa\,T^2}\ .\end{split}\end{eqnarray}
Therefore, we are able to write the scale factor and the metric with a density dependence exclusively
\begin{equation}\label{MetricFRWdensity}
	ds^2=dT^2-a_{0}^2\ \left[\frac{\rho_{d{\scriptscriptstyle 0}}}{\rho_{\scriptscriptstyle dust}(T)}\right]^{2/3}\, \left(dR^2+R^2\ d\Omega^2\right)\ .\end{equation}
Here, $\rho_{d{\scriptscriptstyle 0}}=4/(3\,\kappa\,{T_{0}}^2)$ is the density measured at a particular cosmological time $T_{0}$; the latter usually set according to a scale factor such that $a_{0}=1$.

With the aim of deriving a full picture on how dust influences the observation of gravitational waves, we need to find a coordinate system where dust is described in `static' spherically symmetric coordinates with an origin coincident with the object emitting the gravitational waves. The first step to find this (spherically symmetric) set of coordinates is to impose on the angular element that in FLRW reads as $a(T)^2 R^2 d\Omega^2$, to become simply $r^2d\Omega^2$. The coordinate transformation will be given by
\begin{equation}\label{ChVarDMUnknown}\left\{\,\begin{split}
	\hspace{7pt}T(t,r)&=\text{unknown\, function}\hspace{40pt} \\[2ex]	R(t,r)&=\frac{r}{a(T)}\ .\end{split}\right.\end{equation}

Next we take into account the transformation properties of a rank 2 tensor
\begin{equation}\label{MetricTransformation}
	g_{\mu'\nu'}=\frac{\partial X^\mu}{\partial{x^\mu}'}\,\frac{\partial X^\nu}{\partial{x^\nu}'}\,g_{\mu\nu}\ .\end{equation}
Spherical symmetry is a requirement for both metrics, then the transformation for the two angular variables is characterized by the identity. This means that in the angular diagonal components of the metric the Jacobian corresponds to the identity and the transformation for these components do not bring any new information. Analogous is the case for the angular and off-diagonal elements of the metric.
The transformation is trivial and in both pictures the components of the metric are null. However, there is one off-diagonal element that is not null \textit{per se}; the $g_{tr}=g_{rt}$ element. If a diagonal metric is desired we must impose the vanishing of this element. This translates into the following differential equation for the cosmological time $T$
\begin{equation}\label{DiagonalMetricCondition}
	\partial_r T=-\frac{6\,r\,T}{9T^2-4\,r^2} \qquad\iff\qquad\partial_r\rho_{\scriptscriptstyle dust}= \frac{\kappa\,\rho_{\scriptscriptstyle dust}^2\,r}{1-\frac{\kappa\,\rho_{\scriptscriptstyle dust}}{3}\,r^2}\ .\end{equation}
Of course we already know the structure for the bijection between the cosmological time and the content of the universe, i.e. $\abs{T}\propto1/\sqrt{\rho_{\scriptscriptstyle dust}}$ coming from equation \eqref{MetricFRWdensity}. This allows to write the diagonal metric condition (\ref{DiagonalMetricCondition}) in terms of the non-relativistic matter density only. We have two remaining components of the metric to be transformed: the $TT$ and $RR$--components. If we use only the requirement for the radial transformation \eqref{ChVarDMUnknown} and impose a diagonal structure by \eqref{DiagonalMetricCondition}, the metric reads
\begin{equation}\label{MetricSSSDMUnknown}
	ds^2=\frac{\left(\partial_t\rho_{\scriptscriptstyle dust}\right)^2}{3\,\kappa\,{ \rho_{\scriptscriptstyle dust}}^3} \left[1-\frac{\kappa\,\rho_{\scriptscriptstyle dust}}{3}\,r^2\right]\,\mathrm{d}t^2- \frac{1}{1-\frac{\kappa\,\rho_{ \scriptscriptstyle dust}}{3}\,r^2}\ dr^2 -r^2\ d\Omega^2\ .\end{equation}
Note that $\rho_{\scriptscriptstyle dust}$ is a (scalar) known function of $T$ and therefore it is expected to depend on the new temporal variable $t$; but this dependence is unknown a priori (because $t$ is unknown so far). In any case we see that the new metric necessarily contains a $t$ dependence via $\rho_{\scriptscriptstyle dust}(t)$.

Before continuing, let us reduce the number of dimensionful parameters of the theory. The limit of weak coupling ($\kappa\rightarrow0$) and vanishing matter density ($\rho_{\scriptscriptstyle dust}\rightarrow0$) are essentially the same because both parameters appear together in all the equations --except in the energy momentum conservation which is trivially satisfied. Hence we define: $\tilde{\rho}_{\scriptscriptstyle dust}=\kappa\rho_{\scriptscriptstyle dust}$. Then the solution of \eqref{DiagonalMetricCondition} is
\begin{equation}\label{RelationSolvesODE}
	\frac{6+r^2\,\tilde{\rho}_{\scriptscriptstyle dust}}{{\tilde \rho_{\scriptscriptstyle dust}}^{1/3}}=C(t)\ ;\end{equation}
where $C(t)$ is a constant with respect to $r$, meaning that it can only depends on $t$. Dimensional analysis do the rest of the work. In natural units, $[\tilde\rho_{\scriptscriptstyle dust}]=L^{-2}$, then the units for the `constant' should be $[C(t)]=[\tilde \rho_{\scriptscriptstyle dust}]^{-1/3}=L^{2/3}$. In this picture, $\kappa$ does not belong to the theory, hence there is only one dimensionful parameter to give units to $C(t)$; namely $t$. As $\left[\,t\,\right]=L$, then $C(t)=A\,t^{2/3}$ with $A$ a positive dimensionless parameter to be determined.

It is expected for later times the dust density to be homogeneously diluted. This requirement automatically fixes the remaining free parameter of the theory, $A$. This is translated into the metric \eqref{MetricSSSDMUnknown} to be Minkowskian--like when $t\rightarrow\infty$. This is automatically fulfilled except for the factor in front of the brackets in the $g_{tt}$ element. If a Minkowskian limit is expected, the following enforcement of the prefactor is needed $(\partial_t\rho_{\scriptscriptstyle dust})^2/(3\kappa{\rho_{\scriptscriptstyle dust}}^3)\rightarrow1$. We have a relation for obtaining the temporal variation of the dust density, namely (\ref{RelationSolvesODE}) with the corresponding $C(t)$ function. Therefore, reintroducing $\kappa$ and deriving both sides of the equality, one arrives to
\begin{equation}
	\partial_t(\kappa\rho_{\scriptscriptstyle dust})=-\frac{(\kappa\rho_{\scriptscriptstyle dust})^{4/3}}{1-\frac{\kappa\rho_{\scriptscriptstyle dust}}{3}\,r^2} \,\frac{A}{3\,t^{1/3}}\,.\end{equation}
Squaring the last result, dividing by $3\,(\kappa\rho_{ \scriptscriptstyle dust})^3$, and finally using equation \eqref{RelationSolvesODE} again for replacing $1/t^{2/3}$, the prefactor for the $g_{tt}$ component in the limit of $\rho_{\scriptscriptstyle dust}\rightarrow0$ becomes
\begin{equation}
	\frac{A^2}{3^3\,(\kappa\rho_{\scriptscriptstyle dust})^{1/3}\,t^{2/3}}\hspace{20pt}\longrightarrow\hspace{20pt} \frac{A^3}{3^3\,6}\ .\end{equation}
After all, imposing the flat asymptotic limit to this factor, the constant is fixed to $A=3\sqrt[3]{6}$. Once $A$ is known, $\partial_t\rho_{\scriptscriptstyle dust}$ can be expressed as a function of $\rho_{\scriptscriptstyle dust}$ and $r$ uniquely. Following the same steps as before; replacing $1/t^{2/3}$ from the prefactor by $A/C(t)$ from equation \eqref{RelationSolvesODE}, we get simply
\begin{equation*}
	\frac{\left(\partial_t\rho_{\scriptscriptstyle dust}\right)^2}{3\,\kappa\,{ \rho_{\scriptscriptstyle dust}}^3}= \frac{1}{\left(1+\frac{\kappa\,\rho_{\scriptscriptstyle dust}}{6}\,r^2\right)\left(1-\frac{\kappa\rho_{\scriptscriptstyle dust}}{3}\,r^2\right)^2}\end{equation*}
where the asymptotic limit is shown explicitly. Then the sought for metric turns out ot be
\begin{equation}\label{MetricSSSDM}
	ds^2=\frac{1}{\left(1+\frac{\kappa\,\rho_{\scriptscriptstyle dust}}{6}\,r^2 \right) \left(1-\frac{\kappa\,\rho_{\scriptscriptstyle dust}}{3}\,r^2\right)}\ dt^2 -\frac{1}{1-\frac{\kappa\,\rho_{\scriptscriptstyle dust}}{3}\,r^2}\ dr^2-r^2\ d\Omega^2\ .\end{equation}
This metric has a good Minkowskian limit for $\rho_{\scriptscriptstyle dust}\rightarrow0$. It has similar characteristics to the ones of the SdS metric and it is clearly the relevant metric to describe the Keplerian problem we alluded to in the introduction, in the presence of dust but without cosmological constant, when properly extended with the black hole gravitational potential (see below).

Further discussions concerning dust in these coordinates are included in the appendix. With the full structure for $C(t)$, we present a detailed discussion on the physical solutions to equation \eqref{RelationSolvesODE}. In Appendix \ref{DustDensity} we show that the mere existence of a solution for \eqref{RelationSolvesODE} entails the presence of a horizon in such coordinates. In Appendix \ref{DustEnergyMomentumTensor}, we include the full computation of the \textit{non-diagonal} stress-energy tensor also in these coordinates.

The corresponding change of variables to and from FLRW is given by
\begin{equation}\label{ChVarDMExact}\left\{\begin{split}	
	&\quad t=
	\frac{\left[6 +(\kappa\rho_{d{\scriptscriptstyle 0}})^{2/3}\,(\kappa\rho_{\scriptscriptstyle dust})^{1/3}\,R^2 \right]^{3/2}}{9\,\sqrt{2\,\kappa\rho_{\scriptscriptstyle dust}}}
	\\[2ex]
	&\quad r=\sqrt[3]{\frac{\rho_{d{\scriptscriptstyle 0}}}{\rho_{\scriptscriptstyle dust}}}\ R \end{split}\right.\end{equation}
with $\rho_{\scriptscriptstyle dust}$ given by equation \eqref{ScaleFactorDM}. The metric \eqref{MetricSSSDM} can be linearized for $\rho_{\scriptscriptstyle dust}\to 0$ to
\begin{equation}\label{MetricSSSDMLinear}
	ds^2=\left(1+\frac{\kappa\,\rho_{\scriptscriptstyle dust}}{6}\,r^2\right)\ dt^2-\left(1+\frac{\kappa\,\rho_{\scriptscriptstyle dust}}{3}\,r^2\right)\ dr^2-r^2\ d\Omega^2\ .\end{equation}
Within this linearized approximation it is of course trivial to include the black hole gravitational field. This would give rise to
\begin{equation}\label{MetricSSSDMLinearplusBH}
	ds^2=\left(1-\frac{r_S}{r}+\frac{\kappa\,\rho_{\scriptscriptstyle dust}}{6}\,r^2\right)\ dt^2-\left(1+\frac{r_S}{r}+\frac{\kappa\,\rho_{\scriptscriptstyle dust}}{3}\,r^2\right)\ dr^2-r^2\ d\Omega^2\ .\end{equation}
This expression is valid in the region $r\gg r_S$ and $\kappa\rho_{\scriptscriptstyle dust}\,r^2 \ll 1$. Note that it is not time-independent, unlike its SdS counterpart, but the time dependence enters only via the matter density.

In \eqref{ChVarDMExact} the coordinate transformation is expressed in terms of $\rho_{\scriptscriptstyle dust}$ but the dependence  of $\rho_{\scriptscriptstyle dust}$ on $T$ is known; see equation \eqref{ScaleFactorDM}. We can therefore expand in powers of $\Delta T=T-T_{0}$, where $T_{0}$ is the time where the density of dust is $\rho_{d{\scriptscriptstyle 0}}$. Linearization of the coordinate transformation leads to
\begin{equation}\label{ChangeVariablesLinearDustTrue}\left\{\begin{split}
	&\quad t=T+\frac{R^2}{2}\,\sqrt{\frac{\kappa\rho_{d{\scriptscriptstyle 0}}}{3}}+\Delta T\,R^2\,\frac{\kappa\rho_{d{\scriptscriptstyle 0}}}{12}+\dots\\[2ex]&\quad r=\left(1+\Delta T\,\sqrt{\frac{\kappa \rho_{d{\scriptscriptstyle 0}}}{3}}-\Delta T^2\ \frac{\kappa\rho_{d{\scriptscriptstyle 0}}}{12}\right)\,R+\dots\end{split} \right.\end{equation}
where the scale factor $a_{0}$ is taken equal to $1$. Note that the initial condition $T=T_{0}$ for the cosmological time does not correspond to a unique value for $t$ as the relation does depend on $R$.

\section{Gravitational wave propagation in an analytic $\Lambda$CDM- background}\label{SectionGWLCDM}

Observations show that radiation density is very small in the current epoch; recall the radiation density parameter \eqref{DensityParamR}. If radiation is neglected, the system of Friedmann equations has an analytic solution for the remaining two components: nonrelativistic matter and the cosmological constant.

\subsection{Non-interacting background components}

Another key relation can be found thanks to the Bianchi identities. This is the covariant conservation of the stress-energy tensor
\begin{equation}\label{CovariantConservation}
	\nabla^\mu T_{\mu\nu} = 0\ .\end{equation}
One of these four equations, the temporal one $(\nu=0)$, prides a very useful relation for the evolution of all background energy densities:
\begin{equation}\label{StressTensorConservation}
	\sum_{i=1}\left[\ \dt\rho_i + 3\ \frac{\dt a}{a}\left(\rho_i +P_i\right)\ \right]=0\ ,\end{equation}
where $i$ label each component of the universe. This equation does not provide any new information; indeed is a linear combination of Friedmann equations: deriving the first Friedmann equation and replacing the $\ddt a$ term by the second equation in \eqref{FriedmannEc}, \eqref{StressTensorConservation} is recovered.

If we consider a universe with non-coupled components only, there is no exchange of energy-momentum among them; to wit, each of the addend in the sum \eqref{CovariantConservation} nullifies independently
\begin{equation}
	\nabla_\mu T_{\scriptscriptstyle(\,i\,)}^{\mu\nu}=0\ .\end{equation}
This means that their interaction is purely gravitational and \eqref{StressTensorConservation} holds separately for each individual component $i$. These equations can be rewritten as
\begin{equation}\label{DensityDerivativeEq}
	\frac{d\rho_i}{da} + 3\ \frac{\rho_i}{a}\ \left(1+w_i\right)=0\end{equation}
for each component $i$. This simplification of non-interacting components is very useful since provides one equation for each component apart from the initial two Friedmann equations. These new relations depends only on the state parameter of each component, which obeys
\begin{equation}\label{DensityDependencea}
	\rho_i=\rho_{i{\scriptscriptstyle0}}\ a^{-3(1+\omega_i)} \ .\end{equation}
The initial scale factor at $T_0$ has been fixed to the unity, $a_0=1$. 

If radiation is neglected, the two relevant component for the following discussion are dark matter and the cosmological constant, each of them with the corresponding cosmological parameter already written in \eqref{StateParameters}. For the latter, where $\omega_{\scriptscriptstyle\Lambda}=-1$, equation \eqref{DensityDependencea} becomes
\begin{equation}\label{DensityCC}
	\rho_{\scriptscriptstyle\Lambda}\equiv\rho_{\scriptscriptstyle\Lambda0}\ .\end{equation}
This component has been already discussed. Here energy density of the cosmological constant can be obtained from \eqref{DensityParamCC}.
The second component is to very good approximation a {\it dust-like} component \cite{Frieman}, characterized by $\omega_{\scriptscriptstyle dust}=0$. Therefore, 
\begin{equation}\label{DensityNR}
	\rho_{\scriptscriptstyle dust}(a)=\rho_{d{\scriptscriptstyle0}}\ a^{-3}\ .\end{equation}

Finally, let us consider the combined situation where dust and cosmological constant are both present, and let us derive the evolution of the corresponding cosmological scale factor in FLRW coordinates
The total density is the addition of the density for each component; in this way the first Friedmann equation \eqref{FriedmannEc} can be rewritten as
\begin{equation}\label{DensityFull}
	3\ \left(\ \frac{\dt a}{a}\ \right)^2=\kappa\rho_{\scriptscriptstyle Eff}(T)=\kappa\rho_{\scriptscriptstyle dust}(T)+ \kappa\rho_{\scriptscriptstyle \Lambda}=\kappa\rho_{d{\scriptscriptstyle0}}\, \left[\frac{a_{0}}{a(T)}\right]^3+\rho_{\scriptscriptstyle\Lambda}\ .\end{equation}
$\rho_{d{\scriptscriptstyle0}}$ is the {\it initial} density of "dust" and $\rho_{{\scriptscriptstyle\Lambda}}$ is the (constant) density of "dark energy" when $a(T_{0}=0)= a_{0}=1$. The solution for the latter equation is analytic and gives 
\begin{equation}\label{ScaleFactorDMDE}\begin{split}
	a(T)=\left[\sqrt{1+\frac{\Omega_m}{\Omega_{\scriptscriptstyle\Lambda}}}\,\sinh\left(3H_0\sqrt{\Omega_{\scriptscriptstyle\Lambda}}\ \frac{T}{2}\right)+\cosh\left( 3H_0\sqrt{\Omega_{\scriptscriptstyle\Lambda}}\ \frac{T}{2}\right) \right]^{2/3}\ .\end{split}\end{equation}
In the limit when $\Omega_{d{\scriptscriptstyle 0}}\rightarrow0$ we recover the one component scale factor (\ref{ScaleFactorDE});
or if $\Omega_{\scriptscriptstyle\Lambda}\rightarrow0$, we get \eqref{ScaleFactorDM} as it should be.

\begin{figure}[!h]
	\includegraphics[scale=0.80]{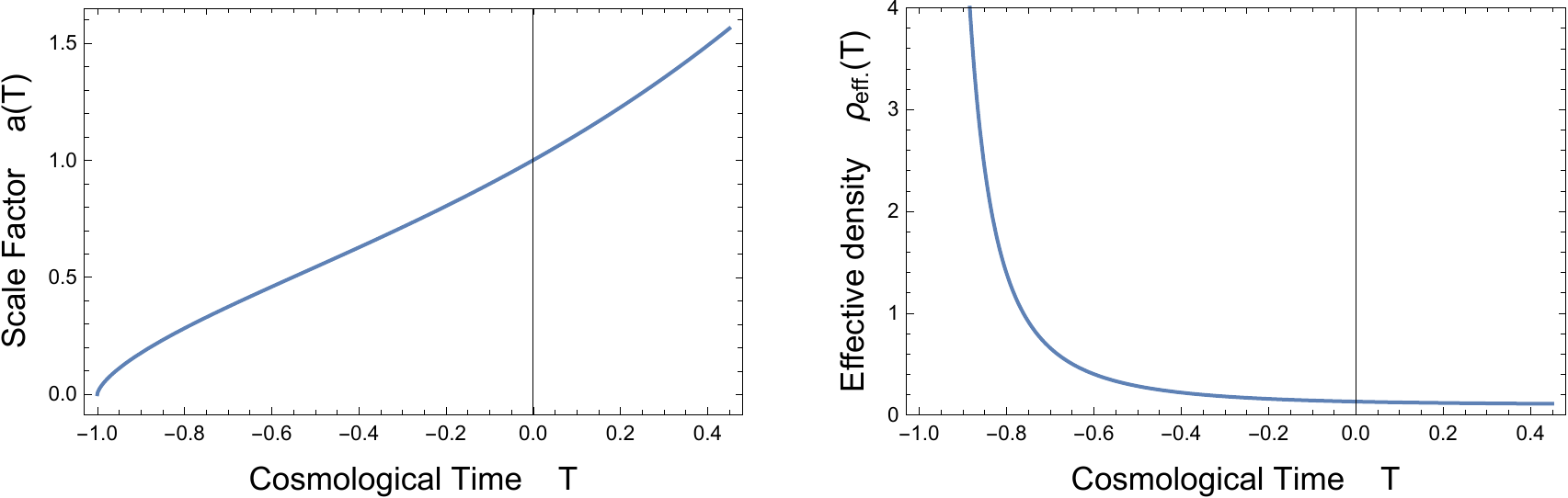}
	\captionsetup{width=0.85\textwidth}
	\caption{\small Scale factor $a(T)$ and the effective energy density $\rho_{\scriptscriptstyle Eff}$ of the 2 main components of the $\Lambda$CDM model. The vertical line corresponds to the value $T_0=0$, where $a(T_0)=a_0=1$.} \label{FigScaleFactorDensity}\end{figure}

Figure \ref{FigScaleFactorDensity} shows the corresponding cosmological scale factor together with the effective energy density. This approximation of gravitationally interacting components is fairly accurate for $a>0.01$ or equivalently for $T>10$ million years with respect to the Big Bang. On the contrary, when radiation is added, this analytic behaviour would not be the case anymore, and there is no analytic solution for the scale factor. 

\vspace{13pt}

\subsection{Non-relativistic matter over an expanding background}

Now that we have combined both components, dust and cosmological constant, let us discuss the same picture but from the point of view where gravitational waves are produced. These components are considered as non-interacting fluids, each with such an equation of state \eqref{EoS}, composing a background over where gravitational waves propagates. As the interaction is purely gravitational, there is no exchange energy-momentum directly among them, i.e. the stress tensor for each component is covariantly conserved separately.

The same picture in cosmological coordinates centered at the gravitational waves source is exceedingly simple in linearized gravity, as previously discussed. The metric will be 
\begin{equation}\label{MetricSSSDMCCLinear}
	ds^2=\left(1+\frac{\kappa\rho_{\scriptscriptstyle dust}}{6}\,r^2- \frac{\Lambda }{3}\,r^2\right)\ dt^2-\left(1+\frac{\kappa\rho_{ \scriptscriptstyle dust}}{3} \,r^2+\frac{\Lambda}{3} \,r^2\right)\ dr^2-r^2\ d\Omega^2\ ;\end{equation}
where the black hole contribution $\sim r_S/r$ has been omitted because is not relevant at large distances. The additive picture permits to recover the individual linearized theory for each component (taking the corresponding limit of the other going to zero). One would expect the same additive behaviour in the perturbations for the change of variables, but if one analyze how the full scale factor (\ref{ScaleFactorDMDE}) expands in Taylor series,
\begin{equation}\label{ScaleFactorDMCCExpanded}\begin{split}
	a(T)&\simeq1+\sqrt{\frac{\Lambda+\kappa\rho_{d{\scriptscriptstyle 0}}}{3}}\ \Delta T+\dots\,,\end{split}\end{equation}
it is clear that the additivity takes place inside the square root. Indeed it is easy to see that the following coordinate transformation does the job of moving from FLRW coordinates to the ones in which the metric \eqref{MetricSSSDMCCLinear} is expressed
\begin{equation}\label{ChVarDMCCLinear}\left\{\begin{split}
	\quad&t\sim T+\frac{R^2}{2}\,\sqrt{\frac{\Lambda+\kappa\rho_{d{ \scriptscriptstyle 0}}}{3}}+\Delta T\,R^2\,\left( \frac{\Lambda}{3} +\frac{\kappa \rho_{d{\scriptscriptstyle 0}}}{12}\right)+\dots\\[2ex] \quad&r=a(T)\ R\sim\left(1+\Delta T\,\sqrt{\frac{\Lambda+\kappa \rho_{d{\scriptscriptstyle 0}}}{3}}\right)\,R+\dots \end{split}\right.\end{equation}
Of course in both limits, $\rho_{d{\scriptscriptstyle 0}}\rightarrow0$ and $\Lambda\rightarrow0$, we recover each $1-$component universe coordinate transformation.

\section{Including relativistic matter}\label{SectionRelMatt}

The contribution of a relativistic background when gravitational waves are propagating through the universe is the weakest among the components of the $\Lambda$CDM model in the current epoch. As discussed along the previous lines, this contribution is almost negligible; the interest of including it is that it provides a generalization to take into account all background energy densities existent as a whole. In this manner, we will find that the results will depend only on the Hubble parameter today, which is a direct measurement of observations, and we would be able to apply our results to a realistic cosmology.

The procedure presented in section \ref{SectionNoRelMatt} can be applied in an analogous manner to a relativistic matter universe. For this single component, characterized by a state parameter $\omega_{\scriptscriptstyle rad}=1/3$, the equation of state \eqref{EoS} is written as
\begin{equation}
	\rho_{\scriptscriptstyle rad.}=3\ P_{\scriptscriptstyle rad.}\end{equation}
and closes the Friedmann system of equations \eqref{FriedmannEc}. In comoving coordinates the scale factor can be written in terms of different parameters
\begin{equation}
	a(T)=a_{0}\ \sqrt{\frac T{T_{0}}}=a_{0}\ \sqrt[4]{ \frac{\rho_{r{\scriptscriptstyle 0}}}{\rho_r(T)}}\end{equation}
as there is a direct link between the energy density and the cosmic time. Radiation gets diluted as
\begin{equation}\label{DensityVsT}
	\rho_r(T)=\frac3{4\,\kappa\,T^2}\ .\end{equation}
The integration constant $\rho_{r{\scriptscriptstyle 0}}=3/(4\kappa T_{ 0}{}^2)$ is the energy density measured today, when the scale factor is $a_0=1$. Let us repeat the steps for non-relativistic but for a universe with this single component. Again it is needed a set of {\it static} coordinates with spherical symmetry such that the FLRW 2-sphere $S^2=a^2R^2d\Omega^2$ becomes a {\it static} 2-sphere $S^2=r^2d\Omega^2$. This requirement fixes the radial change of variables while leave the temporal transformation unknown.  The structure is the same as the one employed in \eqref{ChVarDMUnknown}.

The next step is to change the metric representation from the cosmological coordinates $X^\mu=(T,R,\theta,\phi)$ to the new set $x^\mu= (t,r,\theta,\phi)$ by means of the 2-rank tensor transformation \eqref{MetricTransformation}. Doing the math, it is seen that the diagonal angular components of the metric transform trivially and that any non-diagonal component with at least one angular index remains null. A diagonal metric in the new coordinates is obtained if the following identity is satisfied
\begin{equation}\label{RelationSolvesODERadiation}
	\partial_r\rho_r=\frac{4\,\kappa\rho_r{}^2\,r}{3-\kappa\rho_r\,r^2} \hspace{40pt}\Longrightarrow\hspace{40pt}\frac{6+2\,\kappa\rho_r\,r^2}{\sqrt{\kappa\rho_r}}=C(t)=A\,t\ .\end{equation}
The energy density $\rho_r$ is a function of the new variables $t$ and $r$. According to the first integral of the equation on the left, the integration constant must depend only on the temporal coordinate. Dimensional analysis provides its structure: the units goes as $[C]=[\kappa\rho_r]^{-1/2}=L$; therefore, using equivalent reasons as the ones discusses around equation \eqref{RelationSolvesODE}, any integration constant for $r$ can acquire units only by means of $t$, hence $C(t)=At$. 

The diagonal condition provides a metric which is almost unequivocally defined, except for a global coefficient in the temporal component of the metric involving the temporal derivative of the radiation density
\begin{equation}\label{MetricSSSDMUnknownRadiation}
	ds^2=\frac{3\,\left(\partial_t\rho_r\right)^2}{16\,\kappa\rho_r{}^3}\left(1-\frac{\kappa\rho_r\,r^2}{3}\right)\ dt^2-  \left(1-\frac{\kappa\rho_r\,r^2}{3}\right)^{-1}\ dr^2-r^2\ d\Omega^2\ .\end{equation}
Of course this metric resembles the metric of the non-relativistic case obtained in \eqref{MetricSSSDMUnknown} but with different numerical factors. The temporal derivative can be obtained deriving the right equation in \eqref{RelationSolvesODERadiation}
\begin{equation}\label{DensityDerivativeRad}
	\partial_t\rho_r=-A\ \frac{\sqrt{\kappa\rho_r}\,\rho_r} {3-\kappa\rho_r\,r^2} \ .\end{equation}

One more time, The homogeneously diluted limit for later times, $t\rightarrow\infty$, provides the missing constant $A$. The requirement for the metric to be Minkowskian in this asymptotic regime, $g_{\alpha\beta}\rightarrow\eta_{\alpha \beta}$ while $\rho_{\scriptscriptstyle rad.}\rightarrow0$, is fulfilled for all the components of the metric except for the temporal one. We need to impose $(\partial_t\rho_r )^2/(\kappa\rho_r{}^3)\rightarrow 16/3$ in $g_{tt}$ from equation \eqref{MetricSSSDMUnknownRadiation}. Accordingly, $A=4\sqrt{3}$. After all, we have fixed the density relation, this is 
\begin{equation}
	3+\kappa\rho_r\,r^2=2\sqrt{3\,\kappa\rho_r}\ t\ .\end{equation}

Inserting \eqref{DensityDerivativeRad} in \eqref{MetricSSSDMUnknownRadiation} with the value of $A$, we obtain the metric representation in some {\it static} and spherically symmetric coordinates
\begin{equation}\label{MetricSSSRad}
	ds^2=\left(1-\frac{\kappa\rho_r\,r^2}{3}\right)^{-1}\ dt^2-\left(1- \frac{\kappa\rho_r\,r^2}{3}\right)^{-1}\ dr^2 -r^2 d\Omega^2\ .\end{equation}
The change of variables back to the cosmological coordinate system is given by
\begin{equation}\label{ChVarRadExact}\left\{\begin{split}	
	&\quad t=T+a_{0}{}^2\sqrt{\frac{\kappa\rho_{r{\scriptscriptstyle 0}}}{ 3}}\,\frac{R^2}{2}\\[2ex]&\quad r=a_0\ \sqrt[4]{\frac{\rho_{r{\scriptscriptstyle 0}}}{\rho_r(T)}}\ R\ ,\end{split}\right. \end{equation}
where the dependence of the density on the cosmological time has been written in \eqref{DensityVsT}. Let us mention that time dependence enters only via the energy density of the relativistic matter.

It is enough for our purposes a linearized version of \eqref{ChVarRadExact}. The linearized metric for $\rho_{\scriptscriptstyle rad.}\rightarrow0$ from \eqref{MetricSSSRad} is
\begin{equation}
	ds^2=\left(1+\frac{\kappa\rho_r\,r^2}{3}\right)\ dt^2-\left(1+ \frac{\kappa\rho_r\,r^2}{3}\right)\ dr^2 -r^2 d\Omega^2\end{equation}
and the linear change of variables, expanded in powers of $\Delta T=T-T_0$, results
\begin{equation}\left\{\begin{split}
	&\quad t\sim T+a_0{}^2\sqrt{\frac{\kappa\rho_{r{\scriptscriptstyle 0}}}{3}}\,\frac{R^2}{2}+\dots\\[2ex]
	&\quad r\sim a_0\,\left(1+\sqrt{\frac{\kappa\rho_{r{\scriptscriptstyle 0}}}{3} }\,\Delta T\right)\,R+\dots\end{split} \right.\end{equation}
The magnitudes measured today are related by $1/T_0=2\sqrt{\kappa \rho_{r{\scriptscriptstyle 0}}/3}\,$.

\section{Gravitational wave propagation in a general background}\label{SectionGWH0}

Although the contribution of radiation as a background component is almost negligible and worse, it cannot be included in an analytical expression as found in \eqref{ScaleFactorDMDE}, it is interesting to include its effect anyway for reasons that will be seen below.

At this point where the linearized framework is a fair approximation, the inclusion of radiation in \eqref{MetricSSSDMCCLinear} becomes obvious,
\begin{equation}\label{MetricSdSH0}\begin{split}
	ds^2=&\left[1-\frac{\kappa}{6}\left(2\rho_{\scriptscriptstyle\Lambda}-2\rho_{r}-\rho_{nr}\right)r^2+...\right]dt^2\\&\hspace{60pt}-\left[1+\frac{\kappa}{3}\left(\rho_{\scriptscriptstyle\Lambda}+\rho_{nr}+\rho_{r}\right)r^2+...\right]dr^2-r^2\ d\Omega^2\ .\end{split}\end{equation}
In the later equation we favour the density of dark matter over the cosmological constant. Observations measure the Hubble parameter
\begin{equation}\label{FriedmannEcMinimal}
	H(a)\equiv H_{0}\ \sqrt{\Omega_{dm}\ a^{-3}+\Omega_{ rad}\ a^{-4}+\Omega_{\scriptscriptstyle\Lambda}}\end{equation}
rather than the separate contributions from the cosmological constant and the various matter densities. Its value at a time $T_0$ when $a_0=1$ is a measure of the total energy density of the background on which gravitational waves propagate\footnote{In cosmology, its value is conventionally chosen as $T_0=0$ in the current epoch.}; known as the Hubble constant is
\begin{equation}\label{HubbleConstant}
	H_0{}^2=\frac\kappa3\ (\rho_{\scriptscriptstyle\Lambda}+ \rho_{d{ \scriptscriptstyle 0}}+\rho_{r{\scriptscriptstyle 0}})\ .\end{equation}
--recall the discussion below equation \eqref{HubbleConstantGeneric}.

Well away from the sources, in the limit where $r_S/r\ll\rho_i$, the black hole contribution has been be safely neglected in equation \eqref{MetricSdSH0}, and the linearized picture holds.
In this approach, the gravitational wave would propagate over an expanding spacetime, characterized by a scale factor approximately given by
\begin{equation}
	a(T)\simeq1+H_0\ \Delta T+\dots	\end{equation}
The propagation should take place in coordinates where the expansion is explicit. To change to those cosmological coordinates, is enough to use the linearized change of variables given by
\begin{equation}\label{ChVarDMCCRadLinear}\left\{\begin{split}
	\quad&t\sim T+\frac{R^2}{2}\ H_0+\dots\\[2ex] \quad&r\sim\left(1+\Delta T\ H_0 \right)\,R+\dots\ . \end{split}\right.\end{equation}

In conclusion, the harmonic wave front \eqref{gw1} can be transformed up to linear order in the Hubble parameter, to expanding FLRW coordinates by means of \eqref{ChVarDMCCRadLinear}. Therefore, choosing $T_0=0$ at the time when gravitational waves are emitted, the wave perturbation gets modified to
\begin{equation}\label{GWFRW}
	h_{\mu\nu}\sim\frac{e_{\mu\nu}}{R} \left(1+H_0\, T\right) \cos\Bigl[{w\, (T-R)+w\ H_0\,\Bigl(\frac{R^2}{2}-TR\Bigr)}\Bigr]\ ,\end{equation}
keeping only changes of order $\sqrt{\Lambda}$. $e_{\mu\nu}$ is the transformed polarization tensor that will not be very relevant for the following discussion, although its precise form is essential for precise comparison with observations. From this last expression we can identify the effective frequency and the effective wave number 
\begin{equation}\label{FreqWaveNumb}\begin{split}
	&w_{\scriptstyle eff}= w\,\left(1- H_0\ R\right) \\[3ex]&k_{\scriptstyle eff}= w\, \Bigl(1-H_0\ \frac{R}2\Bigr)\ ;\end{split}\end{equation}
which do not coincide.

In astrophysical observations, it is common to use the cosmological redshift $z$ for characterizing the relative changes between the observed $w_{eff}$ and emitted $w$ frequencies (or wavelengths) of distant objects. This magnitude is defined by
\begin{equation}\label{FrquencyCosmology}
	{w_{eff}}=\frac{w}{1+z}\end{equation}
and at these scales, the redshift is very well approximated by the Hubble relation $z\simeq H_0\,R$. Hence the observed frequency $w_{eff}$ is very well approximated by equation \eqref{Frequency}. We notice that the obtained frequency in \eqref{FreqWaveNumb} agrees with the usual frequency redshift \eqref{Frequency}, or \eqref{FrquencyCosmology}; so this well known result is well reproduced. However the wave number is different. This discrepancy lies at the root of a possible local measurement of $\Lambda$ in PTA, as we will discuss in next chapter.

Before moving on, let us draw some conclusions about the linearized results obtained. Along this chapter, we have discussed extensively how gravitational waves are modified when they propagate over a non-flat background, but over an expanding one. If comoving coordinates are employed during the propagation, we have shown that modifications to the gravitational waves are induced. Simultaneously, the wavelength and the amplitude are increased with the radial coordinate $R$ that measures the distance to the source. There is also a discrepancy between the effective frequency $w_{eff}$ and the effective wave number $k_{eff}$; see the equations in \eqref{FreqWaveNumb}.
The study of the relevant coordinate transformations make us arrive to \eqref{ChVarDMCCRadLinear}, where linear modifications due to the effective background over where waves propagate have been taken into account. The magnitude describing the energy density composition of the background is given by the Hubble parameter today \eqref{HubbleConstant} whose value is dominated by the cosmological constant. This parameter provides the linear corrections to the theory. Measuring these corrections at {\it moderate} values of the redshift $z$, would immediate imply that we are measuring {\it locally} the cosmological constant.

The above calculations have been performed at a linear level, retaining ${\cal O}(H_0\,R)$ terms\footnote{We emphasize once more that at the next, non-linear, order the corrections cannot be all grouped in terms of the Hubble constant, so when we speak of ${\cal O}(H_0{}^2\,R^2)$ corrections this has to be understood in the above sense.}, and considering the corrections of order ${\cal O}(H_0{}^2\,R^2)$ and beyond as subleading. This truncation can be justified based on the smallness of $H_0$ (and of all its ingredients, of course) since we will be interested in sources at distance $R \ll H_0{}^{-1}$.
This approximated picture has a range of validity. Given the current observed value of the cosmological constant \eqref{CosCte}, for events happening at most at a distance of a few Gpc away ($R\lesssim10^{25}$ m) modifications of ${\cal O}(H_0\,R)\lesssim10^{-1}$ appears to be non negligible.
In any case, the expressions could be improved to order $H_0{}^2$ accuracy easily.

\chapter{Local measurements of $\Lambda$ with pulsar timing array}\label{ChapterPTA}

\section{Pulsars as gravitational wave detectors}

A pulsar is a highly magnetized rotating neutron star that emits beams of electromagnetic radiation. They are born in supernova explosions, usually with periods ranging between $10-20$ ms and later on slow down gradually to periods of order $\sim1$ s over millions of years \cite{GoldManchester}. When the beam of emission points towards us, we see an effect much like a lighthouse. Since its first discovery in 1968 \cite{Hewish} radio telescopes have been monitoring repeatedly the corresponding pulsed appearance of emission.

Among the pulsar population, the so-called millisecond pulsars (MSPs) are of special interest to us. They have periods less than 100 ms and a period derivative less than $10^{-17}$. This means that they are very stable and regularly rotating stars. The leading theory for such objects dates from 1982 \cite{Radhakrishnan} and states that the MSPs result from the "recycling" of an old, slowly rotating and probably dead neutron star through accretion from a low-mass companion. This accretion results in a mass transferred from the companion that also carries angular momentum from the orbit to the neutron star, spinning it up and reactivating the pulsar emission process. In some cases the complete ablation of the companion star occurs. This is a viable mechanism for the formation of solitary MSPs. However, globular clusters are much more efficient factories for the production of MSPs.

Owing to their extraordinarily rapid and stable rotation, MSPs are usually used by astronomers as galactic clocks rivaling the stability of the best atomic clocks on Earth. Any factor affecting the arrival time of pulses by more than a few hundred nanoseconds can be easily detected and used to make precise measurements.
The process of pulsar timing is fundamentally dependent on how accurate  the description of everything that affects the times of arrival (ToAs) of the pulsed radiation at the telescope is. Then, there are "timing models" that take into account all orbital, location and motion parameters of the pulsars and of the interstellar medium along where radio waves propagate, to predict the phase of the pulsar's periodic signal at any point in time.
When a gravitational wave passes it disturbs everything around it, and it does it over the observed periods of pulsars. The gravitational waves detection with pulsars depends on the great stability of the MSPs. 

Even the more stable pulsars present intrinsic irregularities in their period at some point. Therefore, observations of an ensemble of MSPs (this is a Pulsar Timing Array) are needed for gravitational wave detection.
The observational method is based on searching for correlated signals among pulsars in a PTA. A disturbance from a passing gravitational wave across the ensemble of pulsars will have the particular quadrupolar spatial signature that corresponds to gravitational waves \cite{Hellings}. If it passes, it will thus be detected.
In practice, the analysis consists in computing the difference between the expected and actual ToA of pulses. This difference is called "timing residuals" and contains information of all unmodeled effects that have not been included in the fit, including gravitational waves \cite{Jenet2005}.

Nevertheless, PTA observations have not yet found convincing evidence of gravitational waves. 
In the next section we will rederive in an alternative way the main results obtained in the previous chapter. This time without moving from different coordinate systems but just computing the wave equation directly in FLRW expanding coordinates, and showing that the wave solution is in fact nothing less than the same solution found before.
As it was pointed out at the end of section \ref{SectionGWH0}, the solution develops some anharmonic effects because propagation takes place through an expanding background. Although globally very small, these anharmonic effect result in a dramatic enhancement of the signals expected in PTA observations. However, the effect only takes place at some particular values of the angle subtended by the source and the observed pulsar.
In section \ref{SectionPTA} we will compute the timing residual numerically and show its main {\it peak-like} characteristics. As can be expected, if calculations are performed in a Minkowskian background the well-known oscillating signal with no enhancement is obtained. However when a dynamical background is considered the anharmonicities appear.
After this, in section \ref{SectionCharacterization} we will characterize the signal; particularly which of the cosmological parameters or characteristic distances modifies the position or width of the peak.
To conclude, we will expose in what manner these results can be tested by the IPTA collaboration. We will propose a way to hopefully detect this very relevant effect; which if present would also provide a way to measure the Hubble constant $H_0$ but at subcosmological distances.

\section{Linearized perturbations in FLRW}

As we have discussed in Chapter \ref{ChapterGW} the gravitational wave propagation takes place not over a flat spacetime, but over an approximately globally de Sitter universe. In \cite{EGR} linearized gravity in curved backgrounds has been discussed but from the point of view of a cosmological observer. Applying the same strategy of linearization on a curved background, we consider small perturbations around a FLRW background metric \eqref{MetricFLRWGenericK0} of the form
\begin{equation}
	g_{\mu\nu}=\tilde{g}_{\mu\nu}^{\scriptscriptstyle (FLRW)} +h_{\mu\nu}\ .\end{equation}
These perturbations are interpreted as gravitational waves propagating on this curved background in analogy to the treatment on a flat spacetime; recall section \ref{SectionLinGrav}.

The linearized equations of motions are provided by the Einstein's equations and are given by the following expansion
\begin{equation}
	G_{\mu\nu}(\tilde{g}+h)=G_{\mu\nu}(\tilde{g})+\frac{\delta G_{\mu\nu}}{\delta g_{\alpha \beta}}\, \biggl\rvert_{\tilde{g}}\ h_{\alpha\beta}+\ .\ .\ .\ = \kappa\left(T_{\mu\nu}^{\scriptscriptstyle\,(0)} +T_{\mu\nu}^{\scriptscriptstyle\,(1)}+ \ .\ .\ .\ \right)\ .\end{equation}
Clearly, these equations are satisfied for the unperturbed FLRW background metric. At zero order we have $G_{\mu\nu}(\tilde{g})=\kappa T_{\mu\nu}^{\scriptscriptstyle\,(0)}$, which in fact gives the well-known Friedmann equations by treating the energy content of the universe as perfect fluids. The energy-momentum tensor includes as a matter of fact all cosmological fluids composing the background, each one described by some equation of state \eqref{EoS}.
Then, at first order, we end up with the following expression as a motion equation of the perturbation tensor
\begin{equation}\label{EEqGenericExpansionInh}
	\frac{\delta G_{\mu\nu}}{\delta g_{\alpha \beta}}\, \biggl\rvert_{\tilde{g}}\ h_{\alpha\beta}= \kappa\ T_{\mu\nu}^{\scriptscriptstyle\,(1)}\ .\end{equation}

Let us make some reasonable assumptions to avoid redundancies under coordinate transformations and go straightforward to the gravitational wave profile in FLRW coordinates. In this new set of coordinates the spatial character of the gravitational wave is not expected to change. Besides, it is reasonable to expect that in spherical coordinates gravitational waves propagate along the radial direction away from the source, along a vector given by $k^\mu=(w,k_{\scriptscriptstyle R},0,0)$. Then transversality to the $R-$coordinate implies that the only non-vanishing $h-$components are the purely angular ones, i.e.
\begin{equation}
	h_{{\scriptscriptstyle T}\alpha}=h_{{\scriptscriptstyle R}\alpha}=0\end{equation}
for $\alpha=(T,R,\theta,\phi)$. Moreover, the traceless nature of gravitational waves should not change; hence the condition $h=\tilde g^{\mu\nu}\,h_{\mu \nu}=0$ relates the diagonal components of the
metric perturbation by 
\begin{equation}
	h_{\phi \phi}=-\sin^2 \theta\ h_{\theta \theta}\ .\end{equation}
Finally, symmetry conditions imply
\begin{equation}
	h_{\theta \phi} = h_{\phi \theta}\ .\end{equation} 	
All in all, there are two nonzero independent components in  agreement with the two degrees of freedom corresponding to the two polarization directions of gravitational waves. 

After imposing these gauge conditions, the two equations for $h_{\phi \phi}$ and $h_{\theta \phi}$ coming from \eqref{EEqGenericExpansionInh} are rewritten up to first order in the perturbations as
\begin{equation}\label{WaveEqFLRW}
	\left[\frac{a}{2}\ \frac{d}{dT}\left(\frac{1}{a}\,\frac{d}{dT} \right)-\frac{R^2}{2a^2}\frac{d}{dR}\left(\frac{1}{R^2}\,\frac{d}{dR}\right)-\frac{1}{R^2\,a^2}-\left(\frac{\dot{a}}{a}\right)^2-\frac{3\,\ddot{a}}{a}\right]\ h_{\mu\nu}=\kappa\,T_{\mu\nu}^{\scriptscriptstyle \,(1)}\ ;\end{equation}
where $\mu,\nu = \theta,\phi$. We notice that the equations for the two independent metric components have exactly the same functional form. The previous discussion is valid far away from the source, where gravitational waves propagates freely along the expanding universe.
Although equations \eqref{WaveEqFLRW} are valid up to order ${\cal O}({\Lambda})$, we will be mostly interested in terms of order ${\cal O}(\sqrt{\Lambda})$ due to the smallness of the cosmological constant value, and also to be able to compare with the previous chapter. Therefore, in the right hand side of the latter equation, we can get rid of the term $\kappa\,T_{\mu\nu}^{\scriptscriptstyle \,(1)}$ because is linear with the densities and therefore of order ${\cal O}({\Lambda})$. Of course there is nothing wrong in keeping this term (we will return to this later).

Gravitational waves are produced by very distant sources. In this manner, in the far away field regime, when waves arrive to our local system, they can be described very approximately by a plane wave front. A more useful coordinate system for doing this analysis is the Cartesian set of coordinates;
\begin{equation}
	ds^2 = dT^2 - a(T)^2 \, \left(dX^2 + dY^2 + dZ^2 \right)\ .\end{equation}
Let us define the radial direction joining the SMBH merger with the Earth as the propagation direction for these waves; let us call it the $Z-$axis of the Cartesian coordinates. For local observers, the usual treatment on linearized gravity applies very well for the current discussion. 
In the transverse ($h_{\mu\alpha}k^\alpha=0$) and traceless ($h_\alpha{}^\alpha=0$) gauge (TT), for a wave propagating with wave vector $k^\mu = (\omega, 0, 0, k_{\scriptscriptstyle Z})$, the nonzero components are $h_{\scriptscriptstyle XX}$, $h_{\scriptscriptstyle YY}$, $h_{\scriptscriptstyle XY}$ and $h_{\scriptscriptstyle YX}$, however only two of the four are independent.
It is a customary to express the solution in terms of the "plus" polarization $h_+=(h_{\scriptscriptstyle XX}+h_{\scriptscriptstyle YY})/2$ and the "cross" polarization $h_\times=h_{\scriptscriptstyle XY}=h_{\scriptscriptstyle YX}$.

In this manner, gravitational waves coming from a very distant source are equivalently described by Cartesian coordinates or correspondingly to spherically symmetric coordinates but in the very small polar angle case, $\theta\approx0$.
The relation between components is given by
\begin{equation}\begin{split}
	&h_{\scriptscriptstyle XX} = + \frac{1}{R^2} \cos(2\phi) \, h_{\theta \theta} - \frac{1}{R^2} \frac{\cos \theta}{\sin \theta} \sin(2\phi) \, h_{\theta \phi}\\[3ex]
	&h_{\scriptscriptstyle YY} = - \frac{1}{R^2} \cos(2\phi) \, h_{\theta \theta} + \frac{1}{R^2} \frac{\cos \theta}{\sin \theta} \sin(2\phi) \, h_{\theta \phi}\\[3ex]
	&h_{\scriptscriptstyle XY} = + \frac{1}{R^2} \sin(2\phi) \, h_{\theta \theta} + \frac{1}{R^2} \frac{\cos \theta}{\sin \theta} \cos(2\phi) \, h_{\theta \phi}\ .
	\end{split}\end{equation}
Note that the apparent singularity in the prefactor of $h_{\theta \phi}$ has to be compensated by the appropriate vanishing of this component.
The characteristic signature of the gravitational waves on the ToAs of radio pulses is a linear combination of these two independent polarizations; see for instance \cite{SazhinDetweilerBertotti}. Nevertheless

By direct substitution of $	h_{\phi \phi}(\propto h_{\theta\theta})$ and $h_{\theta \phi}(=h_{\phi\theta})$ in the small angle limit $\theta\approx0$, we can easily reach the corresponding equations \eqref{WaveEqFLRW} for the Cartesian metric perturbations but written in spherical coordinates, since the above differentials equations do not involve angular derivatives and only the radial derivative plays a role in this coordinate transformation
\begin{equation}\label{EEqWavePlusCross}
	\left[\partial^2_T-\frac{\dot{a}}{a}\,\partial_T-\frac{1}{a^2}\left(\partial^2_R+\frac{2}{R}\,\partial_R \right)\right] h_{ij} = 0\ ;\end{equation}
where now $ij = +,\times$. In the derivation of these equations we have neglected ${\cal O}({\Lambda})$ terms, as we mentioned before. 
We can understand this differential equation as the wave equation for a universe with nonnull matter content; this is
\begin{equation}\label{EEqWaveShperical}
	\Box_{\tilde{g}}h_{\mu\nu}=0\ ,\end{equation}
which yields over the two polarization directions. In the limit of an empty universe, with no cosmological constant or dust, the scale factor $a(T)$ can be taken to be $a_0 = 1$, so equations \eqref{EEqWavePlusCross} reduce to a homogeneous wave equation 
\begin{equation}\label{EEqWaveCartesian}
	\Box_{\eta}h_{\mu\nu}=0\ ,\end{equation}
as expected for the flat spacetime limit.

As expected, it is not \eqref{gw1} the solution of \eqref{EEqWavePlusCross}, because it solves the flat background wave equation \eqref{EEqWaveCartesian}. The solution for \eqref{EEqWavePlusCross} or equivalently to \eqref{EEqWaveShperical} is
\begin{equation}\label{gwfrw3}
	h_{\mu\nu}= \frac{e_{\mu\nu}}{R}\left( 1 + H_0\,T\right)\cos\left[w_{\scriptstyle eff}\,T-k_{\scriptstyle eff}\,R \right]\ ;\end{equation}
where the effective frequency and effective wave number have been introduced in \eqref{FreqWaveNumb}. This solution is nothing more that the previous obtained solution after changing coordinates with \eqref{ChVarDMCCRadLinear}, from the set where waves are produced to the one where they propagate and are measured. This solution has been presented in equation \eqref{GWFRW}. 
As we have seen in the previous chapter, if the universe were static, the coordinate transformation \eqref{ChVarDMCCRadLinear} would simply become $t=T$ and $r=R$, and the corresponding solution is a superposition of harmonic waves of the form written in equation \eqref{gw1}. However now we have derived the anharmonic solution directly in expanding coordinates.
The phase velocity when gravitational waves propagate in expanding coordinates is  $v_{p}\sim 1-H_0\,\Delta T+\mathcal{O}(H_0{}^2)$ \cite{Espriu}. On the other hand, with respect to the ruler distance traveled (computed with $g_{ij}$) the velocity is still $1$.

\section{Relevance for Pulsar Timing Arrays}\label{PTA}\label{SectionPTA}

From the previous discussion it is clear that one has to understand what are the consequences of having $k_{eff} \neq w_{eff}$. As discussed before, PTA observations involve some non-local contributions as the signal is integrated along the line of sight joining the pulsar and the observer. Therefore, the fact that $H_0$ in \eqref{gwfrw3} is nonzero may render these contributions observable at some point.

\begin{figure}[!b]\begin{center}
	\includegraphics[scale=0.65]{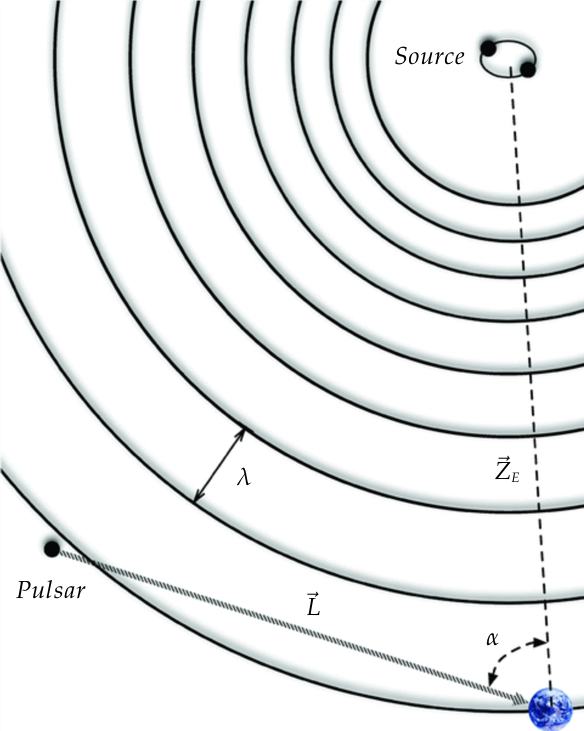}
	\captionsetup{width=0.85\textwidth}\vspace{3pt}
	\caption{\small Relative coordinates of the gravitational wave source ($R=0$), the Earth (at $R=Z_E$ with respect to the source) and the pulsar (located at coordinates $\vec P= (P_X, P_Y, P_Z)$ referred to the source). In FLRW coordinates centered on Earth, the angles $\alpha$ and $\beta$ are the polar and azimuthal angles of the pulsar with respect to the axis defined by the Earth-source direction.} \label{FigSetup} \end{center}\end{figure}

Let us set the experimental framework that might measure this effect. Consider the situation shown in Figure \ref{FigSetup} describing the relative situation of a gravitational wave source (possibly a very massive black hole binary), the Earth and a nearby pulsar.
The pulsar emits pulses with an electromagnetic field phase $\phi_0$ . If this magnitude is measured from the Earth, it will be
\begin{equation}
	\phi(T)=\phi_0\left[T-\tfrac{L}{c}-\tau_0(T)-\tau_{\scriptscriptstyle GW}(T)\right]\ . \end{equation}
The corrections due to the spatial motion of the Earth within the Solar system and the solar system with respect to the pulsar, together with the corrections when the electromagnetic wave propagates through the interstellar medium are taken into account in $\tau_0$. The term $\tau_{\scriptscriptstyle GW}$ involves the corrections owing exclusively to the gravitational wave strength $h_{ij}^{\scriptscriptstyle (FLRW)}$ computed in comoving coordinates. 

In the following we will focus in the phase corrections due to the passage of a gravitational wave; this shift is given by \cite{Deng}
\begin{equation}\label{TimingResidual}
	\tau_{\scriptscriptstyle GW}=-\frac{1}{2}\,\hat{n}^i\hat{n}^j\,{\cal H}_{ij}(T)\ .\end{equation}
The unit vector giving the direction can be written using spherical coordinates as $\hat{n}= (-\sin\alpha \cos\beta; -\sin\alpha \sin\beta; \cos\alpha)$, where $\alpha$ is the angle subtended by the gravitational wave source and the pulsar as seen from the Earth and $\beta$ is the azimuthal angle around the source-Earth axis. ${\cal H}_{ij}$ is the integral of the transverse-traceless metric perturbation along the null geodesic from the pulsar to the Earth. Let us chose the following parametrization: 
\begin{equation}
	\vec R(x)= \vec{P}+L(1+x)\hat{n}\ ,\end{equation}
with $x\in[-1;0]$, $L$ the distance to the pulsar and $\vec{P}$ the pulsar location with respect to the source. Making the dependencies explicit, this magnitude is given by
\begin{equation}\label{tim}
	{\cal H}_{ij}=\frac{L}{c}\int_{-1}^0dx\ h_{ij} \bigl[\,T_E+\tfrac{L}{c}\,x\ ;\ \vec R(x)\,\bigr]\ .\end{equation}
$T_E$ is the time of arrival of the wave that travels a distance $Z_{E}$ to the local system. The speed of light has been restored in this formula.
In the TT--gauge, a gravitational wave propagating through the $\hat Z$ axis has a polarization tensor with only nonzero values in the $X$, $Y$ components; or equivalently in the $+$ and $\times$ components.
Then we can simply write
\begin{equation}\label{Normals}\begin{split}
\hat{n}^i\hat{n}^j\,h_{ij}=\sin^2\alpha\left[(\cos^2\beta-\sin^2\beta)\ h_++2\cos\beta\sin\beta\ h_\times\right]\ , \end{split}\end{equation}
where $h_+$ and $h_\times$ are the two nonnull polarizations of the gravitational wave obtained in \eqref{gwfrw3}. For simplicity, we can also assume $\varepsilon \sim | e_{ij}|$ for $i,j=+,\times$. This approximation does not affect in any significant way the results below, and in fact it can be easily removed.

We have assumed that from the pulsar to the Earth the electromagnetic signal follows the trajectory given by the line of sight and is the distortion created by the intervening gravitational wave what give rise to the timing residual. The real question is whether the observationally preferred small value of the cosmological constant affects this timing residuals from a pulsar at all. This question was answered in the affirmative in \cite{EspriuCC}. 
In order to see the magnitude of the effect we take average values for the parameters involved: $\varepsilon=1.2\times 10^9$ m for a gravitational wave generated by some source placed at $Z_E\sim 0.1 - 1$ Gpc, giving a strength $|h|\sim \varepsilon/R\sim  (10^{-16}-10^{-15})$ which is within the expected accuracy of PTA \cite{jenet}. Monitored MSPs are usually found at a distance of the order of $L\sim0.1-10$ kpc. For the distance to the source we take values in the range $Z = 100$ Mpc $-\,1$ Gpc and we assume the value $w =10^{-8}$ Hz for the frequency. These, together with the preferred value of $H_0$ are all the parameters we need to know. 

In formula \eqref{gwfrw3} enters the modulus of the parametrized trajectory $\vec R(x)$. It is reasonable to assume that $L\ll Z_E$, then
\begin{equation}
	R(x)=|\vec{R}(x)|\sim Z_E+x\,L\ \cos\alpha+\dots\end{equation}
The parametrization for the temporal variable has already been presented in terms of the time of arrival of the wave to the local system $T_{E}$, this is
\begin{equation}\label{Parametrization}
	T=T_E+\tfrac{L}{c}\ x\ .\end{equation}
On geometric grounds, we are always able to choose a plane containing the Earth, the pulsar and the source and set $\beta=0$. This requirement simplifies considerably the contraction \eqref{Normals}. After all these considerations, in order to obtain the timing residual we need to perform the following integral
\begin{equation}\label{TimingResidualH0}\begin{split}
	\tau_{\scriptscriptstyle GW}^{\scriptscriptstyle (\Lambda CDM)}= -&\frac{L \varepsilon}{2c}\ \sin^2 \alpha\ \int_{-1}^0\ dx\ \frac{1+H_0\ (\,T_E+x\,\tfrac{L}{c}\,)} {Z_E+x\,L\ \cos\alpha+\dots}\ \cos (w\,\Theta)\ ;\end{split}\end{equation}
where the argument of the trigonometrical function (less than a global phase) is given by
\begin{equation}\label{Theta}\begin{split}
	\Theta&\equiv x\ \tfrac{L}{c}\,\left(1-\cos\alpha\right)-H_0\ (\,T_E+x\, \tfrac{L}{c}\,\cos\alpha\,)\ \bigl[\,\tfrac{T_E}{2} +x\,\tfrac{L}{c}\, \left(1-\tfrac{\cos\alpha}{2}\right) \,\bigr]\ .\end{split}\end{equation}

Snapshots of the resulting timing residuals as a function of the angle $\alpha$ can be obtained for different values of the Hubble constant  \eqref{HubbleConstant}. 
\begin{figure}[!b]
	\includegraphics[scale=1.02]{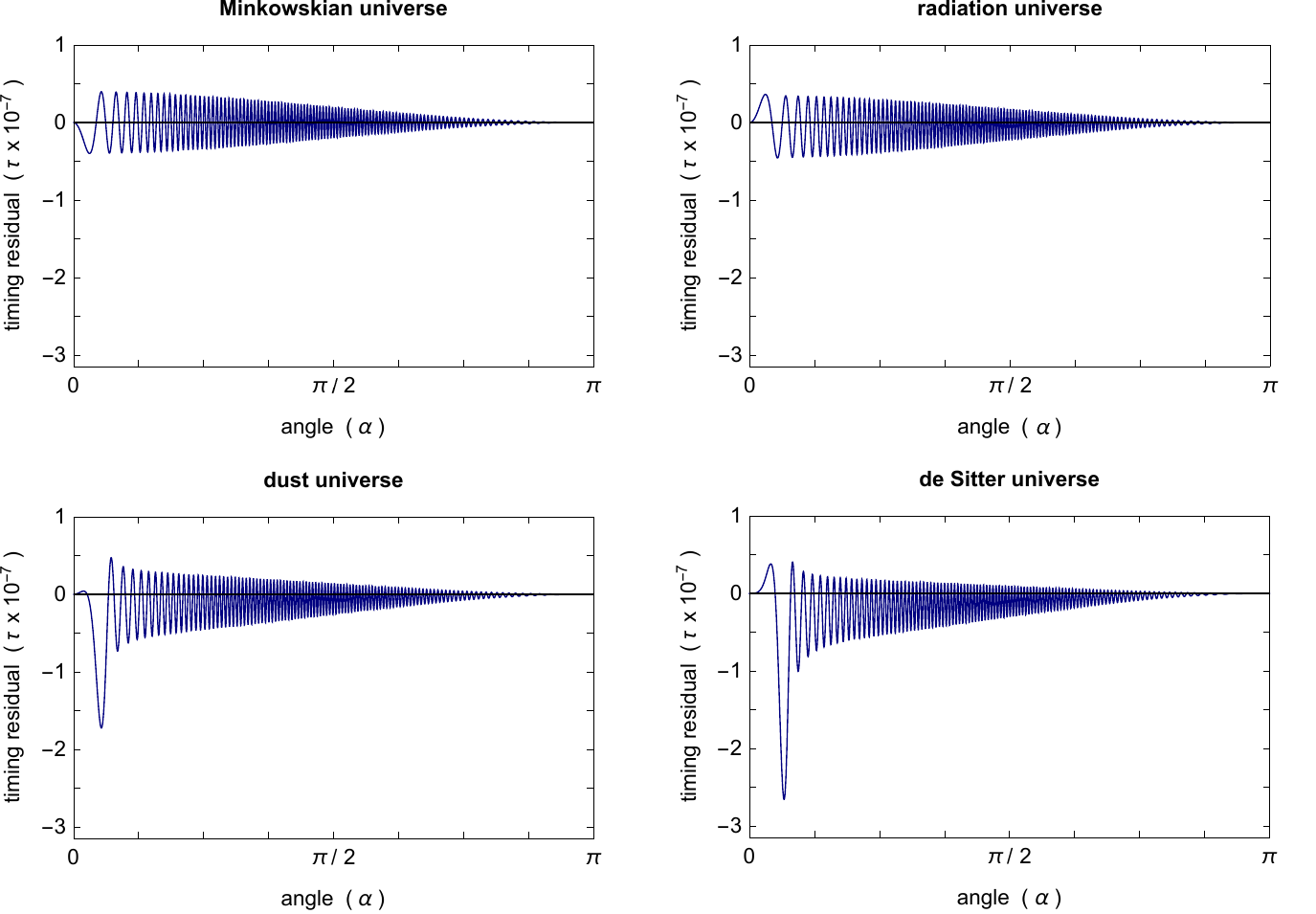}
	\vspace{1pt}\captionsetup{width=0.85\textwidth}\vspace{-13pt}
	\caption{\small Raw timing residual as a function of the angle $\alpha$ subtended by the source and the measured pulsar as seen from the observer. The figures represent a snapshot of the timing residual at a given time $T_E$ in flat space time, in a universe with only relativistic matter, with only non-relativistic matter, and in a de Sitter universe. Clearly, at least in a superficial analysis, it seems plausible to be able to disentangle both contributions. The figures depict the signal at the time $T_E=Z_E/c$. Time evolution or small changes in the parameters will change the phase of the oscillatory signal but not the enhancement clearly visible at relatively low values of the angle $\alpha$. The figures are symmetrical for $\pi\leq\alpha\leq 2\pi$.}\label{FigTimingResidual1Comp}\end{figure}
In Figure \ref{FigTimingResidual1Comp} we compare different universes characterized by various values of $H_0$ (when different densities are taken into account). The distance to the source and the gravitational waves frequency are always chosen to be $Z_E=100$ Mpc and $w=10^{-8}$ Hz respectively, except when specifically indicated.
The figure speaks by itself and it strongly suggests that the angular dependence of the timing residual is somehow influenced by the value of $H_0$, whether this value is due to a cosmological constant, radiation and/or dust. The feature that catches the eye immediately is an enhancement of the signal for a specific small angle $\alpha$, corresponding generally to a source of low galactic latitude or a pulsar nearly aligned with the source (but not quite as otherwise $e^\prime_{ij}\hat n^{i}\hat n^{j}=0$).  In fact, as it is observed, the curves are extremely sensitive to small changes in the $H_0$ parameters.

\begin{figure}[!b]\begin{center}\vspace{15pt}
	\includegraphics[scale=0.65]{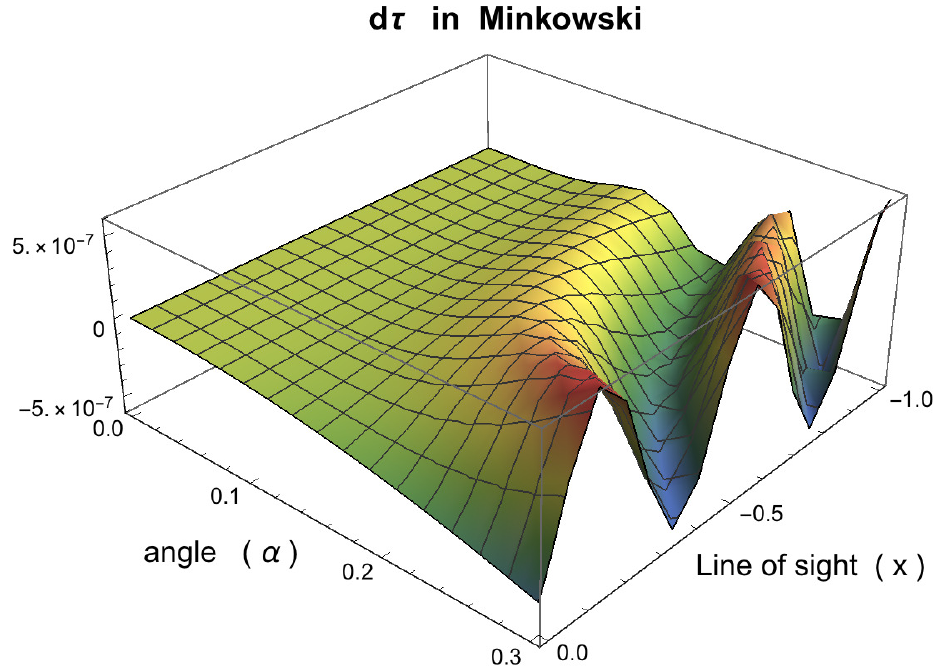}\qquad	\includegraphics[scale=0.65]{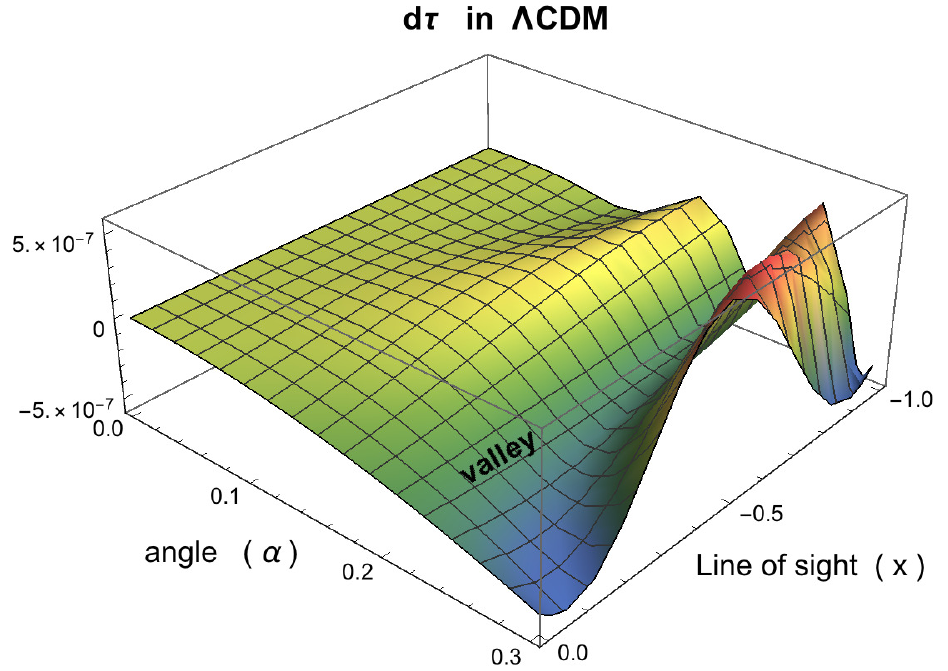}\\[2ex]	\includegraphics[scale=0.85]{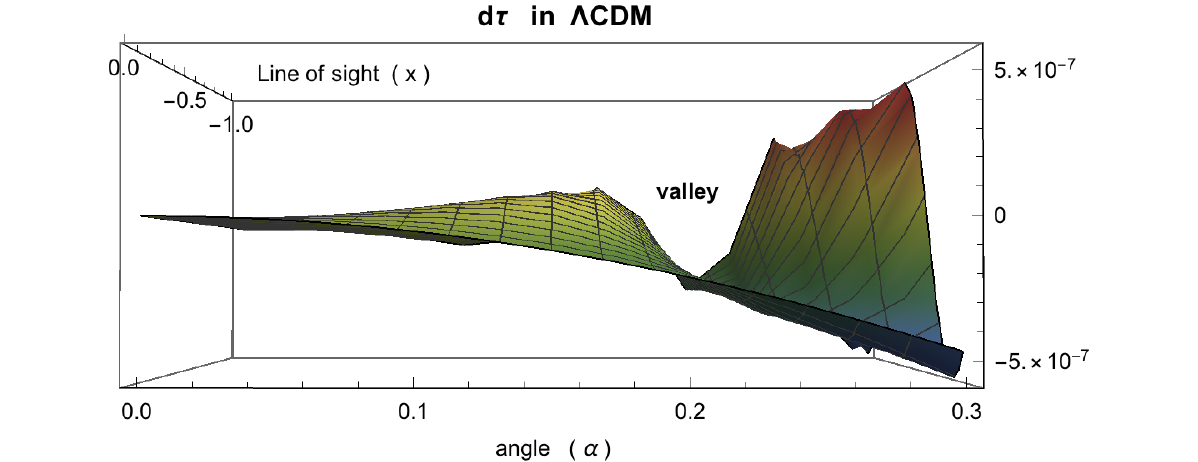}\vspace{5pt}	\captionsetup{width=0.85\textwidth}	\caption{\small In the fist two plots we compare differential timing residual computed in Minkowski and in an expanding background given by $H_0$ in \eqref{HubbleConstant}. We can observe the evolution of the phase along the line of sight from  the pulsar ($x=-1$) to the Earth ($x=0$), for different subtended angles $\alpha$. The code colours goes from the bluish for the more negative values of $d\tau_{\scriptscriptstyle GW}$, to reddish for the higher ones. Around the angle of enhancement $\alpha_{max}\sim0.2$ it can be seen that the signal takes a constant value along the line of sight (in green). The third plot is the same as the second one, but rotated in such a way that the {\sl valley} of stationary phase is obvious.}\label{FigdTau} \end{center}\end{figure}
Surprising as it is, this enhancement is relatively easy to understand after a careful analysis of the integral in equation \eqref{TimingResidualH0} (analogous results as in \cite{Espriu} have been obtained but for the Hubble parameter). In Figure \ref{FigdTau} we plot the differential timing residual in the vertical axis, versus the subtended angle $\alpha$ and versus the line of sight to the pulsar $x$.
We show as a comparison the signal propagating through a flat background on the left and through an expanding universe due to a cosmological constant on the right.
The change occurs around the angle where the enhancement is located; $d\tau_{\scriptscriptstyle GW}$ registers a constant value along all the line of sight. This corresponds to a {\sl valley} along where the phase is nearly constant. The location of this stationary phase path seems to be the responsible of the enhancement at that particular angle $\alpha$. As we will see further on, in the discussion around Figure \ref{FigTimingResidualH0}, the position of this {\sl valley} strongly depends on the value of the Hubble constant.

The effect under discussion is largely degenerated with respect to the details of the source of gravitational waves. Not much depends on the polarization or frequency of the wave front (see e.g. \cite{EspriuCC}). This is so because the enhancement is entirely a consequence of the characteristics of the FLRW metric and it cannot be mimicked by changing the frequency or the amplitude of the gravitational wave, or its polarization. All these changes would enhance or suppress the signal for all angles, not a restricted range of them. The phenomenon described is thus indeed a telltale signal. However, the curves shown in Figure \ref{FigTimingResidual1Comp} are only of theoretical interest, to prove that the effect exists. What one needs to do is to define an observational protocol.

The observable defined in \eqref{TimingResidual} is in fact not necessarily the preferred one in PTA collaborations. Typically, PTA use two-pulsars correlator and use the assumption that the gravitational wave signal in the location in the two pulsars are uncorrelated (see Hellings and Downs \cite{Hellings} and Anholm {\it et al.} \cite{Anholm}). This is so because it is assumed that stochastic gravitational wave prevails and that a single event leading to the direct observable presented here would normally be too small to be observed directly.
Unfortunately, the averaged pulsar-pulsar correlator when supplemented with the uncorrelation hypothesis normally used, likely turns into a local observable. Local observables are not affected by modifications in the wave number. One needs to be able to measure the gravitational waves in two widely separated points for the wave number to be relevant. 
Otherwise only the frequency of the gravitational waves is relevant and as we have seen this is not at all varied with respect to the usual treatment. It is also for this reason that this effect is totally invisible in an interferometric experiment such as LIGO.

It is thus essential to validate the effect observationally to consider single events observations, also in PTA, rather than looking for stochastic gravitational wave background. It is true that the latter may dominate, but the effect reported here brings such a relevant enhancement that the search technique needs to be changed. In the following sections we will provide a detailed characterization of the signal as well as its dependence on various parameters.

\section{Characterization of the signal} \label{SectionCharacterization}

Let us begin by taking into account the influence of all the components in the $\Lambda$CDM model: relativistic and nonrelativistic matter together with the cosmological constant. From its definition in equation \eqref{HubbleConstant}, the Hubble constant $H_0$ represents the current rate of expansion of the universe and contains information of its composition.
Technological progress produced a qualitative change in the observational paradigm in recent decades. The uncertainty decreased significantly, making $H_0$ one of the better measured cosmological parameters. 
However, there is still a great controversy about its actual value.
As we know, there is an apparent incompatibility between `local' measurements using standard candles \cite{Riess} and `cosmological' ones performed with the CMB spectrum \cite{H0Plank}. This fact has generated intriguing questions regarding the local behaviour of the Hubble constant and an intense debate is still ongoing. The discrepancy on $H_0$ is more than $4\sigma$ but less than $6\sigma$; strongly suggesting that it is not dependent on the method, team, or source. In fact, if lucky, our analysis could provide an alternative method of measuring $H_0$ locally, on top of making PTA observations a lot easier.

\subsection{Dependence with the source parameters}

Now we would like to make a more detailed characterization of the possible signal. We analyze how the signal changes when the main characteristics of the gravitational wave front are modified.

\subsubsection*{Dependence with the distance to the source}

\begin{figure}[!b]\begin{center}\vspace{20pt}
	\includegraphics[scale=1.3]{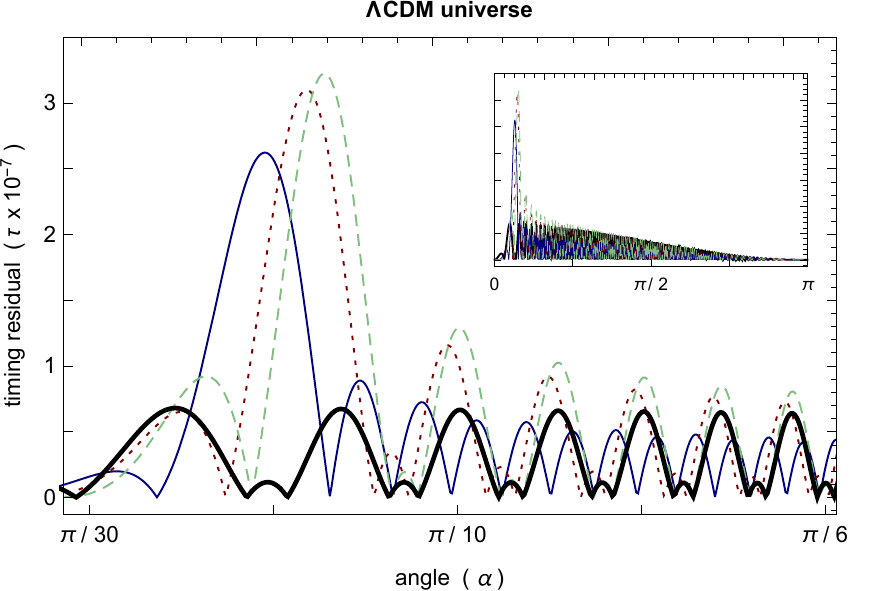}\vspace{5pt}
	\captionsetup{width=0.85\textwidth}	\caption{\small Absolute timing residual $|\tau_{ \scriptscriptstyle GW}|$ for three different values of the Hubble constant: $H_0 = 73.24 \pm 1.74$ km s$^{-1}$ Mpc$^{-1}$ reported in \cite{Riess} (green dashed curve), $H_0 = 67.4 \pm 0.5$ km s$^{-1}$ Mpc$^{-1}$ from \cite{H0Plank} (red dotted curve) and $H_0=60 \pm 6$ km s$^{-1}$ Mpc$^{-1}$ \cite{Tammann} (blue curve). We also include the propagation in Minkowski in thick black line. The remaining cosmological parameters employed for the integration do not vary from the ones used in the computation performed in Figure \ref{FigTimingResidual1Comp}. The small plot contain the four signals shown in their whole angular range for a comparison.} \label{FigTimingResidualH0}\end{center}\end{figure}

In Figure \ref{FigTimingResidualH0} we plot the modulus of the timing residual \eqref{TimingResidualH0} for three different reported values of the Hubble constant. The angular position for the maximum, $\alpha_{max}$, is significantly dependent on the Hubble constant.
In fact, it is possible to derive an approximate formula relating the angle $\alpha$ (the effect is independent of the angle $\beta$) to the value of $H_0$:
\begin{equation}\label{RelationAngleHubble}
	H_0\simeq\frac{2c}{Z_E}\ \sin^2\bigl(\ \frac{\alpha_{max}}{2}\ \bigr)\ .\end{equation}
An analogous relationship has been discussed in \cite{EspriuCC}, but when gravitational waves propagates in a purely de Sitter universe. Taking a fixed distance to the source $Z_E\sim100$ Mpc, the authors verify in this way the validity of an analogous approximate analytical expression to equation \eqref{RelationAngleHubble}, but relating the angular position of the maximum and the (local) value of the cosmological constant, i.e. $\Lambda(\alpha)$. 
This relation is extremely interesting in two ways: on the one side, this prediction could be eventually tested experimentally and may facilitate enormously the detection of gravitational waves produced in SMBH mergers. On the other side, and the most interesting for us, this relation may provide an indirect way to measure the Hubble constant and so achieve a local manner to detect the cosmological constant at subcosmological scales (once the distance to a SMBH binary $Z_E$ and the angular location of the enhancement $\alpha_{max}$ are known). It is particularly interesting to measure $H_0$ as locally as possible as its main component $\Lambda/3$ is assumed to be an intrinsic property of the spacetime, present to all scales, something that needs to be proven.

Obviously, for a fixed value of $H_0$ (assumed to be the one reported in \cite{Riess} from now on), we can test the dependence of the effect on the distance to the source, and also compare the results to equation \eqref{RelationAngleHubble}.
In Table \ref{TableVaryZ} we have tested this relation for a range of validity of the source location in the corresponding approximation. We have selected the pulsar J2033$+$1734 located at $L\sim2$ kpc (all distances to the pulsars used along our study are presented in Table \ref{TablePulsars}). Leaving the pulsar fixed, we vary the distance to the source for three different equally spaced values. We compare the angle of the peak obtained from analogous plots as the one presented in Figure \ref{FigTimingResidualH0}, called $\alpha_{graphical}$, and the same angle computed by the approximate formula \eqref{RelationAngleHubble}, named $\alpha_{theoretical}$. There is a nice agreement between both ways of finding the angular location of the maximum. These results have been checked independently in \cite{AlfaroGamonal}.

\begin{table}[!h]\centering
	\begin{tabular}{|c|c|c|c|c|c|c|c|c|}\hline
	\ Source distance\ &$\hspace{15pt}$Strength$\hspace{15pt}$&$\hspace{20pt}\alpha_{graphical}\hspace{20pt}$&$\hspace{5pt}\alpha_{theoretical}\hspace{5pt}$&$\hspace{10pt}$Deg$\hspace{10pt}$\\\hline\hline 100 Mpc& $1.71\times10^{-6}$&0.2125 / 0.2130&0.2101&$\sim12^{\circ}$\\500 Mpc&$1.60\times10^{-6}$& 0.4716 / 0.4720& 0.4735&$\sim27^{\circ}$\\1 Gpc&$1.67\times 10^{-6}$&0.6748 / 0.6751 &0.6761&$\sim39^{\circ}$\\\hline\end{tabular}\vspace{2pt} \captionsetup{width=0.85\textwidth}\caption{\small Comparison of the angular location of the maximum where the pulsar J2033$+$1734 would experience the enhancement on its timing residual. We compare the angle $\alpha_{max}$ computed in two different ways, graphically by means of numerical integrations as the one performed in Figure \ref{FigTimingResidualH0} and with the formula \eqref{RelationAngleHubble}. As can be seen the accuracy is quite remarkable.}\label{TableVaryZ}\end{table}

\subsubsection*{Dependency on the wave frequency} 

From the previous analysis we have seen that the maximum turns out to be moderately stable under changes of the parameters involved in the gravitational wave production.
Let us now focus on the dependency of the signal when the frequency of the gravitational waves change in the range of the characteristic values where the PTA experiment is expected to be performed. 
\begin{figure}[!b]\begin{center}
	\includegraphics[scale=0.8]{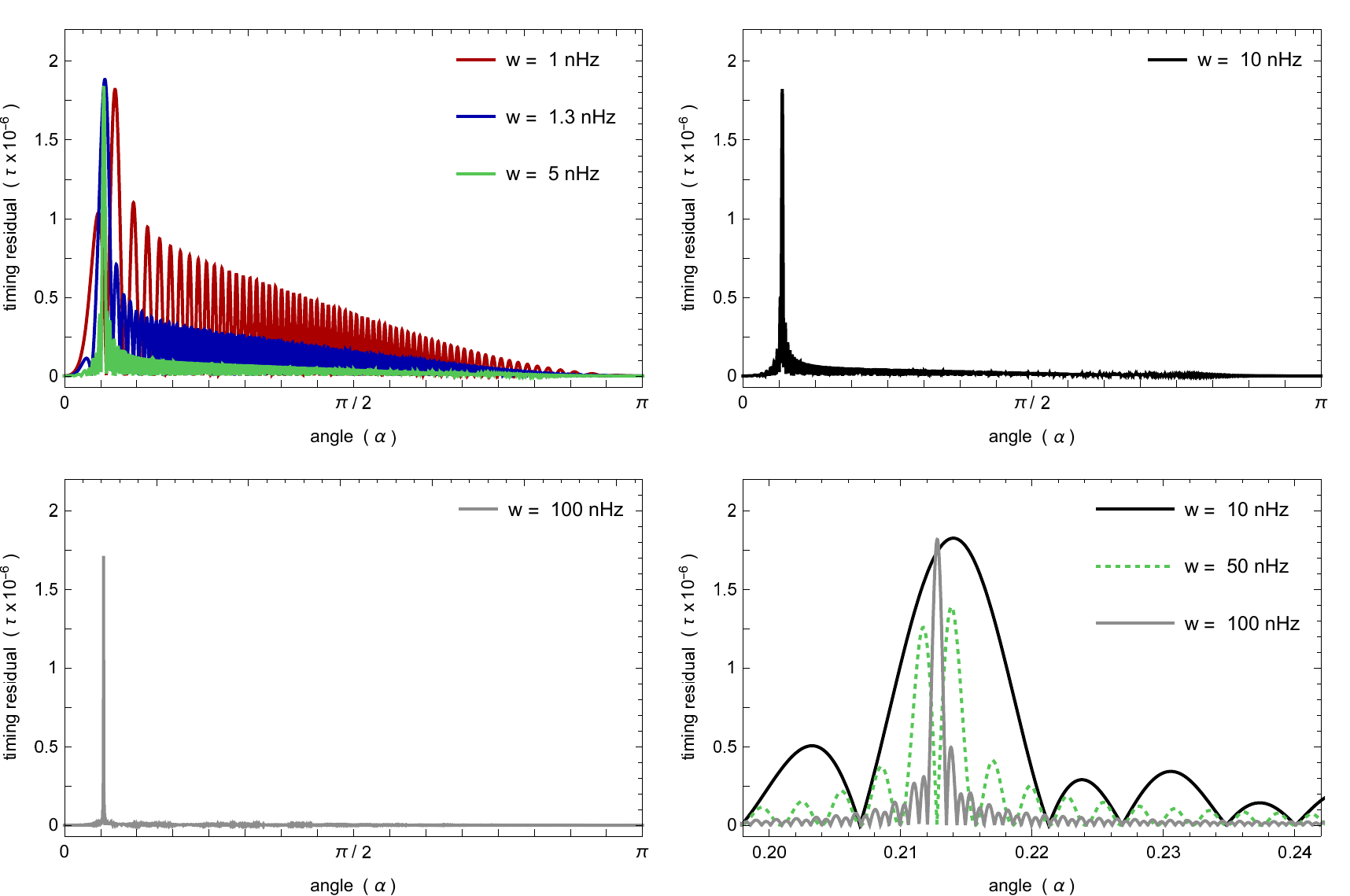}
	\captionsetup{width=0.85\textwidth}\vspace{5pt}
	\caption{\small All curves show monochromatic gravitational waves with the frequency mentioned in the plot. The first graph shows three nearby frequencies. We can see that the signal decreases quite fast its main value when the frequency is increased in some decimals. On the contrary, the peak remains nearly at the same angular position. In the second and third graphs we increase the frequency in 2 orders of magnitude. Accordingly to the first plot, the signal decreases even more while the peak does not. In the last plot we show a zoom of the two previous curves with $w=10$ nHz and $w=100$ nHz, where it is clearly seen that the signal stretches when the frequency is increased. The same compression suffers the peak.}\label{FigMonochromatic}\end{center}\end{figure}
In Figure \ref{FigMonochromatic} the absolute value of the timing residuals for the angular coordinate $\abs{\tau_{\scriptscriptstyle GW}(\alpha)}$ is plotted  for different values of the frequency. This magnitude would be registered by the pulsar J2033$+$1734 located at 2 kpc.

In the fist graph we superpose three different harmonic functions with slightly different values for the frequency: $w=1$ nHz, $w=1.3$ nHz and $w=5$ nHz. For values near  $\sim 1$ nHz the peak already doubles the rest of the signal. If we slightly increase the frequency by a few decimals, the signal falls while the peak remains stable with nearly the same height and location.
We chose these values to show how fast the decay results: increasing only 3 decimals in the frequency the signal falls by half. This pattern remains valid for all frequencies tested. At $w = 5$ nHz the peak is already one order of magnitude higher than the rest of the signal. It is also noticeable that peak becomes narrower as the frequency increases.
The fourth plot shows a zoom on the enhancement region for the latter two figures. Here again it is clear how the signal, peak and oscillations, stretch and fall while the peak remains at the same place with the same height. Sometimes two peaks next to each other appear; however, as can be seen in the signal computed with $w=50$ nHz (green curve), the value of the maximum is still approximately one order of magnitude above the signal. In conclusion, the height of the signal turns out to be nearly constant as the frequency of the gravitational wave varies.

\begin{figure}[!b]\begin{center}
	\includegraphics[scale=0.8]{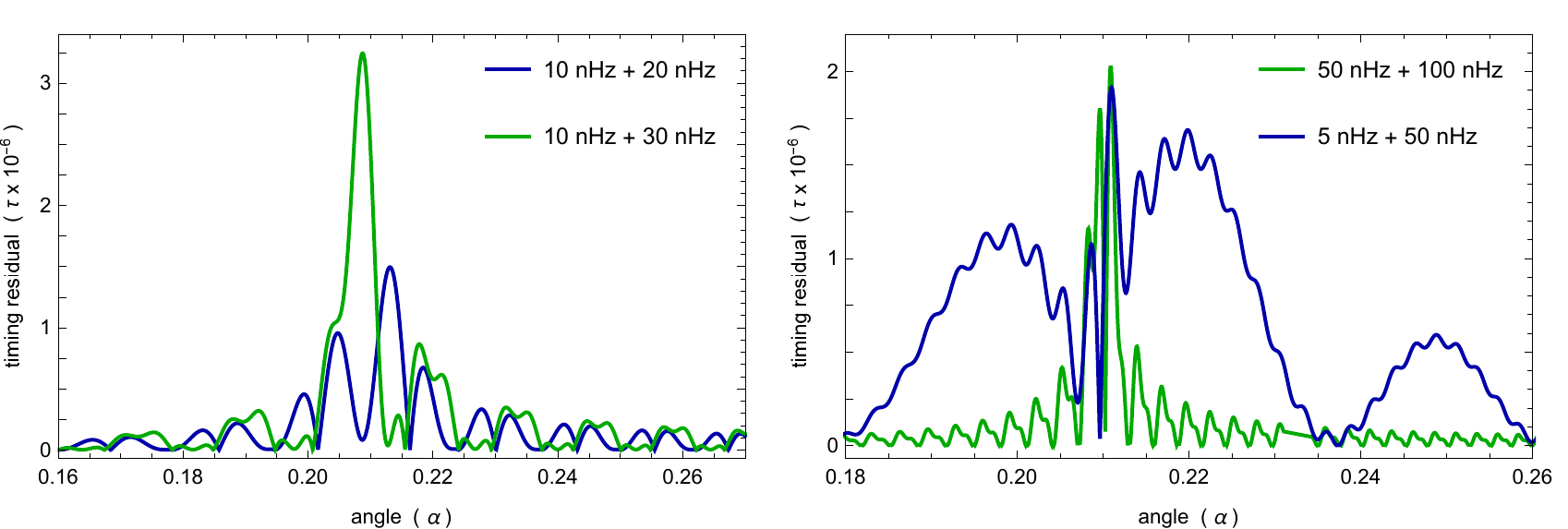}
	\captionsetup{width=0.85\textwidth}\vspace{5pt}
	\caption{\small Absolute timing residual $|\tau_{\scriptscriptstyle GW}(\alpha)|$ for non-monochromatic gravitational waves. In the first plot be combine two frequency with nearer values. It is seen that the peak suffers some modifications if its height. In the second plot, the frequencies differ in an order of magnitude and the enhancement zone becomes wider. The principal peak remains normally in the same position. } \label{FigNonMonochromatic}\end{center}\end{figure}

It is interesting to explore signals consisting in the superposition of several frequencies. In Figure \ref{FigNonMonochromatic} the absolute value of the timing residuals are computed employing bichromatic waves. The main characteristics of the signal remain the same if we plot for the full angular range $\alpha\in[0,\pi]$. We can see an oscillating signal around zero and a remarkable peak always at the same location. It turns out that the peak seems to depend only on the value of the Hubble constant and on the distance to the source.
A closer look reveals that the peak suffers several deformations when two or more frequencies are taken into account, but it is always present. In the figure we show two combinations of two frequencies each; one combines relatively similar frequencies, while the other makes a superposition of rather different ones. In summary, the main peak of the signal remains at the same location when frequencies are varied or combined. However, the combination of frequencies has a desirable effect as it enlarges the range of angles where the enhancement is visible, and this is quite welcome when it comes to a possible detection of the effect.

\subsubsection*{Time dependent amplitudes for the gravitational waves}

Let us assume a varying global factor for the wavefront such that
\begin{equation}\label{dTau}
	d\tau_{\scriptscriptstyle GW}\,\rightarrow\,A(T)\,d\tau_{\scriptscriptstyle GW}\ .\end{equation}
$d\tau_{\scriptscriptstyle GW}$ is the differential timing residual already introduced in the discussion regarding Figure \ref{FigdTau}. 
The next step is to give a structure to the form in which the amplitude varies. Let us impose two different behaviours: a signal modified by a very slowly varying exponential and by means of a linear varying amplitude. For the former, we would have
\begin{equation}\label{Amplitude}
	A(T)=\exp\left\{\frac{c}{L}\ \left[T-\Bigl(T_E-\frac{L}{c}\Bigr)\right] \right\}=\exp\left(1+x\right)\ .\end{equation}
For the second equality we have used the parametrization of the Earth-pulsar path. Then we perform the numerical integration as in \eqref{TimingResidualH0}, with the argument of the trigonometric function given by \eqref{Theta}, but taking into account the new varying amplitude. 
In such manner, when the gravitational wave arrives to the pulsar location, the amplitude equals $1$ in \eqref{Amplitude} (the integration starts at $x=-1$), then it grows while the integration is performed. At the endpoint, the amplitude reaches a value equals to $A(T_E)=e$. While this may not be a realistic change of the wave front, it serves us to test the dependence on $A(T)$. 

The same steps have been done for a linear varying amplitude. We start with a linear structure $A(T)=A_0(T-T_0+b)$ and impose the same boundary conditions. At the start of integration, we want that the amplitude takes a unit value; for this to happen we must impose $b=1/A_0$. We have used again the same parametrization \eqref{Parametrization} and we take $A_0=c/L$. With these considerations, we use 
\begin{equation}
	A(T)=1+(1+x) \end{equation}
to compute the timing residual \eqref{TimingResidualH0}.

\begin{figure}[!h]\vspace{10pt}\begin{center}
	\includegraphics[scale=0.81]{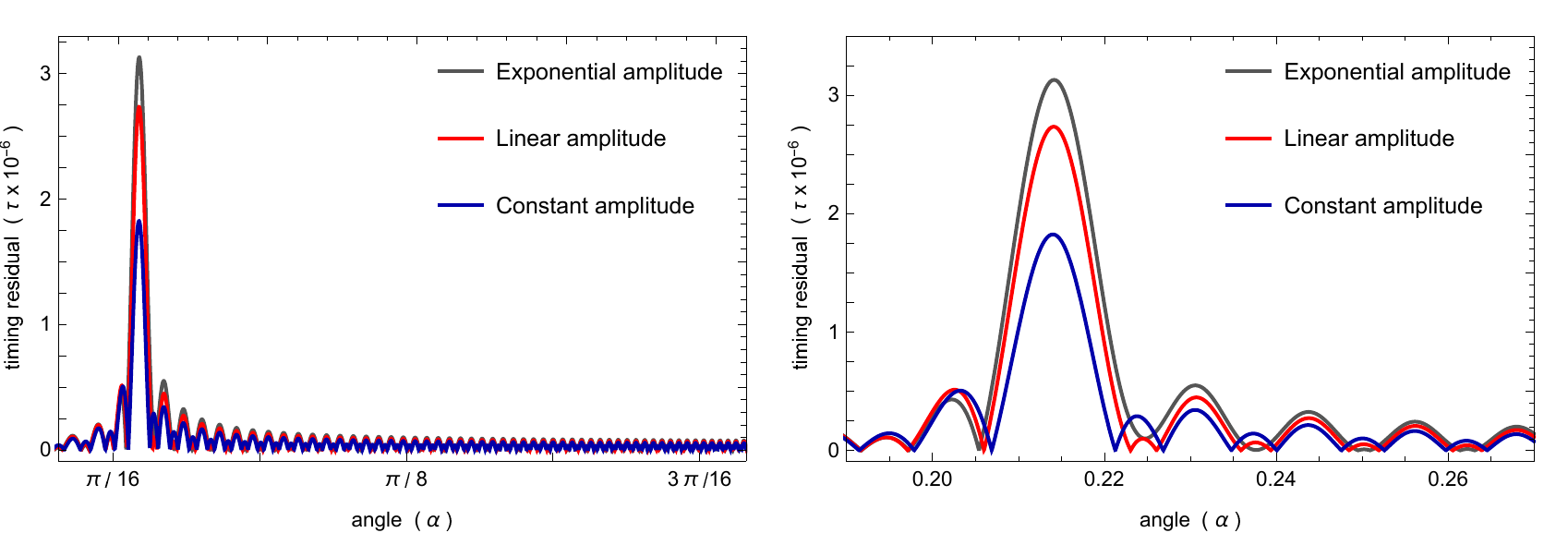}
	\captionsetup{width=0.85\textwidth}\vspace{5pt}
	\caption{\small Absolute timing residual $|\tau_{\scriptscriptstyle GW}(\alpha)|$ for gravitational waves characterized by the two different time varying amplitudes mentioned. As a comparison we also include the timing residual corresponding to a unitary amplitude in blue.}\label{AmplitudeVary}\end{center}\end{figure}

In Figure \ref{AmplitudeVary} we show how the timing residual changes when the above time-dependent amplitudes are 
introduced. The remaining parameters remain fixed with the values taken in the curve in Figure \ref{FigTimingResidualH0} with the Hubble constant value reported by the Planck collaboration. One more time we take the signals coming from the pulsar J2033$+$1734. The timing dependent amplitudes modifies the overall characteristics in the expected way, increasing the whole signal according to the function taken.

\subsection{Dependence with the position of the pulsars}

\begin{table}[!b]\centering\begin{tabular}{|l|r|l|}\hline
		Pulsar&distance (kpc)&C\\\hline\hline J0030$+$0451& \begin{minipage}{0.15\linewidth}\vspace{0.5pt}$0.28^{+0.10}_{-0.06}$ kpc\end{minipage}&N,E\\J0610$-$2100& 3.5$\pm$0.7 kpc &E\\ J0751$+$1807&$0.4^{+0.2}_{-0.1}$ kpc&E\\J1603$-$7202&$1.2\pm 0.2$ kpc&P\\J1802$-$2124&$2.9\pm0.6$ kpc&E\\\hline \end{tabular}\hspace{20pt}\begin{tabular}{|l|r|l|}\hline
		Pulsar&distance (kpc)&C\\\hline\hline J1804$-$2717&$0.8\pm0.2$ kpc&E\\J2033$+$1734&$2\pm0.4$ kpc&E\\J1939$+$2134&$5^{+2} _{-1}$ kpc &E,P\\J1955$+$2908&$4.6\pm0.9$ kpc&N,E\\&&\\\hline \end{tabular}\captionsetup{width=0.85\textwidth}\vspace{2pt}
	\caption{\small Pulsars employed in the analysis \cite{ATNF}. The distances to the pulsars have been taken from the reported values in \cite{Verbiest}. The column C referees to the collaboration monitoring the pulsar: N stands for the NanoGRAV collaboration, E for the EPTA and P for PPTA.}\label{TablePulsars}\end{table}

Let us focus on the dependence of the peak on the distance $L$ to the pulsar. Let us fix the gravitational waves source at $100$ Mpc and use the same values for the parameters as in previous analysis. We plot the absolute value for the timing residual versus the angle subtended by the source and the measured pulsar as seen from the observer, $|\tau_{\scriptscriptstyle GW} (\alpha)|$, for different MSPs from Table \ref{TablePulsars}.

\begin{figure}[!b]\begin{center}
	\includegraphics[scale=1.2]{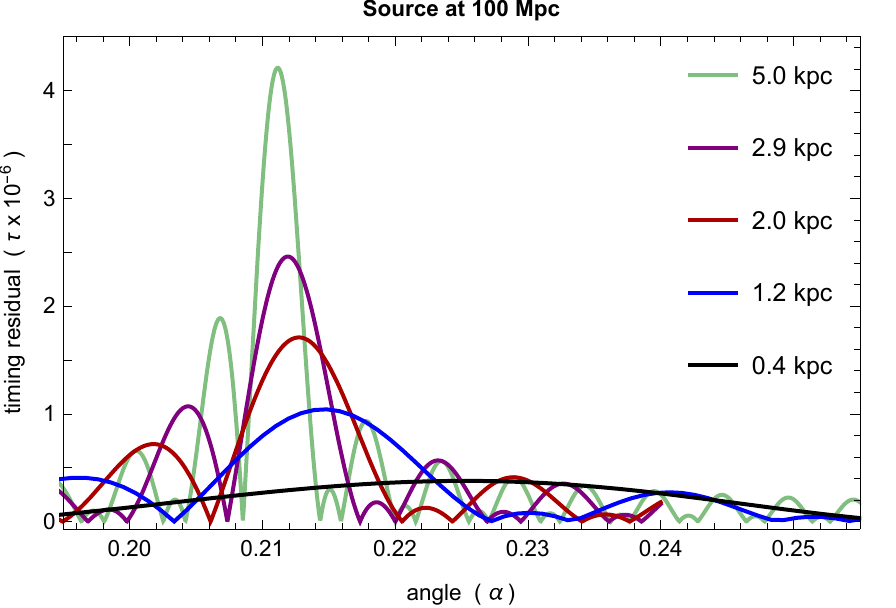}
	\captionsetup{width=0.9\textwidth}\vspace{5pt}
	\caption{\small Zoom of $|\tau_{\scriptscriptstyle GW}|$ on the enhancement region. Each curve represent each of the 5 pulsars chosen with different distances to the Earth. It is worth to note that all peaks are way higher than the rest of the signal. Even the lowest in height, J0751$+$1807 (in black), is significantly bigger (a factor 5) than the rest of the remaining signal. For instance, take as an example the blue curve corresponding to the pulsar J1603$-$7202; this pulsar has nearly the distance employed in Figure \ref{FigTimingResidualH0} (where for the red curve $L$ was fixed to $1$ kpc). The remaining pulsars are: J2033$+$1734 in red, J1802$-$2124 in violet and J1939$+$2134 in green.} \label{FigLdependence}\end{center}\end{figure}

We choose five pulsars with locations halfway apart with respect to the Earth; they are J1939$+$2134, J1802$-$2124, J2033$+$1734, J1603$-$7202 and J0751$+$1807 in a decreasing order of distance from us. In Figure \ref{FigLdependence} we show a zoom of the enhancement region.
As expected, as long as the Hubble parameter as well as the distance to the source are fixed parameters, the angular position where the enhancement peak is located do not vary for the different curves. On the contrary, the strength of the peak grows to the extent that farther away pulsars are used in the calculation. There is a logic behind this, as the accumulative effect of the enhancement becomes larger as it does the optical path.

We emphasize that in all cases the enhancement peak is several orders of magnitude bigger that the rest of the signal. Even for the black curve in Figure \ref{FigLdependence} where at fist sight and because of the scale seems that there is no peak, this reappears when one plots for a wide range of values for $\alpha$.
Note that this pulsar, J0751$+$1907, is located at a characteristic distance very similar to the one used in Figure \ref{FigTimingResidualH0} where $L=0.32$ kpc. In fact, its corresponding plot is very similar to those curves too where it is clear that the peak is still significantly bigger than the whole signal. 

What we also see from these figures is that while the peak becomes higher when the distance to the pulsar grows, it becomes also narrower. So peaks resulting from closer pulsars are broader. Figure \ref{FigLdependence} indicates that pulsars located in a region characterized by $\overline\alpha\in[\alpha_1;\alpha_2]$ register an anomalous enhancement in their timing residual because of the passage of a gravitational wave produced at a distant source. 
This interval is related with the width of the peak; in fact the global maximum is the center of this interval: $\alpha_{max}=(\alpha_1+\alpha_2)/2$. All in all, pulsars placed more near to us will have a bigger range of values $\overline\alpha$ where they would register a significant enhancement in $\tau_{\scriptscriptstyle GW}$.

\begin{figure}[!t]\vspace{10pt}\begin{center}
	\includegraphics[scale=0.95]{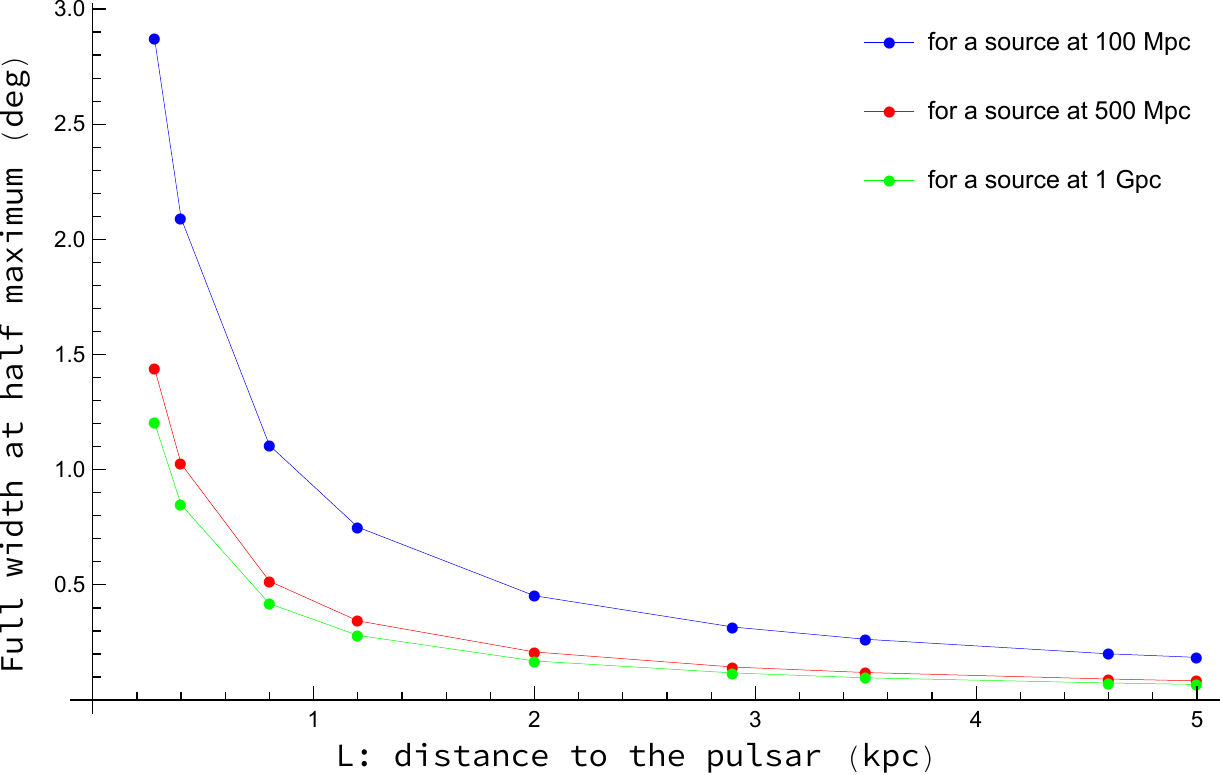}
	\captionsetup{width=0.9\textwidth}\vspace{10pt}
	\caption{\small Each point represents the angular value using the FWHM criterion for obtaining the main peak width of the signal. We compute this magnitude for each of the 9 pulsars in Table \ref{TablePulsars}, order in the $x$-axis by the distance from the pulsar to the Earth. Each curve corresponds as indicated to a different source location.}\label{FigWithVsL} \end{center}\end{figure}

As a more realistic example we will make use of all the real pulsars listed in Table \ref{TablePulsars}. Using three different source locations, we can draw the $\alpha$ profile of the modulus of the timing residuals corresponding to each of the pulsars. Added to this, we need a criterion to link the width of the peak with the interval $\overline{\alpha}$ where the enhancement happens. A conservative election is the {\it full width at half maximum} (FWHM) parameter. It is defined as the distance between points on the curve at which the function reaches half of its maximum value\footnote{Pulsars further away from Earth would register a higher but narrower peak. The FWHM measure can be flexibilized; as the half peak is quite a big threshold that can be reduced significantly in pursuit of increasing the width where the enhancement occurs. This would significantly increase the areas where this effect can be detected in the sky.}. 
In Figure \ref{FigWithVsL} we show the result of computing the FWHM for the 9 pulsars shown in Table \ref{TablePulsars} when the gravitational wave source is fixed, arbitrarily, at 3 different distances (100 Mpc, 500 Mpc and 1 Gpc). Accordingly, we can remark the convenience of employing nearest pulsars and nearer sources as much as possible. As reported in the first data release of the IPTA collaboration in \cite{Verbiest} almost the 25 \% of the pulsars considered are at distances $L \leq 0.5$ kpc from us.

\begin{figure}[!b]\vspace{10pt}\begin{center}
	\includegraphics[scale=0.75]{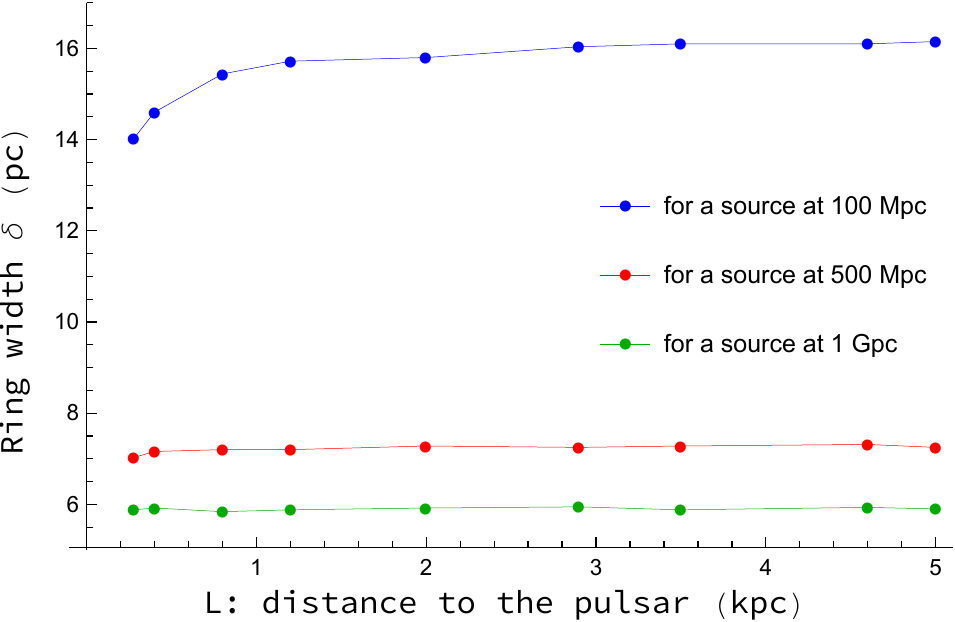}\qquad\includegraphics[scale=0.33]{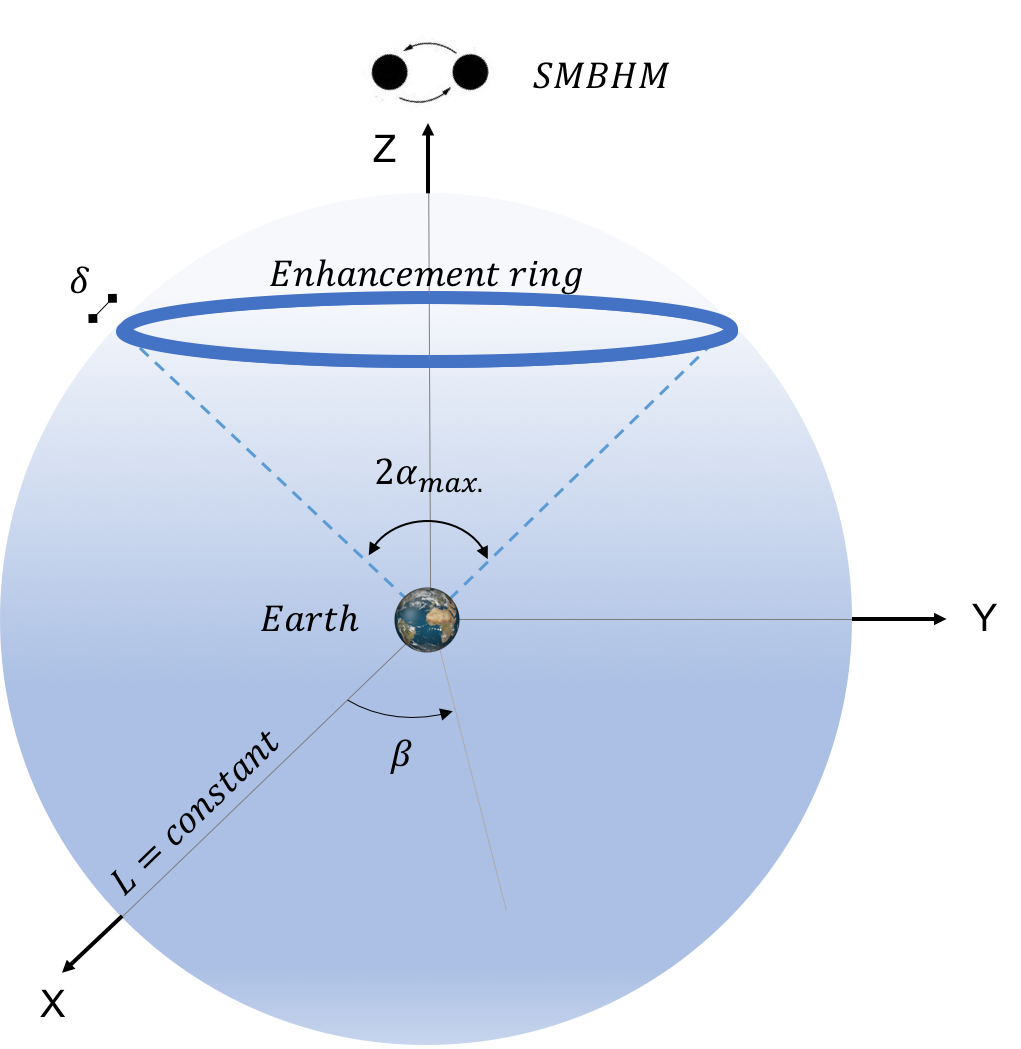} \vspace{10pt}
	\captionsetup{width=0.85\textwidth}	\caption{\small Spatial region (in blue) where any falling pulsar would experience an anomalous enhancement in timing residual $\tau_{\scriptscriptstyle GW}$. This region is shown for only one value of the distance $L$, but appears for all values of $L$ with nearly the same characteristics (the complete region is a truncated cone-like surface). On the left we show the width $\delta$ of the ring that seems to depend only on changes of the source location $Z_E$.} \label{FigRing}\end{center}\end{figure}

We can convert the angular interval where the enhancement takes place to astronomical distances; i.e. $\overline{\alpha}\rightarrow \delta $. Taking into account at which distance the pulsar would be located, we can use trigonometry to calculate the size in parsecs that would correspond to a determined angular interval subtended at that distance $L$. For the 9 pulsars referred previously we obtain the plot in Figure \ref{FigRing}. 
A source located at the referred distance would "light up" a zone characterized by an angular width $\overline{\alpha}$. This interval would correspond to a distance $\delta$ as it is shown in the plot on the left. Any pulsar falling in this interval would experience a significant enhancement in their timing residual. 
The three different curves show that the size of this zone is of a few tens of parsecs and its width is nearly constant with respect to the distance $L$ (the fluctuations of each curve when varying the location $L$ of the pulsar represent more or less 10\% of its value). We can conclude that the width of the peak seems to be highly dependent on the location of the gravitational wave source $Z_E$, as it was for its position $\alpha_{max}$ derived in equation \eqref{RelationAngleHubble}.

There is another important aspect to take into account. According to the definition of the timing residual, the effect is $\beta$ independent and nearly $L$ independent. The $\beta$ independency of the theory is translated into a symmetry of revolution for the enhancement region $\delta$ around the $Z$-axis (the axis coinciding with the direction to the source). This means that a SMBH merger would generate a ring of enhancement at each constant distance $L$ from the Earth characterized by the azimuthal angle ranging between $\beta\in[0;2\pi)$. 
The angular radius of the ring is given by the polar angle $\alpha_{max}$ (recall the results on Table \ref{TableVaryZ} where different angular locations of the enhancement have been obtained by using different distances to the source) and its width $\delta$ is given in parsecs on the plot next to the scheme of Figure \ref{FigRing}, according to the corresponding distance involved. This region where a significant deviation in the timing residual would be registered is shown in the scheme as the blue ring and will be called the {\sl enhancement region}. The invariance with respect to $L$ makes this region a truncated cone-like surface more that a thick ring; however for practical issues that will be discussed in the next section, let us retain the region as a ring-like image such that pulsars placed there would be seen by us as experiencing a notorious deviation on the time of arrival of their pulses. The width of this ring-like enhancement region will be in the 1 to 20 pc range, depending on the distance to the source.

\section{The International Pulsar Timing Array project}

In order to detect correlated signals in observed pulsar timing data, an array of MSPs is monitored with large and sensitive telescopes. Such array called PTA\footnote{The acronym PTA stands for the set of pulsars as well as for the different collaborations monitoring these pulsars.} exploits the great period stability of the MSPs where correlated timing variations among the ToA pulses in the array are searched for \cite{Romani1989FosterBecker1990}.

Three different collaborations has been set up in three different places and have been taking data for more than a decade. Specifically, the Australian Parkes PTA (PPTA) \cite{Manchester2013}, the North-American Nanohertz Observatory for gravitational waves (NANOGrav) \cite{Arzoumanian2015} and the European PTA (EPTA) \cite{Desvignes2016}. Combined, they form the International PTA (IPTA) \cite{Hobbs2010ManchesterIPTA2013}. 
This collaboration monitors around 50 pulsars in the quest for finding footprints of the passage of gravitational waves in the aforementioned correlated signals.

The quest for detection of gravitational waves in PTA follows two main avenues. One proposal focuses on stochastic (isotropic and incoherent) gravitational wave background \cite{Jenet2005,Hellings, FosterBacker1990Kaspi1994} produced by incoherent superposition of radiation from the whole cosmic population. In \cite{Verbiest} a brief review goes through the main sources for this background and define the resulting signal that would (or not) be detected by PTA.

More interesting for our purposes is the search for single or individual sources, if they are sufficiently bright. Clearly, a nearby binary merger would be detectable if and only if the gravitational signal stand out above the root-mean-square value of the aforementioned background \cite{Sesanna2009}. There is controversy on whether a single merger event is detectable by PTA \cite{Seto2009vanHaasteren2010Pshirkov2010Favatara2009}. This debate and its somewhat pessimistic conclusion assumes of course the standard analysis of the signal-to-noise ratio.

As of today there is no conclusive evidence of an indirect measurement of any gravitational radiation in this regime of frequencies wherever they are produced. Single-source limits show that all proposed binary mergers are still below current sensitivity levels: recent limits have been derived in \cite{Babak2015} from the EPTA, in \cite{Arzoumanian2014}  from the NANOGrav data and in \cite{Zhu2014} from the PPTA collaboration. In addition, similar conclusions have been reached for gravitational waves bursts \cite{Wang}. Our analysis reported here may suggest to reconsider the expected signal and how to look for it.

When gravitational waves propagate over flat spacetime the phase effect discussed before is lost as we have shown in the first plot of Figure \ref{FigTimingResidual1Comp} (Minkowskian propagation). Including the familiar redshift in frequency alone as the only way of including the effect of the FLRW metric, without taking in consideration the effect discussed in this work, for each single-source event a ring-like region of enhancement in the sky is possibly missed. The enhancement should be affecting any pulsar in it, irrespective of the distance to the observer $L$. We recall that the enhancement may represent an increase in the signal of several orders of magnitude. From the study presented in the previous sections and the typical values of $L$ we concluded that the width of this ring is of the order of 1 to 20 pc.

\begin{figure}[b!]\begin{center}\vspace{-45pt}
	\includegraphics[scale=0.7]{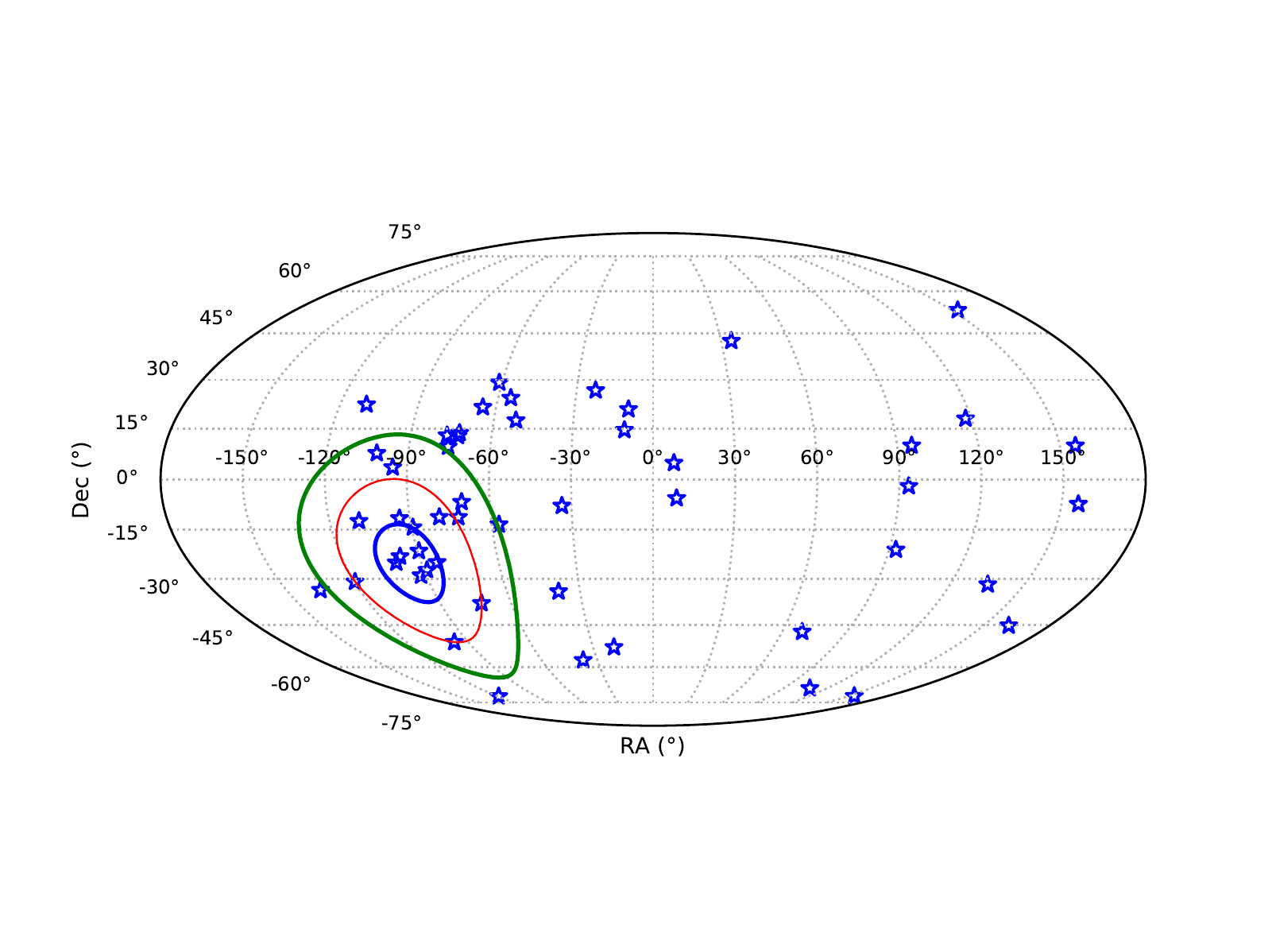}
	\captionsetup{width=0.85\textwidth}	\vspace{-40pt}
	\caption{\small Galactic distribution of all MSPs observed by the IPTA are indicated by blue stars. Enhancement rings correspond to three different sources placed at 100 Mpc in blue and at 1 Gpc in green. The red ellipse would be the enhancement ring if the three further south pulsars are supposed to perceive the enhancement. When we see that at least three pulsars register an anomalous timing residual, we have enough information to know where the effect has its origin. This three points determine an enhancement ring of $\alpha_{max}\simeq25^\circ$ with respect to a center where there should be placed an hypothetical gravitational wave source. According to the relation \eqref{RelationAngleHubble}, the angle of enhancement constrain the spatial location of the source to $Z_E\simeq443$ Mpc. The angular radius subtended between the source and the pulsar are for the other two enhancement rings: $\sim39^\circ$ for the green ellipse and $\sim12^\circ$ for the blue one.}\label{FigMollwide} \end{center}\end{figure}

The image presented in Figure \ref{FigRing}, we reminds us of the coordinates of the pulsars monitored by the IPTA collaboration, irrespective of the value of $L$, the distance between the pulsar and the Earth.
In the graphical representation on Figure \ref{FigMollwide}, for each sphere of the sky at a distance $L$, a SMBH merger would induce a circular ring-like region with a diameter given by twice the angle subtended between the source and the pulsar with respect to the source. In Figure \ref{FigMollwide} we show in a Mollwide projection the location of these pulsars together with the enhancement ring for a source located at $100$ Mpc and at $1$ Gpc. 

Assuming that each collaboration measures the timing residual over an adequate integration period a single-event should represent a simultaneous firing of at least three pulsars. The collaboration should therefore look for such triple signals by making all possible combinations. Each of such triple signals would define a vector pointing towards the source. Once this direction is tentatively determined, a more detailed analysis of the signal effect on other pulsars would be possible.

Note that along this chapter all signals referred to snapshots at a given specified time. Of course, there is a large background that obscures the signal (even taking into account the enhancement) and therefore the signal has to be integrated over an adequate period of time. This issue has not been considered here, but we refer the reader to the work \cite{EspriuCC} where it was preliminarily estimated that observation of the given set of pulsars over a period of three years with an eleven-day period would provide a sufficiently good signal. Needless to say that this is a very crude value that needs careful consideration.

\chapter{Conclusions}

The purpose of this work is to provide some new insights into different aspects of black hole physics, both at a quantum level and at a classical or observational point of view. 
In reference to the former, we discuss its possible internal structure employing some common approximations in the context of quantum mean field theories and give different internal structures based on a graviton condensate structure. With respect to the observable aspects, we discuss how in which manner the fact that we live in an expanding universe would produce any measurable effect over the gravitational waves produced in a black hole merger.

We have started analyzing a novel approach that reformulates the quantum theory of black holes in a language of condensed matter physics. The key point of the theory is to identify the black hole with a Bose--Einstein condensate of gravitons.

In Chapter \ref{ChapterBHasBEC} we have conjectured the set of equations that play the role of the Gross--Pitaevskii non-linear equation that describes the ground state of a quantum system of identical bosons using the Hartree--Fock approximation as an interacting model; they are the field equations derived from the Einstein--Hilbert Lagrangian after adding a chemical potential-like term. We have used a number of different techniques to analyze these equations when the perturbation (i.e. the tentative condensate) has spherical symmetry. The equations appear at first sight rather intractable, but by doing a perturbative analysis around the black hole Schwarzschild metric at quadratic order (i.e. including the leading non-linearities) we found that the chemical potential necessarily vanishes in the exterior of the black hole. On the contrary, in the interior we have found two sets of solutions, one of them has to be discarded as producing a non-normalizable result. The other one leads to a nonzero chemical potential in the interior of the black hole that behaves as $1/r^2$. 
Therefore, there is a finite jump on the chemical potential at the black hole horizon. Surprisingly --or maybe not so-- this solution modifies the coefficient of the $tt$ and $rr$ terms in the Schwarzschild solution, but not its functional form. Of course if there is no chemical potential at all, the modification vanishes, in accordance with well known theorems. However, if the former is nonzero, the modification affecting the metric is also necessarily nonzero.
 
The perturbative analysis triggers a unique consistent physical solution for the non-perturbative (exact) theory. This solution is characterized by a constant density of the wave function for the condensate. 
From the existence and knowledge of this solution, an unambiguous relation between the number of gravitons and the geometric properties of the black hole is obtained. Hence, we find an expression for the Schwarzschild radius that involves an a priori independent and tuneable parameter, the dimensionless chemical potential $X$ (related to the mean-field wave function of the condensate $\varphi$). We find this somewhat strange as this would be a new black hole parameter. Therefore we favour the universal value $X=-1$ as discussed in the text. 

As should be obvious to the reader who has followed our discussion, the approach is somewhat different from the one developed in the initial papers by Dvali, G\'omez and coworkers. We assume from the start the existence of a classical geometry background that acts as confining potential for the condensate. The fact that the functional form of the metric perturbation induced by the condensate is exactly the same as the original background, of course gives a lot of credence to the possibility of deriving the latter from the former in a sort of self-consistent derivation. We have not explored this possibility in detail yet.  

The picture presented in the context of Schwarzschild black holes has been extended to charged black holes (with Reissner--Nordstr\"om metric) and to the de Sitter cosmological horizon. We have found that in all these situations a graviton condensate is possible in the classically inaccessible zone when a chemical potential term is appropriately introduced in Einstein's equations. This seems to suggest that the existence of an horizon is intimately linked to the existence of such condensates.

In order to discern whether this possibility describes in any manner the nature of black holes, in Chapter \ref{ChapterQLE} we analyze how the notion of quasilocal energy in the context of the classical theory of general relativity can be extended so as to encompass the condensate description. Even though there is no total consensus on the definition of a quasilocal energy, we consider the Brown--York prescription \cite{BrownYorkMath} as being adequate for our purposes and physically well-motivated in the present context. If this definition for the energy it chosen, it is found that a graviton Bose--Einstein condensate is energetically favourable for all the studied cases in the regions beyond the horizon.

The possibility of defining a quasilocal energy entails the possibility of deriving a quasilocal potential describing the binding of falling matter to a black hole. Its well-like structure, makes the binding to take place around a fairly thin shell located at the event horizon. This seems to indicate that matter falling into the black hole accumulates at both sides of the horizon, but always in its vicinity. This confers to the black hole a fuzzy boundary. When matter falls and get trapped by this potential, gravitational wave emission is necessarily produced on energy conservation grounds. Part of this energy is emitted towards infinity, however it is expected that a large fraction is trapped in the inner region. This seems to be a very plausible explanation on how the condensate of gravitons behaves and the value of the graviton condensate order parameter $\varphi$ increases. As explained in section \ref{GravitationalBinding}, order of magnitude considerations indicate that practically most of the black hole mass may be stored in the form of a graviton condensate state. Besides, the unceasing capture of {\it gravitons} may indicate that the condensate value should be close to its limit value, $\varphi\rightarrow1$. This statements gives strong plausibility to the original proposal of Dvali and G\'omez; namely that a black hole consist mostly of a BEC of gravitons.

In Chapter \ref{ChapterBHasBECExtension} we show that the extension of the energy considerations to the Reissner--Nordstr\"om case is straightforward. However, is not always possible to define a significant quasilocal potential for every spacetime. For instance, in de Sitter we have been unable to do so, this concluding that while it is energetically viable, a graviton condensate is actually not present.

The validity of the previous picture brings about many interesting points. One of them is that the horizon, while being from the metric point of view, quite well defined, becomes necessarily fuzzy due to quantum effects. A second consequence is that matter falling in the black hole does not get `crunched' by the singularity at $r=0$. When it comes to the quasilocal energy, it is perfectly regular at the origin, and so is the associated potential (it tends to a constant). In the classical theory, the location of the horizon is defined as a limit for timelike world lines to exist. Points where 4-velocity turns null form a `one way' spatial surface. Here, following our interpretation, we see that matter is captured by the quasilocal potential and certainly cannot escape, except for the occasional thermal fluctuation. However, no loss of information would occurs here, except for the usual thermodynamic irreversibility. If the mass of the black hole decreases for some reason and $r_S\to 0$, the potential becomes progressively shallower and matter stored in its potential well can escape. Needless to say that one mechanism whereby the mass of the black hole could decrease is through evaporation. This is an intrinsically quantum mechanism that is accurately described in the usual way. However, if a graviton condensate is present, and we have argued that this is very likely with a value close to the limiting value, one should take into account that Bose--Einstein condensates are not localized objects and therefore some amount of leaking should be present. In their original proposal Dvali and G\'omez sustained the point of view that Hawking radiation could be understood in this way. From the point of view that most of the black hole mass is stored in the form of the condensate, this is a likely possibility, but we have not considered this issue in the present study.

Next, in Chapter \ref{ChapterGW}, we review shortly the main aspects of the $\Lambda$CDM model that describes in a very good agreement the universe over where observational phenomena are aimed to be measured. For instance it is well known that accelerated masses generate gravitational radiation in the form of gravitational waves.
We have reviewed how a nonzero cosmological constant modifies the propagation of a gravitational wave, with an additional contribution to the one that is usually taken into account --the redshift in frequency. The effect is entirely due to the change of coordinate systems between the reference frame where the wave originates (e.g. the merger of two gigantic black holes) and the reference frame where waves are measured in PTA, namely cosmological FLRW coordinates. We have proceeded to extend these results to the case where non-relativistic dust is present, and later on to the combined and analytical situation where non-relativistic matter and vacuum energy are both present. 
Finally the inclusion of radiation motivates a picture dependent on the Hubble constant, including all the components of the background in which gravitational waves propagate. All the results have been derived in a linearized approximation where only the fist deviations are considered. This approximation is however enough for the case of study, as subsequent corrections are seen to be extremely small when the measured values of matter density and cosmological constant (the two main components) are used, combined with the distances involved in the problem.

In the last chapter, we have confirmed the validity of the approximate solution shown in equation \eqref{gwfrw3} by solving directly, up to ${\cal O}(H_0)$, the wave equation for linearized perturbations in a FLRW geometry. The solution has an effective wave number $k_{eff} \neq w_{eff}$, where $w_{eff}$ is the familiar redshifted frequency. This does not mean that gravitational waves propagate subluminically; they do so when the proper `ruler' distance is considered.
This approximate solution is valid to describe gravitational waves originating from sources up to $\sim1$ Gpc, or, equivalently, up to redshift $z=0.2$ approximately. It is important to emphasize that, at this order, all cosmological parameters enter in the form of the Hubble constant $H_0$, but it is not so when higher orders are included.

It came out as a surprise that the gravitational wave solution has a definite impact in observations of gravitational waves involving non-local effects; there is an integration over a path in comoving distances. This is the case of observations made in the framework of the IPTA project (and, incidentally, also in LISA, even though the analysis in this case is yet to be performed). Also in Chapter \ref{ChapterPTA} we have discussed in great detail the dependence of the effect on various parameters influencing the observation: superposition of frequencies, time-varying amplitudes, distance to the source, distance to the intervening pulsars. In all cases the signal should be perfectly visible, well above the usual analysis assumed. The analysis here presented may help to establish a well designed search strategy. We note that one aspect we have not discussed at all are the integration times. In \cite{EspriuCC,E2} an observational strategy was mentioned, but we consider that the concourse of the observational teams is essential in this point.

So far no clear positive measurement of gravitational waves has come out from the existing collaborations. The effect reported here appears to be firmly established, and in conclusion it may provide an opportunity for the IPTA collaboration to detect correlated signals in triplets of pulsars and from there on determine location and other properties of the source of gravitational waves. Needless to say that an independent local determination of $H_0$ (and eventually $\Lambda$) would be of enormous interest.

\chapter*{List of publications}
\addcontentsline{toc}{chapter}{List of publications} 

The research carried out during this PhD thesis has given rise to the following list of papers:

\begin{itemize}
	\item {\it Bose$-$Einstein graviton condensate in a Schwarzschild black hole.}\\
	J. Alfaro, D. Espriu and L. Gabbanelli\\
	Class. Quant. Grav. {\bf 35}, 015001 (2018); doi: 10.1088/1361-6382/aa9771 \\
	arXiv:1609.01639\\

	\item {\it On the propagation of gravitational waves in a $\Lambda$CDM universe.}\\
	J. Alfaro, D. Espriu and L. Gabbanelli\\
	Class. Quant. Grav. {\bf 36}, 025006 (2019); doi:	10.1088/1361-6382/aaf675 \\
	arXiv:1711.08315 \\
	
	\item {\it Condensates beyond the horizons}.\\
	J. Alfaro, D. Espriu and L. Gabbanelli\\
	arXiv:1905.01080\\

	\item {\it Measuring $H_0$ in PTA with gravitational waves}.\\
	D. Espriu, L. Gabbanelli and M. Rodoreda\\
	arXiv:1908.08472

\end{itemize}

\chapter*{Abstract in Spanish}
\addcontentsline{toc}{chapter}{Abstract in Spanish} 

El objetivo de la presente tesis es profundizar en diversos aspectos de la física de los agujeros negros. Tanto en lo que respecta a sus características constitutivas fundamentales, su "estructura" interna, como a la posibilidad de observar o detectar mediante observaciones astrofísicas ciertos efectos producto de su dinámica.

Por un lado, hemos seguido las ideas de Dvali, Gómez {\it et al.} quienes han sugerido la posibilidad de que un agujero negro sea un condensado de Bose--Einstein de gravitones débilmente interactuantes. En nuestro caso hemos estudiado la existencia de este tipo de soluciones sobre diferentes métricas de agujero negro (Schwarzschild y Reissner--Nordstr\"om) que actuarían como potencial confinante para dichos condensados. Un parámetro necesario para ello, es el equivalente a un potencial químico que debe ser incorporado a la relatividad general. Cabe destacar que la solución encontrada puede ser interpretada como la función de campo medio del condensado. Además resulta fuertemente ligada a la estructura clásica de la métrica que la sustenta.

Por otro lado, es bien sabido que la aceleración de cuerpos muy masivos producen perturbaciones de tipo onda en el espaciotiempo. Son de nuestro interés las ondas gravitatorias de baja frecuencia, provenientes de la colisión de agujeros negros supermasivos y que deberían poder ser detectadas mediante sistemas de púlsares (Pulsar Timing Arrays). De acuerdo a una línea de investigación desarrollada por Espriu {\it et al.} la presencian de una constante cosmológica podría tener un efecto en la propagación y por lo tanto en la detección por parte de la colaboración IPTA de estas ondas. En la presente tesis hemos generalizado el método para incluir diferentes tipos de materia (relativista y no relativista) además de la constante cosmológica. Del análisis se deriva que el efecto depende sensiblemente del valor de la constante de Hubble (que engloba todos los tipos de materia presentes). Continuando dicha línea, hemos caracterizado detalladamente el efecto en su dependencia con los parámetros cosmológicas y las distancias involucradas, y cómo podría ser hallado. Esperamos que nuestros resultados puedan contribuir a una definitiva detección por IPTA.

\markboth{}{}

\chapter*{Resume in Spanish}
\addcontentsline{toc}{chapter}{Resume in Spanish} 

Es un hecho que nuestro universo está en expansión. Las mediciones a escalas cosmológicas son consistentes con la presencia de una forma de energía intrínseca uniformemente distribuida. Si bien esta energía es hasta el momento completamente desconocida, se estima que contribuye con aproximadamente un 68 \% del contenido energético del Universo. El modelo teórico más simple en acuerdo con las observaciones es el modelo cosmológico estándar. En este, una constante cosmológica ($\Lambda$) pequeña pero positiva produce un efecto gravitatorio repulsivo, resultando la expansión acelerada. Este modelo tiene sus comienzos en el desarrollo de la teoría general de la relatividad de Albert Einstein y la mejora en las observaciones astronómicas de objetos extremadamente distantes. Una de las predicciones más controversiales de la teoría de la relatividad es la existencia de soluciones de tipo agujero negro. Estas soluciones indican que una concentración de masa lo suficientemente densa en una región acotada del espacio genera un campo gravitatorio tan extremo, que ninguna partícula material, ni siquiera la luz, puede escapar de dicha región.

A un nivel matemático, estos objetos son soluciones de vacío y su estructura puede ser descripta por muy pocos parámetros. Por ejemplo, existe un teorema que afirma que un agujero negro en estado estacionario puede ser descripto completamente  por tres parámetros: su masa, su carga eléctrica y su momento angular. Como consecuencia, dos agujeros negros que comparten los mismos valores para estos parámetros, son indistinguibles entre sí. De acuerdo a las métricas habituales que describen este tipo de objetos estelares, la curvatura del espaciotiempo se vuelve singular. En este lugar, la solución predice que se encuentra localizada toda la masa (en el centro mismo del agujero negro, o si gira, en un anillo infinitamente delgado). Estas singularidades se encuentran envueltas por superficies cerradas denominadas horizontes de sucesos. 

Enfoques recientes desafían esta estructura poco intuitiva y se enfocan en formas de lograr que la materia se extienda en todo el interior en forma consistente con las condiciones de altas presiones debidas a las altas densidades. Claramente, estas condiciones extremas requieren de una teoría cuántica para la gravedad. En los últimos años una novedosa propuesta estableció una conexión entre la información cuántica y la física de los agujeros negros: los agujeros negros podrían entenderse como condensados de Bose--Einstein de gravitones débilmente interactuantes pero densamente empaquetados. De este modo el comportamiento "misterioso" de los agujeros negros no sería más que el producto de efectos colectivos de un sistema cuántico de bosones idénticos.

El objetivo de la presente tesis es llevar esta propuesta al terreno de las soluciones conocidas de la teoría de la relatividad general. Uno de los problemas centrales de esta teoría es que no se tiene una noción local y covariante de la densidad de energía del campo gravitatorio. Por supuesto, esto se encuentra estrechamente vinculado con los problemas históricos para cuantizar este campo. Por lo tanto, el problema de su localidad es objeto de intenso debate. De cualquier manera, la posibilidad de definir este tipo de energía en regiones compactas (que incluso pueden poseer horizontes y singularidades) ha llevado a postular numerosas representaciones cuasilocales de la energía. 
Esto permite asociar a la densidad de energía una densidad de numero de gravitones en estado fundamental conformando el condensado. De acuerdo a la correspondencia propuesta, la radiación semiclásica que Hawking obtuvo al aplicar la teoría cuántica de campos a espacios curvos, no sería otra cosa que la depleción cuántica usual que sufren los condensados de Bose--Einstein. Esta es una propiedad puramente cuántica producto por la cual el condensado se encuentra constantemente {\it chorreando} bosones debido a efectos de {\it scattering} donde un gravitón es excitado fuera de su estado fundamental; esto le permite vencer el potencial de los $N-1$ restantes y abandonarlos.

De acuerdo a estas consideraciones, la depleción es responsable del la variación de los parámetros previamente mencionados, en particular de la densidad de partículas (en un sentido probabilístico). Por lo tanto, es termodinámicamente consistente buscar alguna especie de potencial químico que de cuenta de la variación del número de partículas $N$ del condensado. En el capítulo \ref{ChapterBHasBEC} discutimos cómo introducir este parámetro dentro de la relatividad general.

En el marco de teorías de materia condensada, bajo la aproximación de Hartree--Fock la función de onda del estado fundamental de un sistema de $N$ bosones es descripta por la ecuación de Gross--Pitaevskii. Desde un aspecto geométrico, el potencial químico regularía las componentes de un tensor de orden 2 ($h_{\mu\nu}$) asociado a perturbaciones de una estructura causal clásica de tipo agujero negro. Esta métrica {\it background} haría las veces del potencial confinante y las ecuaciones de campo para $h_{\mu\nu}$ pueden ser interpretadas como las ecuaciones efectivas de tipo Gross--Pitaevskii que rigen al condensado de gravitones. La solución para $h_{\mu\nu}$ posee características fuertemente dependientes de la métrica clásica que sustenta el condensado y por lo tanto puede ser interpretada como la función de campo medio de este condensado. Es interesante remarcar que las propiedades de los agujeros negros que pueden derivarse de esta propuesta resultan siempre parametrizadas únicamente por el radio de Schwarzschild.

Una de las propuestas mejor fundamentadas para definir la energía cuasilocal tiene sus bases en una formulación covariante del formalismo de Hamilton--Jacobi. En el capítulo \ref{ChapterQLE} rediscutimos ciertas consideraciones energéticas que se derivan de esta energía cuasilocal aplicada a los agujeros negros, pero desde el punto de vista de considerarlos como condensados de gravitones. Sorprendentemente la estructura condensada es energéticamente favorable en comparación a las soluciones de vacío. Además, es posible definir un potencial cuasilocal efectivo que permitiría proveer un posible mecanismo para dar cuenta de la absorción y depleción de los constituyentes del condensado. El perfil de este parámetro presenta un pozo de potencial sobre el horizonte de eventos donde la materia podría almacenarse en estado cuántico confiriendo a estos objetos un "borde" difuso, como es bien sabido que sucede para los condensados.

En el capítulo \ref{ChapterBHasBECExtension} mostramos que, de manera contraria a las métricas que presentan horizontes de eventos (por ejemplo de tipo Schwarzschild o Reissner--Nordstr\"om), nuestra formulación no es consistente para métricas con horizontes cosmológicos y por lo tanto no pareciera ser posible justificar una estructura condensada sobre una métrica de tipo de Sitter.

Pasemos ahora al análisis de la dinámica de un tipo particular de agujero negro: los supermasivos, con masas entre $10^7$ y $10^{10}$ masas solares (M$_\odot$). Un sistema binario de este tipo de objetos es la mayor fuente de ondas gravitacionales del Universo; y el perfil del frente de onda es otra de las sólidas predicciones de la teoría de Einstein. En teoría este tipo particular de onda es de baja frecuencia (entre los nHz y los mHz) y es sensible a ser detectada por observaciones astrofísicas de arreglos de pulsares (PTA). En el capítulo \ref{ChapterGW} discutimos por qué en este tipo de observaciones debe tenerse en cuenta el hecho de que la propagación se produce en un universo en expansión (globalmente de tipo de Sitter) y no en un universo plano y estático. Esta expansión haría que las ondas emitidas durante la fase espiral, que pueden ser descompuestas en una suma de ondas armónicas, se vuelvan anarmónicas al llegar al sistema local de coordenadas donde deberían ser detectadas. Esta anarmonicidad es pequeña; sin embargo en el capítulo \ref{ChapterPTA} damos evidencia de que puede ser sensible a las observaciones de PTA. Caracterizamos las posibles modificaciones a la señal que las colaboraciones que monitorean estas estrellas esperan obtener. Analizamos en detalle la estructura de la nueva señal esperada, cómo se relaciona con las características de la fuente de ondas gravitatorias y con las posiciones de los púlsares empleados por la colaboración IPTA para la detección. 

Vale la pena destacar que la locación del efecto encontrado es relativamente sensible al valor de la constante de Hubble, por lo tanto la posibilidad de hallar ondas gravitatorias por este medio podría traer aparejada la medición local de esta constante $H_0$. Es claro que si la constante cosmológica fuera una propiedad intrínseca del Universo, este efecto sería capaz de confirmar la existencia de $\Lambda$ para redshifts tales que $z<1$.


\begin{acknowledgements}
	\addchaptertocentry{\acknowledgementname} 

Dicen algo así como que la suerte te engancha justo donde el mucho esfuerzo te dejó; sin embargo a esta frase le falta un ingrediente clave: las personas que lo recorrieron a tu lado y lo hicieron posible. 

Quiero agradecer en primer lugar a Domènec Espriu con quien compartí a lo largo de todos estos años dos grandes pasiones: hacer física y explorar el mundo. Realmente me encantó haberlo compartido y haber aprendido a su lado. También quiero darle las gracias a Jorge Alfaro por su hospitalidad y su recibimiento en Santiago de Chile. Voy a guardar en mi memoria los encuentros con ambos. A ambos, gracias. 

A Federico Mescia con quien disfruté muchísimo la docencia, otra de las cosas que amo. Gracias a los jurados por aceptar ser parte: Oriol Pujolas, Jorge Russo y Carlos Sopuerta, espero que disfruten leyendo esta tesis tanto como yo disfruté escribiéndola.

Casi al comienzo de mi doctorado, en alguna conferencia, alguna persona dijo algo que cambió mi percepción de la física. Las palabras fueron algo así como "physics is about friendship". Esas palabras me dejaron una huella imborrable. Es así que esta tesis va dedicada a mis colegas alrededor del mundo. A Jorge Ovalle y Adrián Sotomayor, ojalá las casualidades nos sigan reuniendo en los rincones más recónditos. Mientras, estimulemos las causalidades. A Ángel Rincón y Carlitos Rubio Flores, porque hacer cuentas con ellos siempre fue un disfrute y ojalá lo siga siendo.

También quiero dedicársela a les pilares de mi vida. A mi compañera, Guada, quien me enseñó que lo único permanente es el cambio.  A mi hermano, con quien comparto una vida de mucho más que amistad. Siempre son todo lo que está bien y les admiro profundamente. Gracias por estar.

A mi vieja, por su fuerza arrasadora, por sus ganas de vivir cada segundo y por seguirme en los viajes más inesperados. A mi viejo, por su amor incondicional. A Ivo, Agus, Fabi y Gaby, por la perseverancia y por elegir ser libres. A Marisol y Luís, por estar siempre. A la abuela Ilda, a ver si pronto le festejamos el medio milenio :).

A mis amigos de la vida: Javito, Alan, el Pela, el Bishu, Dante, Emito y el Negrus. Me encanta levantarme cada mañana con tres o cuatro ordenes de magnitud de mensajes. Y sí, todavía me parece increíble que no hayan dudado un segundo en venir en banda para acá. Una dedicatoria especial va para SanSan. A él va dedicado todo el capítulo 3 porque escribirlo me llevó devuelta a la mesa redonda de Salguero y los baldes de café. 

A con quienes compartimos kilómetros, con el podio indiscutido para la Colo, Rami, Haize, Ikercito, Jonny y Dieguis. A la inmensa Vane. A la familia extendida: Juli, Gaby, Osqui, y Poli. Al Maicol, Emi y Ro porque cada vez que nos reencontramos es como si el tiempo no pasara. A quienes cruzaron el charco y a quienes no, y a les que siempre están yendo y viniendo: Virgi, Marian, Tomi, Yaguis, Robi y Tati. A Ceci Villegas, la responsable de transformar los desayunos y meriendas. 

A las geniales apariciones; a los conejos de todas las galeras, porque sin querer saber el truco, me quedo con la magia. A Juan, porque descubrí un gran amigo de sopetón. A Eze, slackeros y la gente de Can Batlló con quienes aprendí otra nueva forma de autogestión. A toda la gente colgada, en especial a Kat, Gemma, Joanito, Anna y Clara, por enseñarme a volar. A Marcelo, Darío y Alfonso, agradezco haberme cruzado con ellos para darle vueltas a nuestro increíble instrumento. A la orquesta Latitud y a la fila de bandoneones, es muy emocionante todo lo hecho y seguro todo lo que vendrá. Para todas las personas que construyen mi vida en Barcelona; ustedes son parte fundamental del todo. Por último y muy especialmente, va dedicada a mis cuadrúpedas amigas con quienes compartimos infinitas horas silla de trabajo y estudio. Y a todes con quienes compartimos sonrisas. Gracias por compartir.

\end{acknowledgements}

\begin{appendices}

\chapter{Perturbative aspects of the classical theory} 
\label{AppxClassical}

In the present appendix we will rederive a well known result from \cite{ReggeWheeler}, but in convenient perturbative formalism, useful for constructing a condensate. In this manner, we will study the spacetime curvature part of the field equations, coming from the Einstein tensor.

As we will see, every dimensionful quantity can be expressed as a function of the Schwarzschild radius $r_S$. Let us consider a perturbation on the Schwarzschild metric with the form of \eqref{MetricSchwPert-h}. We will now expand the Einstein tensor up to second order in $h_{\alpha\beta}$. This will allow us to retain the leading non-linearities. Four of the components of the Einstein tensor are nonnull but only three are linearly independent; both angular components are the same. These are $G_t{}^t$, $G_r{}^r$ and $G_\theta{}^\theta$
\begin{equation}\label{EEqSchwPerturbed}\begin{split}
	\hspace{-2pt}-\frac{r-r_S}{r^3}\Big[\frac{r+r_S}{r}h_{rr}+(r-r_S){h_{rr}}'-\frac{(r+2r_S)(r-r_S)}{r^2}{h_{rr}}^2-2\frac{(r-r_S)^2}{r}h_{rr}{h_{rr}}'\Big]&\\[4ex]
	\frac{1}{r^4(r-r_S)^2}\Big[r_Sr^2\left(r-r_S\right)h_{tt}-r\left(r-r_S\right)^3h_{rr}-r^3\left(r-r_S\right)^2{h_{tt}}'+\left(r-r_S\right)^4{h_{rr}}^2&\\+r_Sr^3{h_{tt}}^2-r_Sr\left(r-r_S\right)^2h_{rr}h_{tt}-r^4\left(r-r_S\right)h_{tt}{h_{tt}}'+r^2\left(r-r_S\right)^3h_{rr}{h_{tt}}'&\Big]\\[4ex]
	-\frac{1}{4r^5(r-r_S)^3}\Bigl[r_Sr^3(r-r_S)(2r-r_S)h_{tt}+r_Sr(2r-r_S)(r-r_S)^3h_{rr}\hspace{55pt}&\\+r^4(2r-3r_S)(r-r_S)^2{h_{tt}}'+r^2(2r-r_S)(r-r_S)^4{h_{rr}}'+2r^5(r-r_S)^3{h_{tt}}''&\\-2r_S(2r-r_S)(r-r_S)^4{h_{rr}}^2+2r_Sr^5{h_{tt}}^2-2r_Sr^2(r-r_S)^3h_{rr}h_{tt}&\\+r^5(2r-5r_S)(r-r_S)h_{tt}{h_{tt}}'+r^6(r-r_S)^2{h_{tt}}'^2-2r^3(r - r_S)^4h_{rr}{h_{tt}}'&\\+r_Sr^3(r - r_S)^3h_{tt}{h_{rr}}'-r^4(r-r_S)^4{h_{tt}}'{h_{rr}}'-2r(2r-r_S)(r-r_S)^5h_{rr}{h_{rr}}'&\\+2r^6(r-r_S)^2h_{tt}{h_{tt}}''-2r^4(r-r_S)^4h_{rr}{h_{tt}}''\Bigr]&.
\end{split}\end{equation}
The prime stand for derivative with respect to the radial coordinate $r$. Of course, the zero order vanishes as $G^{\scriptscriptstyle (0)}_{\alpha\beta}(\tilde g_{\alpha\beta})=0$.
With this three components we get the three $O(2)$ vacuum Einstein's equations
\begin{equation}
	G_\alpha{}^\beta(\tilde g_{\alpha\beta}+h_{\alpha\beta})=0\end{equation}
to solve.

It is well known that Birkhoff's theorem \cite{Birkhoff} guarantees that any spherically symmetric solution of the vacuum field equations must be static and asymptotically flat. So this means that the solution of the equations $G_{\alpha\beta}(\tilde g +h)=0$ with $\tilde g_{\alpha\beta}$ being the Schwarzschild's metric, must necessarily be of the Schwarzschild type again. In conclusion, a perturbation of a Schwarzschild metric in vacuum must end in another Schwarzschild metric but taking into account a change in the parameters responsible of inducing the perturbation. Stating this in other words, a spherically symmetric gravitational field should be produced by some massive object at the origin characterized by the Schwarzschild radius, hence a perturbation in itself it can only be produced by an alteration of this radius and therefore a change in its mass. 

Indeed, imposing the following ansatz for the components of the perturbation
\begin{equation}
	h_{tt}=\sum_{n=1}^{\infty} \frac{B^{(n)}}{r^{n}} \,;\qquad h_{rr}=\sum_{n=1}^{\infty} \frac{A^{(n)}}{r^{n}}\end{equation}
and substituting $h_{rr}$ in the corresponding field equation constructed with the first component (temporal) of the Einstein tensor written in \eqref{EEqSchwPerturbed}, is possible to obtain the coefficients for each term of the radial expansion. We obtain
\begin{equation}\begin{split}
	h_{rr}=&\frac{A}{r}+\frac{A^2+2r_SA}{r^2}+\frac{2A^3+3r_SA^2+3r_S{}^2A}{r^3}+\dots\end{split}\end{equation}
where $A=A^{(0)}$. It is obvious that $A=0$ is a solution as implies no perturbation of the metric at all. But if $A\neq 0$ the rest of the coefficients of the expansion are fixed once $A$ is given. The same happens for $h_{tt}$ using the radial equation from \eqref{EEqSchwPerturbed}
\begin{equation}
	h_{tt}=\frac{A}{r}+0+\frac{A^3}{r^3}+\frac{5A^4+8r_SA^3}{12r^4}+\dots\end{equation}
Taking into account the accuracy of the expansion in $h_{\alpha\beta}$ the components of the metric become
\begin{equation}\begin{split}
	&\tilde g_{tt}+h_{tt}\simeq-\left(1-\frac{r_S}{r}\right)+\frac{A}{r} =-\left(1-\frac{2G_N\overbar M}{r}\right)\\[2ex]& {\tilde g_{rr}}+{h_{rr}}\simeq\left(1+\frac{r_S}{r} +\frac{r_S{}^2}{r^2} \right)+\frac{A}{r}+\frac{A^2+2Ar_S}{r^2}=1+\frac{2G_N\overbar M}{r}+\frac{(2G_N\overbar M)^2}{r^2}\ ;\end{split}\end{equation}
$\overbar{M}$ being the new mass associated to the black hole and $\bar r_S=2G_N\overbar M$ the perturbed Schwarzschild radius. Now we will use similar techniques in the `quantum' case and as we will see the results are dramatically different. However it is interesting to note that any perturbation $h_{\alpha\beta}$ (i.e. each {\it graviton}) has associated a certain amount of energy that is reflected in a change of the black hole mass.

\chapter{Action variation}\label{AppxActionVariation}

Here we will shortly review the variation of the action \eqref{ActionGRChemPot} with respect to the perturbation (it is equivalent to compute the variation with respect to  $h^{\alpha\beta}$)
\begin{equation}
	0=\delta S=M_P{}^2\int d^4x\sqrt{-g}\left[\frac{R}{\sqrt{-g}} \frac{\delta\sqrt{-g}}{\delta h^{\alpha\beta}}+\frac{\delta R}{\delta h^{\alpha\beta}}-\frac{ \sqrt{-\tilde g}}{\sqrt{-g}}\ \frac{\mu}{2}\ \frac{\delta(h_{\rho\sigma}h^{\rho\sigma})}{\delta h^{\alpha\beta}} \right]\delta h^{\alpha\beta}\ . \end{equation}
The first two terms corresponding to the Einstein tensor are widely known, let us just clarified a few steps: making use of the chain rule, $\delta/\delta h^{\alpha\beta}=(\delta g^{\rho\sigma}/\delta h^{\alpha\beta})\delta/\delta g^{\rho\sigma}$, we get the Ricci tensor from the second term, $\delta R/\delta g^{\rho\sigma}=R_{\rho\sigma}$, and, with the following identity $\delta{\sqrt{-g}/\delta g^{\rho\sigma}}=-\sqrt{-g}\ g_{\rho\sigma}/2$, with the first term we get both terms of the Einstein tensor. In the third term, the chemical potential term, is useful for the variation to write it as $h_{\rho\sigma}h^{\rho\sigma}=g_{\rho\lambda}g_{\sigma\gamma}h^{\lambda\gamma}h^{\rho\sigma}$.

Since the action principle should hold for any variation $\delta h^{\alpha\beta}$, yields
\begin{equation}
	\left(R_{\rho\sigma}-\frac{R}2\,g_{\rho\sigma}\right)\frac{\delta g^{\rho\sigma}}{\delta h^{\alpha\beta}}-\frac{\mu}{2}\frac{\sqrt{-\tilde g}}{\sqrt{-g}}\left[2g_{\rho\lambda} g_{\sigma\gamma} h^{\lambda\gamma}\frac{\delta h^{\rho\sigma}}{\delta h^{\alpha\beta}} +h^{\lambda\gamma}h^{\rho\sigma}\frac{\delta(g_{\rho\lambda}g_{\sigma\gamma})}{\delta h^{\alpha\beta}}\right]=0\ .\end{equation}
The metric has been defined as a perturbation of the background metric \eqref{MetricPerturbedGeneric}; the background metric is independent with respect to the perturbation, hence gives just Kronecker deltas $\delta(\tilde g^{\rho\sigma}+h^{\rho\sigma})/\delta h^{\alpha\beta}=\delta^\rho_\alpha\delta_\beta^\sigma$ because of the symmetric property of this tensor. 

There is only one remaining term for obtaining the field equations. The final ingredient is the variation of the metric with respect to the inverse metric. For instance, using that $\delta g_{\sigma\gamma}/\delta g^{\alpha\beta}=-(g_{\sigma\alpha}g_{\gamma\beta}+g_{\sigma\beta}g_{\gamma\alpha})/2$ and doing the the corresponding indices contractions, we get the Einstein field equations \eqref{EEqChemPot} with a chemical potential term
\begin{equation}
	G_{\alpha\beta}-\frac{\mu}{2}\ \frac{ \sqrt{-\tilde g}}{\sqrt{-g}} \bigl(2\,h_{\alpha\beta}-2\,h_{\alpha\sigma}h^\sigma{}_\beta\big)=0\ .\end{equation}

\chapter{Extrinsic curvature of $^3B$}\label{AppxExtrinsicCurvature}

In the present appendix some manipulation of the expressions will be presented to make the generalization to general relativity and the derivation of the quasilocal energy for static and spherically symmetric spacetime more tractable.

Let us make use of all the induced coordinates to write the extrinsic curvature \eqref{ExtrinsicCurvature3B} of the three-boundary $^3B$ in terms of the extrinsic and intrinsic geometry of the spacetime foliated into the spacelike hypersurfaces $\Sigma$. 
For doing so, let us write the extrinsic curvature introducing two identity tensors written in terms of the metric tensor defined on $\Sigma$, $\delta_\mu^\alpha=\gamma_\mu^\alpha-u_\mu u^\alpha$. In fact, if we expand $\overline\Theta_{\mu\nu}=\delta_\mu^\alpha\delta_\nu^\beta\overline\,\Theta_{\alpha\beta}$ we obtain
\begin{equation}\label{ExtrinsicCurvature3BSplit}
	\overline\Theta_{\mu\nu}=-\gamma_\mu^\alpha\gamma_\nu^\beta\overline\gamma_\alpha^\rho \nabla_\rho n_\beta-u_\mu u_\nu u^\alpha u^\beta \overline\gamma_\alpha^\rho \nabla_\rho n_\beta+2 \gamma_{(\mu}^\alpha u_{\nu)} u^\beta\overline\gamma_\alpha^\rho \nabla_\rho n_\beta\ ;\end{equation}
where we have used that the extrinsic curvature is a symmetric tensor, because the unit normal $n^\mu$ is surface forming with vanishing vorticity 
\begin{equation}
	\overline\gamma_\alpha^\rho \nabla_\rho n_\beta-\overline\gamma_\beta^\rho \nabla_\rho n_\alpha=0\ .\end{equation}

Let us manipulate each of the terms in \eqref{ExtrinsicCurvature3BSplit}. The orthogonal condition of the unit normals on $^3B$, i.e. $u^\mu n_\mu=0$, makes the projectors on $^3B$ and $\Sigma$ commute; i.e. $\gamma_\mu^\alpha \overline\gamma_\alpha^\rho=\overline\gamma_\mu^\alpha \gamma_\alpha^\rho$. Now the covariant derivative is projected onto $\Sigma$ because of the two projector tensors $\gamma_\nu^\beta\gamma_\alpha^\rho\nabla_\rho n_\beta=D_\alpha n_\nu$. Therefore the first term is simply the extrinsic curvature of $B$ as embedded in $\Sigma$ defined in \eqref{ExtrinsicCurvatureB}
\begin{equation}
	-\overline\gamma_\mu^\alpha D_\alpha n_\nu=-\sigma_\mu^\alpha D_\alpha n_\nu=	K_{\mu\nu}\ .\end{equation}
Both spacetime tensors $K_{\mu\nu}$ and $D_\alpha n_\nu$ are tangent to $\Sigma$.
The second term can be manipulated according to the differentiation of the orthogonal condition between normal vectors $u^\beta\nabla_\rho n_\beta+n_\beta\nabla_\rho u^\beta=0$ to be proportional to the acceleration $a^\beta=u^\rho\nabla_\rho u^\beta$ of the timelike hypersurface normal; this is
\begin{equation}
	u_\mu u_\nu u^\alpha\overline\gamma_\alpha^\rho n_\beta\nabla_\rho u^\beta=u_\mu u_\nu n_\beta\, a^\beta\ .\end{equation}
Finally, the last term in eq. \eqref{ExtrinsicCurvature3BSplit} is related to the extrinsic curvature of $\Sigma$. In fact, using these relationships, $\overline\gamma_\alpha^\rho u^\beta\nabla_\rho n_\beta=-\overline\gamma_\alpha^\rho n^\beta\nabla_\rho u_\beta$ and $\gamma_\mu^\alpha\overline\gamma_\alpha^\rho= \sigma_\mu^\alpha\gamma_\alpha^\rho$ on $^3B$ we obtain for the second term in \eqref{ExtrinsicCurvature3BSplit}  an expression related to the extrinsic curvature defined in \eqref{ExtrinsicCurvatureSigma}
\begin{equation}
	-2\gamma_{(\mu}^\alpha u_{\nu)}  \overline\gamma_\alpha^\rho n^\beta\nabla_\rho u_\beta = -2\sigma_{(\mu}^\alpha u_{\nu)} n^\beta\gamma_\alpha^\rho\nabla_\rho u_\beta =2\sigma_{(\mu}^\alpha u_{\nu)} n^\beta \Theta_{\alpha\beta}\ .\end{equation}
Plugging everything together, the extrinsic curvature can be written as
\begin{equation}\label{ExtrinsicCurvatureDecomposition}
	\overline\Theta_{\mu\nu}=K_{\mu\nu}+u_\mu u_\nu n_\beta a^\beta+2\sigma_{(\mu}^\alpha u_{\nu)} n^\beta \Theta_{\alpha\beta}\ . \end{equation}
The first term gives the projection onto $B$; the second term, the projection along the normal $u^\mu$; and the two symmetrized last terms give the off-diagonal projection of $\overline\Theta_{\mu\nu}$. This components can be written as the off-diagonal projection of the hypersurface extrinsic curvature $\Theta_{\mu\nu}$ making use of the relationship $\sigma_{\mu}^\alpha  n^\beta \Theta_{\alpha\beta}= -\sigma_\mu^\alpha  u^\beta\overline\Theta_{\alpha\beta}$.
From these decomposition the trace is directly obtained,
\begin{equation}
	\overline\Theta=\overline\gamma^{\mu\nu}\ \overline\Theta_{\mu\nu}=K-n_\beta a^\beta\ .\end{equation}

It is also useful to employ the decomposition for the extrinsic curvature of $^3B$ in \eqref{ExtrinsicCurvatureDecomposition} to rewrite the boundary momentum $\pi^{ij}$ in \eqref{MomentumTensorsPi}. Replacing one identifies the gravitational momentum $P^{ij}$ conjugate to $\gamma_{ij}$ and finally write
\begin{equation}\label{MomentumBoundaryDecomposition}
	\pi^{ij}=\frac{N\sqrt{\sigma}}{2\kappa}\left[K^{ij}+(n\cdot a)\sigma^{ij}-K\,\overline\gamma^{\,ij}\right]-\frac{2N\sqrt{\sigma}}{\sqrt{\gamma}}\sigma^{(i}_\alpha u^{j)} P^{\alpha\beta}\, n_\beta\ .\end{equation}
Let us remark the difference in the tensor indices; in the latter expression $i,j$ refer to coordinates on $^3B$ while $\alpha,\beta$ are related to coordinates in the slices $\Sigma$.

\chapter{Cosmological horizon for a dust universe}
\label{DustDensity}

Let us make a short comment on the equation that relates the density of dust and the variables $t$ and $r$ in \eqref{RelationSolvesODE}. Once the temporal function $C(t)=A\,t^{3/2}$ is fixed, the following change of variables
\begin{equation}\label{Defux}
	u^3 = r^2\,\kappa\rho\, ,\qquad x=A\,\sqrt[3]{\frac{t^2}{r^2}}\end{equation}
makes the relation \eqref{RelationSolvesODE} to correspond with the roots of an extended function
\begin{equation}
	\label{fEq} f(u)=u^3-xu+6\ .\end{equation}
The procedure goes as follows: at any particular time $t$, the density profile is determined at each point $r$. The density takes a value such that the function $f$ vanishes, $f(u=\sqrt[3]{r^2\, \kappa\rho})=0$. Or in other words, at a certain point of the spacetime, the relation between the magnitudes $u$ and $x$ is given by the roots of the function $f(u)$.

The new function accomplish $f(-\infty)=-\infty$ and $f(0)=6$, hence at least one of the roots of $f(u)$ has a real and negative value, $u_1<0$. This root of \eqref{fEq} has no physical interpretation because as it has been defined in \eqref{Defux}, $u$ must be positive. Therefore, we know that a positive solution must exist; as
$f(\infty)=\infty$ there must be at least one double root (or  two more roots). To find it (them), let's analyze the function $f(u)$. The critical points are obtained by the roots of the first derivative of $f(u)$
\begin{equation}
	f'(u_c)=3\,u_c^2-x=0 \qquad\Longrightarrow\qquad u_{c_\pm}=\pm\sqrt{\frac{x}{3}}\ .\end{equation}
The second derivative classifies whether the critical points are a maximum, a minimum or a saddle point
\begin{equation}
	f''(u_c)=6u_c\ .\end{equation}
Therefore, being also $x>0$, $u_{c_+}=\sqrt{x/3}$ is a local minimum and  $u_{c_-}=-\sqrt{x/3}$ is a local maximum. We are interested in $u>0$. The next step is to locate the image of the positive critical point $u_{c_+}$ (the minimum), because this value will determine the existence of the wanted remaining root(s)
\begin{equation}
	f(\,u_{c_+}\,)=u_{c_+}^3-3\,u_{c_+}^3+6\leq0\ .\end{equation}
We are searching for positive real roots of \eqref{fEq}, hence the value of the minimum must be zero or negative (one double root or two simple roots). It is interesting to note that the latter bound can be written as $3^{5/3} \leq x$. Bringing the value of $x$ back, we will see that the previous constraint reflects the presence of a horizon. The mere existence of a positive solution for $f(u)=0$ with $u\propto\rho$ entails the presence of a horizon in such coordinates. With the value of the constant $A=3\sqrt[3]{6}$, the cosmological horizon equation is written as
\begin{equation}\label{HorizonBound}
	\frac{r}{t}\,\,\leq\,\,\sqrt{\frac{2}{3}}\ .\end{equation}

\chapter{Energy momentum tensor for dust in static coordinates}
\label{DustEnergyMomentumTensor}

In this appendix we will give the components of the stress-energy tensor explicitly for a dust universe in the new set of coordinates. Dust is a pressureless perfect fluid with an energy-momentum tensor given by \eqref{EMTensorDust}; this definition is coordinate independent. Once the coordinates are fixed one can find the explicit stress-energy tensor by means of Einstein's equations \eqref{EEq}.
The Einstein tensor is straightforward once the metric is chosen; from \eqref{MetricSSSDMUnknown} for instance the computations are particularly simple --it would be equivalent to use the metric \eqref{MetricSSSDM}, at this point $\partial_t\rho_{\scriptscriptstyle dust}$ is already know. 

The product of the two nonnull off-diagonal components of the Einstein tensor lead us to
\begin{equation}\label{EinEqLHS}
	G_t{}^r\,G_r{}^t=-\frac{(\kappa\rho_{\scriptscriptstyle dust})^3\,r^2} {3\left(1-\frac{\kappa\rho_{\scriptscriptstyle dust}}{3}\,r^2\right)^2}\ .\end{equation}
Notice that $G_t{}^r\neq G_r{}^t$. The right hand side of Einstein's equations equals the previous relation to
\begin{equation}\label{EinEqRHS}
	\kappa^2\,T_t{}^r\,T_r{}^t=(\kappa\rho_{\scriptscriptstyle dust})^2\,U_tU^t\,U_rU^r\,.\end{equation}
We aim at solving the 4--velocity and we have another equation to do so: the normalization condition over the 4--velocity that yields
\begin{equation}\label{Norm4Vel}
	g_{\mu\nu}\,U^\mu U^\nu=U_tU^t+U_rU^r=1\ .\end{equation}
The system of two equations: \eqref{EinEqLHS} $=$ \eqref{EinEqRHS} and \eqref{Norm4Vel} has the following solution
\begin{align}\label{VelocityTemp}
	&U_tU^t=\frac{1}{1-\frac{\kappa\rho_{ \scriptscriptstyle dust}}{3}\,r^2}\,;\\[3ex] &U_rU^r=-\frac{\kappa\rho_{\scriptscriptstyle dust}}{3}\,\frac{r^2}{1-\frac{\kappa\rho_{ \scriptscriptstyle dust}}{3}\,r^2}\,.\end{align}
We have chosen the solution noticing that $U_tU^t>0$ and $U_rU^r<0$. The next step is to isolate each component of the velocity; using the prescription $U_\mu{}^2=U_\mu U^\mu\,g_{\mu\nu}$, for $\mu=t,r$ we obtain
\begin{align}
	&U_t=\frac{\abs{\partial_t\rho_{ \scriptscriptstyle dust}}}{(3\,\kappa\rho_{ \scriptscriptstyle dust}^3)^{1/2}}\,;\\[3ex] &U_r=\sqrt{\frac{\kappa\rho_{ \scriptscriptstyle dust}}{3}}\,\frac{r}{1-\frac{\kappa\rho_{ \scriptscriptstyle dust}}{3}\,r^2}\ .\end{align}

A relevant comment is that if the temporal Einstein's equation
\begin{equation}
	G_t{}^t=\kappa\,\frac{r\,\partial_r\rho_{ \scriptscriptstyle dust}+3\,\rho}{3}=\kappa\,\rho\,U_tU^t\end{equation}
is combined with the corresponding velocity in equation \eqref{VelocityTemp}, and then one solves for the radial derivative 
one gets \eqref{DiagonalMetricCondition}.

Finally,
\begin{equation}\begin{aligned}
	&T_t{}^t=\frac{3\,\rho_{\scriptscriptstyle dust}}{3-\kappa\rho_{ \scriptscriptstyle dust}\,r^2}&&&&\qquad T_t{}^r=-\frac{ \abs{\partial_t\rho_{ \scriptscriptstyle dust}}}{3}\,r\\[3ex]
	&T_r{}^t=\frac{9\,\kappa\rho_{ \scriptscriptstyle dust}^3}{ \abs{\partial_t\rho_{ \scriptscriptstyle dust}}\left(3-\kappa\rho_{ \scriptscriptstyle dust}\,r^2\right)^2}\,r&&&&\qquad T_r{}^r= -\frac{\kappa\rho_{ \scriptscriptstyle dust}^2}{3-\kappa\rho_{ \scriptscriptstyle dust}\,r^2}\,r^2\ .\end{aligned}\end{equation}

\end{appendices}


\end{document}